\documentclass[11pt,a4paper]{article}
 \pdfoutput=1
 \usepackage{cite}

\usepackage[dvips]{graphics}
\usepackage[pdftex]{graphicx}
\usepackage[pdftex]{xcolor}
\usepackage{amsmath, amsthm}
\usepackage{amssymb}
\usepackage{fancyvrb}
\usepackage[affil-it]{authblk}

\usepackage{verbatim}	

\usepackage{syntonly} 

\usepackage{empheq}
\usepackage[pdftex]{hyperref}
\hypersetup{colorlinks=false}

\title{Dreaming Awake: Disentangling the Underlying Physics in Case of a SUSY-like Discovery at the LHC}

\author[1]{Konstantin T.~Matchev}

\author[2,3]{Filip Moortgat}

\author[2]{Luc Pape}

\affil[1]{Institute for Fundamental Theory, Physics Department, University of Florida, Gainesville, FL 32611, USA}
\affil[2]{CERN, Geneva, Switzerland}
\affil[3]{Ghent University, Ghent, Belgium}

\date{\today}

\begin{document}

\newcommand{\swsq}{\ensuremath{\sin^2\theta_W}}
\newcommand{\cwsq}{\ensuremath{\cos^2\theta_W}}
\newcommand{\tanb}{\ensuremath{\tan\beta}}
\newcommand{\tanbsq}{\ensuremath{\tan^{2}\beta}}
\newcommand{\sidb}{\ensuremath{\sin 2\beta}}
\newcommand{\alpS}{\ensuremath{\alpha_S}}
\newcommand{\alpt}{\ensuremath{\tilde{\alpha}}}
\newcommand{\QL}{\ensuremath{Q_L}}
\newcommand{\sQ}{\ensuremath{\tilde{Q}}}
\newcommand{\sQL}{\ensuremath{\tilde{Q}_L}}
\newcommand{\ULC}{\ensuremath{U_L^C}}
\newcommand{\sUC}{\ensuremath{\tilde{U}^C}}
\newcommand{\sULC}{\ensuremath{\tilde{U}_L^C}}
\newcommand{\DLC}{\ensuremath{D_L^C}}
\newcommand{\sDC}{\ensuremath{\tilde{D}^C}}
\newcommand{\sDLC}{\ensuremath{\tilde{D}_L^C}}
\newcommand{\LL}{\ensuremath{L_L}}
\newcommand{\sL}{\ensuremath{\tilde{L}}}
\newcommand{\sLL}{\ensuremath{\tilde{L}_L}}
\newcommand{\ELC}{\ensuremath{E_L^C}}
\newcommand{\sEC}{\ensuremath{\tilde{E}^C}}
\newcommand{\sELC}{\ensuremath{\tilde{E}_L^C}}
\newcommand{\sEL}{\ensuremath{\tilde{E}_L}}
\newcommand{\sER}{\ensuremath{\tilde{E}_R}}
\newcommand{\sFer}{\ensuremath{\tilde{f}}}
\newcommand{\sQua}{\ensuremath{\tilde{q}}}
\newcommand{\sUp}{\ensuremath{\tilde{u}}}
\newcommand{\suL}{\ensuremath{\tilde{u}_L}}
\newcommand{\suR}{\ensuremath{\tilde{u}_R}}
\newcommand{\sDw}{\ensuremath{\tilde{d}}}
\newcommand{\sdL}{\ensuremath{\tilde{d}_L}}
\newcommand{\sdR}{\ensuremath{\tilde{d}_R}}
\newcommand{\sTop}{\ensuremath{\tilde{t}}}
\newcommand{\stL}{\ensuremath{\tilde{t}_L}}
\newcommand{\stR}{\ensuremath{\tilde{t}_R}}
\newcommand{\stone}{\ensuremath{\tilde{t}_1}}
\newcommand{\sttwo}{\ensuremath{\tilde{t}_2}}
\newcommand{\sBot}{\ensuremath{\tilde{b}}}
\newcommand{\sbL}{\ensuremath{\tilde{b}_L}}
\newcommand{\sbR}{\ensuremath{\tilde{b}_R}}
\newcommand{\sLep}{\ensuremath{\tilde{l}}}
\newcommand{\sLepC}{\ensuremath{\tilde{l}^C}}
\newcommand{\sEl}{\ensuremath{\tilde{e}}}
\newcommand{\sElC}{\ensuremath{\tilde{e}^C}}
\newcommand{\seL}{\ensuremath{\tilde{e}_L}}
\newcommand{\seR}{\ensuremath{\tilde{e}_R}}
\newcommand{\snL}{\ensuremath{\tilde{\nu}_L}}
\newcommand{\sMu}{\ensuremath{\tilde{\mu}}}
\newcommand{\sNu}{\ensuremath{\tilde{\nu}}}
\newcommand{\sTau}{\ensuremath{\tilde{\tau}}}
\newcommand{\Glu}{\ensuremath{g}}
\newcommand{\sGlu}{\ensuremath{\tilde{g}}}
\newcommand{\Wpm}{\ensuremath{W^{\pm}}}
\newcommand{\sWpm}{\ensuremath{\tilde{W}^{\pm}}}
\newcommand{\Wz}{\ensuremath{W^{0}}}
\newcommand{\sWz}{\ensuremath{\tilde{W}^{0}}}
\newcommand{\sWino}{\ensuremath{\tilde{W}}}
\newcommand{\Bz}{\ensuremath{B^{0}}}
\newcommand{\sBz}{\ensuremath{\tilde{B}^{0}}}
\newcommand{\sBino}{\ensuremath{\tilde{B}}}
\newcommand{\sHino}{\ensuremath{\tilde{h}}}
\newcommand{\Zz}{\ensuremath{Z^{0}}}
\newcommand{\sZino}{\ensuremath{\tilde{Z}^{0}}}
\newcommand{\sGam}{\ensuremath{\tilde{\gamma}}}
\newcommand{\Hone}{\ensuremath{H_{d}}}
\newcommand{\sHone}{\ensuremath{\tilde{H}_{d}}}
\newcommand{\Htwo}{\ensuremath{H_{u}}}
\newcommand{\sHtwo}{\ensuremath{\tilde{H}_{u}}}
\newcommand{\sHig}{\ensuremath{\tilde{H}}}
\newcommand{\sHa}{\ensuremath{\tilde{H}_{a}}}
\newcommand{\sHb}{\ensuremath{\tilde{H}_{b}}}
\newcommand{\sHpm}{\ensuremath{\tilde{H}^{\pm}}}
\newcommand{\hz}{\ensuremath{h^{0}}}
\newcommand{\Hz}{\ensuremath{H^{0}}}
\newcommand{\Az}{\ensuremath{A^{0}}}
\newcommand{\Hpm}{\ensuremath{H^{\pm}}}
\newcommand{\chiz}{\ensuremath{\tilde{\chi}^{0}}}
\newcommand{\chip}{\ensuremath{\tilde{\chi}^{+}}}
\newcommand{\chim}{\ensuremath{\tilde{\chi}^{-}}}
\newcommand{\chipm}{\ensuremath{\tilde{\chi}^{\pm}}}
\newcommand{\chimp}{\ensuremath{\tilde{\chi}^{\mp}}}
\newcommand{\sGra}{\ensuremath{\tilde{G}}}
\newcommand{\mtil}{\ensuremath{\tilde{m}}}
\newcommand{\mtau}{\ensuremath{m_\tau}}
\newcommand{\Em}{\ensuremath{E\!\!\!/}}
\newcommand{\Mm}{\ensuremath{M\!\!\!\!\!/\;}}
\newcommand{\ETm}{\ensuremath{{E\!\!\!/}_T}}
\newcommand{\PTm}{\ensuremath{{P\!\!\!/}_T}}
\def\ga{\mathrel{\rlap{\raise.6ex\hbox{$>$}}{\lower.6ex\hbox{$\sim$}}}}
\def\la{\mathrel{\rlap{\raise.6ex\hbox{$<$}}{\lower.6ex\hbox{$\sim$}}}}
\newcommand{\eps}{\ensuremath{\epsilon}}
\newcommand{\epss}{\ensuremath{\epsilon_{sel}}}
\newcommand{\epsh}{\ensuremath{\epsilon_{h}}}
\newcommand{\epszh}{\ensuremath{\epsilon_{0h}}}
\newcommand{\epsth}{\ensuremath{\epsilon_{\tau h}}}
\newcommand{\epse}{\ensuremath{<\!\epsilon_{e}\!>}}
\newcommand{\epsm}{\ensuremath{<\!\epsilon_{\mu}\!>}}
\newcommand{\epspe}{\ensuremath{<\!\epsilon'_{e}\!>}}
\newcommand{\epspm}{\ensuremath{<\!\epsilon'_{\mu}\!>}}
\newcommand{\epsze}{\ensuremath{<\!\epsilon_{0e}\!>}}
\newcommand{\epszm}{\ensuremath{<\!\epsilon_{0\mu}\!>}}
\newcommand{\epspeu}{\ensuremath{<\!\epsilon'_{e1}\!>}}
\newcommand{\epsed}{\ensuremath{<\!\epsilon_{e2}\!>}}
\newcommand{\epspmu}{\ensuremath{<\!\epsilon'_{\mu 1}\!>}}
\newcommand{\epsmd}{\ensuremath{<\!\epsilon_{\mu 2}\!>}}
\newcommand{\epsesq}{\ensuremath{<\!\epsilon_{e}\!>^2}}
\newcommand{\epsmsq}{\ensuremath{<\!\epsilon_{\mu}\!>^2}}
\newcommand{\epsem}{\ensuremath{<\!\epsilon_{e}\!> <\!\epsilon_{\mu}\!>}}
\newcommand{\epspesq}{\ensuremath{<\!\epsilon^{'}_{e}\!>^2}}
\newcommand{\epspmsq}{\ensuremath{<\!\epsilon^{'}_{\mu}\!>^2}}
\newcommand{\epspem}{\ensuremath{<\!\epsilon^{'}_{e}\!> <\!\epsilon^{'}_{\mu}\!>}}
\newcommand{\epszesq}{\ensuremath{<\!\epsilon^2_{0e}\!>}}
\newcommand{\epszmsq}{\ensuremath{<\!\epsilon^2_{0\mu}\!>}}
\newcommand{\epszem}{\ensuremath{<\!\epsilon_{0e} \epsilon_{0\mu}\!>}}
\newcommand{\epspeusq}{\ensuremath{<\!\epsilon'_{e1}\!>^2}}
\newcommand{\epsedsq}{\ensuremath{<\!\epsilon_{e2}\!>^2}}
\newcommand{\epspmusq}{\ensuremath{<\!\epsilon'_{\mu 1}\!>^2}}
\newcommand{\epsmdsq}{\ensuremath{<\!\epsilon_{\mu 2}\!>^2}}
\newcommand{\epspemu}{\ensuremath{<\!\epsilon'_{e1}\!> <\!\epsilon'_{\mu 1}\!>}}
\newcommand{\epsemd}{\ensuremath{<\!\epsilon_{e2}\!> <\!\epsilon_{\mu 2}\!>}}
\newcommand{\epste}{\ensuremath{<\!\epsilon_{\tau e}\!>}}
\newcommand{\epstm}{\ensuremath{<\!\epsilon_{\tau \mu}\!>}}
\newcommand{\epstesq}{\ensuremath{<\!\epsilon_{\tau e}\!>^2}}
\newcommand{\epstmsq}{\ensuremath{<\!\epsilon_{\tau \mu}\!>^2}}
\newcommand{\epstem}{\ensuremath{<\!\epsilon_{\tau e}\!> <\!\epsilon_{\tau \mu}\!>}}
\newcommand{\epszte}{\ensuremath{<\!\epsilon_{0\tau e}\!>}}
\newcommand{\epsztm}{\ensuremath{<\!\epsilon_{0\tau \mu}\!>}}
\newcommand{\epsztesq}{\ensuremath{<\!\epsilon_{0\tau e}^2\!>}}
\newcommand{\epsztmsq}{\ensuremath{<\!\epsilon_{0\tau \mu}^2\!>}}
\newcommand{\epsztem}{\ensuremath{<\!\epsilon_{0\tau e} \epsilon_{0\tau \mu}\!>}}
\newcommand{\Be}{\ensuremath{B_{e}}}
\newcommand{\Bm}{\ensuremath{B_{\mu}}}
\newcommand{\Bt}{\ensuremath{B_{\tau}}}
\newcommand{\Bpe}{\ensuremath{B'_{e}}}
\newcommand{\Bpm}{\ensuremath{B'_{\mu}}}
\newcommand{\Bpesq}{\ensuremath{B^{'2}_{e}}}
\newcommand{\Bpmsq}{\ensuremath{B^{'2}_{\mu}}}
\newcommand{\Rl}{\ensuremath{R_{l}}}
\newcommand{\Rh}{\ensuremath{R_{h}}}
\newcommand{\Bze}{\ensuremath{B_{0e}}}
\newcommand{\Bzm}{\ensuremath{B_{0\mu}}}
\newcommand{\Bzt}{\ensuremath{B_{0\tau}}}
\newcommand{\Bztlh}{\ensuremath{B_{0\tau lh}}}
\newcommand{\Rzl}{\ensuremath{R_{0l}}}
\newcommand{\Rzh}{\ensuremath{R_{0h}}}
\newcommand{\Rddl}{\ensuremath{R_{22l}}}
\newcommand{\Rdul}{\ensuremath{R_{21l}}}
\newcommand{\Rdzl}{\ensuremath{R_{20l}}}
\newcommand{\Rddh}{\ensuremath{R_{22h}}}
\newcommand{\Rduh}{\ensuremath{R_{21h}}}
\newcommand{\Rdzh}{\ensuremath{R_{20h}}}
\newcommand{\Rtdl}{\ensuremath{R_{32l}}}
\newcommand{\Rtul}{\ensuremath{R_{31l}}}
\newcommand{\Rtzl}{\ensuremath{R_{30l}}}
\newcommand{\Rtdh}{\ensuremath{R_{32h}}}
\newcommand{\Rtuh}{\ensuremath{R_{31h}}}
\newcommand{\Rtzh}{\ensuremath{R_{30h}}}
\newcommand{\Rqdl}{\ensuremath{R_{42l}}}
\newcommand{\Rqul}{\ensuremath{R_{41l}}}
\newcommand{\Rqzl}{\ensuremath{R_{40l}}}
\newcommand{\Rqdh}{\ensuremath{R_{42h}}}
\newcommand{\Rquh}{\ensuremath{R_{41h}}}
\newcommand{\Rqzh}{\ensuremath{R_{40h}}}
\newcommand{\Pc}{\ensuremath{P_{\chipm_1}}}
\newcommand{\Pn}{\ensuremath{P_{\chiz_2}}}
\newcommand{\Pcc}{\ensuremath{P_{\chipm_1\chipm_1}}}
\newcommand{\Pcn}{\ensuremath{P_{\chipm_1\chiz_2}}}
\newcommand{\Pnn}{\ensuremath{P_{\chiz_2\chiz_2}}}
\newcommand{\flasum}{\ensuremath{\Sigma}}

\newcommand{\Mtil}{\ensuremath{\tilde{M}}}
\newcommand{\Mtilp}{\ensuremath{\tilde{M'}}}

\newcommand{\myemph}[1]{\textbf{\emph{***** #1}}}%

\maketitle

\begin{abstract}

The purpose of this review is to investigate what kind of physics can be extracted at the LHC,
assuming a discovery is made in events with missing transverse momentum,
as generically expected in supersymmetry (SUSY) with R-parity conservation.
To set the scene, we first discuss the collider phenomenology of the six possible 
electroweakino benchmark scenarios, as they provide valuable insight into what one might be facing at the LHC.
We review the existing methods for mass reconstruction from measured kinematic endpoints in the distributions of 
suitable variables, e.g., the invariant masses of various sets of visible decay products, 
as well as the $M_{T2}$ and the $M_2$ types of variables.
We propose to extend the application of these methods to the various topologies 
of fully hadronic final states, possibly with hadronically reconstructed massive bosons ($W$, $Z$ or $h$).
We test the idea with a simplified simulation of events in the main electroweakino benchmark scenarios.
We find that the fully hadronic events allow the complete determination of the relevant mass spectrum.
For comparison, we also review the potential of the standard kinematic endpoint methods 
for final states involving leptons from the decays of (on-shell or off-shell) sleptons.
We find that with $300 \; {\rm fb}^{-1}$, the statistics for the leptonic events 
is very marginal and they look less promising than the fully hadronic channels.
This corresponds to a complete reversal of the usual paradigm, where leptonic events comprised the gold-plated SUSY channels.
Finally, we put together all available information and summarize what level of understanding of the underlying physics can be achieved.
We show that, as a by-product of the mass reconstruction, 
it is also possible to determine the production cross sections and decay branching ratios, which
in turn enable us to pinpoint the underlying model.

\end{abstract}

\newpage
\tableofcontents

\section{Introduction}
\label{sect:intro}

The Large Hadron Collider (LHC) at CERN is currently operating at a $13$ TeV centre of mass energy, with 
a plan to increase the energy to 14 TeV and collect an integrated luminosity of $300 \; {\rm fb}^{-1}$ before the end of 2023.
This period will be followed by an upgrade to the High Luminosity LHC (HL-LHC),
expected to start around 2025 in view of collecting $3000 \; {\rm fb}^{-1}$ at the same energy.
Another very promising prospect would be an increase in energy to the HE-LHC of 25-30 TeV, still in the same LHC tunnel,
or a game-changing jump to the 100 TeV FCC (Future Circular Collider) in a new circular tunnel of 100 km circumference \cite{Golling:2016gvc}.

This timeline gives the LHC and its successors a reasonable shot for discovering new physics beyond the Standard Model (SM). 
A well-studied and theoretically sound paradigm for a possible new physics scenario is
low energy supersymmetry (SUSY) \cite{Ramond:1971gb,Golfand:1971iw,Neveu:1971rx,Volkov:1972jx,Wess:1973kz,Wess:1974tw,Fayet:1974pd}, which encompasses a variety of models extending the SM to include 
superpartners of the known particles, whose gauge and Yukawa interactions are fixed by supersymmetry, 
and whose masses are of order the electroweak scale. 
One of the many attractive features of SUSY is the fact that it introduces new neutral weakly interacting particles 
which can be potential dark matter candidates \cite{Jungman:1995df}. 
The two possibilities are: the Lightest Supersymmetric Particle (LSP) is a spin 0 sneutrino, which is the superpartner of a SM neutrino,
or a spin $1/2$ neutralino, which is a mixture of the superpartners of the neutral gauge bosons and Higgs bosons. 
In what follows, we shall focus on the latter option, since it is less constrained experimentally and more 
readily realized in specific models.

In order to account for the cosmological dark matter, the neutralino must be long lived on cosmological time scales.
The simplest way to achieve this is to envoke a new symmetry, a $Z_2$ parity known as $R$-parity, which assigns
parity $+1$ to the SM particles and $-1$ to their superpartners \cite{Farrar:1978xj}. There exists a minimal version of SUSY, the so-called
Minimal Supersymmetric Standard Model (MSSM)\footnote{It is not the purpose of this report to review the MSSM,
for which excellent articles exist in the literature, see, 
e.g., \cite{Martin:1997ns,Nilles:1983ge,Haber:1984rc,Chung:2003fi,Pape:2006ar,MSSMBaer} 
and references therein. In the following, it will be assumed that the reader is knowledgeable about the basics 
of the MSSM collider phenomenology.}, in which one more Higgs doublet is added to the SM before
imposing supersymmetry and introducing superpartners. However, SUSY itself does not predict the masses of the 
superpartners --- those masses must arise mostly from SUSY breaking effects, since no superpartners has been found so far. 
In principle, the superpartner masses could be arbitrarily large, but arguments based on naturalness and 
grand unification suggest that (at least some of) the superpartners are not too far away from the electroweak scale.
In the years leading up to the LHC, this hint was taken perhaps too literally (and perhaps somewhat too optimistically) \cite{Feng:2013pwa}, 
and as a result, most SUSY studies done in those days relied on benchmark points with relatively light SUSY mass spectra
\cite{Battaglia:2001zp,Allanach:2002nj}. Once the LHC became operational, the negative results from the many SUSY searches quickly eliminated
this ``low hanging fruit", while still leaving open the parameter space regions with heavier superpartners \cite{AbdusSalam:2011fc,Baer:2013ula,Bechtle:2015nta}.
Our main goal in this report is to revisit the collider phenomenology of SUSY at the LHC, 
taking into account these new realities:
\begin{itemize}
\item {\em Higher superpartner mass scale.} We already mentioned that the current SUSY limits already point in the direction 
of a mass scale which is perhaps heavier than previously thought. However, there are additional factors supporting this expectation:
\begin{enumerate}
\item {\em The Higgs boson mass.} In the meantime, a SM-like Higgs boson state $h$ has been discovered and its mass was
measured to be $m_h=125$ GeV \cite{Aad:2015zhl}. In order to accommodate such a large value for $m_h$ in the MSSM, 
one has to rely on the radiative corrections from top squark (stop) loops. 
This in turn requires the stops to be relatively heavy, in the multi-TeV range \cite{Bagnaschi:2014rsa,Hajer:2015gka,Craig:2016ygr}. 
Given that in typical models, the top squarks are expected to be among the lightest squarks, due to the 
large top Yukawa coupling effects in the RGE evolution, the other squark states should be even heavier.
\item {\em The case for Higgsino or Wino dark matter.} Many of the original SUSY benchmark points 
incorporated the assumption of gaugino mass unification, whereby the three gaugino masses, $M_1$ for the Bino,
$M_2$ for the Winos and $M_3$ for the gluino, are exactly unified at the grand unification theory (GUT) scale.
With this assumption, the subsequent RGE evolution guaranteed that the lightest superpartner, 
and therefore the dark matter candidate, is the Bino, the superpartner of the hypercharge gauge boson of the SM.
The Bino, however, has suppressed couplings --- first, due to the smallness of the hypercharge gauge coupling $g_1$,
and second, due to the small numerical values for the hypercharges of the SM fermions, most notably the left-handed squarks.
Furthermore, the Bino does not couple to any SM gauge bosons. All these factors lead to a suppression in the 
Bino annihilation cross-section in the early universe. Consequently, to account for all of the dark matter, 
the Bino must be relatively light, on the order of a few hundred GeV. This is why models with gaugino mass unification
always preferred a light LSP, which would suggest a light superpartner spectrum overall. However, 
the Bino is only one of several possibilities for the identity of the lightest neutralino --- 
it is also very possible that the lightest neutralino is a Higgsino, a Wino, or some mixture of the two.
In contrast to the Bino case, the annihilation cross-sections for Wino-like and Higgsino-like neutralinos are not suppressed,
and thus the preferred mass range for a Wino (Higgsino) dark matter is $2-3$ TeV (1 TeV), much higher than the Bino case.
Since the dark matter candidate is the lightest superpartner, all other superpartners must be even heavier.
\item {\em The subjectiveness and the danger of naive fine-tuning arguments.} By ensuring the exact cancellation 
of the quadratic divergences in the Higgs mass radiative corrections, supersymmetry neatly solves the technical hierarchy problem.
However, logarithmic divergences still remain, thus heavy superpartners tend to destabilize the hierarchy, albeit only 
logarithmically. By adopting a numerical tolerance for the amount of fine-tuning in the Higgs potential,
one could ``predict" the superpartner mass scale \cite{Barbieri:1987fn,Anderson:1994tr,Chankowski:1997zh}. 
Over the last 10-15 years, this line of thought has come under
further scrutiny. First, the notion of ``allowed amount of fine-tuning" is very subjective, and may have been 
underestimated in the past \cite{Barbieri:1998uv}. Perhaps more importantly, all known methods of calculating a measure of 
fine-tuning neglect to take into account possible correlations at the GUT or Planck scale \cite{Feng:1999mn,Horton:2009ed,Younkin:2012ui,Yanagida:2013ah}. Examples exist when 
such correlations would tend to reduce the apparent fine tuning from the low energy point of view, 
and therefore allow much heavier mass spectra \cite{Feng:2013pwa}.
\end{enumerate}
\item {\em Increased mass gaps among the superpartners.} An immediate consequence from the previous point 
is the fact that as the overall superpartner mass scale increases, so do the mass gaps between the different superpartners.
As the mass gaps grow, new decay channels, which were previously kinematically forbidden, will open up.
In this paper, we shall focus on the electroweakino sector, and in particular on the opening of the two-body 
decays of various electroweakino states to on-shell $W$, $Z$ and Higgs bosons.
\item {\em The presence of massive SM bosons in the SUSY decay chains.} In order to perform 
precision measurements post-discovery, many previous studies have focused on the 
classic SUSY decay chain, which consists of a sequence of two-body decays, emitting a light SM quark or a SM lepton
at each vertex. To an excellent approximation, the masses of those SM quarks and leptons in the final state 
can be safely neglected, which leads to significant simplifications in the analysis and the corresponding formulae.
Once the two-body decays to $W$, $Z$ and Higgs bosons become relevant, we cannot ignore their masses any more, and
the analysis becomes more complicated. To the best of our knowledge, there are no existing studies 
of SUSY-like decay chains with massive reconstructed SM particles in the final state. This is why, another one of our goals here is
to demonstrate how to generalize the existing analyses to the case with massive visible final state particles.
\item {\em New experimental conditions.} It goes without saying that there will be challenges on the experimental side as well ---
the higher integrated luminosity or energy also put severe constraints on the detector and the physics analysis.
Better methods will need to be developed to mitigate the effect of increased pile-up;
it will be very challenging to unravel the contents of highly boosted objects like top quarks, 
$W$, $Z^0$ or $h^0$, which require excellent two-track resolution in the tracker system.
An unprecedented precision of the tracker will be required for momentum measurements of tracks in the multi-TeV regime.
\end{itemize}

In summary, an increase of the SUSY mass scale will generally enrich the SUSY signatures with
on-shell $W$, $Z$ and Higgs bosons. All three of these particles have both hadronic and leptonic decay modes. 
The advantage of the hadronic decay modes is the higher branching fraction, and the fact that 
the mother particle can be fully reconstructed\footnote{Contrast this with the case of a leptonically decaying $W$-boson, where a neutrino goes missing.},
as demonstrated by several existing LHC searches, e.g., for the case of hadronic top quarks 
\cite{Aad:2012ans,Sirunyan:2017uhk,Aaboud:2017ayj,Sirunyan:2017pjw,Sirunyan:2018ryr}, 
hadronic $W$ and $Z$-bosons \cite{Sirunyan:2016cao,Aaboud:2017eta,Sirunyan:2017acf,Sirunyan:2018iff,Sirunyan:2018ivv,Sirunyan:2018hsl}
and hadronic $h$-bosons \cite{Khachatryan:2015axa,Khachatryan:2015bnx,Sirunyan:2017wto}.

Given all these new developments, it becomes likely that if an excess over the SM background is observed, 
this excess {\em will first be detected in the fully hadronic final states}
and the immediate question which it raises is: {\em what next}?
Indeed, it is not unreasonable to expect that in this case some excesses will also appear in other channels,
in particular leptonic final states, but this may require a considerable increase in integrated luminosity,
since the leptonic decay branching ratios are typically small.
In this report, we will examine how to unravel the underlying physics by reconstructing the sparticle masses, 
estimating the production cross sections and decay branching fractions and identifying the type of underlying physics model.
For this purpose, we shall make crucial use of the hadronic channels with fully reconstructed gauge and Higgs bosons, which
have largely been overlooked until now.

In what follows, we shall focus mainly on the relevant features of the {\em signal}.
At the time of the discovery most backgrounds are estimated using largely data-driven methods, which in turn
allow the simulation to be checked.
As the mass reconstruction can only be undertaken after a sufficient amount of excess events has been collected
(much more than for a discovery),
it can be expected that the experiment, and in particular the simulation, are thoroughly understood at that time.
The SM backgrounds can then be estimated directly from simulation,
without the need of data-driven methods.
For simplicity, we will assume that these backgrounds have been subtracted from the distributions or eliminated by the experimental cuts
and hence in the following we will ignore the SM background altogether.
However, it is likely that in every topology studied below,
the main backgrounds will be due to {\em other MSSM processes}
and those will be much harder to evaluate or reject.
Fortunately, the mass reconstruction requires only the observation and measurement of endpoints in distributions.
They may still be observable, even above an unknown, but {\em smooth}, background distribution.

\begin{figure}[htb]
 \begin{center}
 \includegraphics[width=0.6\textwidth]{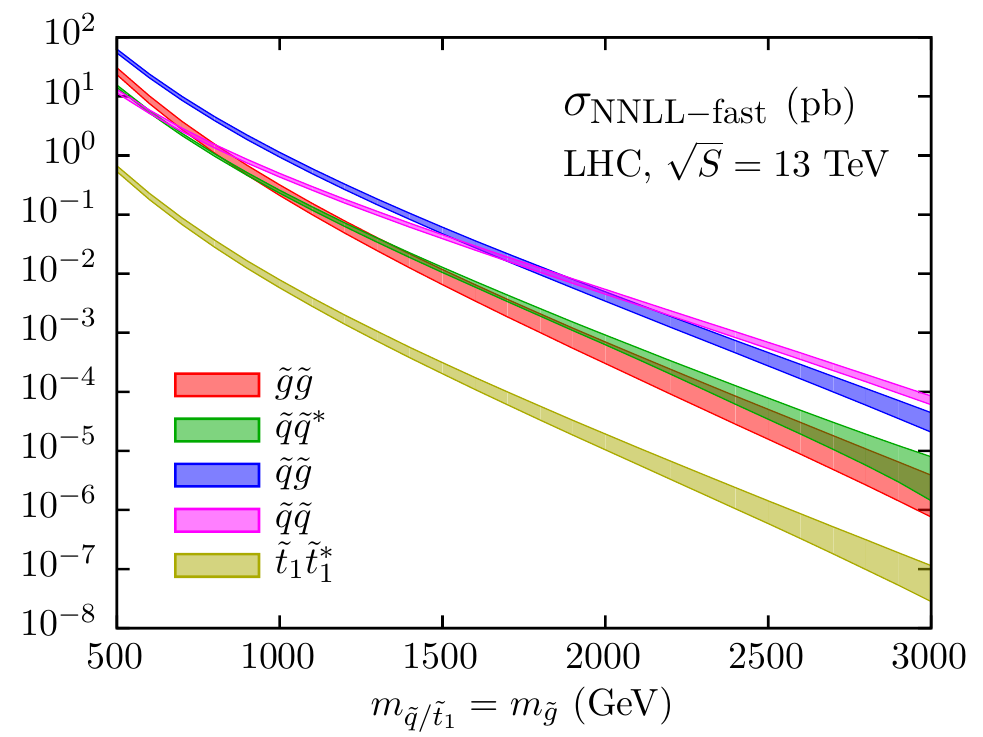}
 \caption{ Cross section predictions for squark and gluino production at the LHC with CM energy $13$ TeV
at NNLOApprox +NNLL accuracy, including Coulomb resummation and bound state effects (taken from Ref.~\cite{Beenakker:2016lwe}). 
The error bands denote the theoretical uncertainty due to scale variation and the combined pdf $+ \alpha_s$  error.
 }
 \label{fig.xsec.vsmass}
 \end{center}
\end{figure}

At the LHC, the most likely process for a discovery is the strong pair-production of colored sparticles, i.e., gluinos and/or squarks.
At leading order (LO), the relevant production processes are:
\begin{subequations}
\begin{eqnarray}
 \sGlu \sGlu : & q_i \bar{q}_i &\rightarrow \sGlu \sGlu  \\
 			& g g  &\rightarrow \sGlu \sGlu  \\
 \sGlu \sQua : & g q_i  &\rightarrow \sGlu \sQua_i  \qquad \mathrm{and \; c.c.} \\
 \sQua \bar{\sQua} : & q_i \bar{q}_j &\rightarrow \sQua_i \bar{\sQua}_j  \\
 			& g g  &\rightarrow \sQua_i \bar{\sQua}_i  \\
 \sQua \sQua : & q_i q_j &\rightarrow \sQua_i \sQua_j \qquad \mathrm{and \; c.c.}
\label{eq:intro.processes}
\end{eqnarray}
\end{subequations}
where the indices designate the quark/squark flavor.
The dependence of the cross sections on the corresponding sparticle mass is shown in Fig.~\ref{fig.xsec.vsmass}.
It is observed that for the same mass, the cross section between gluino and squark differs by at least a factor of 10.
As discussed in \cite{Hubisz:2008gg}, such a ratio is easily expected from the difference in spin between the two states.

The produced sparticles will subsequently decay independently of each other.
If gluinos are produced, they will first decay to (on-shell or off-shell) squarks and quarks through strong interactions.
The squarks will subsequently decay via electroweak interactions to charginos and/or neutralinos.
The latter may have several possible decay modes, depending on the mass hierarchy and the chargino/neutralino composition,
i.e., the relative fractions of gauginos and of higgsinos in the chargino/neutralino physical mass eigenstates.
Altogether, each event contains two decay chains, each producing some detectable particles and terminating
with the emission of an unseen stable neutral particle, the LSP.
After a discovery is made, the most pressing question becomes {\em what underlying physics is leading to the excess}.
This can be at least partly answered by reconstructing the sparticle mass spectrum.
Further constraints on the model may come from the measurement of the production cross section and the decay branching fractions.
As, so far, no hints of Supersymmetry or other Beyond the SM physics have been discovered, 
this ambitious program indeed deserves the ``dreaming awake" clarification in the title.

Numerous methods for mass reconstruction have appeared in the literature \cite{Barr:2010zj}.
Here we will deliberately concentrate on only two types of methods available for the mass reconstruction {\em from endpoints},
as this minimalistic approach suffices to cope with all situations.
\begin{itemize}
\item {\em Invariant mass distributions from a single decay chain.} 
First, we may analyze the successive steps (made of sequential two-body decays) in the decay chain of a single mother sparticle.
In this case, the relevant variable to use is the {\em invariant mass} of two or more visible systems (jets and/or leptons) of the decay chain.
As discussed in Appendix~\ref{sect:invmass},
the endpoints of the distributions provide relations between the masses of the sparticles involved in the decay.
Historically, this was the first proposed method of SUSY mass reconstruction~\cite{Bachacou:1999zb}.
\item {\em (Transverse) invariant mass variables utilizing both decay chains.}
The second method applies when we consider together both decay chains of the two mother sparticles originally produced in the event.
Then, several variables have been proposed, the most popular of which is the Cambridge $M_{T2}$ variable, 
first proposed in \cite{MT2variable}. 
The distribution of the $M_{T2}$ variable (or that of its 3-dimensional analogue, $M_2$, \cite{Cho:2014naa}) also exhibits an endpoint
whose measurement places a constraint on the masses of the sparticles involved in the decay chain. 
The $M_{T2}$ ($M_2$) method is presented in Appendix~\ref{sect:MT2}.
\end{itemize}
Sometimes (especially for the case of sufficiently long decay chains), both types of methods can be applied, 
which may result in over-constrained determination of the masses.

As already mentioned, it is likely that a signal for new physics will first appear in the fully hadronic final states.
As will be illustrated below, these channels will already allow the extraction of a vast amount of physics from the data.
The leptonic events, whose contributions may become visible only later, will bring very valuable additional information
and should allow more precise determination of the masses.
The leptonic channels also present the unique possibility to discover sleptons, the superpartners of the leptons.

It should be emphasized that the mass reconstruction from endpoints always requires a hypothesis to be specified for the decay chains,
defining the nature of the successive sparticles and particles involved.
Some guidance is provided by the event topology, the number and identity of jets and/or leptons,
and possible special features, like the presence of b-jets, $W^\pm$ or $Z^0$ gauge bosons, or the $h^0$ boson.
Frequently, this procedure does not lead to an unambiguous identification of the decay chains,
as several hypotheses may give rise to the same final state topology.
If this occurs, the various possibilities will all need to be tested and their results may or may not allow a ``best hypothesis" to be determined.
For example, this will be the case for the production of $\sTop$ or $\sBot$, whose decay chains are not always easy to distinguish from each other,
nor will it be easy to identify which of the two is the parent.

In the following, we will concentrate on and use particle names from the MSSM. However, the reader should keep in mind that 
the reconstruction techniques, relying purely on kinematics, are model independent,
and would equally well apply to the Universal Extra Dimensions (UED) model 
with KK-parity conservation \cite{Appelquist:2000nn,Cheng:2002iz,Cheng:2002ab, Rizzo:2001sd}
and to the Littlest Higgs model with T-parity \cite{ArkaniHamed:2002qy,Schmaltz:2005ky,Perelstein:2005ka, Belyaev:2006jh}, for example.
The only requirement is that the new particles are produced in pairs and that their decays end 
with the emission of an undetected neutral particle, the LSP.
Note that the $\sGlu$ and $\sQua$ of the MSSM can be distinguished from their equivalent particles in these other models by their spins:
the $\sGlu$ is spin 1/2 and the $\sQua$ is spin 0, while the analogous particles in UED are spin 1 and 1/2, respectively.

This report is structured as follows.
First, in Section \ref{sect:scenarios} we will discuss some extreme physics scenarios to provide some insight into what we may be facing and to set the scene for what follows.
Appendices \ref{sect:invmass} and \ref{sect:MT2} summarize the relevant results
(some of which were previously unpublished) for mass reconstruction from endpoints, using invariant masses or $M_{T2}$ ($M_2$).
In Section \ref{sect:had}, these methods will be applied to the various topologies of fully hadronic final states.
As rather little has been published on this subject, the potential of the methods will be tested using a simplified simulation of the events
in a selected scenario.
It will be seen that the masses of all sparticles participating in the decays can, in principle, be determined.
Section \ref{sect:leptslept} reviews the (mostly published) potential of these methods for events involving leptons,
under the assumption that the decays involve a charged slepton $\sLep$ or a sneutrino $\sNu$.
The case where no sleptons are involved has not been covered in the literature,
and will be tested in Section \ref{sect:lept} using a simplified simulation of the events
in the same scenario as for the hadronic decays\footnote{The results for the remaining electroweakino benchmark scenarios can be found in \cite{Hadr} and \cite{Lept}, respectively.}. It appears that the leptonic channels may be powerful if 
the sleptons are on-shell, and the decays to them have sufficient branching fractions, otherwise
they turn out to be much less useful than hadronic decays.
Finally, Sections \ref{sect:correl}-\ref{sect:BRs} put everything together and summarize 
what can be achieved in terms of understanding the new physics responsible for the signal.
It will be seen that, in addition to the mass reconstruction, it is also possible to determine the production cross sections and decay branching ratios.
The production cross sections, together with the known masses, then enable us to identify the underlying model
by comparing the measurements to the theoretical calculations, which give very different results depending on the spin.
Hence, this enables us to pinpoint the underlying model as the MSSM as opposed to one of its twin scenarios \cite{Cheng:2002ab}.

\section{Benchmark Physics Scenarios in the MSSM}
\label{sect:scenarios}

As already mentioned in the introduction, strong production dominates the SUSY cross-sections at the LHC.
The experimental signatures are determined by the corresponding decay mode patterns. The decays of gluinos always involve a squark:
$\sGlu \rightarrow q \sQua$, where the squark $\sQua$ is either on-shell or off-shell, depending on whether the gluino $\sGlu$
is heavier or lighter than the squark. For concreteness, here we choose the gluino to be heavier than the squarks, 
and pick the masses according to the CMS test point HL2 \cite{PTDR2}, with
$m_{\tilde g}= 1785$ GeV and $m_{\tilde q}=1656$ GeV. 
These masses are on the order of 2 TeV, and still just beyond the reach obtained with the current data set.
The squarks $\sQua$ will subsequently decay to neutralinos or charginos, whose decays in turn will determine the final signal event topology. 

Following Refs.~\cite{Konar:2010bi,Francescone:2014pza,Gainer:2015zna,Altunkaynak:2015kia} we shall investigate several potential MSSM scenarios, 
differing by the relative ordering of the 
electroweak superpartners. In each case, the mass spectra and decay branching fractions will be computed using {\sc Susyhit}
(which comprises of {\sc Suspect}, {\sc Sdecay} and {\sc Hdecay}) \cite{Djouadi:2006bz}.

\subsection{The six possible electroweakino hierarchies}
\label{sect:scenarios.first}

To simplify the picture, we first consider scenarios where all sfermions are heavy and do not affect the 
electroweakino\footnote{Following the standard CMS jargon, the Higgsino, Wino and Bino states will be collectively called "electroweakinos".} 
branching fractions. In particular, the sleptons will be taken to be degenerate with the squarks, 
which prevents any leptonic decays of charginos or neutralinos to sleptons. Furthermore, to prevent 
gluino decays $\tilde g \to t \sTop$ to the lightest top squark mass eigenstate $\sTop$, 
the stop mass parameters will be chosen to be sufficiently large, $\sim 3$ TeV.\footnote{The case 
with a light stop will be discussed separately in Section \ref{sect:scenarios.sfermions} below.}
Throughout this paper, we adopt $tan \beta = 10$, although this choice plays a very minor role.
The light Higgs boson mass is required to be at about 125 GeV, which is achieved 
by adjusting the mixing parameter $A_t$ of the stop,
whereas the heavy Higgs bosons are fixed at 1 TeV and do not participate in the decays of charginos and 
neutralinos.\footnote{For alternative scenarios with the heavy Higgs bosons being observable in sparticle decay chains, see Ref.~\cite{Datta:2003iz}
and references therein.}

As pointed out in \cite{Konar:2010bi}, the relevant SUSY collider signatures depend {\em mostly} on the 
mass {\em ordering} of the superpartners and to a much lesser extent on the absolute mass scale.
The physical masses in the electroweakino sector are determined in terms of three input parameters:
the Bino mass parameter $M_1$, the Wino mass parameter $M_2$, and the higgsino mass parameter $\mu$.\footnote{Since $\mu$ can have either sign, the quantity relevant for the mass hierarchy is $|\mu|$.} 
Therefore, there are $3!=6$ possible cases which are worth considering, each obtained from a unique
ordering of the values of $M_1$, $M_2$ and $|\mu|$. This in turn will change the composition of the 
charginos and neutralinos, i.e., modify the nature of the four neutralino eigenstates $\tilde \chi^0_i, (i=1,2,3,4)$
and the respective two chargino mass eigenstates $\tilde\chi^\pm_i, (i=1,2)$. Note that the two neutral 
higgsino states and the charged higgsino state form an almost degenerate triplet with mass splittings of only a few GeV, 
while the neutral Wino and the charged Wino are extremely degenerate as well, with a mass gap typically smaller than a GeV
\cite{Cheng:1998hc,Feng:1999fu,Gherghetta:1999sw,Ibe:2012sx}.
For concreteness, we fix the numerical values of the mass parameters of the electroweakino sector to be
400 GeV, 800 GeV and 1200 GeV, and obtain the different scenarios from the 
six possible permutations in the set of three variables $\{M_1,M_2,|\mu|\}$,
as illustrated in Figs.~\ref{fig.scenarios.spectra} and \ref{fig.scenarios.spectra1}.
\begin{figure}[t]
\center
\begin{picture}(470,150) 
\put(28, 150){\line(1,0){410}}
\put(28, 10){\line(1,0){410}}
\put(28, 150){\line(0,-1){140}}
\put(128,150){\line(0,-1){140}}
\put(228,150){\line(0,-1){140}}
\put(328,150){\line(0,-1){140}}
\put(438, 150){\line(0,-1){140}}
 \put(0, 140){Mass}
  \put(0, 110){2000}
 \put(0, 80){1200}
\put(0, 50){800}
 \put(0, 20){400}
 \put(60, 140){Bino}
 \put(30, 110){$\sQua_{L,R}$}
 \put(60, 110){\line(1,0){20}}
 \put(30, 80){$\sHig_{u, d}$}
 \put(60, 85){\line(1,0){20}}
 \put(60, 75){\line(1,0){20}}
 \put(90, 80){$\chiz_{3,4} / \chipm_2$}
 \put(30, 50){$\sWino$}
 \put(60, 50){\line(1,0){20}}
 \put(90, 50){$\chiz_2 / \chipm_1$}
 \put(30, 20){$\sBino$}
 \put(60, 20){\line(1,0){20}}
 \put(90, 20){$\chiz_1$}
 \put(160, 140){Wino}
 \put(130, 110){$\sQua_{L,R}$}
 \put(160, 110){\line(1,0){20}}
  \put(130, 80){$\sHig_{u, d}$}
 \put(160, 85){\line(1,0){20}}
 \put(160, 75){\line(1,0){20}}
 \put(190, 80){$\chiz_{3,4} / \chipm_2$}
 \put(130, 50){$\sBino$}
 \put(160, 50){\line(1,0){20}}
 \put(190, 50){$\chiz_2$}
 \put(130, 20){$\sWino$}
 \put(160, 20){\line(1,0){20}}
 \put(190, 20){$\chiz_1 / \chipm_1$}
  \put(260, 140){Mixed}
 \put(230, 110){$\sQua_{L,R}$}
 \put(260, 110){\line(1,0){20}}
 \put(230, 80){$\sWino$}
 \put(260, 80){\line(1,0){20}}
 \put(290, 80){$\chiz_4 / \chipm_2$}
 \put(230, 50){$\sHig_{u, d}$}
 \put(260, 55){\line(1,0){20}}
 \put(260, 45){\line(1,0){20}}
 \put(290, 50){$\chiz_{2,3} / \chipm_1$}
 \put(230, 20){$\sBino$}
 \put(260, 20){\line(1,0){20}}
 \put(290, 20){$\chiz_1$}
 \put(350, 140){Higgsino}
 \put(330, 110){$\sQua_{L,R}$}
 \put(360, 110){\line(1,0){20}}
 \put(330, 80){$\sWino$}
 \put(360, 80){\line(1,0){20}}
 \put(390, 80){$\chiz_4 / \chipm_2$}
 \put(330, 50){$\sBino$}
 \put(360, 50){\line(1,0){20}}
 \put(390, 50){$\chiz_3$}
  \put(330, 20){$\sHig_{u, d}$}
 \put(360, 25){\line(1,0){20}}
 \put(360, 15){\line(1,0){20}}
 \put(390, 20){$\chiz_{1,2} / \chipm_1$}
\end{picture}
\caption{Qualitative representation of the SUSY mass spectra for the four main scenarios.}
\label{fig.scenarios.spectra}
\end{figure}

In what follows, we shall focus primarily on the four scenarios depicted in Fig.~\ref{fig.scenarios.spectra}, since
they lead to distinct phenomenology. As discussed in Sections~\ref{sec:mixed} and \ref{sec:higgsino} below,
the remaining two scenarios from Fig.~\ref{fig.scenarios.spectra1} share
the main features of the Mixed and Higgsino scenarios of Fig.~\ref{fig.scenarios.spectra}, correspondingly.
Therefore, they will not be considered in detail and are only mentioned here for completeness\footnote{Strictly speaking, 
the Bino and Wino scenarios will turn out to be very similar as well, but we chose to consider them separately because of their ubiquity in the literature.}.

\begin{figure}[t]
\begin{centering}
\begin{picture}(200,150) 
\put(28, 150){\line(1,0){200}}
\put(28, 10){\line(1,0){200}}
\put(28, 150){\line(0,-1){140}}
\put(128,150){\line(0,-1){140}}
\put(228,150){\line(0,-1){140}}
 \put(0, 140){Mass}
  \put(0, 110){2000}
 \put(0, 80){1200}
\put(0, 50){800}
 \put(0, 20){400}
 \put(40, 140){Inverted Mixed}
 \put(30, 110){$\sQua_{L,R}$}
 \put(60, 110){\line(1,0){20}}
 \put(30, 80){$\sBino$}
 \put(60, 80){\line(1,0){20}}
 \put(90, 80){$\chiz_4$}
 \put(30, 50){$\sHig_{u, d}$}
 \put(60, 55){\line(1,0){20}}
 \put(60, 45){\line(1,0){20}}
 \put(90, 50){$\chiz_{2,3} / \chipm_2$}
 \put(30, 20){$\sWino$}
 \put(60, 20){\line(1,0){20}}
 \put(90, 20){$\chiz_1 / \chipm_1$}
 \put(135, 140){Inverted Higgsino}
 \put(130, 110){$\sQua_{L,R}$}
 \put(160, 110){\line(1,0){20}}
  \put(130, 80){$\sBino$}
 \put(160, 80){\line(1,0){20}}
 \put(190, 80){$\chiz_4$}
 \put(130, 50){$\sWino$}
 \put(160, 50){\line(1,0){20}}
 \put(190, 50){$\chiz_3 / \chipm_2$}
 \put(130, 20){$\sHig_{u, d}$}
 \put(160, 25){\line(1,0){20}}
 \put(160, 15){\line(1,0){20}}
 \put(190, 20){$\chiz_{1,2} / \chipm_1$}
 \end{picture}
\caption{Qualitative representation of the SUSY mass spectra for the two additional scenarios which we call ``inverted" due to the switching of the roles of the Bino and Wino mass parameters 
$M_1$ and $M_2$ (compare to the corresponding ``Mixed" and ``Higgsino" scenarios from Fig.~\ref{fig.scenarios.spectra}).}
\label{fig.scenarios.spectra1}
\end{centering}
\end{figure}

The relevant branching fractions for the four main scenarios from Fig.~\ref{fig.scenarios.spectra}
are listed in Table~\ref{tab.scenarios.BR}.
\begin{table}[htb]
\begin{center}
\resizebox{\columnwidth}{!}{%
\begin{tabular}{|ccc|ccc|ccc|ccc|} \hline
                 & Bino     &   					&  &Wino	& 							     &  & Mixed	& 							& & Higgsino   & \\
\multicolumn{3}{|c|}{$\chiz_1 < \chiz_2 = \chipm_1$}  &  \multicolumn{3}{|c|}{$\chiz_1 =  \chipm_1 < \chiz_2$} 	&  \multicolumn{3}{|c|}{$\chiz_1 < \chiz_2 =  \chiz_3 = \chipm_1$}
&  \multicolumn{3}{|c|}{$\chiz_1 = \chiz_2 = \chipm_1 <  \chiz_3$}	   \\
\hline
$\sQua_L$ & $\rightarrow q \chipm_1$  & 65\% & $\sQua_L$ & $\rightarrow q \chipm_1$  & 66\% &$\sQua_L$ & $\rightarrow q \chipm_2$ & 60\% &$\sQua_L$ & $\rightarrow q \chipm_2$ & 63\%   \\
		& $\rightarrow q \chiz_2$  & 32\% & &   $\rightarrow q \chiz_1$  & 33\% & &   $\rightarrow q \chiz_4$  & 30\% & &   $\rightarrow q \chiz_4$  & 31\%    \\
		& $\rightarrow q \chipm_2$  & 1\% & & $\rightarrow q \chiz_2$  & 1\% 				& & $\rightarrow q \chipm_1$ & 4\% &  & $\rightarrow q \chiz_3$  & 2\%    \\
		& $\rightarrow q \chiz_1$  & 1\% & & & 										& & $\rightarrow q \chiz_2$ & 2\% &  &  $\rightarrow q \chipm_1$  & 3\%   \\
		& & & & & 																& & $\rightarrow q \chiz_1$ & 4\% &  &  &   \\
$\sQua_R$ & $\rightarrow q \chiz_1$  & 100\% & $\sQua_R$ & $\rightarrow q \chiz_2$  & 100\% &$\sQua_R$ & $\rightarrow q \chiz_1$ & 100\% &$\sQua_R$ & $\rightarrow q \chiz_3$ &99\%   \\
$\chipm_1$ & $\rightarrow W \chiz_1$  & 100\% &  &  & 								 &$\chipm_2$ & $\rightarrow Z \chipm_1$ & 26\% &$\chipm_2$ & $\rightarrow h \chipm_1$ &26\%   \\
		& & & & & 																& & $\rightarrow h \chipm_1$ & 25\% &  & $\rightarrow Z \chipm_1$ &25\%   \\
		& & & & & 																& & $\rightarrow W \chiz_2$ & 25\% &  & $\rightarrow W \chiz_2$ &24\%   \\
		& & & & & 																& & $\rightarrow W \chiz_3$ & 24\% &  & $\rightarrow W \chiz_1$ &24\%   \\
$\chiz_2$ & $\rightarrow h \chiz_1$  & 94\% & $\chiz_2$ & $\rightarrow W \chipm_1$  & 65\% &$\chiz_4$ & $\rightarrow W \chipm_1$ & 50\% &$\chiz_4$ & $\rightarrow W \chipm_1$ &49\%   \\
		& $\rightarrow Z \chiz_1$  & 6\%   &			 & $\rightarrow h \chiz_1$  & 33\% &			 & $\rightarrow h \chiz_2$ &24\% &			 & $\rightarrow Z \chiz_2$ &22\%   \\
		&   &   &								 & $\rightarrow Z \chiz_1$  & 2\%   &			 & $\rightarrow Z \chiz_3$ &24\% &			 & $\rightarrow h \chiz_1$ &21\%   \\
		&   &   &					&   &   									&	$\chipm_1$ & $\rightarrow W \chiz_1$  & 100\%  &	 & $\rightarrow h \chiz_2$ &4\%   \\
		&   &   &					&   &   									&	$\chiz_2$ & $\rightarrow h \chiz_1$  & 92\%   	&	& $\rightarrow Z \chiz_1$ &4\%   \\
		&   &   &					&   &   									&			 & $\rightarrow Z \chiz_1$  & 8\%   &			  & &    \\
		&   &   &					&   &   									&	$\chiz_3$ & $\rightarrow Z \chiz_1$  & 93\%   	& $\chiz_3$  & $\rightarrow W \chipm_1$ &52\%   \\
		&   &   &					&   &   									&			 & $\rightarrow h \chiz_1$  & 7\%   	& 		  & $\rightarrow Z \chiz_2$ &23\%   \\
		&   &   &					&   &   									&			 & 	  & 	   					& 		  & $\rightarrow h \chiz_1$ &23\%   \\
 \hline
\end{tabular}%
}
\caption{Branching fractions for the four main scenarios from Fig.~\ref{fig.scenarios.spectra}.}
\label{tab.scenarios.BR}
\end{center}
\end{table}
For the first two generations of squarks, the Yukawa couplings are negligible, hence the decays are governed by gauge couplings only.
The dominant decay modes are then easily understood in terms of the quantum numbers of the SUSY partners.
For example, the right-handed squarks $\sQua_R$ are not charged under $SU(2)_W$ and thus decay 100\%
of the time to the respective Bino state $\sBino$.
On the other hand, the left-handed squarks $\sQua_L$ decay predominantly to the Wino states for two reasons: first, 
the $SU(2)_W$ gauge coupling $g_2$ is larger than the hypercharge gauge coupling $g_Y$; and second, the 
$q_L$ hypercharge is relatively small, $Y_{q_L}=\frac{1}{6}$. Due to the presence of two charged Wino states, $\tilde W^+$
and $\tilde W^-$, and only one neutral Wino state, $\tilde W^0$, the $\sQua_L$ branching fraction to a Wino-like chargino 
is higher by a factor of 2 relative to the $\sQua_L$ branching fraction to a Wino-like neutralino:
\begin{equation}
\frac{B(\sQua_L\to \tilde W^\pm)}{B(\sQua_L\to \tilde W^0)}=2.
\label{eq2:1}
\end{equation} 
We shall now briefly describe the typical signatures for 
each of the four main scenarios in Fig.~\ref{fig.scenarios.spectra}.

\subsubsection{The Bino scenario with $M_1<M_2<|\mu|$}
\label{sec:bino}

The \underline{\em Bino scenario} (left-most columns in Fig.~\ref{fig.scenarios.spectra} and Table~\ref{tab.scenarios.BR}) 
is the one most commonly tested so far, since it is encountered in most of the parameter space of the 
mSUGRA (CMSSM) model \cite{Chamseddine:1982jx,Barbieri:1982eh,Hall:1983iz}. Here, the Higgsinos are the heaviest electroweakinos, comprising the states 
$\tilde\chi^0_4$, $\tilde\chi^0_3$ and $\tilde\chi^\pm_2$.
Due to the smallness of the Yukawa couplings, these states are essentially bypassed in the 
decays of the light flavor squarks, and thus do not appear in the left column of Table~\ref{tab.scenarios.BR}.
The lightest electroweakino is the Bino LSP $\tilde\chi^0_1$, and the right-handed squarks $\sQua_R$ always decay directly to it.
In contrast, the left-handed squarks $\sQua_L$ decay primarily to the Wino-like states 
$\tilde\chi^\pm_1$ and $\tilde\chi^0_2$ in the proportion $2:1$, as predicted by eq.~(\ref{eq2:1}).
The Winos, in turn, decay further to the LSP: the decay of the charged Wino $\chipm_1$ produces a $W$ boson, while 
the decay of the neutral Wino $\chiz_2$ yields typically a Higgs boson $h$ instead of a $Z$ boson,
due to an additional neutralino mixing suppression.\footnote{Since the SM has no triple gauge boson vertices involving three neutral gauge bosons,
the decay of a gaugino-like neutralino to another gaugino-like neutralino necessarily requires higgsino-gaugino mixing and is therefore suppressed.
The decay to a Higgs boson $h$ requires a Higgsino-Gaugino-Higgs boson vertex and a single mixing suppression, while the decay to a $Z$ boson
requires a Higgsino-Higgsino-Gauge Boson vertex and is doubly suppressed \cite{Djouadi:2001kba}. }
Note that in this scenario, there would be no sign in the data of the existence of the heavier Higgsino-like chargino $\chipm_2$ and neutralinos $\chiz_{3}$ and $\chiz_{4}$.
The number of bosons ($W$, $Z$ or $h$) per squark decay chain is either 0 or 1. 
The latter case is due to the decay of a left-handed squark, $\sQua_L$,
whose relative branching fractions are approximately $B(\sQua_L\to q'W \chiz_1):B(\sQua_L\to qZ \chiz_1):B(\sQua_L\to qh \chiz_1+X) \simeq 2:0:1$.
Depending on the further decays of the SM bosons, we will obtain events with a certain number of jets and leptons.
In fully hadronic events, the largest number of jets (at leading order) will be an 8-jet topology, due to $\sGlu \sGlu$ production,
followed by $\sGlu \to \sQua_L$ and hadronic decays of the SM bosons. 
On the other hand, the largest number of leptons per event, 4, will be obtained when the gauge boson on each side is a $Z$-boson decaying leptonically.

\subsubsection{The Wino scenario with $M_2<M_1<|\mu|$}
\label{sec:wino}

In the \underline{\em Wino scenario}, shown in the second columns in Fig.~\ref{fig.scenarios.spectra} and Table~\ref{tab.scenarios.BR},
the lightest neutralino $\chiz_1$ and the lightest chargino $\chipm_1$ are nearly mass degenerate. The chargino $\chipm_1$ 
is slightly heavier than the neutralino $\chiz_1$,
and will decay promptly, but its decay products will be too soft to be detected reliably.
Thus, from an experimental point of view, the chargino $\chipm_1$ behaves like its Wino-like brother $\chiz_1$, which is the true LSP.
This implies that in a standard hadronic analysis the decay to a $\chipm_1$ may be mistakenly interpreted as being to a $\chiz_1$.
Compared to the Bino-like scenario of Sec.~\ref{sec:bino}, the roles of the left-handed and right-handed squarks are now reversed:
it is now the left-handed squarks $\sQua_L$ which decay directly to the (effective) Wino LSP in the proportion predicted by (\ref{eq2:1}).
The right-handed squarks $\sQua_R$, in turn, decay to the Bino-like second-lightest neutralino $\chiz_2$, 
and the experimental signatures are determined by its branching fractions.
Table~\ref{tab.scenarios.BR} shows that the neutral $\chiz_2$ state decays to the lightest chargino $\chipm_1$
twice as often as to the lightest neutralino $\chiz_1$:
\begin{equation}
\frac{B(\chiz_2\to \chipm_1)}{B(\chiz_2\to \chiz_1)}=2,
\label{eq2:2}
\end{equation} 
in analogy to (\ref{eq2:1}). The decays to $h^0$ and $Z^0$ involve the same vertices as in the Bino-like scenario
from Section~\ref{sec:bino}, and are similarly suppressed: the decay to $h$ ($Z$) requires single (double) gaugino-higgsino mixing suppression.
Also similarly to the Bino scenario from Section~\ref{sec:bino}, there would be no sign in the data of the heavier higgsino-like neutralinos
$\chiz_3$ and $\chiz_4$ and chargino $\chipm_2$,
since they are bypassed in the squark cascade decays.
The number of bosons ($W$, $Z$ or $h$) per decay chain originating from a squark $\sQua$ is either 0 or 1, and in the latter case the
relative branching fractions are again approximately given by
$B(\sQua_R\to q'W \chiz_1):B(\sQua_R\to qZ \chiz_1):B(\sQua_R\to qh \chiz_1) \simeq 2:0:1$.
At leading order, the largest number of jets in fully hadronic events would be the 8-jet topology (from $\sGlu \sGlu$ production), while
the largest lepton multiplicity per event would be 4.

\begin{figure}[!htb]
 \begin{center}
 \includegraphics[width=0.4\textwidth]{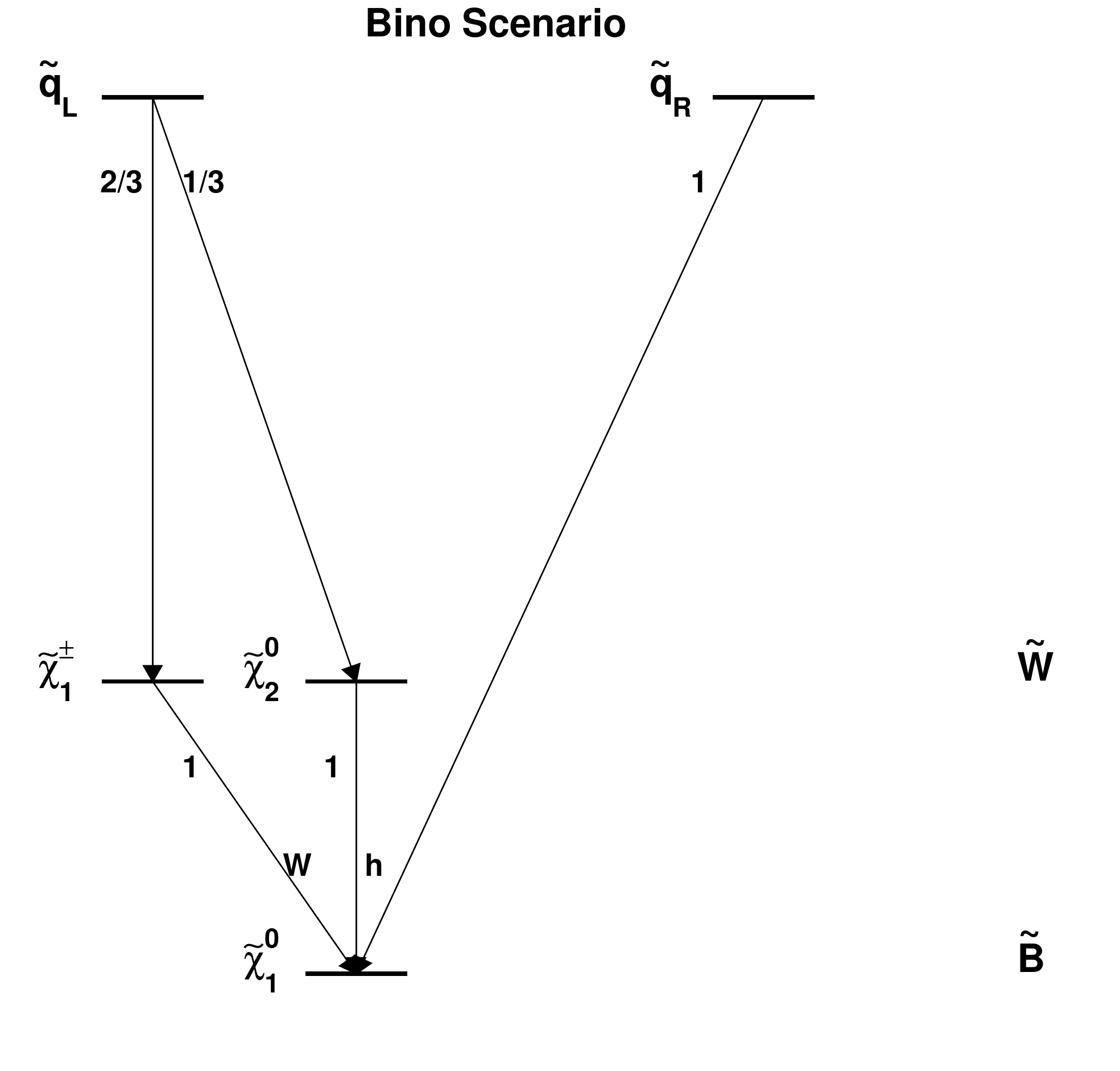}
 \hspace*{1.5cm}
 \includegraphics[width=0.4\textwidth]{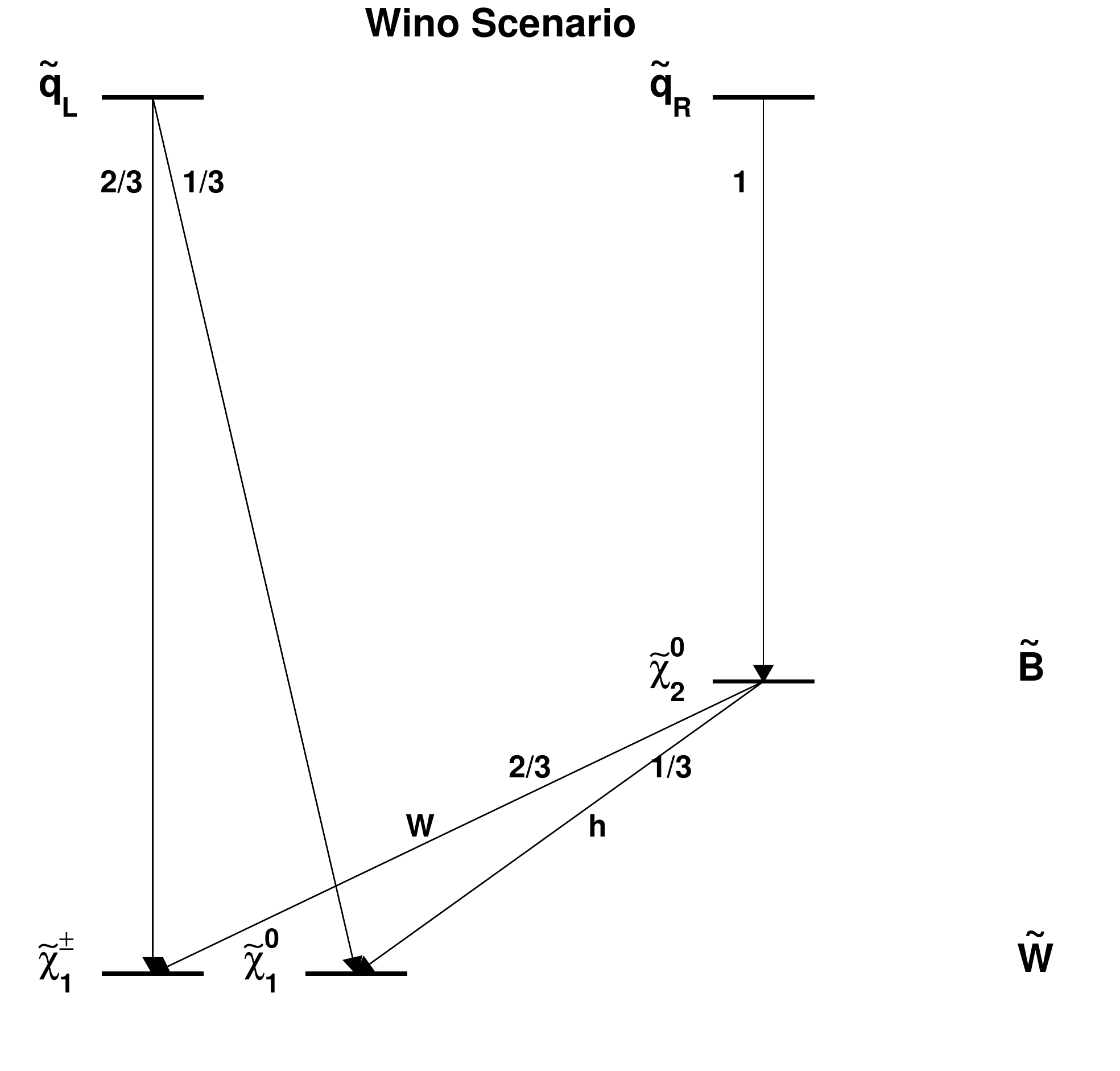}
\caption{Decays and branching fractions for
the Bino-like scenario (left) and the Wino-like scenario (right).
}
 \label{fig.distscenarios.BWino}
 \end{center}
\end{figure}

As seen from Table~\ref{tab.scenarios.BR} and the previous discussion,
the Bino and Wino scenarios can be easily confused with each other.
This is illustrated pictorially in Fig.~\ref{fig.distscenarios.BWino}, which depicts the dominant decay modes and their branching fractions in the two scenarios.

Each squark decay chain may produce a single jet; a jet accompanied by a $W$; or a jet accompanied by a Higgs boson;
with relative branching fractions given in Table~\ref{tab:binowinoBR}.

\begin{table}[!htb]
\begin{center}
\resizebox{\columnwidth}{!}{%
\begin{tabular}{||c|ccc|cc||} 
\hline
\multicolumn{6}{||c||}{Bino scenario}\\
\hline
No &
\multicolumn{3}{|c|}{Decays } &
\multicolumn{2}{c||}{Branching fractions}\\
\hline
 & & & & & \\
 &$\sQua_L$	& $\rightarrow q \chipm_1$ 	& $\rightarrow q W \chiz_1$  & $= \frac{2}{3} \times  1$  & $= \frac{2}{3}$ \\
$N_{\tilde q_L}$    & & & & & \\
 &			& $\rightarrow q \chiz_2$ 		& $\rightarrow q h \chiz_1$  & $= \frac{1}{3} \times  1$  & $= \frac{1}{3}$ \\
 & & & & & \\
\hline
 & & & & & \\
 & $\sQua_R$	& $\rightarrow q \chiz_1$ 		&					& 					& $= 1$ \\
$N_{\tilde q_R}$    & & & & & \\
 & & & & & \\
 & & & & & \\
 \hline
\end{tabular}
\begin{tabular}{||c|ccc|cc||} 
\hline
\multicolumn{6}{||c||}{Wino scenario}\\
\hline
No &
\multicolumn{3}{|c|}{Decays } &
\multicolumn{2}{c||}{Branching fractions}\\
\hline
 &  & & & & \\
 & $\sQua_R$	& $\rightarrow q \chiz_2$ 		& $\rightarrow q W \chipm_1$  & $= 1 \times \frac{2}{3}$	& $= \frac{2}{3}$ \\
 $N_{\tilde q_R}$&  & & & & \\
 &			& 				 		& $\rightarrow q h \chiz_1$	& $= 1 \times \frac{1}{3}$  &  $= \frac{1}{3}$ \\
 & & & & & \\
\hline
 & & & & & \\
 & $\sQua_L$	& $\rightarrow q \chipm_1$ 	&   $\rightarrow q \pi^\pm \chiz_1 $ &	& $= \frac{2}{3}$   \\
$N_{\tilde q_L}$    & & & & & \\
 &			& $\rightarrow q \chiz_1$ 		&   					&	& $= \frac{1}{3}$ \\
  & & & & & \\
\hline
\end{tabular}
}
\caption{Squark decays and branching fractions for the Bino and Wino scenarios.}
\label{tab:binowinoBR}
\end{center}
\end{table}

The starting numbers for each type of squarks, $N_{\tilde q_L}$ for the left-handed squarks $\sQua_L$ and $N_{\tilde q_R}$ 
for the right-handed squarks $\sQua_R$, are {\em a priori} unknown. In the Bino scenario, there are $N_{\tilde q_R}$
cascades with zero bosons, $\frac{2}{3}N_{\tilde q_L}$ cascades with a $W$-boson and $\frac{1}{3}N_{\tilde q_L}$ cascades
with a $h$ boson. On the other hand, in the Wino scenario, there are $N_{\tilde q_L}$
cascades with zero bosons, $\frac{2}{3}N_{\tilde q_R}$ cascades with a $W$-boson and $\frac{1}{3}N_{\tilde q_R}$ cascades
with a $h$ boson. We see that the final state multiplicities for the two scenarios can be mapped into each other by interchanging 
the values of $N_{\tilde q_L}$ and $N_{\tilde q_R}$, which are {\em a priori} unknown. 

This brings up the problem of how to distinguish experimentally the Bino and Wino scenarios, without any theoretical assumptions. 
One important difference is that in the Bino scenario, the decay of the Wino-like chargino, $\chipm_1$, to its Wino-like neutralino cousin, $\chiz_2$, is 
enormously suppressed by phase space and never takes place --- instead, the chargino always prefers to decay directly to the LSP, the Bino-like $\chiz_1$.
On the other hand, in the Wino scenario, the Wino-like chargino $\chipm_1$ has no other choice but to decay to its Wino-like neutralino cousin, which this time is
$\chiz_1$ itself. Due to the extreme degeneracy of the two Wino-like states, the experimental signature depends on the exact value of the mass splitting.
If the mass difference $M(\chipm_1) - M(\chiz_1) \leq m_{\pi}$, the chargino decay would be $\chipm_1 \rightarrow \chiz_1 e^{\pm} \nu$ yielding a soft
lepton.\footnote{In extreme cases, the chargino can be long lived, which motivates a dedicated class of searches for disappearing tracks
\cite{Aaboud:2017mpt,Sirunyan:2018ldc}.}
However, due to radiative corrections \cite{Cheng:1998hc,Feng:1999fu,Gherghetta:1999sw,Ibe:2012sx}, the mass splitting is more likely to be
$M(\chipm_1) - M(\chiz_1) \geq m_{\pi}$, in which case the decay is prompt, $\chipm_1 \rightarrow \chiz_1 + \pi^\pm$.
As the mass difference is at most some 200 MeV, the decay products will be difficult to detect.

\subsubsection{The Mixed scenario with $M_1<|\mu|<M_2$}
\label{sec:mixed}

The \underline{\em Mixed scenario} is motivated by models in which the lightest neutralino can be a good dark matter candidate
due to just the right amount of Bino-Higgsino mixing. The ``focus-point" region of mSUGRA is a well-known such example
\cite{Feng:2000gh,Feng:2000bp,Feng:2011aa} (in a model-independent setup this case is sometimes referred to as 
``the well-tempered neutralino" \cite{ArkaniHamed:2006mb}). As seen from the third column of Table~\ref{tab.scenarios.BR},
the collider phenomenology is quite distinctive and may give rise to some rather long decay chains, 
e.g., $\sQua_L \rightarrow q \chiz_4 \rightarrow q (W^{\pm} \chimp_1) \rightarrow q (W^{\pm} W^{\mp} \chiz_1)$
or $\sQua_L \rightarrow q \chiz_4 \rightarrow q (h \chiz_2) \rightarrow q (h h \chiz_1)$.
The Higgsino-like states $\chiz_2$, $\chiz_3$ and $\chipm_1$ are nearly mass degenerate and so are the Wino-like states $\chiz_4$ and $\chipm_2$.
Note that due to the smallness of the Yukawa couplings, the left-handed squarks $\sQua_L$ decay to the heavier Winos, 
rather than the lighter Higgsinos, although the decays to Higgsinos are favored by phase space.
Both Wino-like states, $\chipm_2$ and $\chiz_4$, decay to the respective Higgsino states plus a boson, $Z^0$, $h^0$, or $W$, in proportions $1:1:2$.
The Wino states do not decay directly to the Bino, $\chiz_1$, since there is no direct Wino-Bino coupling at tree level.
The typical number of bosons ($W$, $Z$, or $h$) in a squark decay chain is
either zero (for right-handed squarks $\sQua_R$) or 2 (for left-handed squarks $\sQua_L$).
On rare occasions ($\sim 6\%$) a left-handed squark will bypass the Winos and decay directly to a higgsino state, 
in which case there will be a single boson per decay chain.
In this Mixed scenario, the largest number of jets in fully hadronic events is found in a 12-jet topology (from $\sGlu \sGlu$ production), while
the largest lepton multiplicity would be 8 (when there are two $Z$ bosons in each decay chain).

\subsubsection{The Inverted Mixed scenario with $M_2<|\mu|<M_1$}
\label{sec:mixedinv}

\begin{figure}[!htb]
 \begin{center}
 \includegraphics[width=0.4\textwidth]{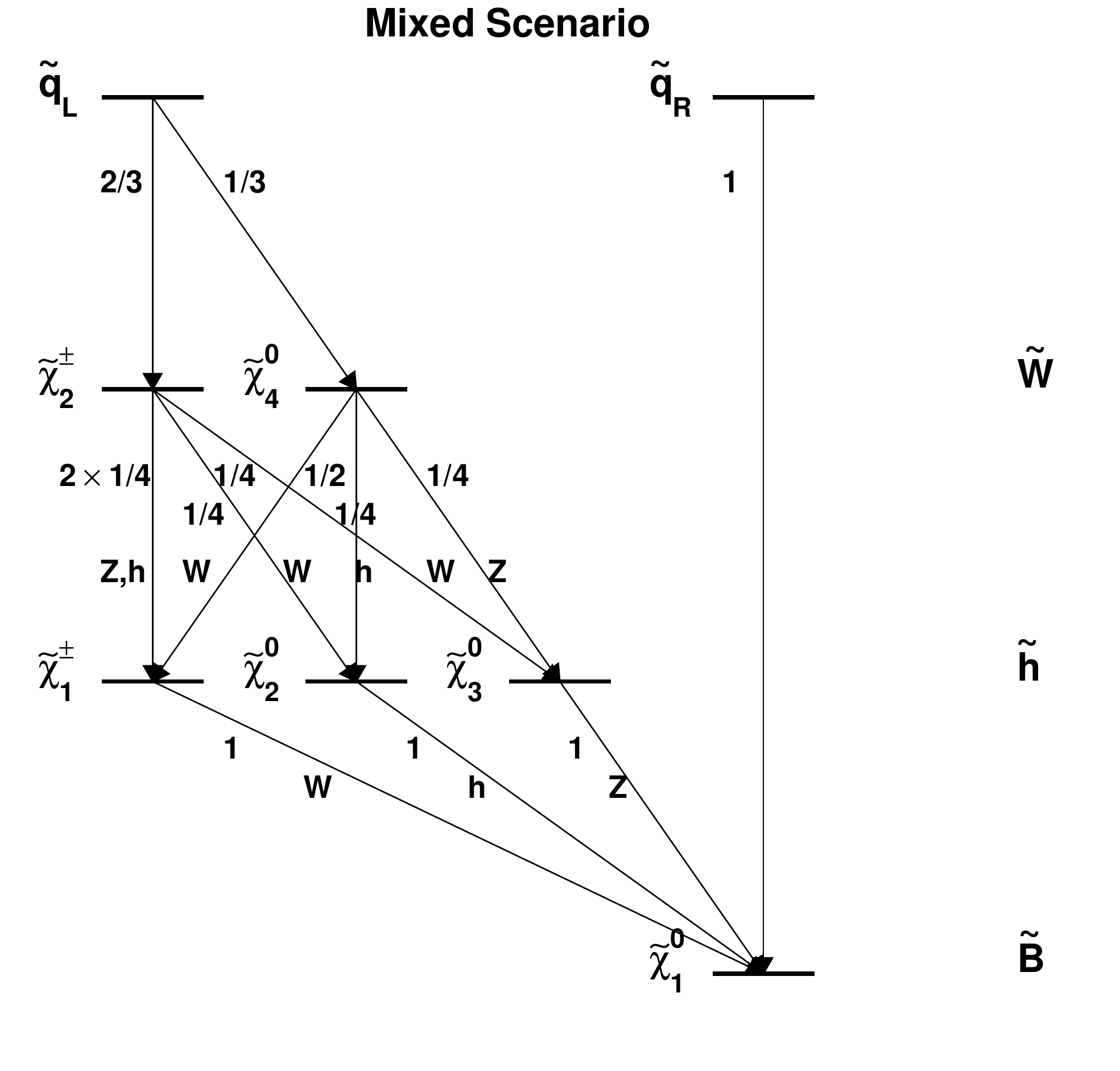}
 \hspace*{1.5cm}
 \includegraphics[width=0.4\textwidth]{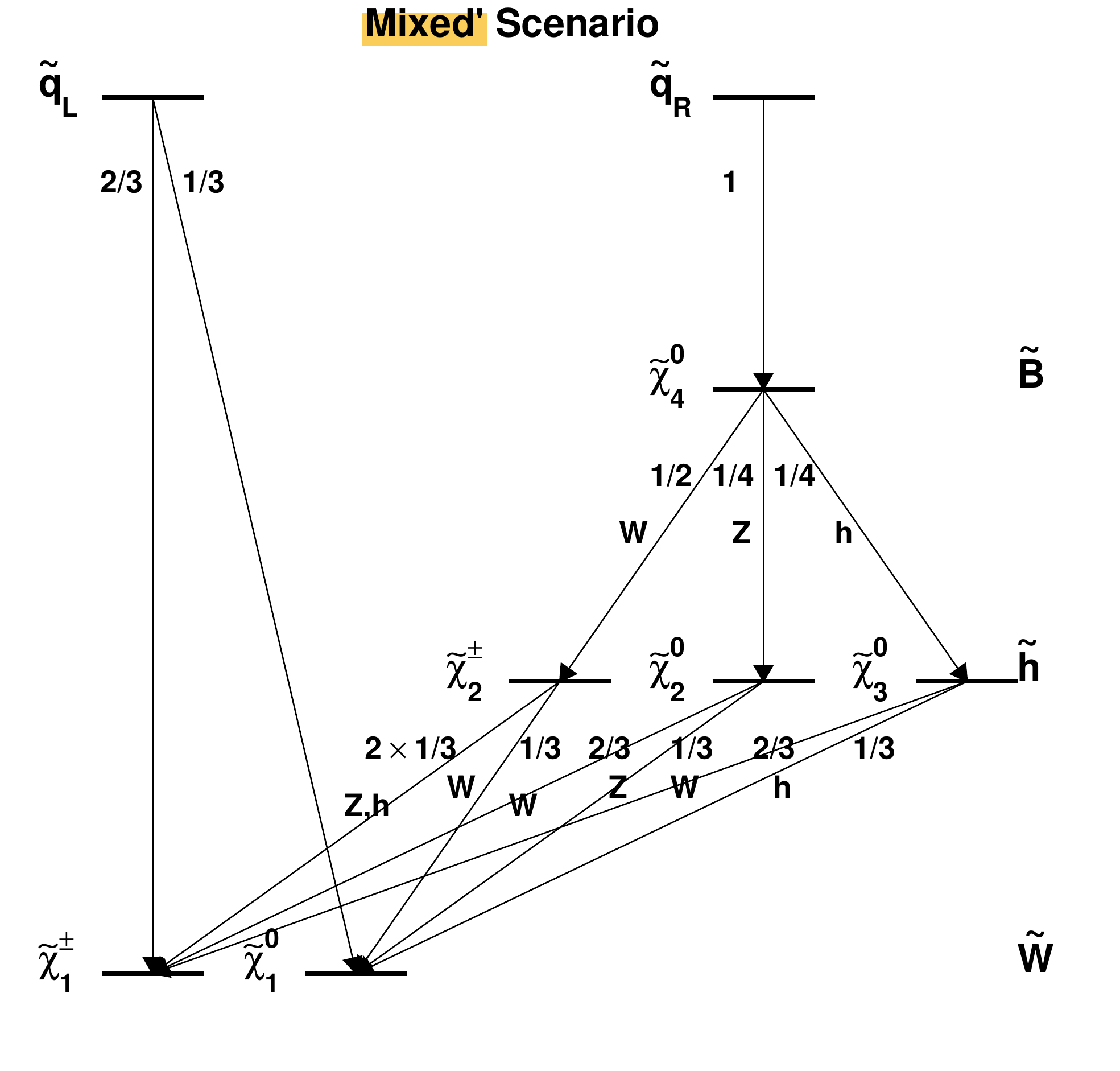}
\caption{Decays and branching fractions for the Mixed scenario of Sec.~\ref{sec:mixed}
(left), and the Inverted Mixed scenario of Sec.~\ref{sec:mixedinv} (right).
}
 \label{fig.distscenarios.Mixed}
 \end{center}
\end{figure}

The next logical possibility is the \underline{\em Inverted Mixed scenario}, illustrated in the left column of Fig.~\ref{fig.scenarios.spectra1}.
This case is similar to the Mixed scenario of Sec.~\ref{sec:mixed}, only now the roles of the Bino and Wino have been reversed, and the mass ordering is
$M_2<|\mu|<M_1$. Correspondingly, the LSP, $\chiz_1$, is now a neutral Wino and the heaviest neutralino, $\chiz_4$, is Bino-like.
The decays and branching ratios for the last two scenarios, Mixed and Inverted Mixed, are illustrated in Fig.~\ref{fig.distscenarios.Mixed}.

Fig.~\ref{fig.distscenarios.Mixed} demonstrates that the LHC signatures of the two scenarios are very similar. 
In each case, one of the two squark species ($\sQua_R$ in the Mixed scenario and $\sQua_L$ in the Inverted Mixed scenario)
decay directly to the LSP with the emission of a jet. The other species of squarks ($\sQua_L$ in the Mixed scenario and $\sQua_R$ in the Inverted Mixed scenario)
decay predominantly to the heaviest electroweakinos, initiating a longer decay chain with two bosons. 
The branching ratios for the different final states can be readily calculated and are shown in Table~\ref{tab:mixedBR}.

\begin{table}[!htb]
\begin{center}
\resizebox{\columnwidth}{!}{%
\begin{tabular}{||c|cccc|cc||} 
\hline
\multicolumn{7}{||c||}{Mixed scenario}\\
\hline
No &
\multicolumn{4}{|c|}{Decays } &
\multicolumn{2}{c||}{Branching fractions}\\
\hline
&&  & & & & \\
& $\sQua_L$	& $\rightarrow q \chipm_2$	& $\rightarrow q Z \chipm_1$ 	& $\rightarrow q Z W \chiz_1$  & $= \frac{2}{3} \times \frac{1}{4} \times  1$  & $= \frac{1}{6}$ \\
&&  & & & & \\ 
& 			&						& $\rightarrow q h \chipm_1$ 	& $\rightarrow q h W \chiz_1$  & $= \frac{2}{3} \times \frac{1}{4} \times  1$  & $= \frac{1}{6}$ \\
&&  & & & & \\
& 			&						& $\rightarrow q W \chiz_2$ 	& $\rightarrow q W h \chiz_1$  & $= \frac{2}{3} \times \frac{1}{4} \times  1$  & $= \frac{1}{6}$ \\
&&  & & & & \\
 $N_{\tilde q_L}$& 			&			& $\rightarrow q W \chiz_3$ 	& $\rightarrow q W Z \chiz_1$  & $= \frac{2}{3} \times \frac{1}{4} \times  1$  & $= \frac{1}{6}$ \\
&&  & & & & \\
&			& $\rightarrow q \chiz_4$		& $\rightarrow q W \chipm_1$ 	& $\rightarrow q W W \chiz_1$  & $= \frac{1}{3} \times \frac{1}{2} \times  1$  & $= \frac{1}{6}$ \\
& &  & & & & \\
& 			&						& $\rightarrow q h \chiz_2$ 	& $\rightarrow q h h \chiz_1$  & $= \frac{1}{3} \times \frac{1}{4} \times  1$  & $= \frac{1}{12}$ \\
& &  & & & & \\
& 			&						& $\rightarrow q Z \chiz_3$ 	& $\rightarrow q Z Z \chiz_1$  & $= \frac{1}{3} \times \frac{1}{4} \times  1$  & $= \frac{1}{12}$ \\
& &  & & & & \\
 \hline
&&  & & & & \\
& $\sQua_R$	& $\rightarrow q \chiz_1$		&					 	&  					  &  & $=  1$   \\
 $N_{\tilde q_R}$&&  & & & & \\
& &  & & & & \\
& &  & & & & \\
\hline 
\end{tabular}
\begin{tabular}{||c|cccc|cc||} 
\hline
\multicolumn{7}{||c||}{Inverted Mixed scenario} \\
\hline
No &
\multicolumn{4}{|c|}{Decays } &
\multicolumn{2}{c||}{Branching fractions}\\
\hline
& &  & & & & \\
& $\sQua_R$	& $\rightarrow q \chiz_4$		& $\rightarrow q W \chipm_2$ 	& $\rightarrow q W Z \chipm_1$  & $= 1 \times \frac{1}{2} \times \frac{1}{3}$  & $= \frac{1}{6}$ \\
& &  & & & & \\
& 			&						& 					 	& $\rightarrow q W h \chipm_1$  & $= 1 \times \frac{1}{2} \times \frac{1}{3}$  & $= \frac{1}{6}$ \\
&  &  & & & & \\
& 			&						& 					 	& $\rightarrow q W W \chiz_1$    & $= 1 \times \frac{1}{2} \times \frac{1}{3}$  & $= \frac{1}{6}$ \\
&  &  & & & & \\
 $N_{\tilde q_R}$ & 			&						& $\rightarrow q Z \chiz_2$ 	& $\rightarrow q Z W \chipm_1$  & $= 1 \times \frac{1}{4} \times \frac{2}{3}$  & $= \frac{1}{6}$ \\
&   &  & & & & \\
& 			&						& 					 	& $\rightarrow q Z Z \chiz_1$  	  & $= 1 \times \frac{1}{4} \times \frac{1}{3}$  & $= \frac{1}{12}$ \\
&  &  & & & & \\
&			& 						& $\rightarrow q h \chiz_3$ 	& $\rightarrow q h W \chipm_1$  & $= 1 \times \frac{1}{4} \times \frac{2}{3}$  & $= \frac{1}{6}$ \\
&  &  & & & & \\
& 			&						& 					 	& $\rightarrow q h h \chiz_1$  	  & $= 1 \times \frac{1}{4} \times \frac{1}{3}$  & $= \frac{1}{12}$ \\
& &  & & & & \\
\hline
& &  & & & & \\
&  $\sQua_L$	& $\rightarrow q \chipm_1$	&  $\rightarrow q \pi^\pm \chiz_1 $ & 					  &   & $=   \frac{2}{3}$  \\
 $N_{\tilde q_L}$&  &  & & & & \\
 &			& $\rightarrow q \chiz_1$		&					 	&  					  &   & $=   \frac{1}{3}$  \\
 & &  & & & & \\
  \hline
\end{tabular}
}
\caption{The same as Table~\ref{tab:binowinoBR}, but this time comparing the Mixed and Inverted Mixed scenarios.}
\label{tab:mixedBR}
\end{center}
\end{table}

We notice that Table~\ref{tab:mixedBR} is very similar to Table~\ref{tab:binowinoBR} in the sense that there is an almost perfect duality 
between the experimental signatures of the two considered scenarios, which is realized by interchanging the number of squarks of each type,
$N_{\tilde q_L}\leftrightarrow N_{\tilde q_R}$. The relative rates of the different diboson final states are the same,
and are uniquely fixed in terms of combinatorics factors. For completeness, in the table we kept $W$, $Z$ and $h$ as distinct 
objects, although, as we shall see later, the di-jet mass resolution may not be sufficient to separate the hadronic $W$'s, hadronic $Z$'s
and hadronic $h$'s. However, even at this somewhat idealized level, the two scenarios match almost identically and will be very difficult to distinguish experimentally.
Just like in the case of the Bino and Wino scenarios discussed earlier, the only difference between the two scenarios appears in the decay of the 
Wino-like chargino $\chip_1$ in the Inverted Mixed scenario, where one can attempt to look for a long-lived chargino or disappearing tracks
(see discussion at the end of Sec.~\ref{sec:wino}).

\subsubsection{The Higgsino scenario with $|\mu|<M_1<M_2$}
\label{sec:higgsino}

The final two out of the six possible mass hierarchies for the electroweakinos are characterized with the Higgsino mass parameter $|\mu|$ being 
smaller than the gaugino masses $M_1$ and $M_2$, and as a result, the LSP is Higgsino-like.\footnote{This situation is typical of several classes of
``natural" SUSY models \cite{Chan:1997bi,Kitano:2006ws,Baer:2011ec,Papucci:2011wy,Baer:2012uy,Baer:2012cf}.}
In fact, the two lightest neutralinos, $\chiz_1$ and $\chiz_2$, and the lightest chargino, $\chipm_1$, are nearly mass degenerate and all of them act,
from an experimental point of view, as if they were the LSP. The small mass differences ($< 10$ GeV) between the Higgsino states produce 
very soft objects ($X$) which will escape the standard analyses.

The specific scenario considered in this subsection is the \underline{\em Higgsino scenario}
with $|\mu|<M_1<M_2$ depicted in the fourth column of Fig.~\ref{fig.scenarios.spectra}.\footnote{The {\em Inverted Higgsino scenario} with $|\mu|<M_2<M_1$ is the subject of the next Section~\ref{sec:higgsinoinv}.}
The decay modes follow the same patterns as described in previous scenarios, with the branching fractions listed in the last column of Table~\ref{tab.scenarios.BR}.
As before, the left-handed squarks $\sQua_L$ preferentially decay to the two Wino states $\chiz_4$ and $\chipm_2$, while the right-handed squarks $\sQua_R$
decay directly to the Bino state $\chiz_3$. In turn, the Wino and Bino states all decay directly to the Higgsino states 
$\chipm_1$, $\chiz_2$, and $\chiz_1$.
The number of bosons ($W$, $Z$ or $h$) per squark decay chain is always 1, with 
relative branching fractions 
$B(\sQua\to q'W \chiz_1):B(\sQua\to qZ \chiz_1):B(\sQua\to qh \chiz_1) \simeq 2:1:1$.
Therefore squark pair production will give no di-jet events, except for the invisible decays of $Z^0$, with $Z^0 \rightarrow \nu \bar{\nu}$.
The largest number of jets in fully hadronic events is given by the 8-jet topology (from $\sGlu \sGlu$ production), while the 
largest lepton multiplicity per event would be 4.

In this scenario, the left-handed squarks $\sQua_L$ and the right-handed squarks $\sQua_R$ give the same final state signatures.
Nevertheless, there will be kinematic differences between $\sQua_L$ events and $\sQua_R$ events, 
which can be explored in the endpoint analyses described below in order to 
prove that there are two different types of decay chains present in the data.

\subsubsection{Inverted Higgsino scenario with $|\mu|<M_2<M_1$}
\label{sec:higgsinoinv}

\begin{figure}[!htb]
 \begin{center}
 \includegraphics[width=0.4\textwidth]{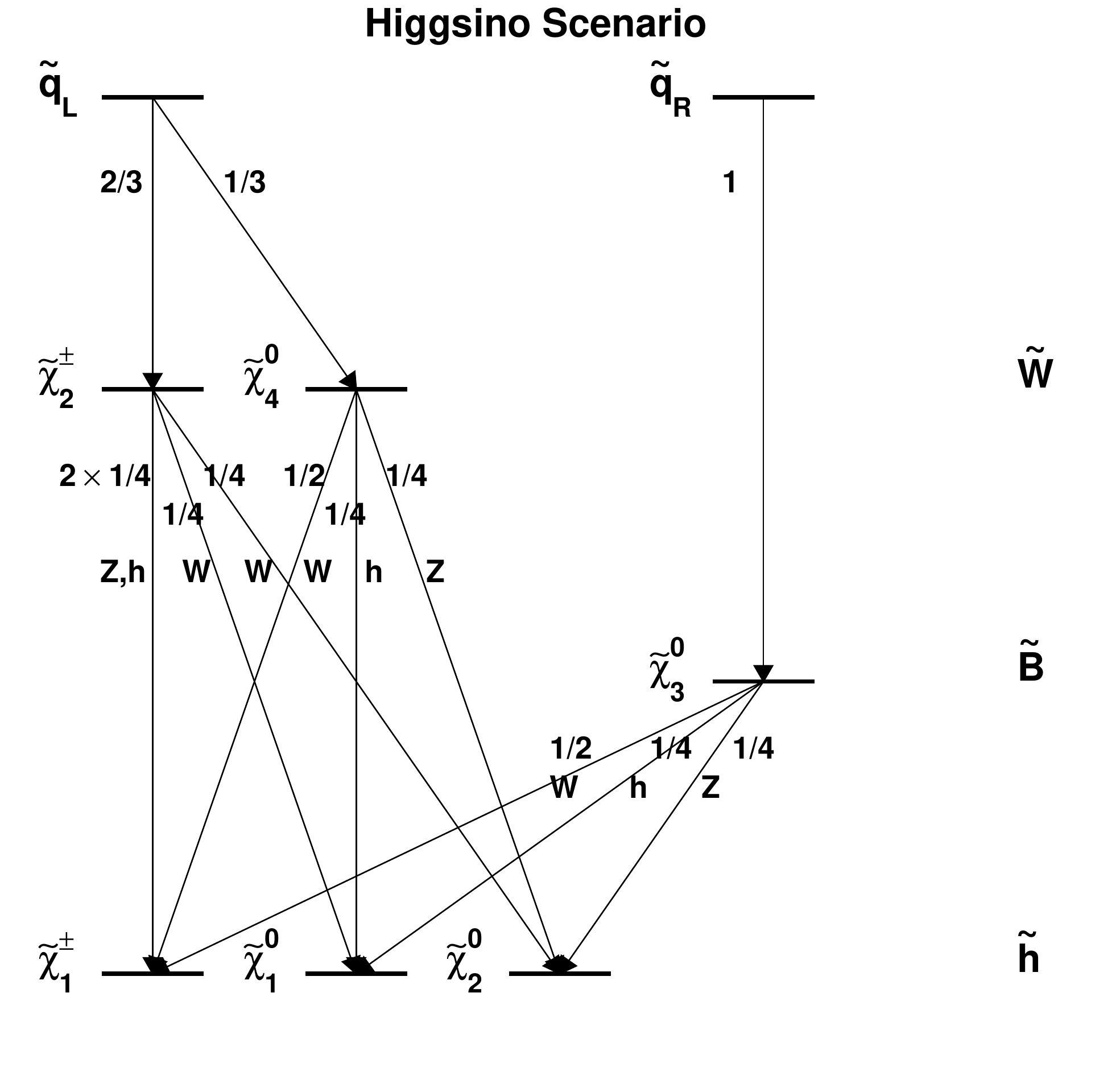}
 \hspace*{1.5cm}
 \includegraphics[width=0.4\textwidth]{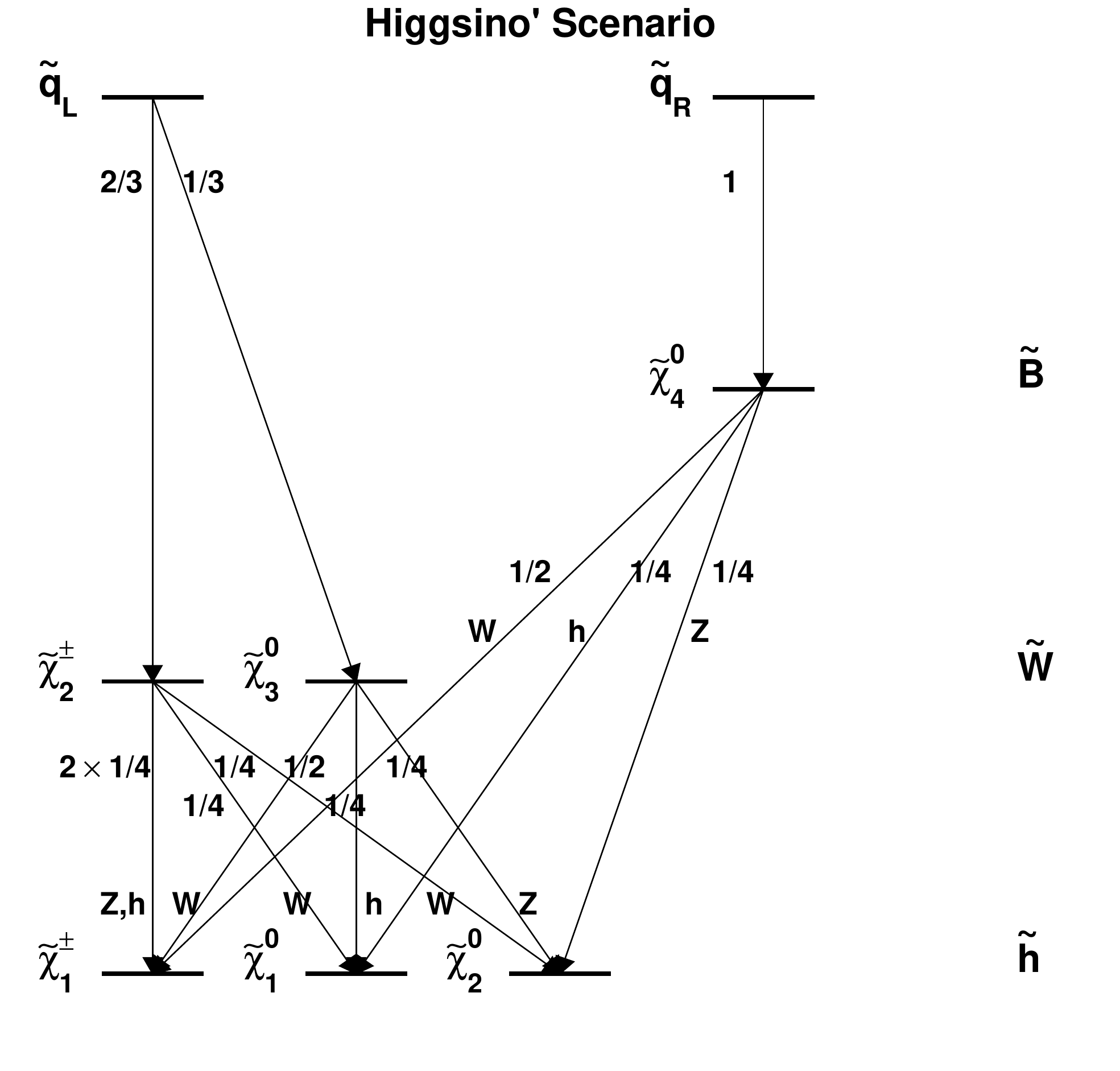}
\caption{Decays and branching fractions for the Higgsino scenario of Sec.~\ref{sec:higgsino}
(left), and the Inverted Higgsino scenario of Sec.~\ref{sec:higgsinoinv} (right).}
 \label{fig.distscenarios.Hino}
 \end{center}
\end{figure}

 The last scenario left to consider is the \underline{\em Inverted Higgsino scenario}, with $|\mu|<M_2<M_1$,
 shown in the right column of Fig.~\ref{fig.scenarios.spectra1}. Compared to the Higgsino scenario from the previous subsection,
 the roles of the two gauginos have been switched, and now $M_2<M_1$. The main decays (with their corresponding branching fractions) 
 for the two scenarios are compared in Fig.~\ref{fig.distscenarios.Hino}.
For completeness, in Table~\ref{tab:higgsinoBR} we list the branching ratios for the different final states, in analogy to 
Tables~\ref{tab:binowinoBR} and \ref{tab:mixedBR}.
\begin{table}[!htb]
\begin{center}
\resizebox{\columnwidth}{!}{%
\begin{tabular}{||c|cccc|cc||} 
\hline
\multicolumn{7}{||c||}{Higgsino scenario}\\
\hline
No &
\multicolumn{4}{|c|}{Decays } &
\multicolumn{2}{c||}{BFs}\\
\hline
&&  & & & & \\
& $\sQua_L$	& $\rightarrow q \chipm_2$	& $\rightarrow q Z \chipm_1$ 	& $\rightarrow q ZX\chiz_1 $  & $= \frac{2}{3} \times \frac{1}{4}$  & $= \frac{1}{6}$ \\
&&  & & & & \\ 
& 			&						& $\rightarrow q h \chipm_1$ 	& $\rightarrow q hX\chiz_1$  & $= \frac{2}{3} \times \frac{1}{4}$  & $= \frac{1}{6}$ \\
&&  & & & & \\
& 			&						& $\rightarrow q W \chiz_2$ 	& $\rightarrow q WX\chiz_1$  & $= \frac{2}{3} \times \frac{1}{4} $  & $= \frac{1}{6}$ \\
&&  & & & & \\
 $N_{\tilde q_L}$& 			&			&  $\rightarrow q W\chiz_1$  	& 						& $= \frac{2}{3} \times \frac{1}{4} $  & $= \frac{1}{6}$ \\
&&  & & & & \\
&			& $\rightarrow q \chiz_4$		& $\rightarrow q W \chipm_1$ 	& $\rightarrow q W X \chiz_1$  & $= \frac{1}{3} \times \frac{1}{2} $  & $= \frac{1}{6}$ \\
& &  & & & & \\
& 			&						& $\rightarrow q Z \chiz_2$ 	& $\rightarrow q Z X \chiz_1$  & $= \frac{1}{3} \times \frac{1}{4} $  & $= \frac{1}{12}$ \\
& &  & & & & \\
& 			&						& $\rightarrow q h \chiz_1$ 	& 						 & $= \frac{1}{3} \times \frac{1}{4}$  & $= \frac{1}{12}$ \\
& &  & & & & \\
 \hline
&&  & & & & \\
& $\sQua_R$	& $\rightarrow q \chiz_3$		& $\rightarrow q W \chipm_1$ 	& $\rightarrow q W X \chiz_1$  & $= 1 \times \frac{1}{2} $  & $= \frac{1}{2}$ \\
& &  & & & & \\
 $N_{\tilde q_R}$& 			&						& $\rightarrow q Z \chiz_2$ 	& $\rightarrow q Z X \chiz_1$  & $= 1 \times \frac{1}{4} $  & $= \frac{1}{4}$ \\
& &  & & & & \\
& 			&						& $\rightarrow q h \chiz_1$ 	& 						 & $= 1 \times \frac{1}{4}$  & $= \frac{1}{4}$ \\
& &  & & & & \\
\hline 
\end{tabular}
\begin{tabular}{||c|cccc|cc||} 
\hline
\multicolumn{7}{||c||}{Inverted Higgsino scenario} \\
\hline
No &
\multicolumn{4}{|c|}{Decays } &
\multicolumn{2}{c||}{BFs}\\
\hline
&&  & & & & \\
& $\sQua_L$	& $\rightarrow q \chipm_2$	& $\rightarrow q Z \chipm_1$ 	& $\rightarrow q ZX\chiz_1 $  & $= \frac{2}{3} \times \frac{1}{4}$  & $= \frac{1}{6}$ \\
&&  & & & & \\ 
& 			&						& $\rightarrow q h \chipm_1$ 	& $\rightarrow q hX\chiz_1$  & $= \frac{2}{3} \times \frac{1}{4}$  & $= \frac{1}{6}$ \\
&&  & & & & \\
& 			&						& $\rightarrow q W \chiz_2$ 	& $\rightarrow q WX\chiz_1$  & $= \frac{2}{3} \times \frac{1}{4} $  & $= \frac{1}{6}$ \\
&&  & & & & \\
 $N_{\tilde q_L}$& 			&			&  $\rightarrow q W\chiz_1$       & 					    & $= \frac{2}{3} \times \frac{1}{4} $  & $= \frac{1}{6}$ \\
&&  & & & & \\
&			& $\rightarrow q \chiz_3$		& $\rightarrow q W \chipm_1$ 	& $\rightarrow q W X \chiz_1$  & $= \frac{1}{3} \times \frac{1}{2} $  & $= \frac{1}{6}$ \\
& &  & & & & \\
& 			&						& $\rightarrow q Z \chiz_2$ 	& $\rightarrow q Z X \chiz_1$  & $= \frac{1}{3} \times \frac{1}{4} $  & $= \frac{1}{12}$ \\
& &  & & & & \\
& 			&						& $\rightarrow q h \chiz_1$      & 						& $= \frac{1}{3} \times \frac{1}{4}$  & $= \frac{1}{12}$ \\
& &  & & & & \\
\hline
&&  & & & & \\
& $\sQua_R$	& $\rightarrow q \chiz_4$		& $\rightarrow q W \chipm_1$ 	& $\rightarrow q W X \chiz_1$  & $= 1 \times \frac{1}{2} $  & $= \frac{1}{2}$ \\
& &  & & & & \\
 $N_{\tilde q_R}$& 			&						& $\rightarrow q Z \chiz_2$ 	& $\rightarrow q Z X \chiz_1$  & $= 1 \times \frac{1}{4} $  & $= \frac{1}{4}$ \\
& &  & & & & \\
& 			&						& $\rightarrow q h \chiz_1$ 	& 						 & $= 1 \times \frac{1}{4}$  & $= \frac{1}{4}$ \\
& &  & & & & \\
  \hline
\end{tabular}
}
\caption{The same as Tables~\ref{tab:binowinoBR} and \ref{tab:mixedBR}, but now comparing the Higgsino and Inverted Higgsino scenarios.
Here $X$ generically denotes the soft decay products resulting from transitions among the relatively degenerate Higgsino states,
$\chipm_1\to X\chiz_1$ and $\chiz_2\to X\chiz_1$.}
\label{tab:higgsinoBR}
\end{center}
\end{table}

Fig.~\ref{fig.distscenarios.Hino} and Table~\ref{tab:higgsinoBR}
reveal that the decay patterns of the two scenarios are identical. 
This time, the decays of both $\sQua_L$ and $\sQua_R$ result in the same final states, so 
it appears unlikely that the two scenarios can be distinguished, and furthermore, 
we would not know which state is $\sQua_L$ and which state is $\sQua_R$.

\subsection{Further variations on the six main electroweakino scenarios}
\label{sec:variations}

When introducing the six main electroweakino scenarios in the previous Section~\ref{sect:scenarios.first}, for simplicity we kept
all scalars (squarks, sleptons and non-SM-like Higgs bosons $H^\pm$, $H^0$ and $A^0$) heavy, so that they do not impact the 
electroweakino decays and branching fractions. By eliminating the role of the scalars, we were able to concentrate on the salient 
features of the electroweakino sector, without additional complications. Nevertheless, there are well motivated SUSY models in which
(some) of the scalars can be lighter and thus play a major role.

\subsubsection{Scenarios with some light sfermions}
\label{sect:scenarios.sfermions}

One such possibility, considered in Section~\ref{sect:leptslept}, is to take the sleptons (but not the squarks) to have masses 
of the same order as the electroweakinos themselves. This scenario is motivated by models in which the superpartner masses 
are generated through a mechanism utilizing the SM gauge interactions, e.g., RGE running or radiatively induced SUSY breaking
in the visible sector. Under those circumstances, it is plausible that states with similar quantum numbers have a similar mass scale,
in particular, that all the non-colored superpartners are similar in mass.

In Section~\ref{sect:leptslept} we consider a light slepton scenario, where for definiteness we introduce
the sleptons between the two lowest lying electroweakino states, so that the heavier one is kinematically allowed to decay 
to a charged slepton $\sLep$ or a sneutrino $\sNu$.
Such decays to sleptons are mediated by gauge interactions and are not suppressed
(the decay modes listed in Table \ref{tab.scenarios.BR} still exist but are suppressed and therefore subdominant).
This light slepton scenario will be considered only in relation to final states with leptons, which is the subject of Section~\ref{sect:leptslept}.

\subsubsection{Gluino decays with a light stop}
\label{sect:scenarios.stop}

In principle, one could also design a similar scenario where instead of sleptons, some squark states are introduced at a lower mass scale.
First, this will open up the phase space (and therefore increase the branching fractions) for decays of the gluino to such lighter squarks, 
and second, this will allow the heavier electroweakinos to decay to these squark states by emitting a jet (provided the decays are open).
For brevity, we shall not discuss in detail generic situations with light squarks, and instead only consider the case where the top squark (the stop)
is the only one which is significantly lighter than the others. There are several reasons why a top squark mass eigenstate can be expected to be lighter than the others \cite{Baer:1991cb,Demina:1999ty} --- 
the effects of tree-level $L-R$ mixing are largest in the stop sector, pushing the lighter mass eigenstate lower, 
RGE running from higher scales suppresses the squark masses proportionally to their Yukawa couplings, of which the top Yukawa is the largest, 
and in certain models the third generation squark masses could be lighter to begin with.

In our case specifically, the stop mass was reduced to $\sim 1576$ GeV, so that the gluino has a sizable branching fraction to $\sTop_1$
(see Table~\ref{tab.simul.BR} below). The first generation squark masses were unchanged. The results for the relevant branching fractions 
are shown in Table \ref{tab.scenarios.BRstop}.
\begin{table}[tb]
\begin{center}
\resizebox{\columnwidth}{!}{%
\begin{tabular}{|ccc|ccc|ccc|ccc|} \hline
                 & Bino     &   					&  &Wino	& 							     &  & Mixed	& 					& & Higgsino	&   \\
\multicolumn{3}{|c|}{$\chiz_1 < \chiz_2 = \chipm_1$}  &  \multicolumn{3}{|c|}{$\chiz_1 =  \chipm_1 < \chiz_2$} 	&  \multicolumn{3}{|c|}{$\chiz_1 < \chiz_2 =  \chiz_3 = \chipm_1$}
&  \multicolumn{3}{|c|}{$\chiz_1 = \chiz_2 = \chipm_1 <  \chiz_3$}	   \\
\hline
$\sTop_1$ & $\rightarrow b \chipm_2$  & 28\%		& $\sTop_1$ & $\rightarrow b \chipm_2$  & 25\% &$\sTop_1$ & $\rightarrow t \chiz_2$ & 30\% &$\sTop_1$ & $\rightarrow b \chipm_1$ & 32\%   \\
		& $\rightarrow t \chiz_4$  & 26\% 		& &  $\rightarrow t \chiz_4$  & 24\% 				& &   $\rightarrow b \chipm_1$  & 29\% 	& &   $\rightarrow t \chiz_1$  & 29\%    \\
		& $\rightarrow t \chiz_1$  & 14\% 		& &   $\rightarrow b \chipm_1$  & 21\% 			& &   $\rightarrow t \chiz_3$  & 24\% 		& &   $\rightarrow t \chiz_2$  & 28\%    \\
		& $\rightarrow t \chiz_3$  & 13\% 		& &   $\rightarrow t \chiz_3$  & 14\% 				& &   $\rightarrow t \chiz_1$  &8\%		 & &    $\rightarrow t \chiz_3$  & 5\%    \\
		& $\rightarrow b \chipm_1$  & 12\% 		& &   $\rightarrow t \chiz_1$  & 10\% 				& &   $\rightarrow b \chipm_2$  &6\% 	& &    &     \\
		& $\rightarrow t \chiz_2$  & 6\% 		& &  $\rightarrow t \chiz_2$  & 7\% 				& &  $\rightarrow t \chiz_4$  &3\% 		& &    &     \\
$\chipm_2$ & $\rightarrow Z \chipm_1$  & 30\% 	&  $\chipm_2$ & $\rightarrow h \chipm_1$  & 31\%	& $\chipm_2$ & $\rightarrow Z \chipm_1$  & 25\%	 & & &     \\
		& $\rightarrow h \chipm_1$  & 29\% 		& & $\rightarrow Z \chipm_1$  & 31\% 			& &  $\rightarrow W \chiz_2$  & 25\% 	&  &  &   \\
		& $\rightarrow W \chiz_2$  & 30\% 		& & $\rightarrow W \chiz_1$  & 30\% 			& &  $\rightarrow h \chipm_1$  & 25\% 	&  &  &   \\
		& $\rightarrow W \chiz_1$  & 11\% 		& & $\rightarrow W \chiz_2$  & 8\%				& & $\rightarrow W \chiz_3$  & 24\%		 &  &  &   \\
$\chiz_4$ & $\rightarrow W \chipm_1$  & 59\% 	& $\chiz_4$ & $\rightarrow W \chipm_1$  & 60\% 	&   $\chiz_4$ & $\rightarrow W \chipm_1$  & 50\%&  & &  \\
		 & $\rightarrow h \chiz_2$  & 28\% 		& & $\rightarrow h \chiz_1$  & 27\% 				& &  $\rightarrow h \chiz_2$  & 25\%		&  & &  \\
		& $\rightarrow h \chiz_1$  & 11\%   		& & $\rightarrow h \chiz_2$  & 8\%				 & &  $\rightarrow Z \chiz_3$  & 25\%	 & &  &   \\
$\chiz_3$	& $\rightarrow W \chipm_1$  & 58\% 	& $\chiz_3$ & $\rightarrow W \chipm_1$  & 61\%   	& $\chiz_3$ & $\rightarrow Z \chiz_1$ & 93\% & $\chiz_3$ & $\rightarrow W \chipm_1$  & 52\%   \\
		& $\rightarrow Z \chiz_2$  & 29\% 		& & $\rightarrow Z \chiz_1$  & 27\%  			&   	 & $\rightarrow h \chiz_1$  & 7\%  	 & & $\rightarrow Z \chiz_2$  & 23\%     \\
		& $\rightarrow Z \chiz_1$  & 11\% 		& & $\rightarrow Z \chiz_2$  & 8\%				& &   & 						   	&  & $\rightarrow h \chiz_1$  & 22\%	   \\
$\chipm_1$& $\rightarrow W \chiz_1$  & 100\%  	& &   &   									& $\chipm_1$ & $\rightarrow W \chiz_1$ & 100\% & &  &  \\
$\chiz_2$ & $\rightarrow h \chiz_1$  & 94\% 		& $\chiz_2$ & $\rightarrow W \chipm_1$  & 65\%   	&$\chiz_2$ & $\rightarrow h \chiz_1$ & 92\% & & &    \\
		& $\rightarrow Z \chiz_1$  & 6\% 		& &   $\rightarrow h \chiz_1$  & 33\% 			& &   $\rightarrow Z \chiz_1$ & 8\% 		& &  &   \\
 \hline
\end{tabular}%
}
\caption{Relevant branching fractions in the four main scenarios in the presence of a light stop.}
\label{tab.scenarios.BRstop}
\end{center}
\end{table}
Some important differences appear compared to Table \ref{tab.scenarios.BR}.
First, due to the large Yukawa coupling of the stop, many decays of  $\sTop_1$ are predominantly to  Higgsino states in all four scenarios.
Second, as the stop is strongly mixed, we observe decays typical of left as well as of right states for $\sTop_1$.
Except for the higgsino-like scenario, the heavier electroweakinos also participate in the decays and could be experimentally observed.
All decays of  $\sTop_1$ give rise to a b-quark, hence the b-tagging or b-veto allow the separation between the first two generation squarks and the stop.
The number of $W / Z / h$ per decay chain, including the $W$ of the top decay, is always larger for the direct production of $\sTop_1$ than for the first two generation squarks.
In the Bino scenario, the number of $W / Z / h$ per decay chain can be 1, 2 or even 3 (giving no di-jet events).
In the Wino scenario, the number of $W / Z / h$ per decay chain can be 0 (giving di-jet events), 1 or 2 or 3.
In the Mixed scenario, the number of $W / Z / h$ per decay chain can be 1, 2 or 3  (giving no di-jet events).
In the Higgsino scenario, the number of $W / Z / h$ per decay chain can be 0 (giving di-jet events), 1 or 2.
If  $\sTop_1$ comes from the decay of a $\sGlu$, it is accompanied by a top-quark, leading to an additional $W$ and b-quark.
The largest jet multiplicity in fully hadronic events from gluino pair production would amount to 20 jets (including the hadronic top decays)
in the Bino-like scenario, very crowded events indeed.
The largest lepton multiplicity per event would now be 12, assuming that both top quarks decay leptonically.

\subsubsection{Gluino lighter than the squarks}
\label{sect:scenarios.lightg}

The cases considered so far already represent a large variety of scenarios.
There exists, however, yet another potentially interesting one.
Previously, it was always assumed that the gluino $\sGlu$ was heavier than the squarks $\sQua$,
but this is by no means guaranteed.
If the $\sGlu$ is substantially lighter than the squarks, its pair production may dominate all other production processes, as seen in Fig.~\ref{fig.xsec.vsmass}.
The $\sGlu$ will decay via an off-shell $\sQua$ (plus a $q$), followed by decay modes of the $\sQua$ similar to the ones presented above.
Although heavier than the gluino, the squarks $\sQua$ may still be produced directly, then they will decay to an on-shell $\sGlu$ (plus a $q$),
and their direct decays to electroweakinos will be strongly suppressed. This scenario will generally lead to higher jet multiplicities --- for example, di-jet events, 
originating from $\sQua$ pair production followed by the decay $\sQua \rightarrow q \chiz_1$, would be absent.
In what follows, we shall not consider the light gluino scenario any further, and instead focus on the cases discussed in the previous two subsections
\ref{sect:scenarios.sfermions} and \ref{sect:scenarios.stop}.

\section{Fully Hadronic Final States}
\label{sect:had}

\subsection{Mass determinations from kinematic endpoints}
\label{sect:massdet}

As already mentioned, the masses of the sparticles appearing in a cascade decay chain 
can in principle be determined by kinematic methods by measuring the kinematic endpoints of suitable distributions.
If we can isolate a single decay chain, the classic method for extracting sparticle masses 
makes use of the invariant masses of combinations of visible particles in the final state \cite{Baer:1994nr,Hinchliffe:1996iu,Paige:1997xb}.
We shall not review the method here and instead refer the interested reader 
to the literature \cite{Barr:2010zj} (a collection of relevant formulas can be found in Appendix \ref{sect:invmass}). 
On the other hand, SUSY models with conserved $R$-parity are characterized by the {\em pair production}
of sparticles, whereby each event contains {\em two} cascade decay chains. Under those circumstances, 
it is often beneficial to use kinematic variables which utilize both decay chains simultaneously.
Among the well known such examples are the Cambridge $M_{T2}$ variable \cite{MT2variable} or 
the $M_2$ variables \cite{Mahbubani:2012kx,Cho:2014naa}. By measuring enough kinematic endpoints of these variables, one can again 
extract the underlying SUSY mass spectrum (the necessary formulas are collected in Appendix \ref{sect:MT2}).
With both methods, one tries to relate the endpoints of the kinematic distributions to the masses of the sparticles involved
and then solve for the masses.

In order to apply these reconstruction methods, one generally has to face a combinatorial problem 
in trying to identify which particles belong to one and which to the other decay chain.
Multi-jet events can be subdivided in two "hemispheres" by using, for example, the technique presented in \cite{PTDR2}.
It consists in collecting all particles in two groups, called the hemispheres, such that the sum of the invariant masses squared of the two hemispheres is minimal.
Ideally, each hemisphere is supposed to contain the decay products of one of the two originally produced parent SUSY particles.
While the hemisphere algorithm was originally developed for multi-jet events, it can also be applied to generic events 
containing some collection of jets, leptons, photons, etc.

\subsection{Simulation test of fully hadronic events}
\label{sect:simul}

As so far almost no simulation has been published for the reconstruction of masses in purely hadronic final states,
it is not really known how well the mass reconstruction will behave in such topologies.
Here, a crude simulation was performed to establish a proof of principle for the approach outlined above.
We could have chosen some of the Simplified Models already available in CMS or ATLAS and used to set limits.
However, the downside of the simplified model approach is that it focuses entirely on a single production channel with branching fractions assumed to be 100\%.
As a result, one is missing combinatorial backgrounds coming from other signal processes.
This is why here we preferred to generate a ``soup" of events from several production channels and with more realistic branching fractions.

For $\sGlu$ and $\sQua$ masses, the CMS test point HL2 \cite{PTDR2} was chosen with $m(\sGlu) = 1785$ and $m(\sQua) = 1656$ GeV.
These masses are still just beyond the reach obtained with the current data set.
For these masses, the NLO production cross sections computed with Prospino2 \cite{Beenakker:1996ed} and summed over 4 squark flavours 
are listed in Table \ref{tab.simul.xsec}.
\begin{table}[t]
\begin{center}
\begin{tabular}{|c|cccc|c|} \hline
					&				&				&					&				&	\\
					& $\sGlu \sGlu$		& $\sGlu \sQua$	& $\sQua \bar{\sQua}$	& $\sQua \sQua$	& Total	 \\
\hline
 cross section (fb)		& 4.9				& 10				& 1.6					& 10				& 26.5	\\
 Events in 100 ${\rm fb}^{-1}$	& 490			& 1000			& 160				& 1000			& 2650	\\
 Events in 300 ${\rm fb}^{-1}$	& 1470			& 3000			& 480				& 3000			& 7950	\\
 \hline
\end{tabular}
\caption{Cross sections for the various production modes at a 13 TeV LHC: $\sGlu \sGlu$, $\sGlu \sQua$, $\sQua \bar{\sQua}$ and $\sQua \sQua$, respectively.
The $\sGlu$ ($\sQua$) mass is 1785 GeV (1656 GeV).
}
\label{tab.simul.xsec}
\end{center}
\end{table}
Numbers of events produced with 100 ${\rm fb}^{-1}$, which might be reached around the end of 2017,
and 300 ${\rm fb}^{-1}$, which is expected before the long shutdown at the end of 2023, are also given.
Direct production of $\sTop \sTop$ is neglected, as its cross section is usually much lower.

A complete mass spectrum and the corresponding decay branching fractions were computed with {\sc Susyhit} 
for the ``Mixed" scenario discussed in Section~\ref{sec:mixed}.
The light Higgs mass was forced to be 125 GeV by adjusting the stop parameter $A_t$ to 2750 GeV,
which also fixes the mass splitting between the two physical stop states ---
the soft stop mass parameters were taken to be $m(\sTop_L) = m(\sTop_R) = 1656$ GeV,
the same as for the other squark species,
leading to $m(\sTop_1) = 1576$ and $m(\sTop_2) = 1796$ GeV.
As a consequence, the gluino $\sGlu$ can decay to $\sTop_1$ but not to $\sTop_2$.
\begin{table}[htb]
\begin{center}
\begin{tabular}{|c|c|c|c|cc|cc|} \hline
$\sGlu \rightarrow \sQua q$ 	& $\sGlu \rightarrow \sBot_1 b$& $\sGlu \rightarrow \sBot_2 b$ &	$\sGlu \rightarrow \sTop_1 t$ 		 \\
\hline
55\%						&  7\%					& 7\%					& 31\%			\\
 \hline
\end{tabular}
\caption{Branching fractions for the gluino decays.
}
\label{tab.simul.BR}
\end{center}
\end{table}
Table~\ref{tab.simul.BR} lists the actual gluino branching fractions.
We observe that gluinos decay 31\% to $(t \sTop_1)$ and 69\% to the other 5 squark flavours.
Decays of $\sGlu$ to $\sBot_1$, with mass of 1714 GeV, and $\sBot_2$, with mass of 1724 GeV, are also included.
The slepton masses were kept at 2 TeV to avoid any influence on the electroweakino decays,
as we are interested here in the purely hadronic decay modes.
Following the discussion in Section~\ref{sect:scenarios.first}, the electroweakino mass parameters were 
taken at the values $M_1 = 400$, $M_2 = 1200$ and $\mu = 800$ GeV.
For definiteness, the value of $\tan \beta$ was chosen to be 10.
The resulting mass spectrum is shown in Table \ref{tab.simul.masses}.
\begin{table}[htb]
\begin{center}
\begin{tabular}{|ccccccccccc|} \hline
$\sGlu$	& $\sQua$	& $\sBot_2$	& $\sBot_1$	& $\sTop_1$	& $\chiz_4$	& $\chiz_3$	& $\chiz_2$	& $\chiz_1$	& $\chipm_2$	& $ \chipm_1$		 \\
 1785	& 1656		& 1724		& 1714		& 1576		& 1230		& 809		& 802		& 396		& 1230		& 800			\\
 \hline
\end{tabular}
\caption{Masses in GeV for the sparticles in the Mixed scenario.}
\label{tab.simul.masses}
\end{center}
\end{table}

Once the masses and stop mixing were fixed, they were input to the {\sc Susyhit} program \cite{Djouadi:2006bz} which computes the branching fractions for all SUSY particles.
The relevant results were already displayed in Tables \ref{tab.scenarios.BR} and \ref{tab.scenarios.BRstop}.

Events were generated at leading order according to sequential two-body phase space decays ignoring spin correlations 
and were subsequently analyzed {\em at parton level}, but including the detector resolution effects. In particular,
parton momenta were smeared according to the momentum resolution of the CMS Particle Flow algorithm \cite{CMS-PRF-14-001}.
Experimental cuts were applied by requesting jets with transverse momentum $p_T \geq 30$ GeV and pseudorapidity $|\eta| \leq 2.4$.
Events with leptons ($e$ or $\mu$) from boson decays satisfying $p_T \geq 10$ GeV and $|\eta| \leq 2.4$ were rejected.
As an example of the resulting event statistics for $300\ {\rm fb}^{-1}$, of the 7950 events generated, 5998 events remained after these experimental selection cuts.

The results summarized below are based on detailed analyses of the resulting signatures in the hadronic \cite{Hadr}
and leptonic channels \cite{Lept}. 
The summary presented here is meant as a roadmap illustrating the basic analysis techniques.
In the current section, we discuss the purely hadronic final states, demonstrating the feasibility of the mass reconstruction. 
In the next two sections, Section~\ref{sect:leptslept} and \ref{sect:lept}, we shall discuss the corresponding leptonic channels, both in the presence and in the absence of light sleptons in the spectrum.

The remainder of this section is organized by jet multiplicity. For simplicity, we concentrate on ``symmetric" events, where 
the hemisphere algorithm produces equal number of jets in each hemisphere. Therefore, when referring to $n$-jet events below, 
it should be understood that they correspond to symmetric $(n/2, n/2)$ topologies with $n/2$ jets in each hemisphere.
In particular, we shall not discuss $n$-jet events with $n$ being odd.

\subsection{Setting the overall mass scale: the $M_1 = \sqrt{\hat{s}}_{min}$ variable}

Having accumulated a sizable sample of signal events, the first order of business is to estimate the mass scale of the newly produced particles.
For this purpose, several variables have been proposed in the literature, 
most recently the variable $M_1 \equiv \sqrt{\hat{s}}_{min}$ defined as \cite{Konar:2008ei}
\begin{eqnarray}
 M_1 \equiv \sqrt{\hat{s}}_{min} = \sqrt{E^2 - p_z^2} + \sqrt{\Mm^2 + \PTm^2},
 \label{eq:smin}
\end{eqnarray}
where $E$ is the energy and $p_z$ is the momentum z-component of the visible system, while
$\PTm$ is (the magnitude of) the missing transverse momentum and $\Mm$ is the total mass of the invisible particles in the final state.
Note that the variable (\ref{eq:smin}) depends on a single unknown parameter, $\Mm$, and can only be evaluated once we make an ansatz for the value of $\Mm$.
It was shown \cite{Konar:2008ei} that, with the correct ansatz, the $M_1$ distribution should, at least in principle, 
peak at around the mass threshold for the pair production of the parent sparticles,
thus leading to an estimate of the SUSY mass scale. However, the value of $\Mm$ is {\em a priori} unknown,  
so the measurement will have to be repeated for different values of $\Mm$, leading to a {\em relationship} between the 
SUSY mass scale and the invisible LSP mass. Since the latter cannot be less than zero, 
the peak in the $M_1$ distribution provides a useful {\em lower bound} on the SUSY mass scale.

\begin{figure}[htb]
 \begin{center}
 \includegraphics[width=0.49\textwidth]{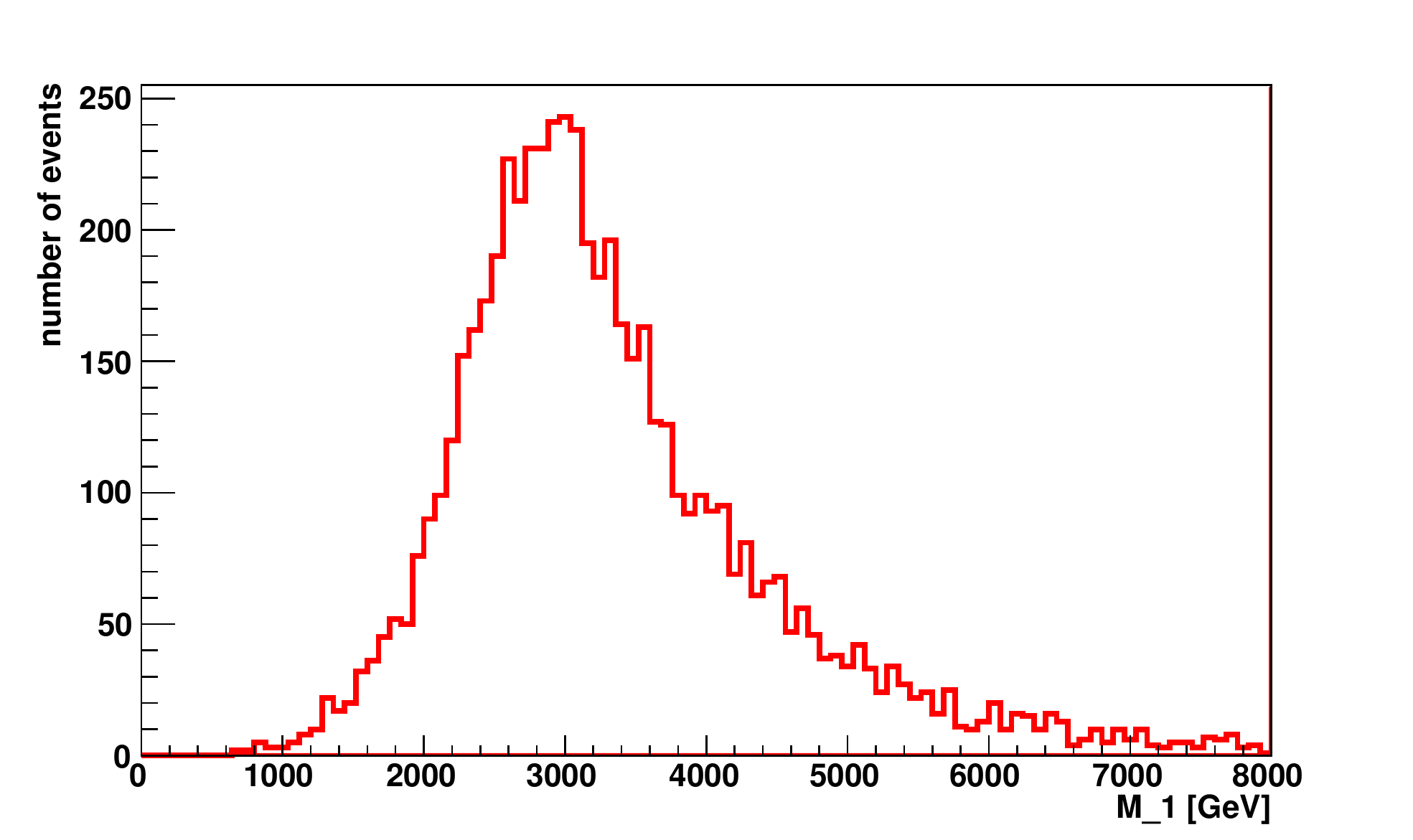}
 \includegraphics[width=0.49\textwidth]{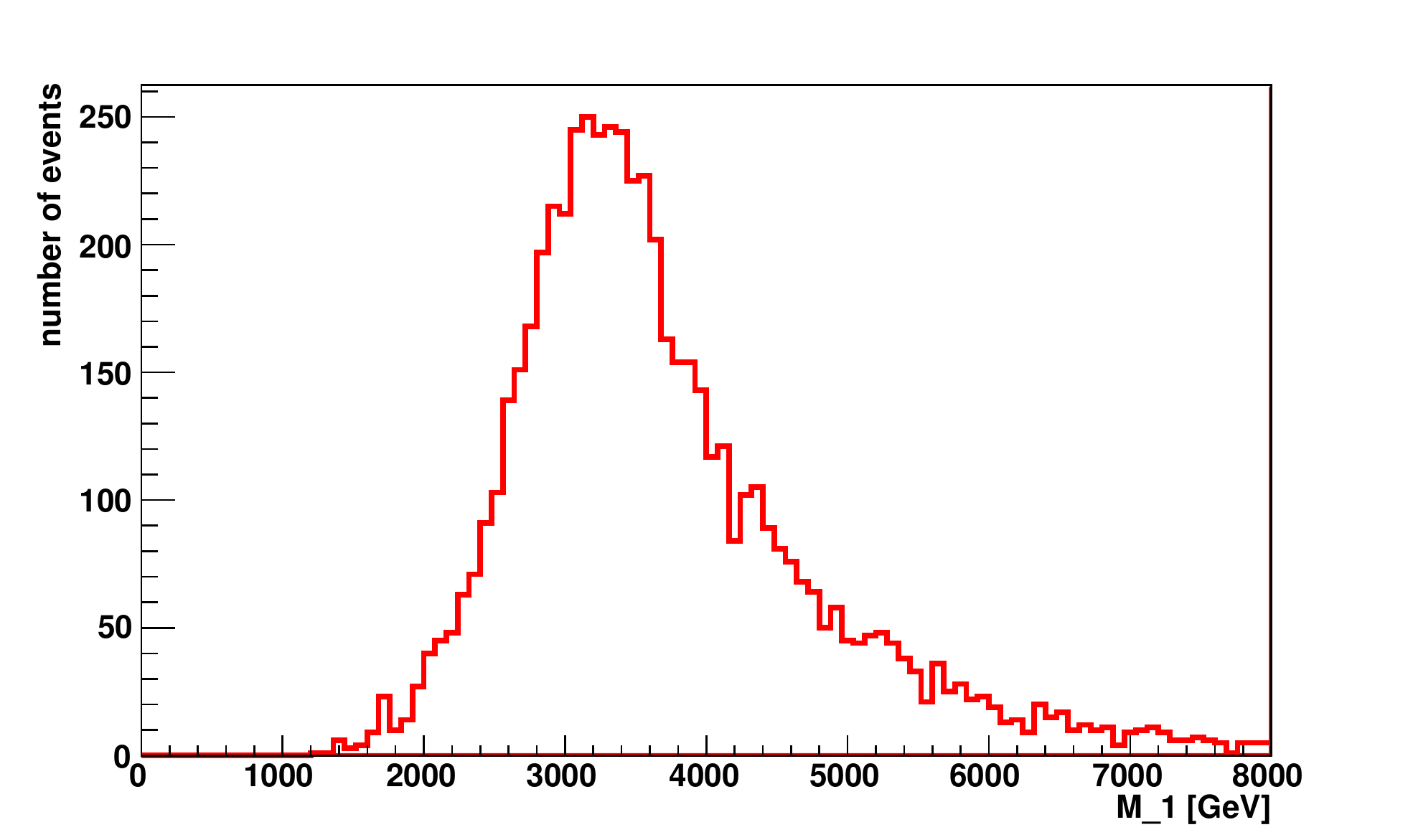}
 \caption{Distribution of $M_1$ for zero test mass, $\Mm=0$, (left) and for the correct test mass of $\Mm=2m_{\tilde\chi^0_1}$ (right).}
 \label{fig.simul.stats.M1}
 \end{center}
\end{figure}

The above procedure is illustrated in Fig.~\ref{fig.simul.stats.M1}, which shows 
distributions of $M_1 = \sqrt{\hat{s}}_{min}$ for a zero test mass, $\Mm=0$, (left panel) and for the correct 
value of $\Mm=2 m_{\tilde\chi^0_1}$ (right panel). In either case, the $M_1$ peak is clearly visible. 
For zero test mass (left panel), the peak is observed near $3000$ GeV, providing a 
useful lower bound on the colored sparticle mass scale of around $1500$ GeV. 
With the correct test mass of $\Mm=2m_{\tilde\chi^0_1}$ (right panel), the peak position shifts to about $3200-3400$ GeV, 
indicating a mass scale of order $1600-1700$ GeV. This is reasonable, given our input values of the 
squark masses ($1656$ GeV) and gluino mass ($1785$ GeV).
One should keep in mind that these are only leading order results, and at next-to-leading order
the distribution of $\sqrt{\hat{s}}_{min}$ will be affected by initial state radiation (ISR), contributions from underlying events, pile-up etc.
\cite{Papaefstathiou:2009hp,Papaefstathiou:2010ru,Konar:2010ma,Robens:2011zm}.

\subsection{Hadronic $W/Z/h$ selection}
\label{sec:hadronicselection}

In Section~\ref{sec:mixed} we observed that in the Mixed scenario a large number of decay chains involve 
not one, but {\em two} SM bosons ($W$, $Z$ or $h^0$), see Fig.~\ref{fig.distscenarios.Mixed} and Table~\ref{tab:mixedBR}.
Since the hadronic branching fractions of all three SM particles, $W$, $Z$ and $h^0$, are rather large, this motivates
an analysis which targets the $W/Z/h$ systems inclusively through their hadronic decay modes. 
To the best of our knowledge, such analysis has not been performed in the literature on SUSY LHC phenomenology.

\begin{figure}[!htb]
 \begin{center}
 \includegraphics[width=0.49\textwidth]{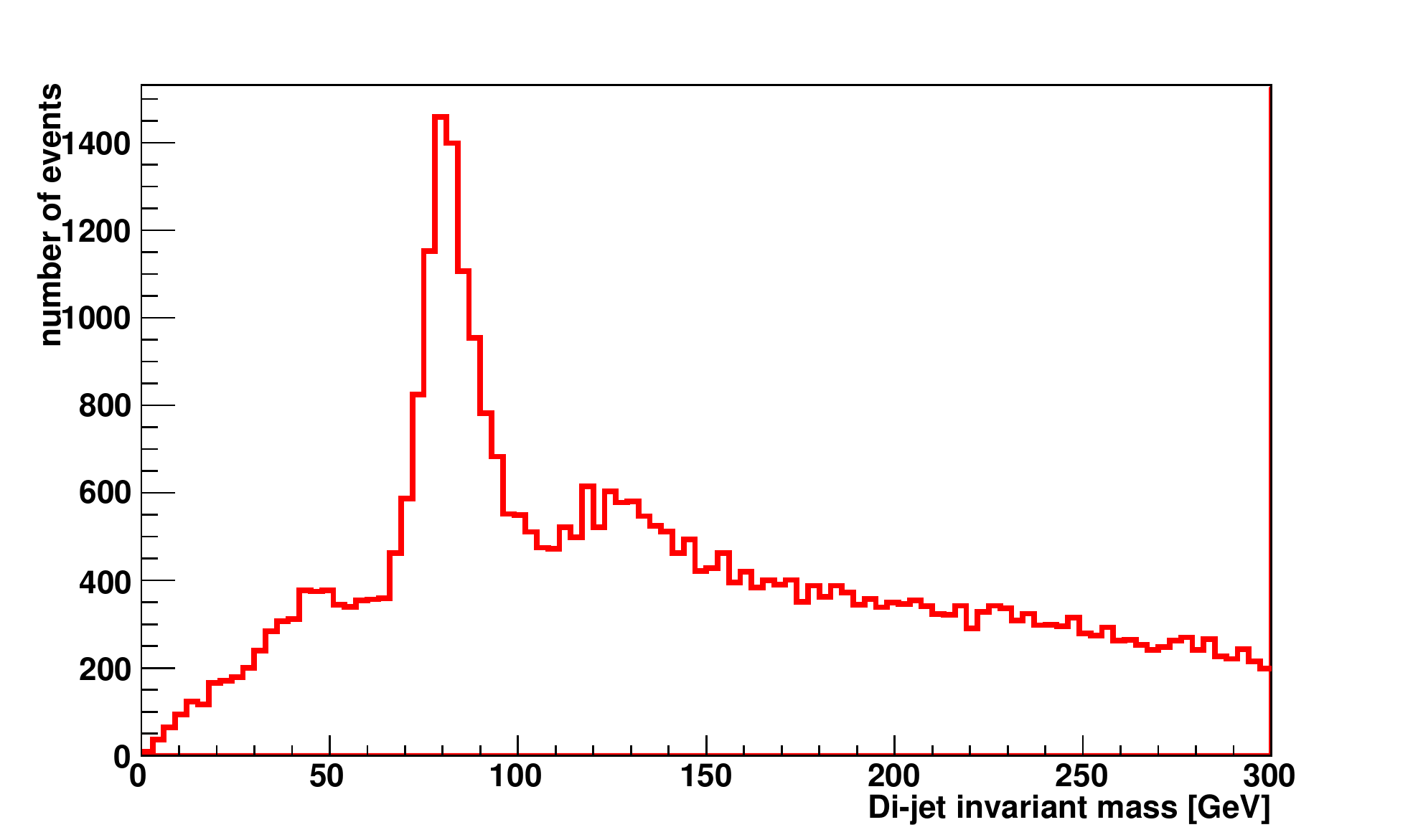}
 \includegraphics[width=0.49\textwidth]{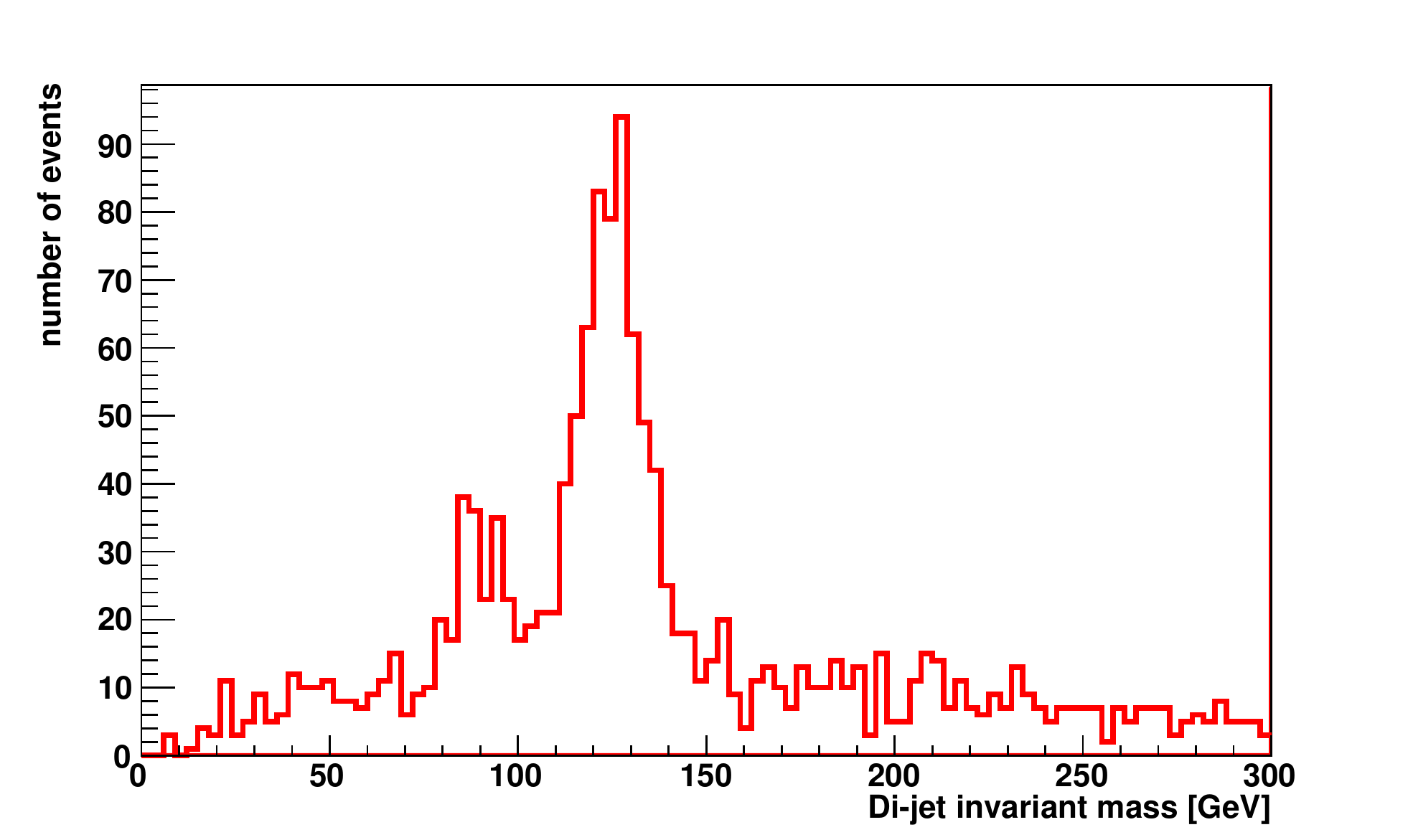}
\caption{Di-jet invariant mass for jet combinations without a $b$-jet (left panel) and with at least one $b$-jet (right panel).}
 \label{fig.simul.WZh.massbnob}
 \end{center}
\end{figure}

As a first step, we study the di-jet invariant mass distributions for jets within the same hemisphere.
In the left panel of Fig.~\ref{fig.simul.WZh.massbnob} we take pairs of jets, neither of which carries a $b$-tag, while in the right
panel of Fig.~\ref{fig.simul.WZh.massbnob} we demand that at least one of the jets in the pair is $b$-tagged.
With this selection, we would expect to see a well-identified $W/Z$ peak in the case of no $b$-tags,
and a $h^0$ peak for the $b$-tagged case. This expectation is confirmed by Fig.~\ref{fig.simul.WZh.massbnob}:
in the left panel we see a clean $W/Z$ peak, plus a very slight excess near the Higgs mass, due to the imperfect $b$-tagging efficiency.
Conversely, in the right panel we observe a well-defined peak at the Higgs mass, accompanied by a much smaller peak 
near the $Z$ mass due to $Z\to b\bar{b}$ decays.
The results shown in Fig.~\ref{fig.simul.WZh.massbnob} in principle allow to estimate the number of 
$W$-bosons, $Z$-bosons and Higgs bosons in the total distribution 
and thus test the relative branching fractions theoretically predicted for the Mixed scenario.

Once the hadronic $W/Z/h$ peaks are identified, we can proceed with an exclusive selection of 
hadronic SM bosons. Note that the plots in Fig.~\ref{fig.simul.WZh.massbnob} are inclusive, 
i.e., for a given event, there could be more than one combination within the same hemisphere which
falls within the area of the peak, leading to a combinatorial ambiguity. In order to resolve this, 
in what follows we shall always choose the di-jet solution with invariant mass closest 
to the nominal $W$, $Z$ or $h^0$ mass.

\subsection{Kinematic variables with hadronic $W/Z/h$ candidates}

As discussed in Section~\ref{sec:mixed}, the Mixed scenario is characterized by the presence of rather long decay chains 
initiated by left-handed light flavor squarks --- in that case, we get two SM bosons per decay chain, see Fig.~\ref{fig.distscenarios.Mixed}. 
On the other hand, decay chains initiated by right-handed light flavor squarks are relatively short, and give no SM gauge bosons.
As a result, we expect to have three types of events\footnote{This picture tends to be
complicated by the decays through stops, since they bring about additional $W$'s from top decays.}: 
a) with 4 SM gauge bosons, 2 in each hemisphere; 
b) with 2 SM gauge bosons, both of them within the same hemisphere; c) with no SM gauge bosons.
Therefore, the kinematic analysis will depend on whether the identified hadronic $W/Z/h$ candidates
are within the same hemisphere or not. We shall consider each case in turn.

\subsubsection{$W/Z/h$ candidates in the same hemisphere}

We first focus on events with two distinct $W/Z/h$ candidates in the same hemisphere. 
The observation of a statistically significant number of such events is a clear indication of the 
{\em presence of heavier charginos or neutralinos} in the decay chain, see Fig.~\ref{fig.distscenarios.Mixed}.
In the simulation, we found 2270 hemispheres with two $W/Z/h$ candidates in $300 \; {\rm fb}^{-1}$.
The large number of such hemispheres confirms that two-step decays are involved
and therefore that there are higher chargino/neutralino states appearing in the decay chains.

\begin{figure}[htb]
 \begin{center}
 \includegraphics[width=0.49\textwidth]{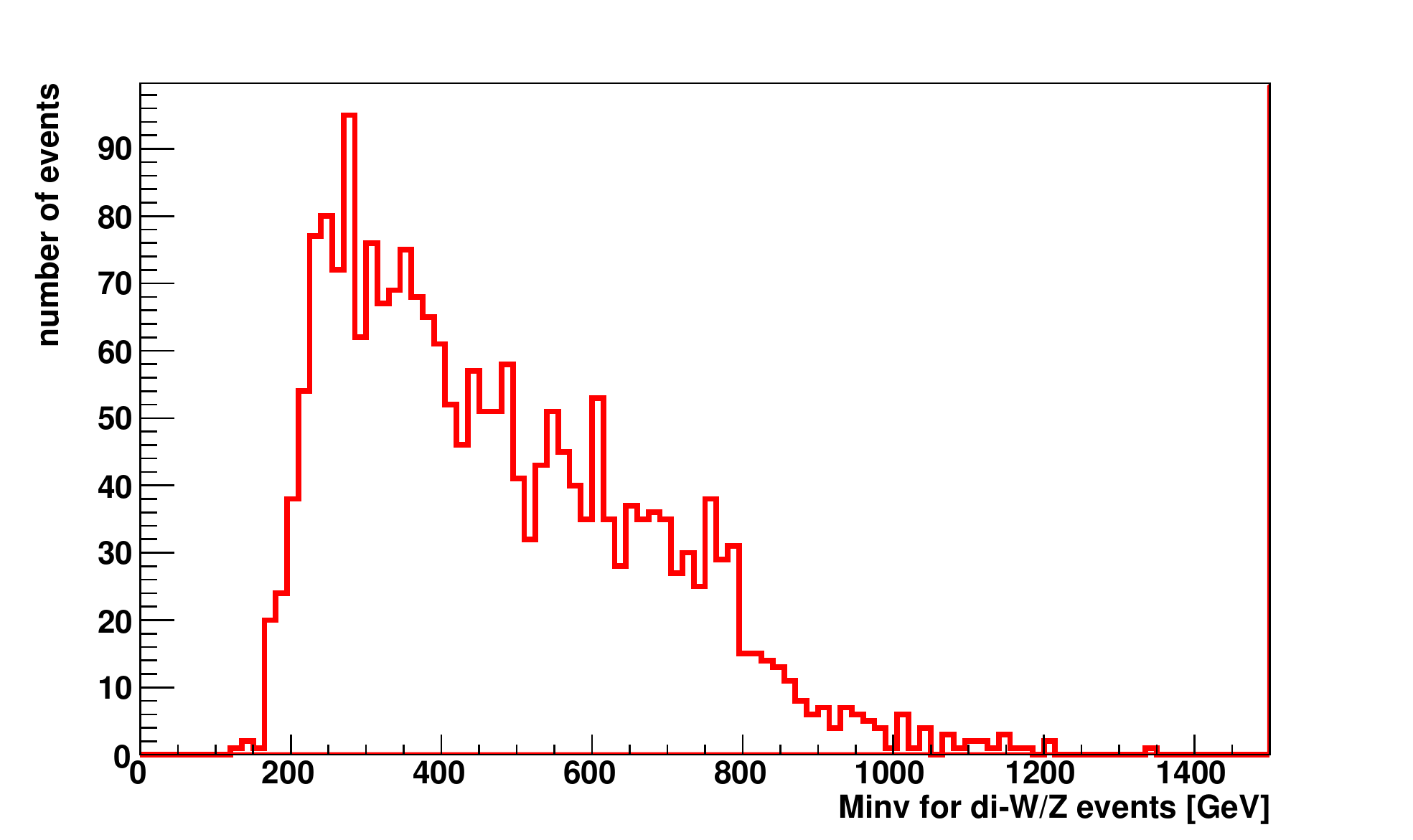}
 \includegraphics[width=0.49\textwidth]{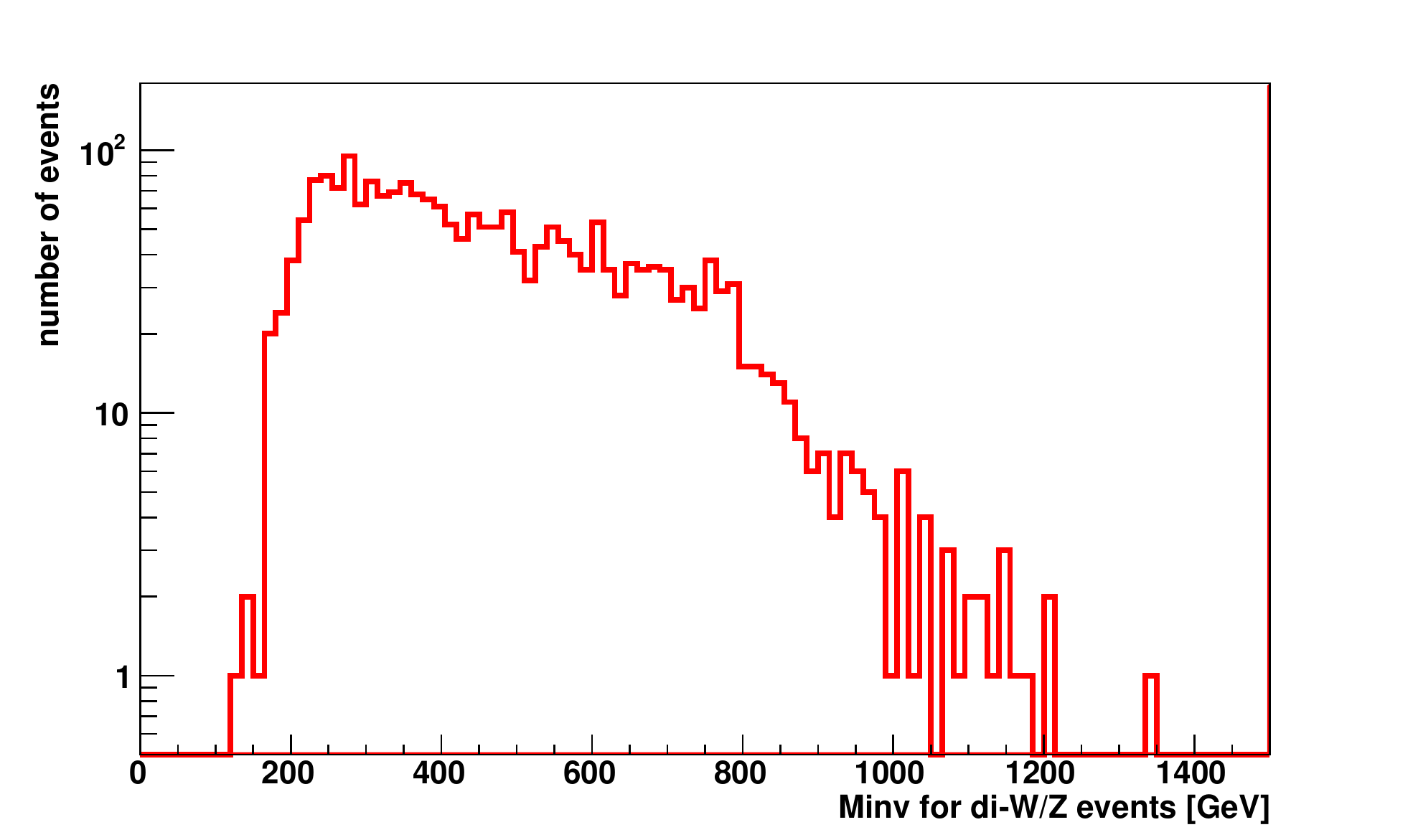}
\caption{Invariant mass distribution of all di-$W / Z / h$ candidate combinations per hemisphere:
in linear scale (left panel), or in semi-log scale (right panel).
}
 \label{fig.simulmultijetWZh.Minvin2WZbest}
 \end{center}
\end{figure}

Generalizing the classic invariant mass endpoint analysis, we consider the invariant mass 
of all pairs of $W / Z / h$ candidates in the same hemisphere and show the result in Fig.~\ref{fig.simulmultijetWZh.Minvin2WZbest}.
The distribution exhibits an endpoint around 800 GeV, which is 
in agreement with the expected value of 813 GeV for the decay $\chipm_2 \rightarrow \chipm_1 \rightarrow \chiz_1$.
This measured endpoint supplies one relation between the masses of $\chipm_2$, $\chipm_1$ and $\chiz_1$.

\begin{figure}[htb]
 \begin{center}
 \includegraphics[width=0.49\textwidth]{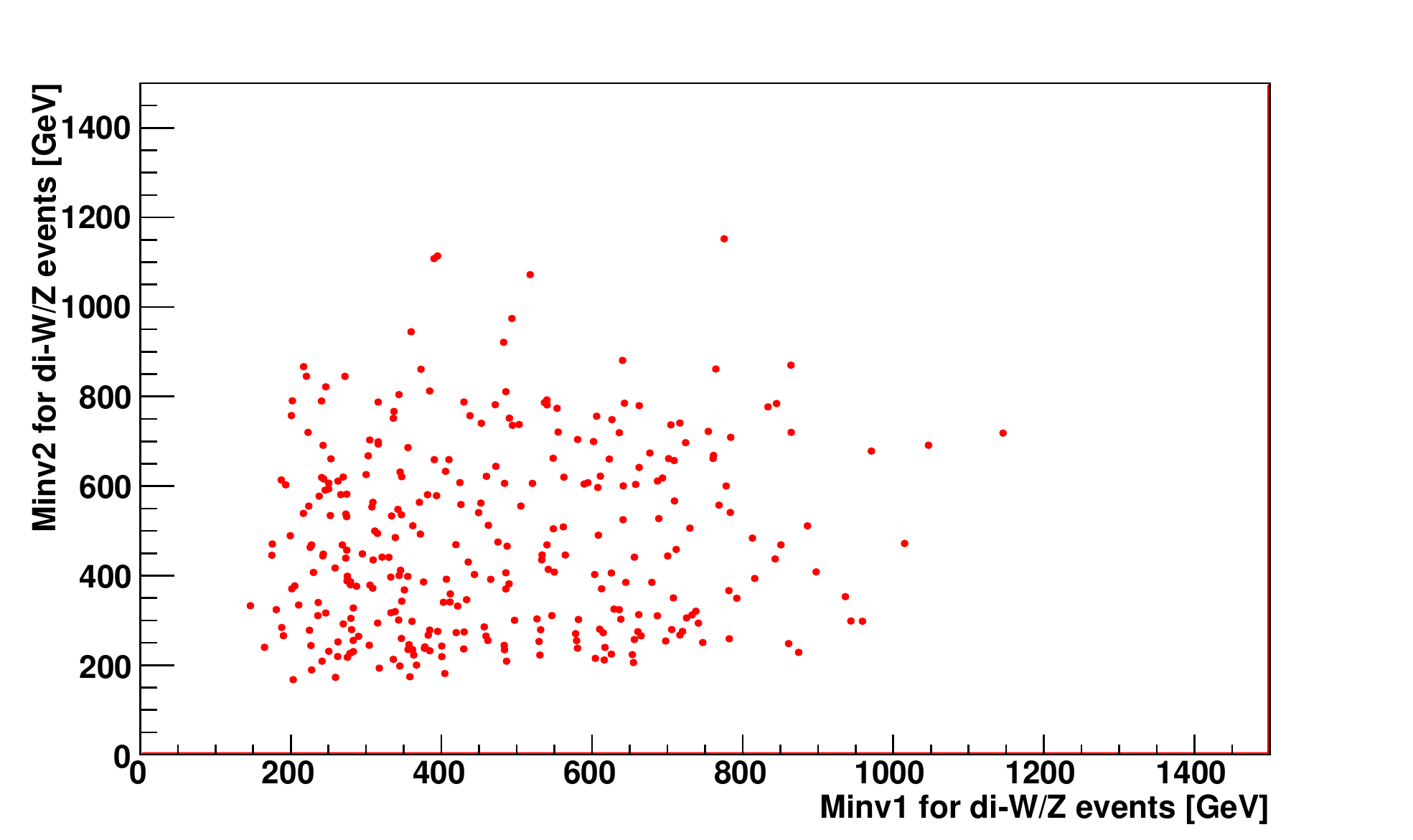}
\caption{Dalitz-like plot of the invariant mass distribution of all di-$W / Z / h$ candidate combinations per hemisphere.
}
 \label{fig.simulmultijetWZh.Minvin2WZbest2D}
 \end{center}
\end{figure}

There is an alternative method which allows a cleaner extraction of the endpoint, albeit at the cost of reduced statistics.
The idea is to select events in which two boson candidates were reconstructed within {\em each} hemisphere.
Then one can produce a Dalitz-like plot of the invariant masses of the two diboson candidates
in the two hemispheres \cite{Bisset:2005rn}. The resulting distribution, depicted in Fig.~\ref{fig.simulmultijetWZh.Minvin2WZbest2D},
exhibits a clear drop in density around 800 GeV, and has less combinatorial background above the kinematic endpoint.

\subsubsection{$W/Z/h$ candidates in different hemispheres}
\label{WZhInDifferentHemispheres}

\begin{figure}[!htb]
 \begin{center}
 \includegraphics[width=0.49\textwidth]{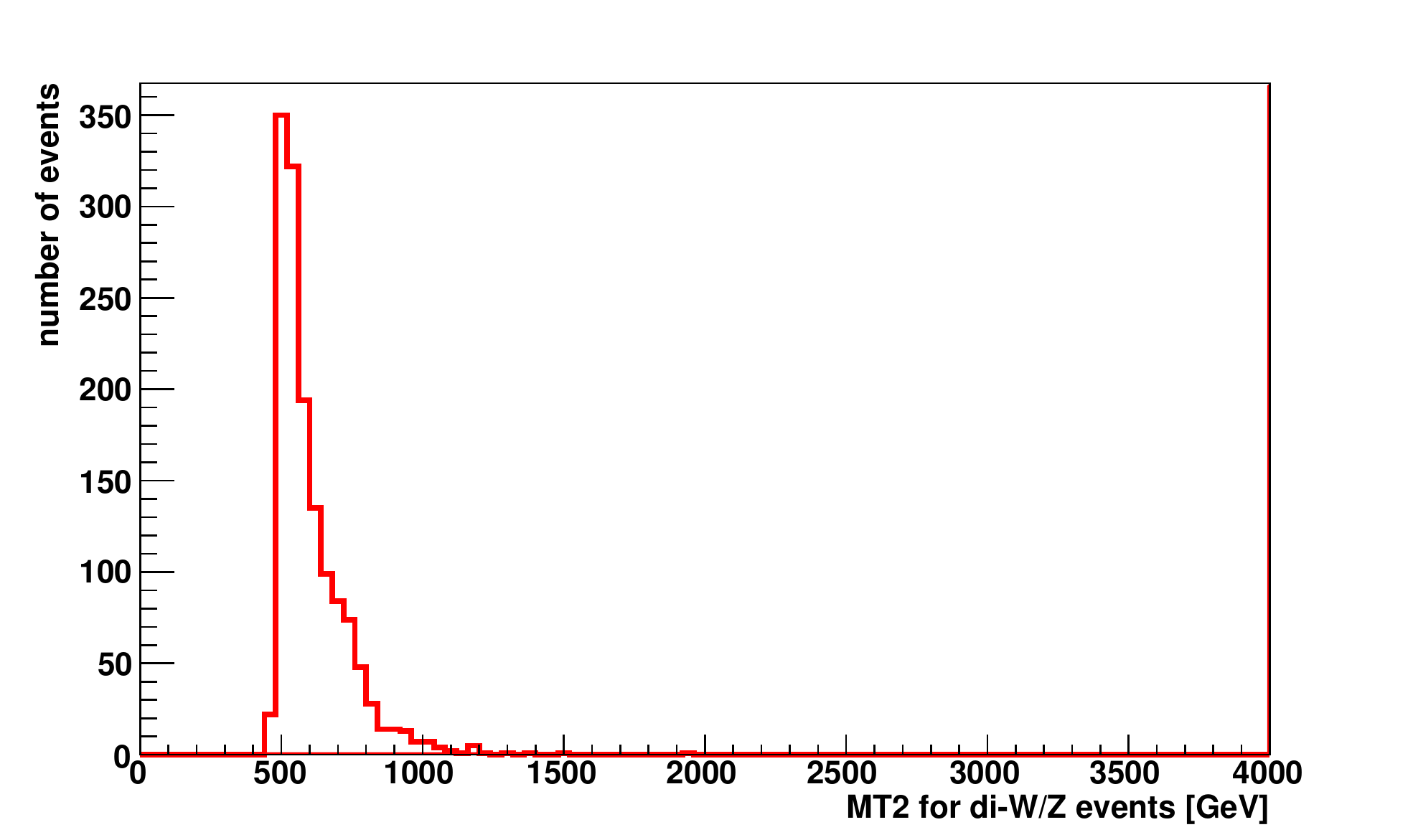}
 \includegraphics[width=0.49\textwidth]{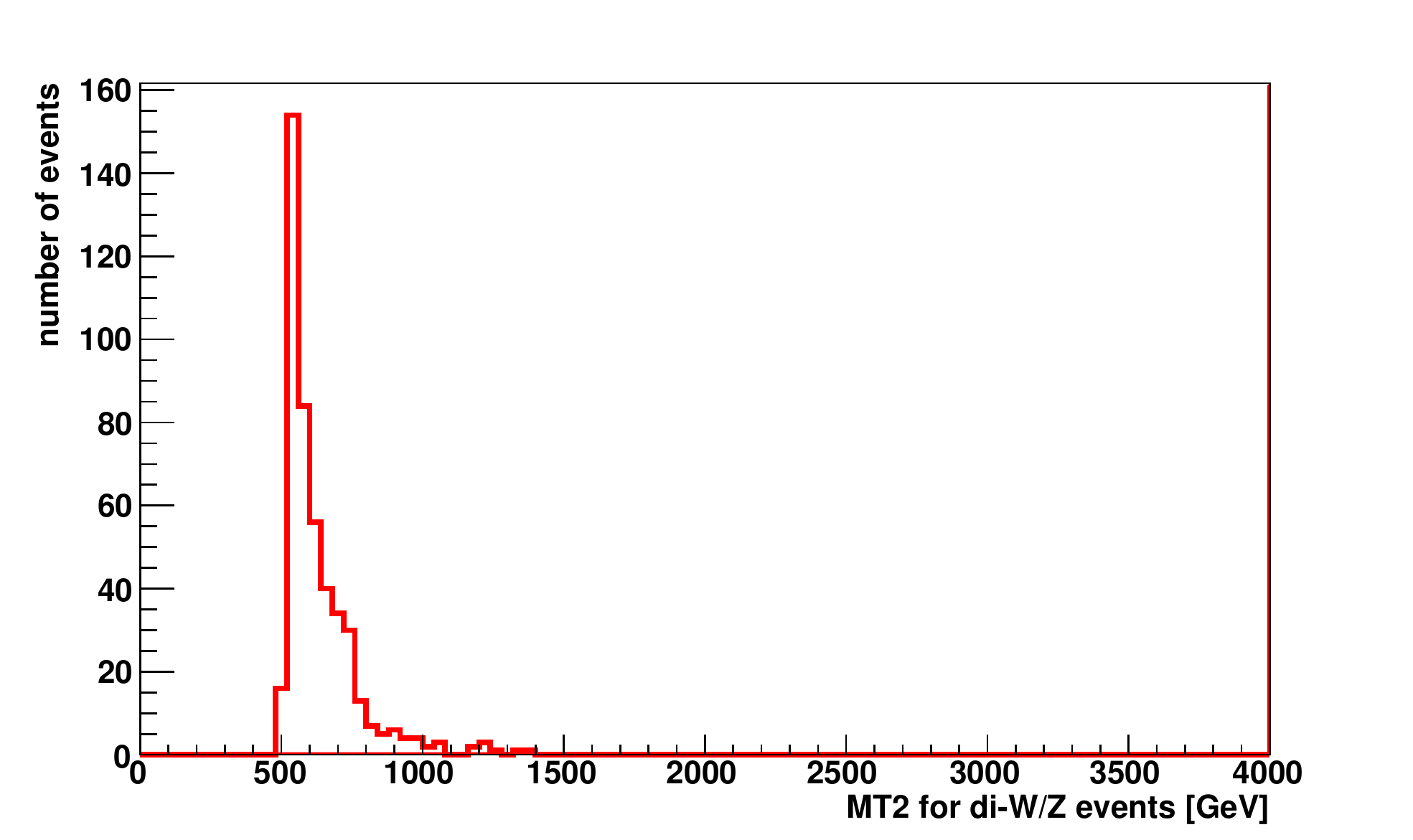}
 \includegraphics[width=0.49\textwidth]{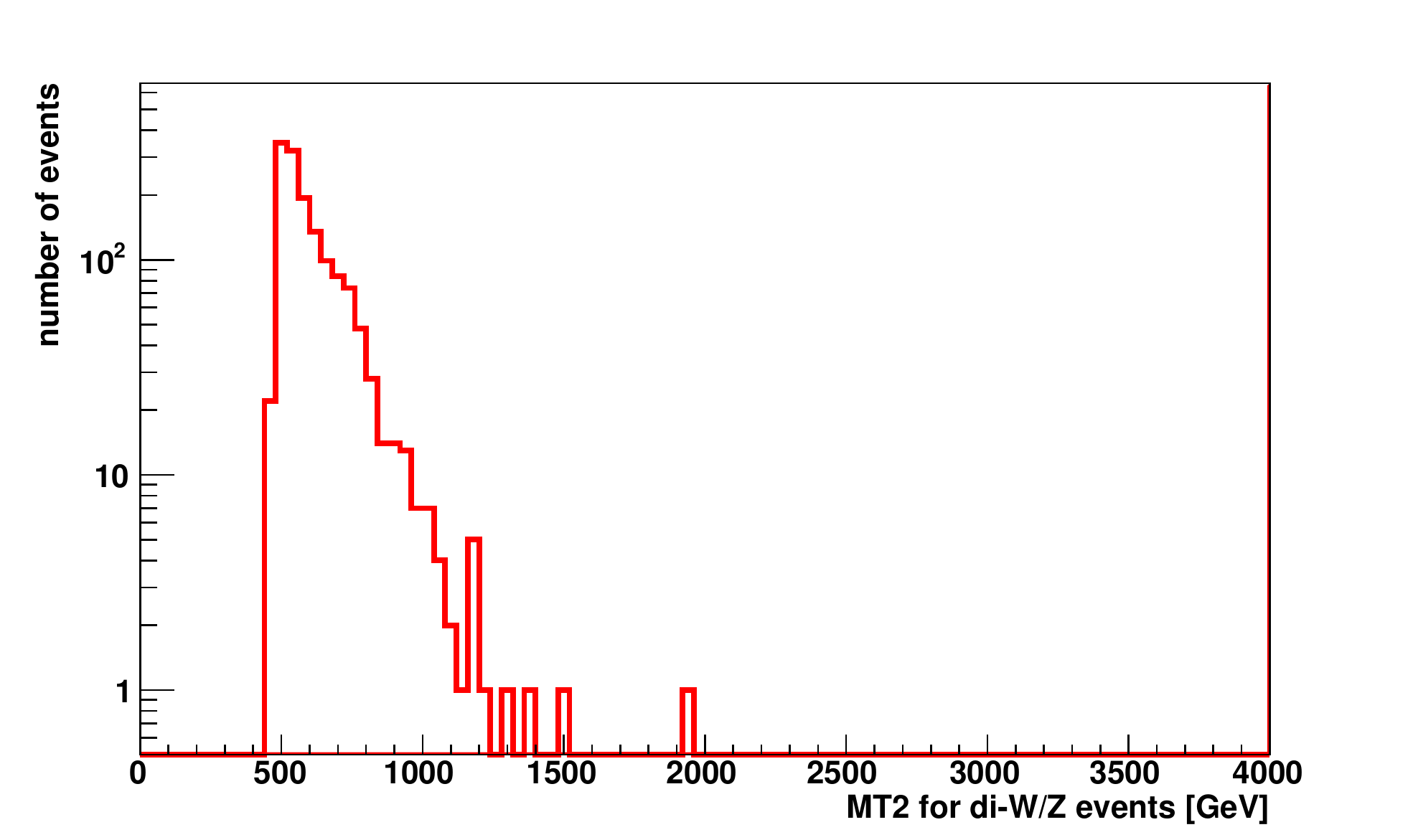}
 \includegraphics[width=0.49\textwidth]{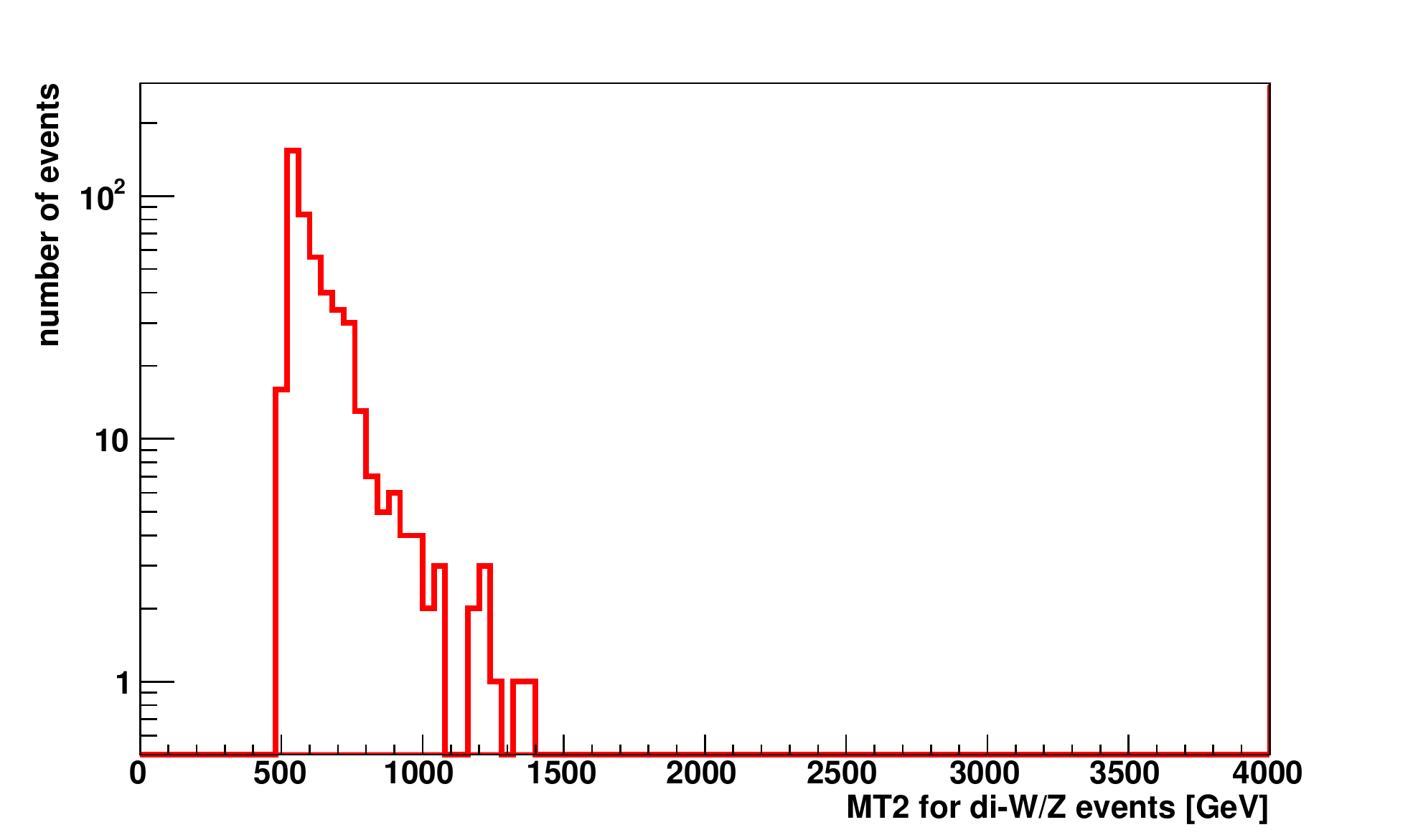}
\caption{$M_{T2}$ distributions for events with $W/Z/h$ di-jet candidates, one in each hemisphere, when no pair is b-tagged (left panel) and with at least 1 $b$-tag (right panel).
The plots in the upper (lower) row are on a linear (semi-log) scale. 
}
 \label{fig.simul.WZh.MT2WZWh}
 \end{center}
\end{figure}

If there exist $W/Z/h$ candidates in {\em each} hemisphere, this opens the possibility of constructing 
$M_{T2}$-type kinematic variables. Fig.~\ref{fig.simul.WZh.MT2WZWh} shows the $M_{T2}$ distributions
for events with two $W/Z/h$ di-jet candidates, one in each hemisphere. The plots in the upper (lower) row are on a linear (semi-log) scale.
In the left panels, we select events where no di-jet combination is $b$-tagged, while in the right panels, there is at least one $b$-tag.
In each case, we use the correct value of the LSP mass, which sets the lower kinematic endpoint of the $M_{T2}$ distribution.
The interesting physics information is encoded in the upper kinematic endpoint, which is seen around $800$ GeV, with a small tail of events extending up to around 1400 GeV.
While this endpoint is consistent with the mass of the intermediate higgsino states in the Mixed scenario, its
interpretation is complicated by the fact that the reconstructed dijet candidate is equally likely to be produced in a higgsino or in a Wino decay, 
see Fig.~\ref{fig.distscenarios.Mixed}.

Nevertheless, encouraged by this result, we take the analysis one step further, by investigating the 
dependence of the kinematic endpoint on the upstream $p_T$ \cite{Barr:2007hy,Burns:2008va}.
The true value of the test LSP mass could be determined by requiring that the $M_{T2}$ kinematic endpoint $M_{T2}^{max}$ is independent of the upstream $P_T$.
A convenient procedure for testing for $P_T$ dependence was outlined in \cite{Konar:2009wn}. As a preliminary step, one first studies the
$M_{T2\perp}$ distribution for some fixed value of the test LSP mass. Its endpoint $M_{T2\perp}^{max}$ is $P_T$-independent by construction, so
this measurement of $M_{T2\perp}^{max}$ can be used to predict the value of $M_{T2\perp}^{max}\equiv M_{T2}^{max} ( P_T = 0)$ for any other value of the test LSP mass via a simple
analytical expression. In the second step, one studies the actual $M_{T2}$ distribution for different choices of the test LSP mass and records the number of events 
whose $M_{T2}$ values exceed the corresponding $M_{T2\perp}^{max}$ predicted for that test mass. This number will be minimized at the true value of the LSP mass, since in general
$M_{T2}^{max} ( P_T) \ge M_{T2}^{max} ( P_T = 0)$, and the equality is reached only at the true value of the test mass.

\begin{figure}[tb]
 \begin{center}
 \includegraphics[width=0.49\textwidth]{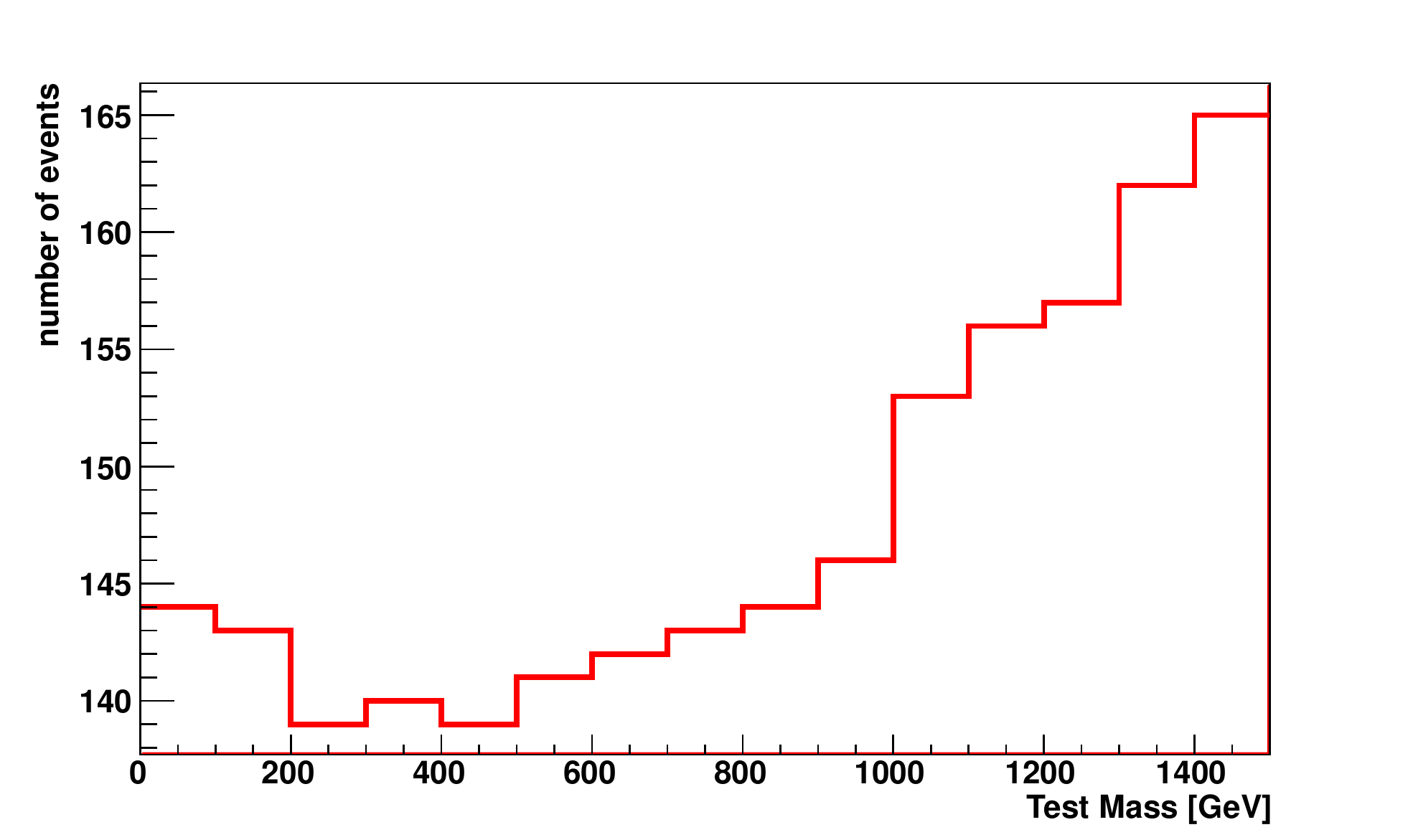}
\caption{Number of events with $M_{T2}$ larger than the endpoint $M_{T2}^{max} ( P_T = 0)$ as a function of the test mass.
}
 \label{fig.simul.WZh.Endptdiff}
 \end{center}
\end{figure}

The result from this procedure, when applied to our diboson sample above, is shown in Fig. \ref{fig.simul.WZh.Endptdiff}. In the figure we show 
the distribution of the number of events with $M_{T2}$ values larger than the endpoint $M_{T2}^{max} ( P_T = 0)$, as a function of the test LSP mass.
A minimum is indeed observed in the region between 200 and 600 GeV, which contains the true value, 396 GeV, of the $\chiz_1$, but with a large
uncertainty \cite{Konar:2009wn,Cohen:2010wv}.
Once we have determined the value of the test mass, we then obtain the mass of the parent particle $\chipm_1$ using the measured $M_{T2\perp}^{max}$ endpoint.
The variation of the $M_{T2\perp}^{max}$  endpoint with the test mass is usually rather weak, so the mass of the parent may be determined with reasonable precision.

The same measurement  of the masses of the parent $\chipm_1$ and child particle $\chiz_1$ can be performed in a different way, by combining the $M_{T2\perp}^{max}$
measurement with a measurement of $M_{T2}^{max} ( P_T)$ at a conveniently chosen value of $P_T$ (to maximize statistics, one can use a value near the maximum of the $P_T$ distribution).
From these two measurements, and using the $P_T$-dependent formulae of \cite{Burns:2008va}, 
one can in principle again determine the masses of the parent and the child.

\subsection{Results for different jet multiplicities} 

In what follows, we shall analyze the different types of events, categorized by jet multiplicity. As already mentioned, we only consider events in which the hemisphere algorithm returns equal number of jets in 
the two hemispheres.

\subsubsection{Di-jet events} 
\label{sec:dijets}

\begin{figure}[!htb]
 \begin{center}
 \includegraphics[width=0.49\textwidth]{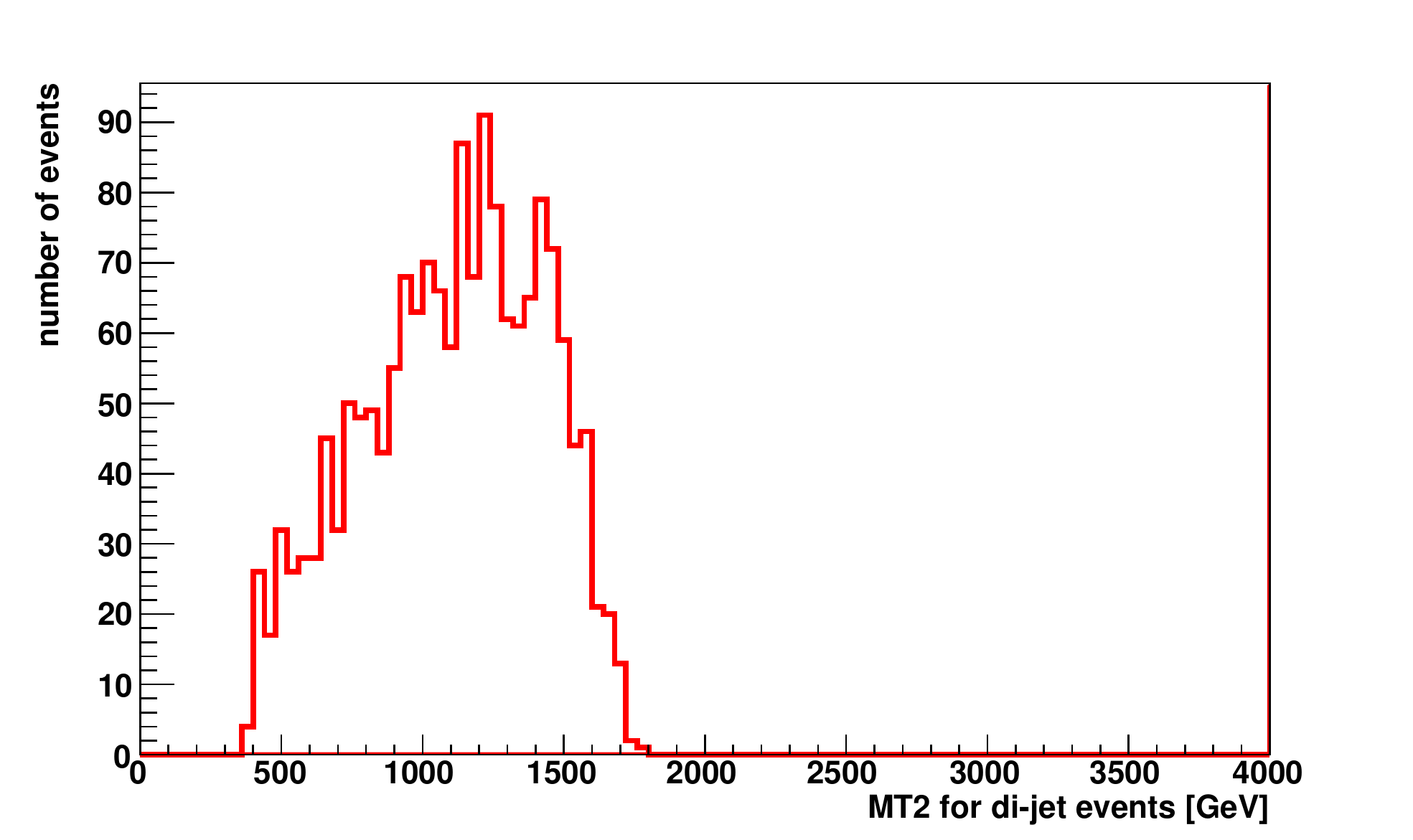}
 \caption{$M_{T2}$ distribution for di-jet events using the true value for the test mass.
}
 \label{fig.simul.dijet.MT2}
 \end{center}
\end{figure}

The simplest case is that of di-jet events, where there is a single jet in each hemisphere.
In the presence of a $b$-tag veto, such events are likely to originate from the 
\begin{eqnarray}
 \sQua_R \rightarrow q \chiz_1 
\label{eq:dijet.sqtoqX21}
\end{eqnarray}
decay chain with $\sQua_R$ being any of the first two generation right-handed squarks.
This is the classic event topology which motivated the introduction of the $M_{T2}$ variable in the first place.
The corresponding $M_{T2}(q)$ distribution is shown in Fig. \ref{fig.simul.dijet.MT2}.
It exhibits a clear endpoint which supplies one relation between the masses of $\sQua_R$ and $\chiz_1$.
However, at this point the individual masses remain undetermined, and there has been considerable interest in
developing methods for additional measurements which could pinpoint the overall mass scale 
\cite{Cho:2007qv,Gripaios:2007is,Barr:2007hy,Cho:2007dh,Cheng:2008hk,Burns:2008va,Barr:2009jv,Matchev:2009ad,Matchev:2009fh,Konar:2009wn,Cho:2010vz,Cheng:2011ya}.
These additional measurements tend to be very subtle and are more likely to be successful in events with leptons rather than jets.
Nevertheless, the measured $M_{T2}(q)$ endpoint from Fig.~\ref{fig.simul.dijet.MT2} provides a useful check on the masses determined in higher multiplicity events (see below).

It should be noted that, provided the sneutrino $\sNu$ is light enough, another decay mode,
\begin{eqnarray}
 \sQua_L \rightarrow q \chiz_2 \rightarrow q (\nu \sNu )
\label{eq:dijet.sqtoqX2}
\end{eqnarray}
would also give rise to the same di-jet final state.
However, since the masses of the charged left-handed slepton $\sLep_L$ and the sneutrino $\sNu$ should be near each other (by $SU(2)$-symmetry),
the decay via $\sLep_L$ should also be observed in events with 2 or 4 additional leptons.

The same analysis can also be applied to di-jet events  with b-tags, which can be interpreted as sbottom pair-production.
In that case, the mass of $\sBot_1$ could similarly be determined as a function of the $\chiz_1$ mass.

\subsubsection{Four-jet events} 
\label{sec:4jet}

\begin{figure}[!htb]
 \begin{center}
 \includegraphics[width=0.49\textwidth]{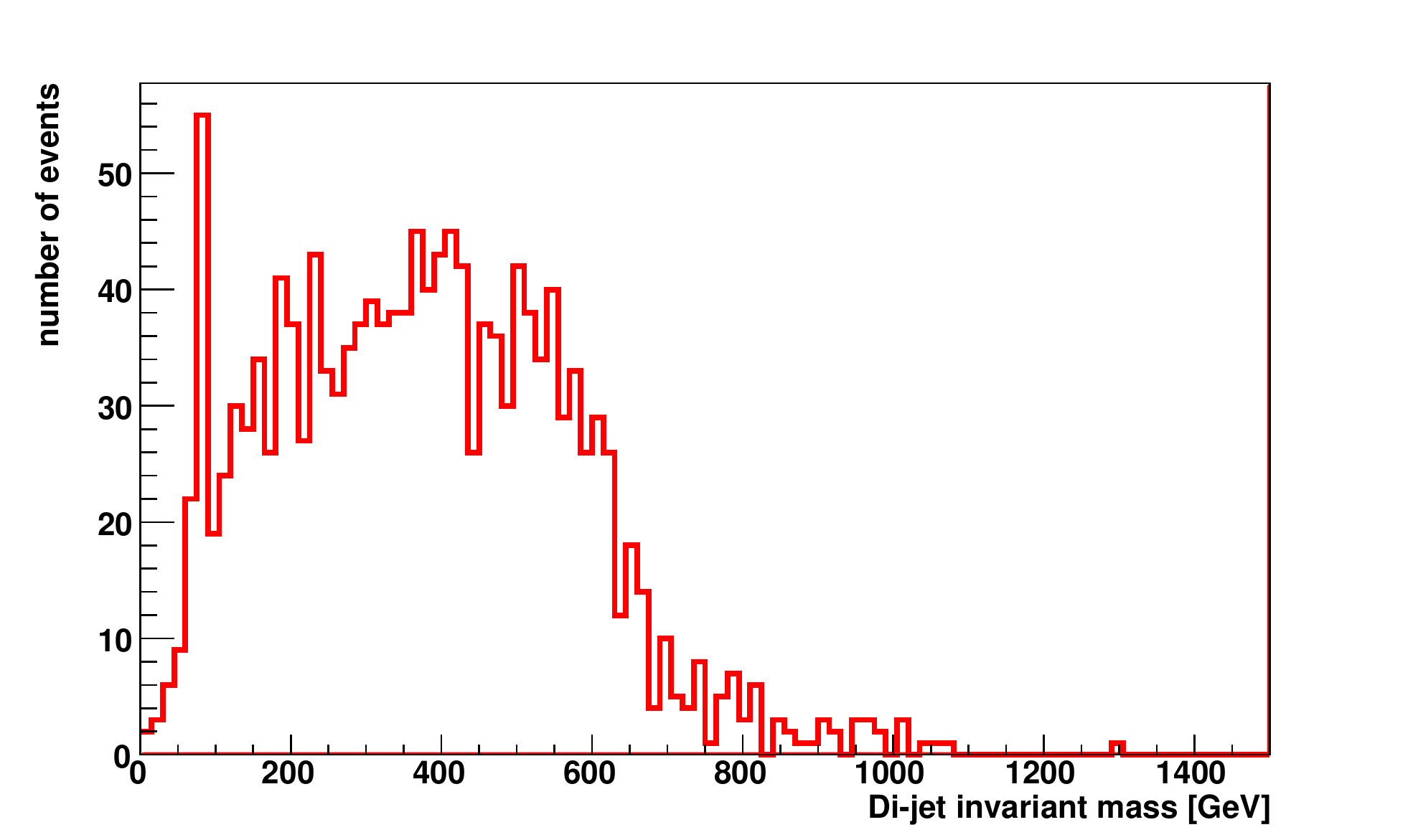}
  \includegraphics[width=0.49\textwidth]{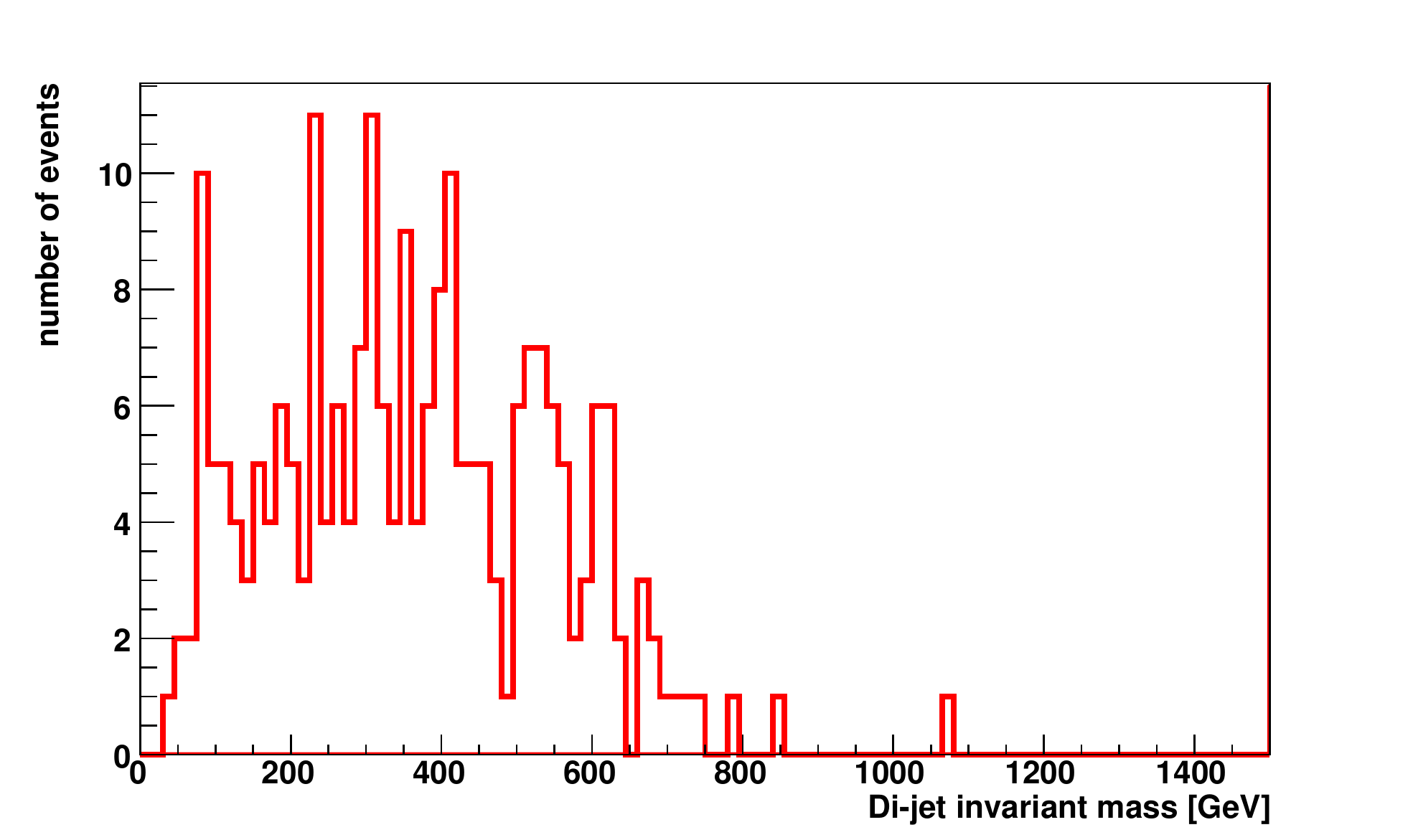}
\caption{Di-jet invariant mass distribution for all events with at least one di-jet hemisphere  
(left) and for four-jet events only (right).}
 \label{fig.simul.fourjet.Minv}
 \end{center}
\end{figure}

The next simplest case is that of four-jet events, with 2 jets in each hemisphere. We first look at the di-jet invariant mass distribution 
for jets belonging to the same hemisphere.
The left panel in Fig.~\ref{fig.simul.fourjet.Minv} shows the di-jet invariant mass for all events with at least one di-jet hemisphere (including ``asymmetric" events), 
while the right panel in Fig.~\ref{fig.simul.fourjet.Minv} shows the di-jet invariant mass combinations in four-jet events only.
While the distribution in the left panel does have a small peak at low invariant mass which can be attributed to $W/Z/h$,
such events are largely due to backgrounds from events with charginos and neutralinos, misreconstructed as four-jet events. 
The di-jet invariant mass distribution in the right panel does not indicate a strong presence of $W/Z/h$, 
lending support to the hypothesis that the decay chains are primarily initiated by gluinos
\begin{eqnarray}
 \sGlu \rightarrow q_1 \sQua_R \rightarrow q_1 ( q_2 \chiz_1 ).
 \label{eq:fourjet.sgtosq}
\end{eqnarray} 
For this decay chain, the invariant mass distribution of the two jets\footnote{Note that one of the two final state quarks in (\ref{eq:fourjet.sgtosq}) is actually an anti-quark. 
However, for notational simplicity we shall not 
indicate which one is the antiquark jet, and will omit the bar over the corresponding $q$.}, $q_1$ and $q_2$, is expected to have an upper kinematic endpoint 
whose value is a function of the masses of $\sGlu$, $\sQua_R$ and $\chiz_1$, in this case 647 GeV (see eq.~(\ref{eq.invmass.2step.Mffmax}) in Appendix \ref{sect:invmass}).
This is confirmed by Fig.~\ref{fig.simul.fourjet.Minv} --- each dijet mass distribution has a reasonably clear endpoint near the expected value, 
with a tail due to MSSM background events.

\begin{figure}[!htb]
 \begin{center}
 \includegraphics[width=0.49\textwidth]{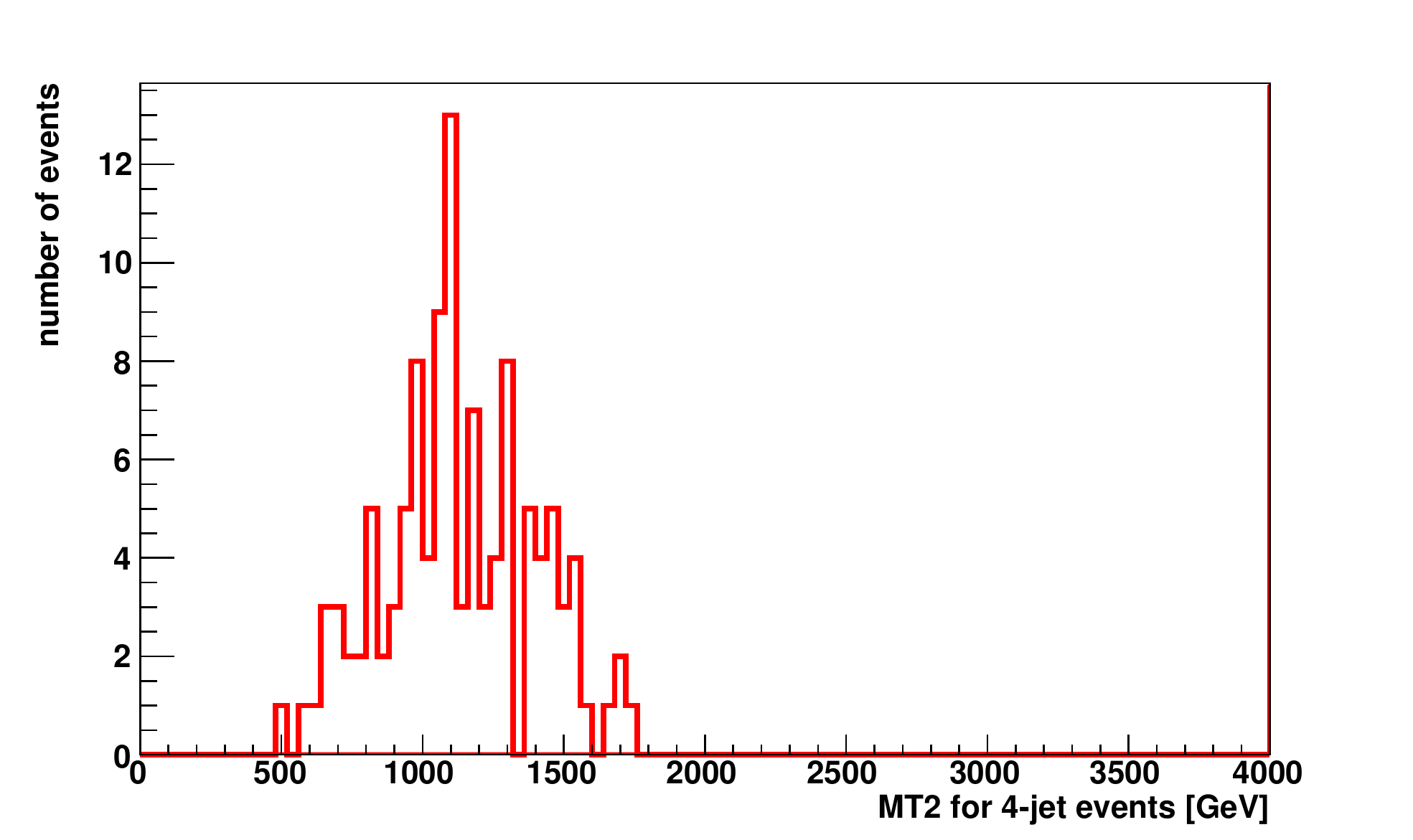}
 \includegraphics[width=0.3\textwidth]{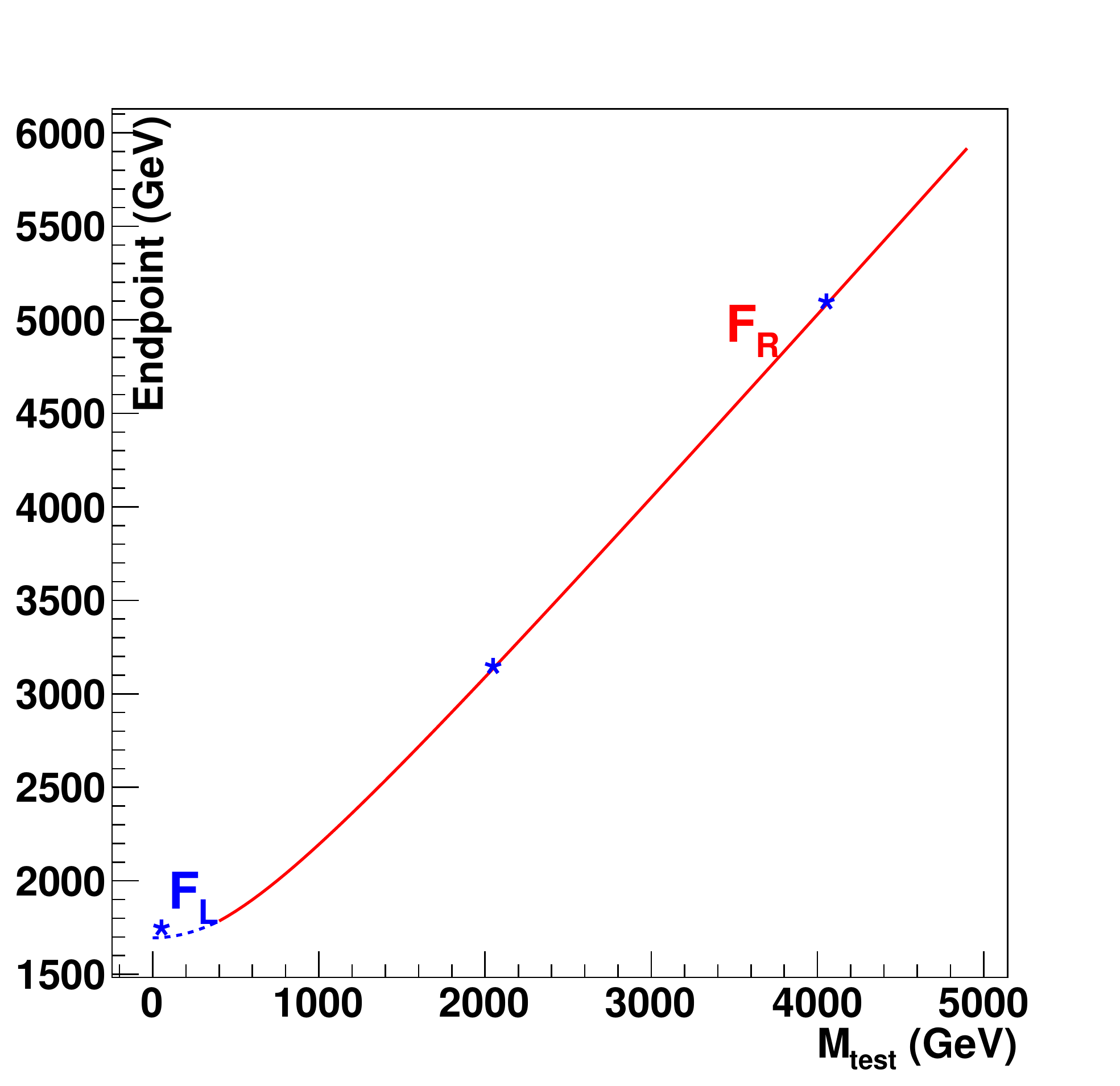}
\caption{Left panel: $M_{T2}(qq)$ distribution in four-jet events, using the true value for the test $\chiz_1$ mass.
Right panel: dependence of the $M_{T2}(qq)$ endpoint on the input test mass. The blue dotted (red solid) line 
corresponds to the left (right) branch given by eq.~(\ref{FL2step}) (eq.~(\ref{FR2step})) of Appendix~\ref{sect:MT2}.
The star symbols indicate the results from measurements at test mass values of 0, 2000 and 4000 GeV, respectively.
}
 \label{fig.simul.fourjet.MT2}
 \end{center}
\end{figure}
Gluino pair-production events with the gluinos decaying as in (\ref{eq:fourjet.sgtosq}) can be suitably studied with the 
Cambridge $M_{T2}$ variable, where each di-jet pair belongs to a separate branch. The distribution of the resulting
variable, $M_{T2}(qq)$, is shown in the left panel of Fig.~\ref{fig.simul.fourjet.MT2}. 
A clear endpoint is observed near 1700 GeV, to be compared with the input $\sGlu$ mass of 1785 GeV.
However, this $M_{T2}(qq)$ endpoint only gives the $\sGlu$ mass as a function of the $\chiz_1$ mass,
which must be supplied as an input to the $M_{T2}$ computation (in the left panel of Fig.~\ref{fig.simul.fourjet.MT2}
we used the true value of the $\chiz_1$ mass, which is of course a priori unknown).
The dependence of the $M_{T2}(qq)$ endpoint on the input test mass is illustrated in the right panel of Fig.~\ref{fig.simul.fourjet.MT2}.
The lines correspond to the theoretically predicted relation, eq.~(\ref{FL2step}) (blue dotted line) and eq.~(\ref{FR2step}) (red solid line) of Appendix~\ref{sect:MT2}.
It has been shown \cite{Burns:2008va} that measuring the $M_{T2}(qq)$ endpoint at three different values of the test mass is sufficient in order to 
extract all three unknown masses (for details, see Sec.~\ref{sect:MT2.MT2.2step}).
If we apply the same idea here, we obtain the three measurements denoted with 
star symbols in the right panel of Fig.~\ref{fig.simul.fourjet.MT2}. The three results are found to be in perfect agreement with the expected theoretical curve,
which means that the correct mass values would be extracted from such analysis.
However, the uncertainties are expected to be large, given the large uncertainties 
on the endpoint position.\footnote{There is an additional complication ---  there remains a twofold ambiguity in the $\sQua_R$ mass determination \cite{Burns:2008va}.
The true value of the $\sQua_R$ mass of 1656 GeV will be accompanied by another estimate of 431 GeV.
The difference between the two alternatives is large enough, and the ambiguity should be easily solved by using additional information,
e.g., from the measured $M_{T2}(q)$ endpoint (see Sec.~\ref{sec:dijets}).}

Before concluding our discussion of four-jet events, we shall investigate a question which to the best of our knowledge has not been 
addressed in the existing literature. We already saw that the natural variable for di-jet events was $M_{T2}(q)$ (see Sec.~\ref{sec:dijets}),
while for four-jet events it makes sense to use $M_{T2}(qq)$ (see Fig.~\ref{fig.simul.fourjet.MT2}). Nevertheless, one may still ask the question
whether the {\em single jet} $M_{T2}$ variable, $M_{T2}(q)$, can be usefully applied to four-jet events as well. At first glance,
$M_{T2}(q)$ seems incompatible with four-jet events, for two reasons. First, there is a four-fold ambiguity in picking a single jet out of each 
di-jet hemisphere, and only two of those selections make physics sense --- when the selected jets are the ones emitted first ($q_1$) in the cascade (\ref{eq:fourjet.sgtosq}),
or when they are the ones emitted last ($q_2$); let us call the corresponding $M_{T2}$ variables $M_{T2}(q_1)$ and $M_{T2}(q_2)$.
Second, the input to the calculation of $M_{T2}$ is different in these two cases: for $M_{T2}(q_1)$,
the momentum of the second jet $q_2$ needs to be added to the missing $\vec{p}_T$ and the true value of the test mass is the mass of the intermediate resonance in (\ref{eq:fourjet.sgtosq}), $\sQua_R$;
while for $M_{T2}(q_2)$, the momentum of the first jet $q_1$ has to be added to the upstream $\vec{P}_T$
and the test mass is interpreted as the mass of the LSP $\chiz_1$.

\begin{figure}[!htb]
 \begin{center}
 \includegraphics[width=0.49\textwidth]{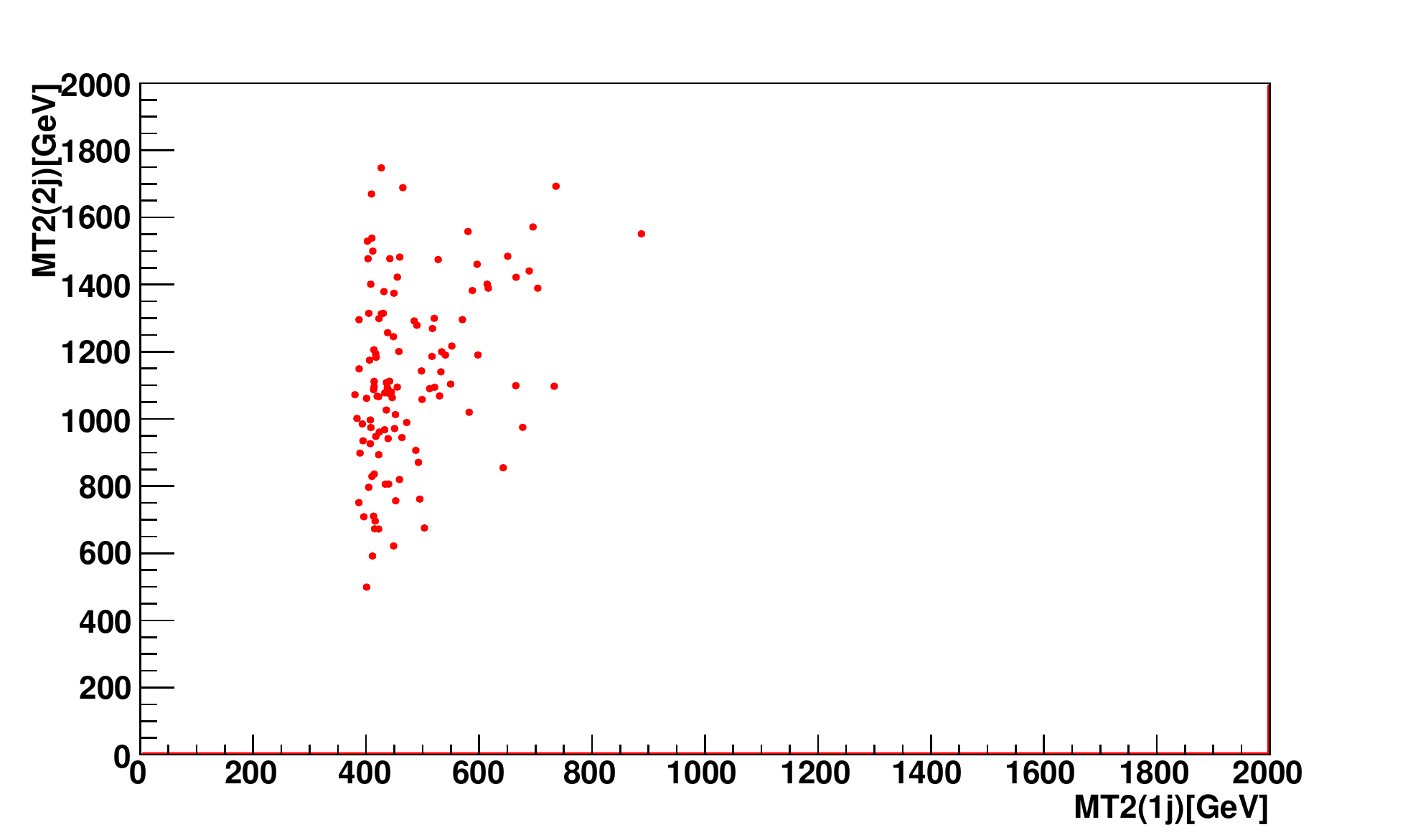}
  \includegraphics[width=0.49\textwidth]{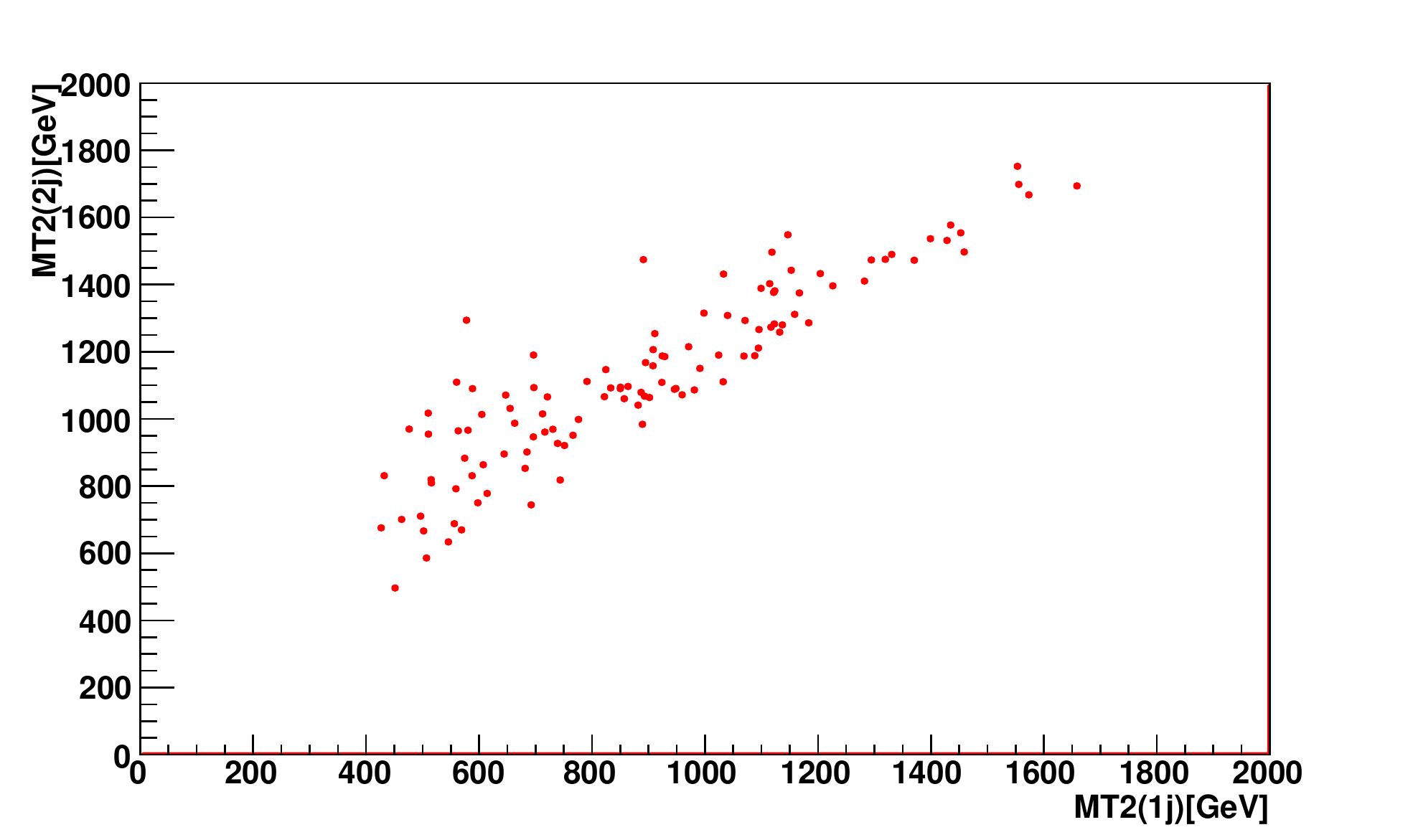}
\caption{$M_{T2}(q_2)$ distributions in four-jet events (selecting one-jet per hemisphere), plotted against $M_{T2}(qq)$.
The remaining two jets are treated as upstream momentum, and we use the true $\chiz_1$ mass in the $M_{T2}$ calculation.
In the left (right) panel we choose the smallest (largest) $M_{T2}(q_2)$ value per event.}
 \label{fig.simul.fourjet.MT21j2nd2D}
 \end{center}
\end{figure}

We propose to overcome those difficulties by utilizing the kinematic correlations with the di-jet $M_{T2}(qq)$ variable in order to
to separate the combinations where the jets are from the first decay ($q_1$) from the ones where the jets are from the second decay ($q_2$). 
The main idea is that for $M_{T2}(q_2)$, the endpoint is reached in the same kinematical configuration as for $M_{T2}(qq)$.
If we plot one against the other, the value of the $M_{T2}(q_2)$ should increase at the same time as the value of $M_{T2}(qq)$.
This behavior is verified in the right panel of Fig.~\ref{fig.simul.fourjet.MT21j2nd2D}, where we plot the
largest of the four $M_{T2}(q_2)$ values per event versus the value of $M_{T2}(qq)$. 
We see that the extreme values of $M_{T2}(qq)$ are reached for the same events which also
maximize the largest $M_{T2}(q_2)$ value per event.

\begin{figure}[!htb]
 \begin{center}
  \includegraphics[width=0.49\textwidth]{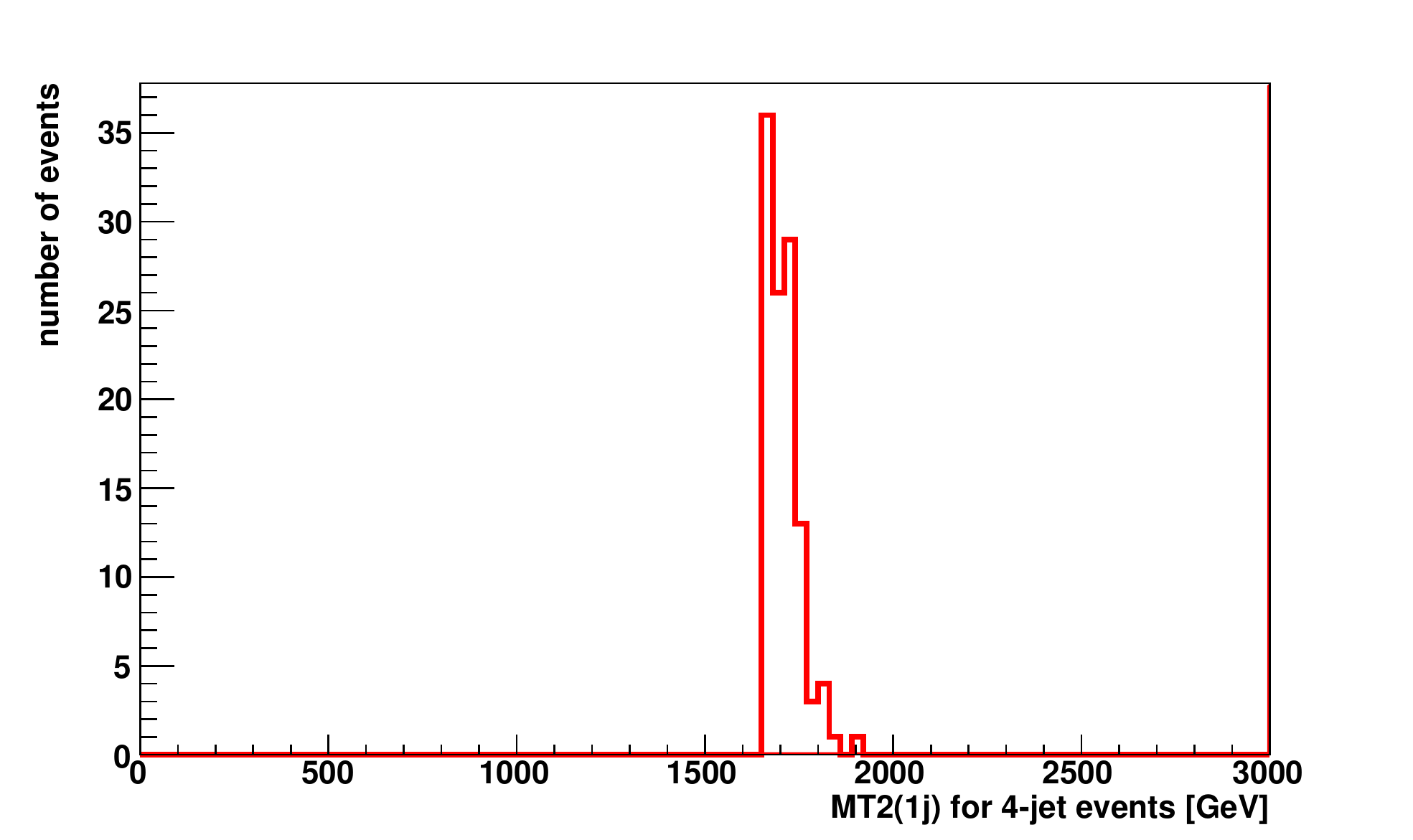}
 \includegraphics[width=0.49\textwidth]{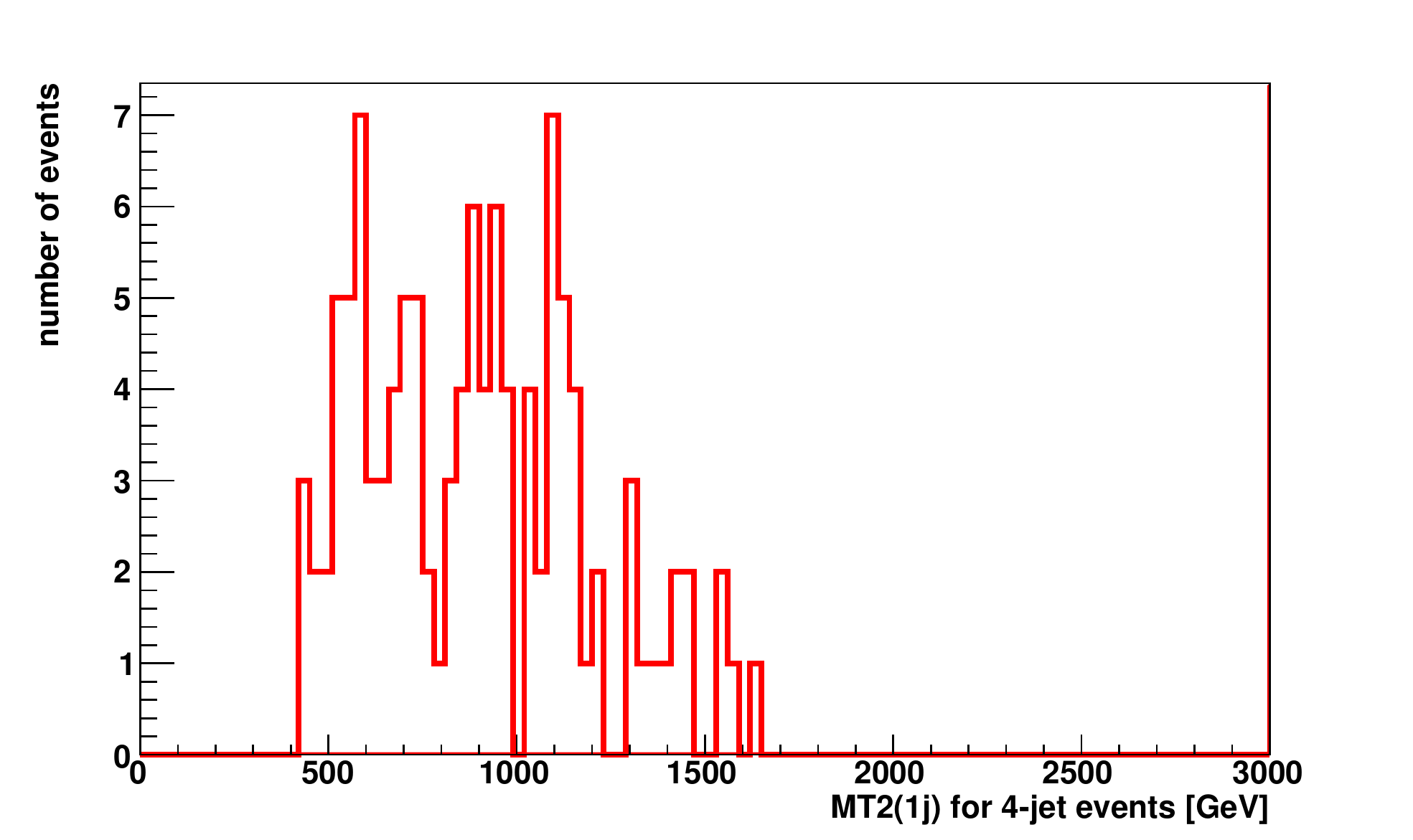}
\caption{$M_{T2}(q)$ distributions for one-jet combinations per hemisphere in four-jet events.
In the left (right) panel we show the smallest $M_{T2}(q_1)$ (largest $M_{T2}(q_2)$) value per event.}
 \label{fig.simul.fourjet.MT21j1st}
 \end{center}
\end{figure}

The above discussion suggests that the single-jet $M_{T2}$ variable can be of interest in four-jet events as well. 
There are actually two variants, $M_{T2}(q_1)$ and $M_{T2}(q_2)$, depending on our hypothesis about the selected jets,
and in each case the remaining two jets are treated accordingly. Fig.~\ref{fig.simul.fourjet.MT21j2nd2D} showed that
one useful kinematic endpoint is obtained by taking the largest value, $\max(M_{T2}(q_2))$, of $M_{T2}(q_2)$.
As for the $M_{T2}(q_1)$ variable, in order to preserve the upper kinematic endpoint, we conservatively take the 
smallest of the four possible values per event, $\min(M_{T2}(q_1))$.
The resulting distributions of $\min(M_{T2}(q_1))$ and $\max(M_{T2}(q_2))$ are plotted in the left and right panels of Fig.~\ref{fig.simul.fourjet.MT21j1st}, respectively.
In both distributions, an endpoint is observable. 
The $M_{T2}(q_1)$ endpoint is at 1850 GeV, reasonably near the expected value of the $\sGlu$ mass of 1785 GeV.
The endpoint is given by a single function of the sparticle masses, 
hence it yields a relation between the masses of $\sGlu$ and $\sQua_R$. 
The $M_{T2}(q_2)$ endpoint in the right panel of Fig.~\ref{fig.simul.fourjet.MT21j1st} is around 1600 GeV, near the expected value of 1656 GeV for the $\sQua_R$ mass.
Repeating the measurement with three suitably chosen values of the test mass again allows a determination of all three sparticle masses. 
This shows that from the four-jet events the masses of $\sGlu$, $\sQua_R$ and $\chiz_1$ can be over-determined. 
In particular, it can be verified that their values satisfy the mass relation obtained from the di-jet events.

\begin{figure}[!htb]
 \begin{center}
 \includegraphics[width=0.49\textwidth]{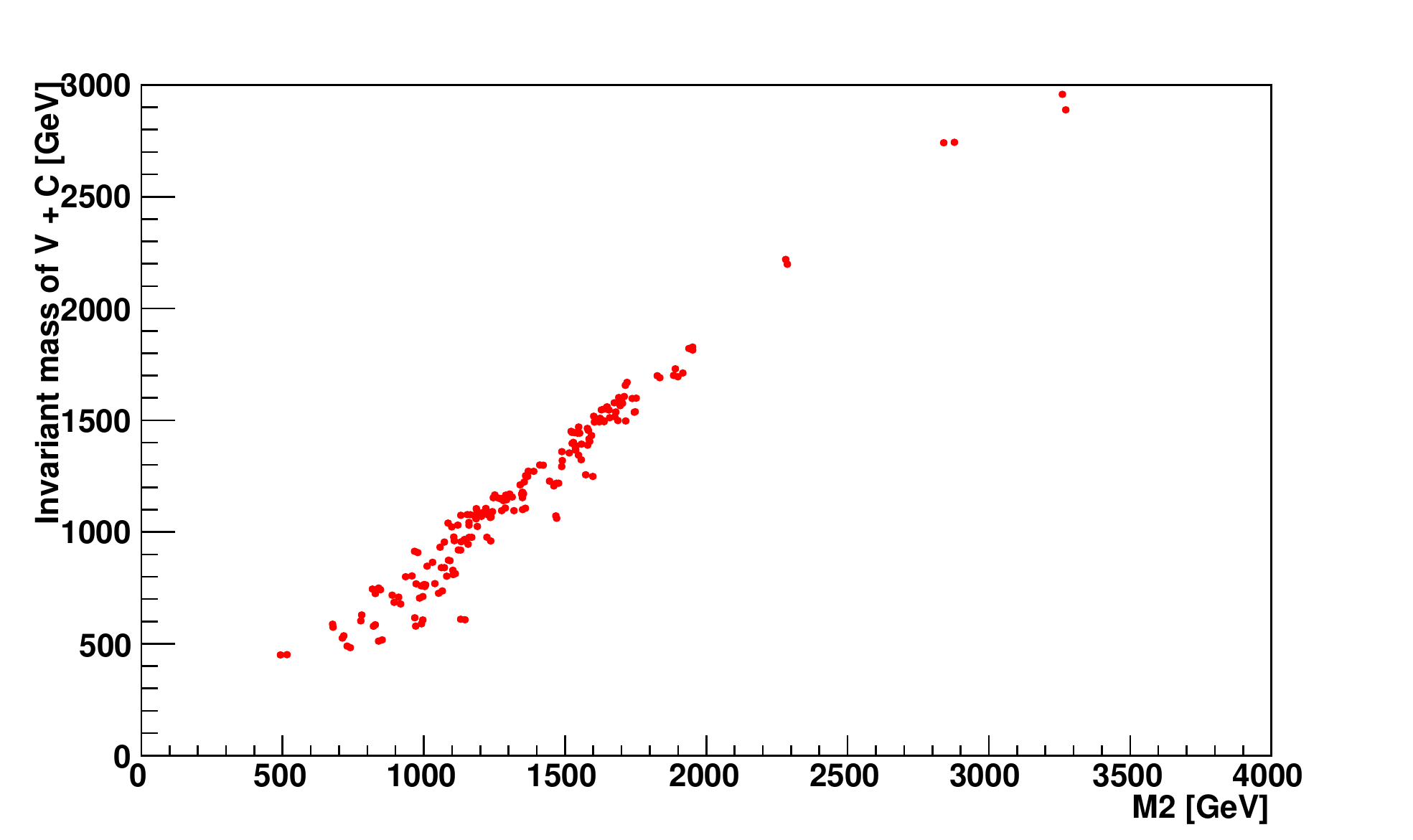}
\caption{A scatter plot of $M_2(qq)$ ($x$-axis) versus the reconstructed $\sQua_R$ mass ($y$-axis), using the invisible momenta obtained in the $M_2(qq)$ calculation.}
 \label{fig.simul.fourjet.M2invM}
 \end{center}
\end{figure}

A similar analysis can be carried out using the $M_2$ variables \cite{Cho:2014naa,Cho:2015laa} instead of $M_{T2}$. The advantage of the $M_2$ variables is that they provide an ansatz for the full
3-momenta of the unseen neutrals, which opens the door for direct mass reconstruction of the intermediate resonances.
For example, when constructing $M_2(qq)$, which is the analogue of $M_{T2}(qq)$, one imposes an additional mass constraint during the 
minimization, namely the equality of the two intermediate $\sQua_R$ masses. In a jetty channel like the one considered here, 
this brings up the usual combinatorial problem of the ordering of the jets as $q_1$ and $q_2$ in each decay chain.
Applying the same idea as in the case of $M_{T2}(q_2)$ above, we consider all four possibilities and then choose the largest of the four $M_2(qq)$ values.
As shown in \cite{Kim:2017awi} (see related discussion in Section~\ref{sect:MT2.M2}),
the correct mass value for the intermediate $\sQua_R$ resonance can be estimated by examining the 2D plot of the reconstructed 
$\sQua_R$ invariant mass (using the invisible momenta obtained in the minimization) versus $M_2(qq)$.
The result is shown in Fig.~\ref{fig.simul.fourjet.M2invM}, which indicates a $\sQua_R$ mass around 1750 GeV, to be compared with the nominal value of 1656 GeV.

Finally, if the $\sQua_R$ mass is heavier than the $\sGlu$ mass, the decay (\ref{eq:fourjet.sgtosq}) proceeds as a direct 3-body decay.
This would be the case in, e.g.,  a (mini-)split SUSY model \cite{Wells:2003tf,ArkaniHamed:2004fb,Giudice:2004tc,Wells:2004di}.
As discussed in Sec.~\ref{sect:MT2.MT2.3body}, the two unknown masses, of the $\sGlu$ and of the $\chiz_1$, can still be measured in this scenario as well.

\subsubsection{Six-jet and eight-jet events} 

In the mixed scenario, squark decays yield either 2 or 0 bosons, see Fig.~\ref{fig.distscenarios.Mixed}. Therefore, in order to have a total of 6 or 8 jets at the generator level in an event, 
the event must be asymmetric, involving a $\sQua_L$ in one branch and a $\sQua_R$ in the other. In that case, however, we would expect to end up with a different number of jets in each 
hemisphere\footnote{This logic would be affected by the additional presence of ISR jets, which we are ignoring in our analysis.}. Thus we conclude that in order to obtain a 
symmetric six-jet or eight-jet event of the type considered here, i.e., with $3+3$ or $4+4$ jets per hemisphere at the reco level, the event must have migrated from a higher 
multiplicity topology, where some jets were lost or have been merged. This expectation was explicitly checked at the generator level, confirming that most such events are indeed
combinatorial background events.

It should be noted that in the other scenarios from Section~\ref{sect:scenarios.first}, namely, the Bino and Higgsino scenarios (as well as their twin scenarios) 
there should be plenty of six-jet and eight-jet channels with sizable statistics,
which should in principle allow the masses to be measured, just like we do for the 10-jet and 12-jet events considered below.

\subsubsection{Ten-jet events} 
\label{sec:10jet}

Ten-jet events (with 5 jets in each of the two hemispheres) are most naturally interpreted as left-handed squark ($\sQua_L$) pair production, followed by, e.g.,
\begin{eqnarray}
 \sQua_L \rightarrow q \; \chipm_2 \rightarrow q \; Z^0/h^0 \; \chipm_1 \rightarrow q \; Z^0/h^0 \; W \; \chiz_1
\label{eq:tenjet.sqtoqX2}
\end{eqnarray}
or similar decay chains involving two mass levels of electroweakinos, see Fig.~\ref{fig.distscenarios.Mixed}.
In this case, clear $W/Z/h$ peaks are again observed in the dijet mass distribution for a given hemisphere and, after selecting the best {\em two} di-jet solutions, $V_1$ and $V_2$,
the following invariant masses can be constructed: $M(V_1V_2)$, $M(qV_1V_2)$, $M(qV_1 \oplus qV_2)$ and $M(qV_1\cup qV_2)$, where
$V$ stands for any $W/Z/h$ boson. Here $M(qV_1 \oplus qV_2)$ is defined as\footnote{Eq.~(\ref{MqVRMS}) is a special case ($\alpha=1$) of the 
more general symmetrized variable $$M_{qV(s)}(\alpha) \equiv \left[M(qV_1)^{2\alpha}+M(qV_2)^{2\alpha}\right]^{1/\alpha}$$ introduced in \cite{Matchev:2009iw}.}
\begin{equation}
M(qV_1 \oplus  qV_2) \equiv \sqrt{M(qV_1)^2+M(qV_2)^2},
\label{MqVRMS}
\end{equation}
providing an additional useful measurement while avoiding the combinatorics problem \cite{Matchev:2009iw},
while $M(qV_1\cup qV_2)$ is simply the combined $M(qV)$ distribution with two entries per event. 

\begin{figure}[!htb]
 \begin{center}
 \includegraphics[width=0.49\textwidth]{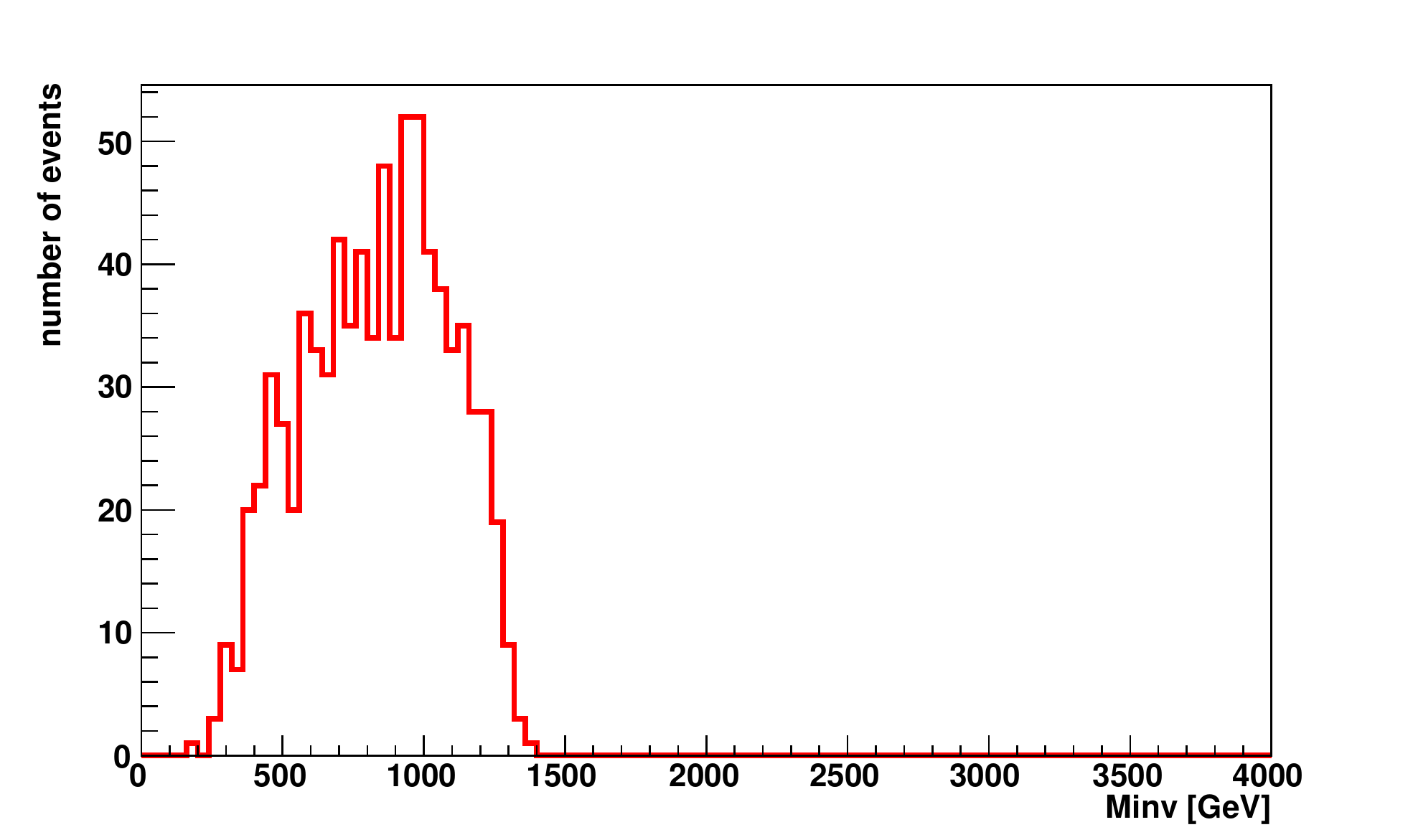}
 \includegraphics[width=0.49\textwidth]{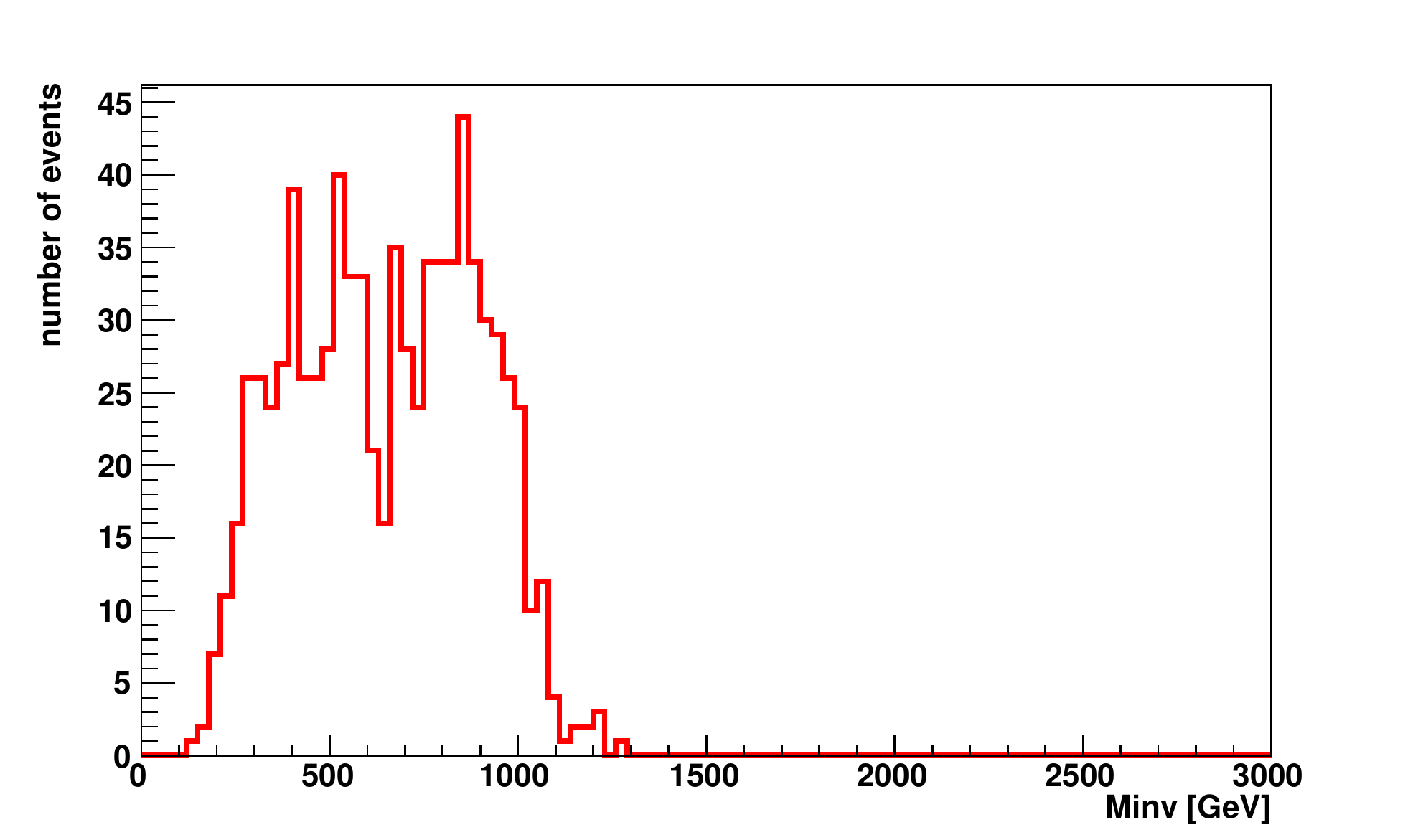}
\caption{Distributions of $M(qV_1V_2)$ (left panel) and $M(qV_1 \oplus  qV_2)$ (right panel) in ten-jet events.
}
 \label{fig.simul.10jet.Minv10jqWW}
 \end{center}
\end{figure}

One might start by investigating the 5-jet invariant mass in each hemisphere, i.e., the
invariant mass of the left over jet with both bosons, $M(qV_1V_2)$.\footnote{In order to increase the statistics, we consider all events with a 5-jet hemisphere.}
This distribution is shown in the left panel of Fig.~\ref{fig.simul.10jet.Minv10jqWW}
and it exhibits a clear endpoint around 1300 GeV, close to the numerically computed expected value of 1260 GeV.
The distribution of $M(qV_1 \oplus qV_2)$ is shown in the right panel of Fig.~\ref{fig.simul.10jet.Minv10jqWW}
and also has an upper kinematic endpoint, albeit at a slightly lower value, due to the identity
\begin{equation}
M^2(qV_1 \oplus qV_2) = M^2(qV_1V_2)-\left[M^2(V_1V_2)-M^2(V_1)- M^2(V_2)   \right].
\end{equation}

\begin{figure}[!htb]
 \begin{center}
 \includegraphics[width=0.49\textwidth]{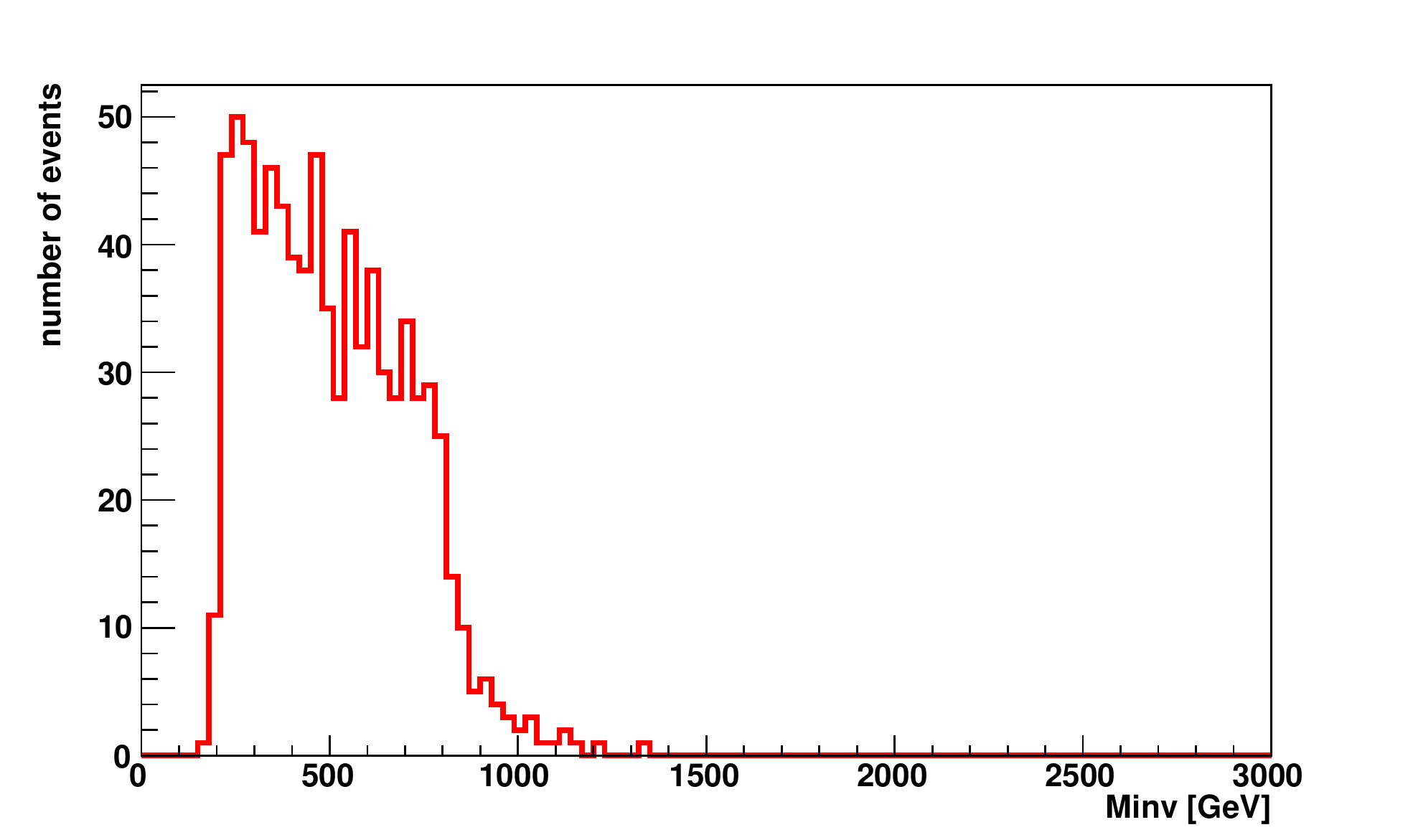}
\caption{Invariant mass distribution of $M(V_1V_2)$.
}
 \label{fig.simul.10jet.Minv10jWW}
 \end{center}
\end{figure}

Another useful kinematic endpoint is provided by the invariant mass distribution of $M(V_1V_2)$
shown in Fig.~\ref{fig.simul.10jet.Minv10jWW}.
The distribution looks clean and an endpoint could be located in the region 800-900 GeV, 
where the distribution drops off significantly. The theoretical value of this endpoint
(computed numerically, as the $V$'s are massive) is expected at 812 GeV, in reasonable agreement with the measured value.

\begin{figure}[!htb]
 \begin{center}
  \includegraphics[width=0.49\textwidth]{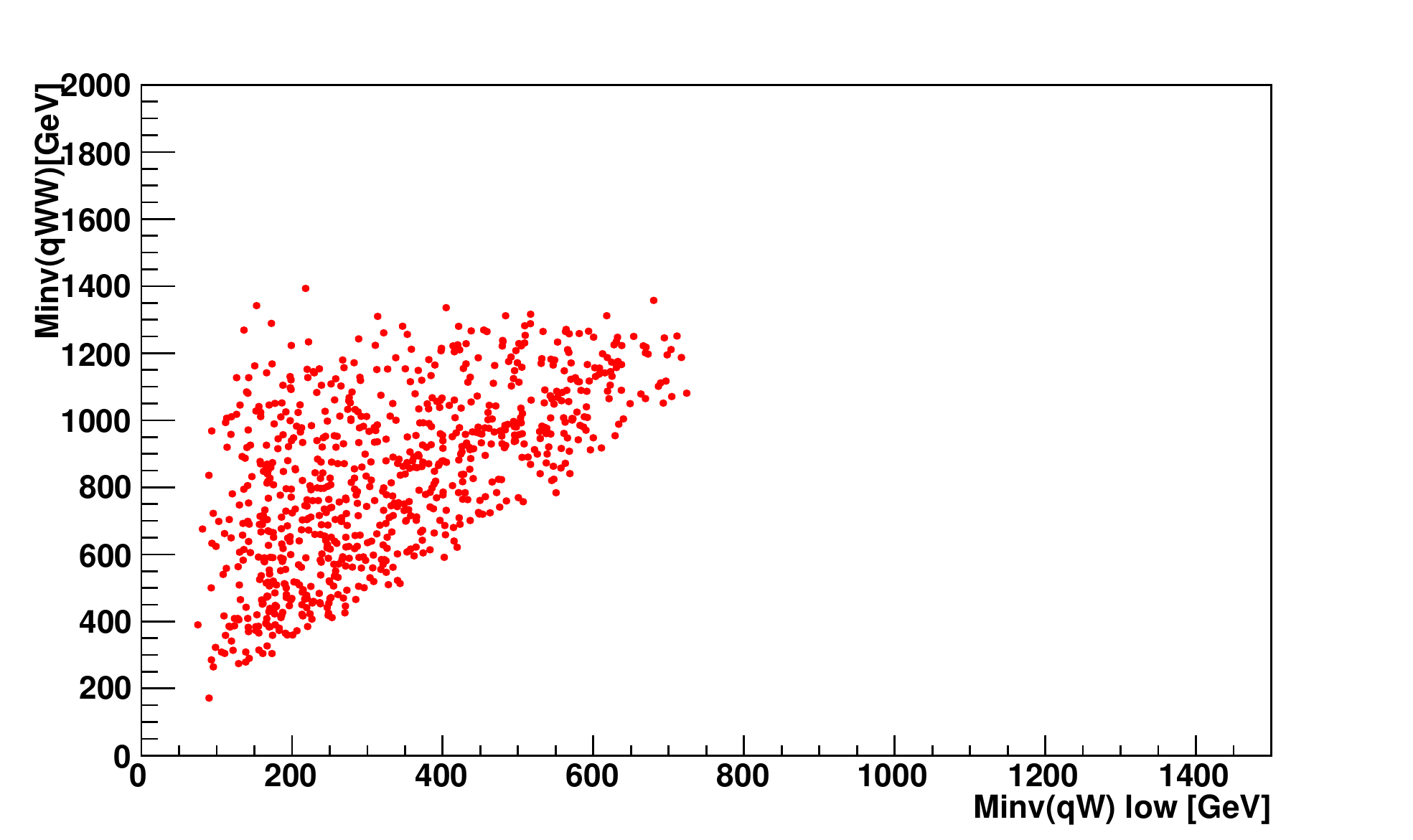}
  \includegraphics[width=0.49\textwidth]{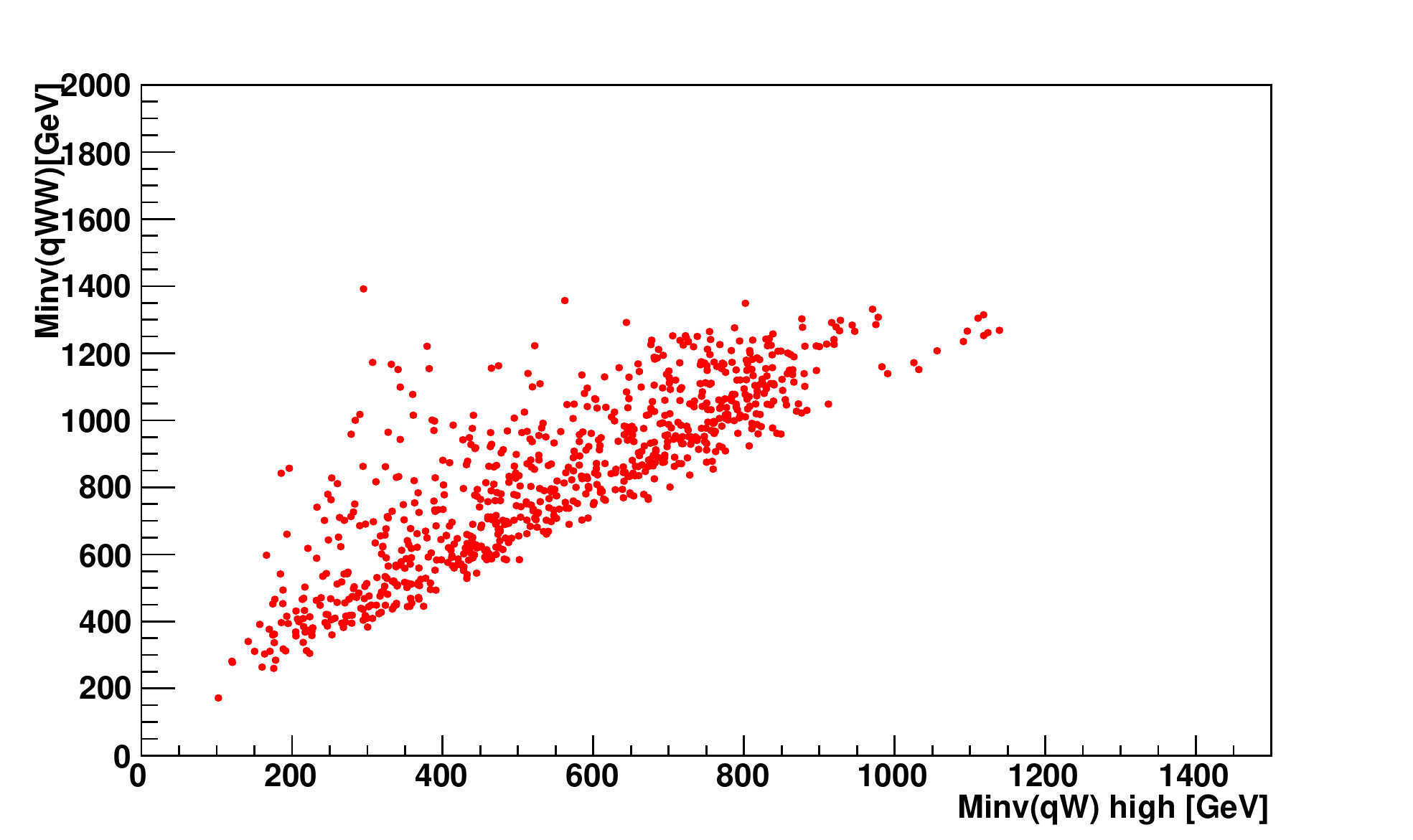}
\caption{Correlation plots of $M(qV_1V_2)$ versus ${\min}_{i=1,2}\{M(qV_i)\}$ (left panel)  and versus $\max_{i=1,2}\{M(qV_i)\}$ (right panel).
}
 \label{fig.simul.10jet.Minv10jqWlowhig2D}
 \end{center}
\end{figure}

\begin{figure}[!htb]
 \begin{center}
  \includegraphics[width=0.49\textwidth]{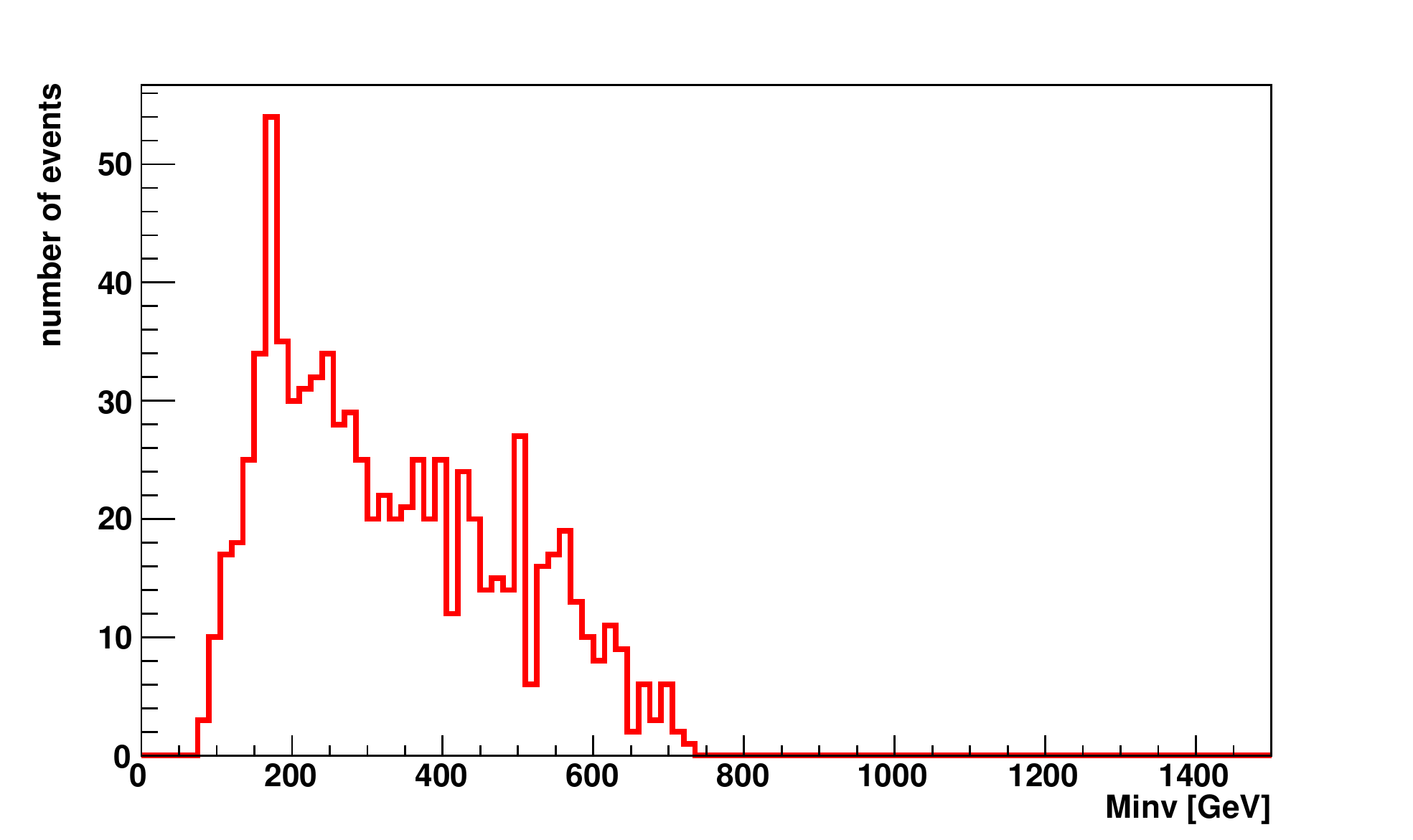}
  \includegraphics[width=0.49\textwidth]{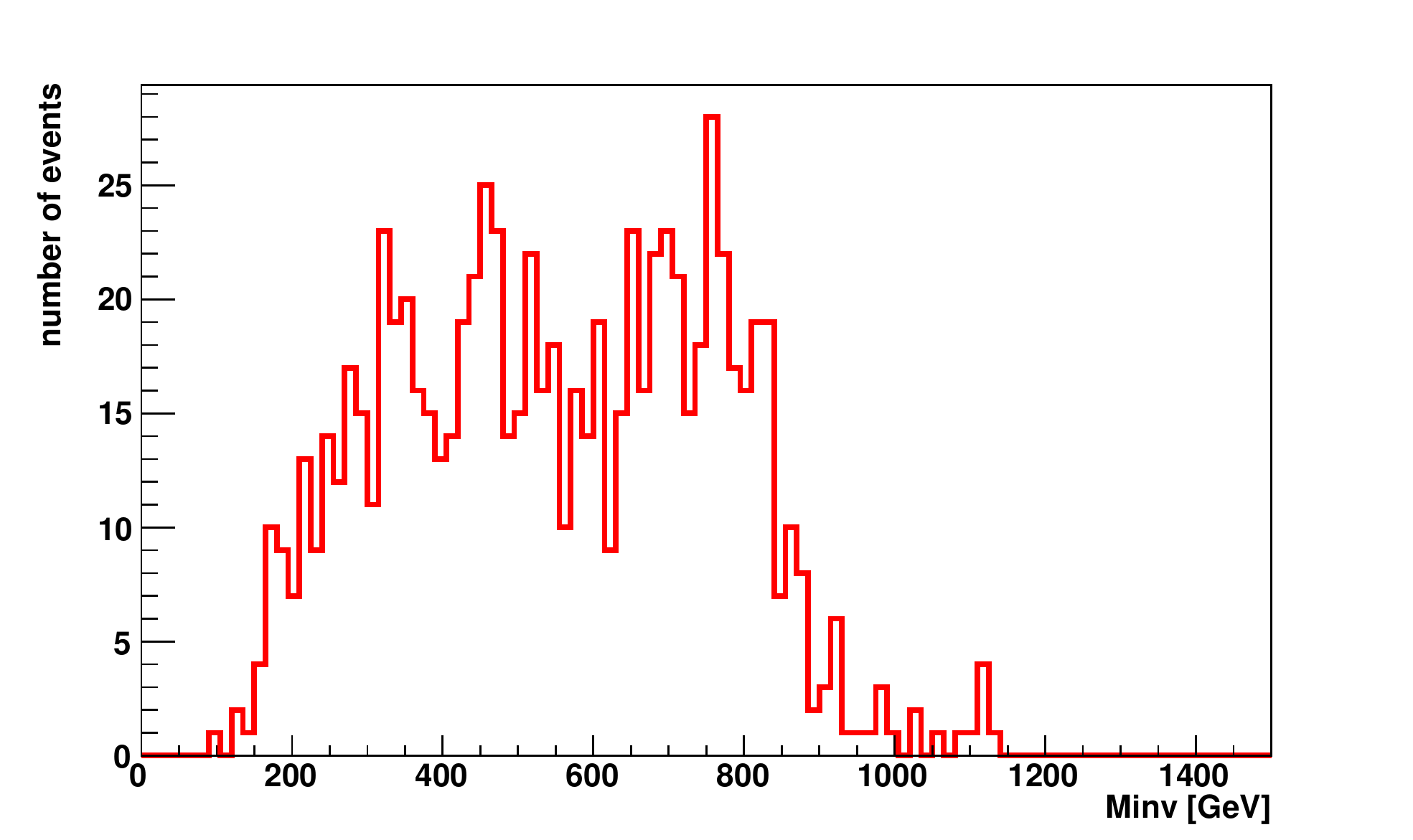}
\caption{Invariant mass distribution of ${\min}_{i=1,2}\{M(qV_i)\}$ (left panel)  and $\max_{i=1,2}\{M(qV_i)\}$ (right panel).
}
 \label{fig.simul.10jet.Minv10jqWlowhig}
 \end{center}
\end{figure}

For $M(qV_1\cup qV_2)$, we have two entries per hemisphere and we do not know which 
one corresponds to the first ($V_1$) and which to the second($V_2$) boson in the decay sequence.
However, following up on the idea from Section~\ref{sec:4jet}, we can study the correlation 
of the two $M(qV)$ entries with $M(qV_1V_2)$. Just as before, we would expect a  
strong correlation for $V_2$, and a weaker one for $V_1$.
This is investigated in Fig.~\ref{fig.simul.10jet.Minv10jqWlowhig2D}, where we plot
$M(qV_1V_2)$ versus ${\min}_{i=1,2}\{M(qV_i)\}$ (left panel)  and versus $\max_{i=1,2}\{M(qV_i)\}$ (right panel).
It is seen that the strongest correlation is for the high invariant mass solution, $\max_{i=1,2}\{M(qV_i)\}$, 
indicating correctly that this is the one corresponding to the case of $V_2$.
The individual ordered invariant mass distributions are shown in Fig.~\ref{fig.simul.10jet.Minv10jqWlowhig}.
For the lower invariant mass, an endpoint is present at about 700-750 GeV;
for the higher invariant mass an endpoint is seen around 900 GeV; 
both are in reasonable agreement with the expected values (computed numerically).

From the endpoints discussed above, four relations can be derived for the four unknown masses
for the particles $\sQua_L$, $\chipm_2$ (or $\chiz_{3, 4}$),
$\chipm_1$ (or $\chiz_2$) and $\chiz_1$.
In principle, this provides a complete determination of all four masses, on the basis
of invariant mass distributions alone.

\begin{figure}[!htb]
 \begin{center}
 \includegraphics[width=0.45\textwidth]{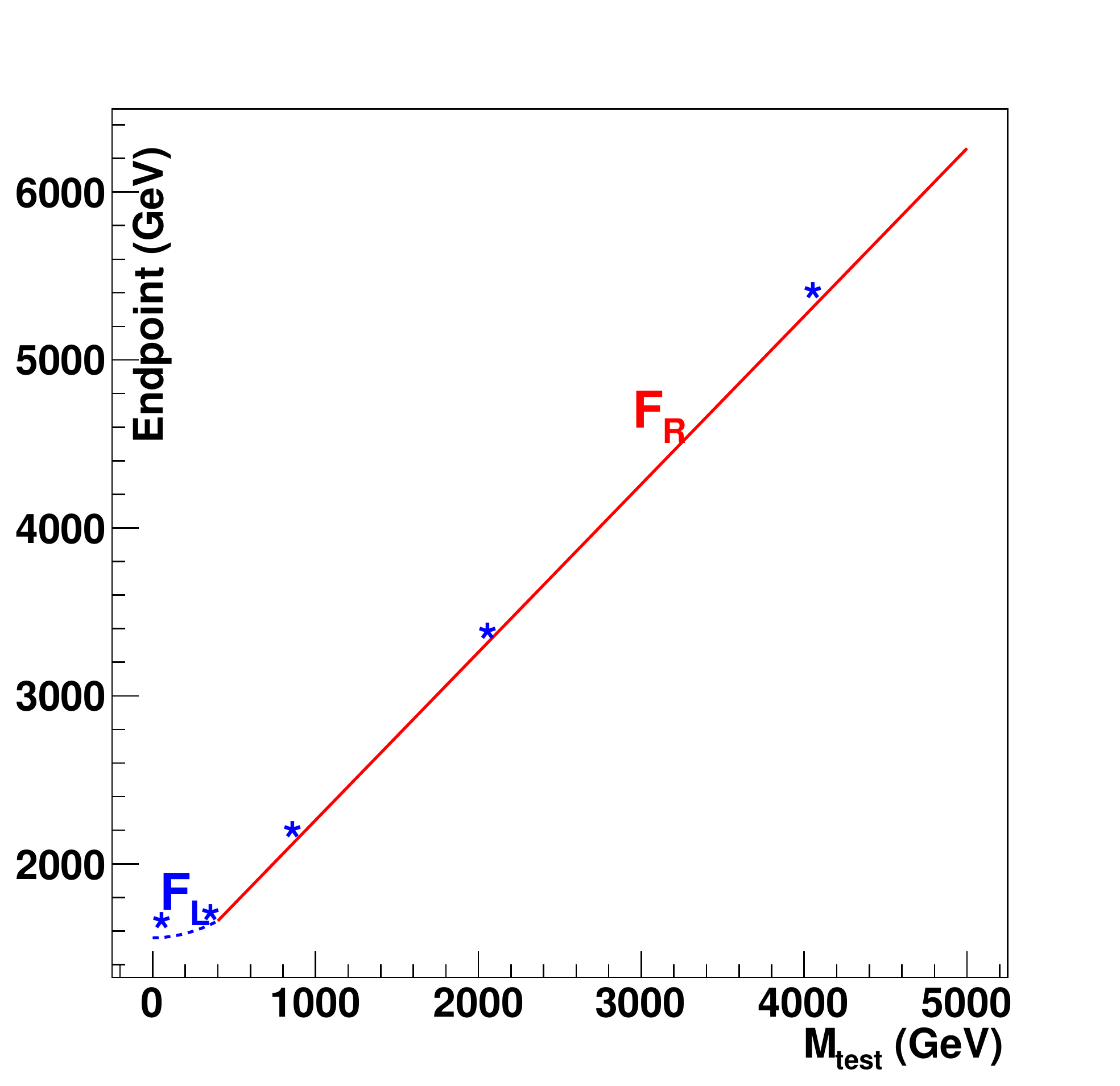}
\caption{Dependence of the $M_{T2}(qV_1V_2)$ endpoint on the test $\chiz_1$ mass and
five sample measurements for test mass values of 0, 300, 800, 2000 and 4000 GeV.
}
 \label{fig.simul.10jet.MT2endptbest}
 \end{center}
\end{figure}

Since 10-jet events have two well identified decay chains (one per hemisphere), one could also study the corresponding $M_{T2}$ and $M_2$ distributions,
and obtain useful information from their endpoints as well. The downside is that requiring a second 5-jet hemisphere in the event significantly reduces the available statistics.
Nevertheless, the distribution of $M_{T2}(qV_1V_2)$, for example, does exhibit a clear endpoint. In analogy to Fig.~\ref{fig.simul.fourjet.MT2},
one could then study the dependence on the observed endpoint to the input test mass for $\chiz_1$, as demonstrated in Fig.~\ref{fig.simul.10jet.MT2endptbest},
where the red dotted (blue solid) line corresponds to the theoretical prediction for test masses smaller (larger) than the true $\chiz_1$ mass.
The blue $\star$ symbols represent the results from repeated measurements of the endpoint, done for different values of the test mass. 
A similar analysis can be performed for the endpoints of the $M_{T2}(q)$ and $M_{T2}(V_1V_2)$ variables (and the corresponding $M_2$ counterparts), although the statistics becomes marginal.

In conclusion, we find that in 10-jet events, there is in principle enough kinematical information for the independent determination of
all four unknown masses in the decay chain (\ref{eq:tenjet.sqtoqX2}).

\subsubsection{Twelve-jet events} 

The twelve-jet event category (with 6 + 6 jets per hemisphere) is very complex and may have several origins. For example, it may be due to gluino pair-production, with
\begin{eqnarray}
 \sGlu \rightarrow q \sQua_L,
\end{eqnarray}
followed by the same decay as in (\ref{eq:tenjet.sqtoqX2}) for ten-jet events. 
It may also arise in events involving top squark decay chains, since they will produce SM top quarks, etc. 

Ignoring any $b$-tag information, the simplest invariant mass distributions within a 6-jet hemisphere (considering all events) are: $M (qqV_1V_2)$, $M(V_1V_2)$, $M(qq)$ and $M(qV_1V_2)$, 
where once again we select the best solutions for the bosons $V_1$ and $V_2$. 
The first three variables do not suffer from any combinatorial issues, and their distributions have clearly observed endpoints, 
providing three mass relations between the sparticles involved. 
In the case of $M(qV_1V_2)$ there are 2 possible combinations per hemisphere.
In principle, each combination will lead to an endpoint, but with our choice of mass spectrum the two endpoints ended up too close to be separately observed.
These measured endpoints, together with the results obtained previously in ten-jet events, allow the determination of the mass of the gluino $\sGlu$.

Just like in the case of 10-jet events, here one could also consider the $M_{T2}$ variables formed from (objects in) the two hemispheres. 
Unfortunately, the available statistics with $300\ {\rm fb}^{-1}$ integrated luminosity is insufficient for the useful extraction of any kinematic endpoints.

A variation of the previously described invariant mass analysis can be obtained if we make use of $b$-tagging, which would allow for the exclusive selection of a Higgs boson 
$h$ instead of the generic boson $V$. As before, we may consider the invariant mass distributions of $M(qqhV)$, $M(hV)$, $M(qq)$ and $M(qhV)$,
and the measured four endpoints will impose four mass constraints on the unknown particle masses. 
Weaker constraints, due to ambiguities in their interpretation, can be imposed from the endpoints of $M(qh)$, $M(qqh)$, $M(qV)$ and $M(qqV)$. 
Nevertheless, they could still be used in an overall fit of all hadronic channels.

\begin{figure}[!htb]
 \begin{center}
 \includegraphics[width=0.49\textwidth]{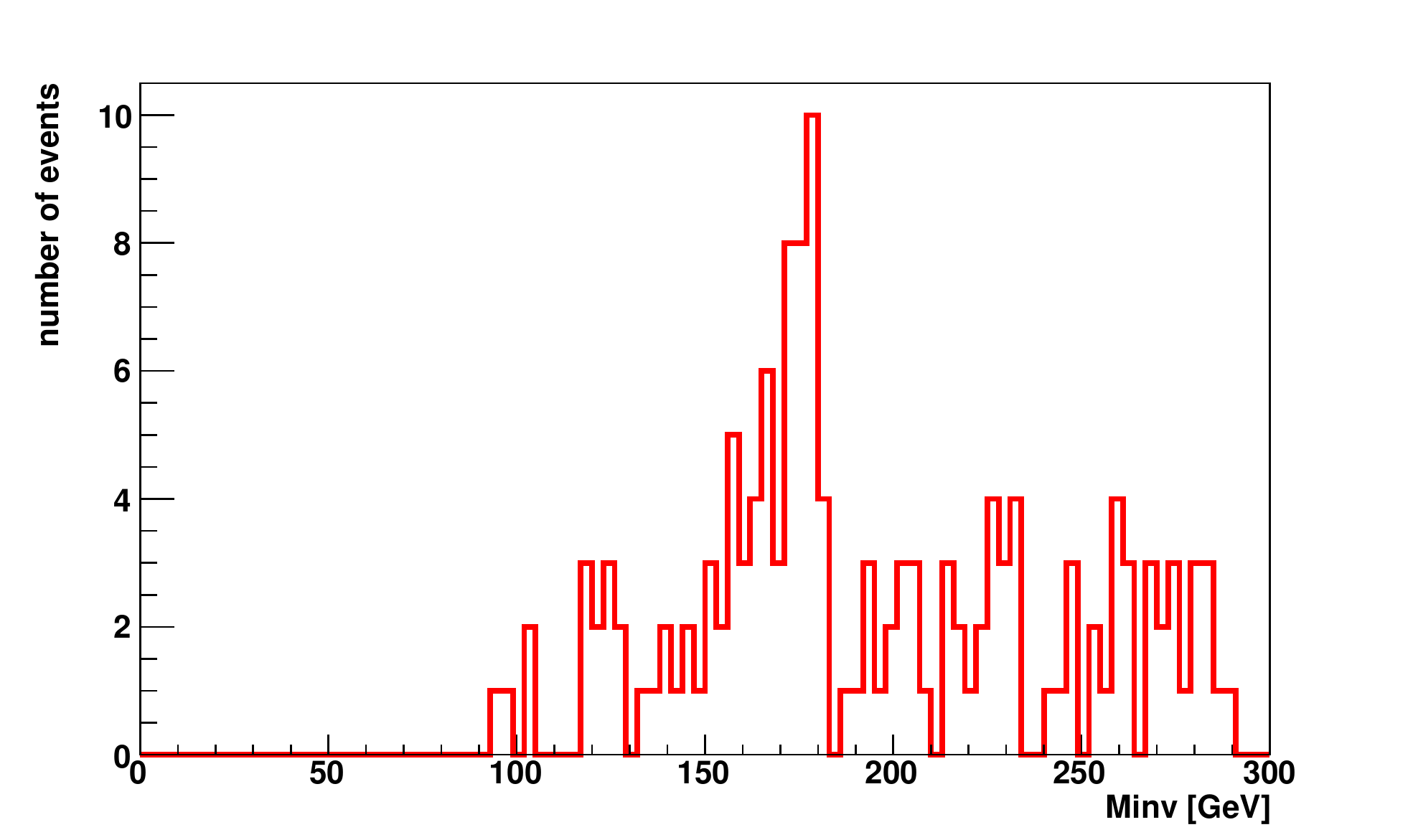}
\caption{Invariant mass $M(b V)$ in 12-jet events.
}
 \label{fig.simul.12jet.Minv12jWb}
 \end{center}
\end{figure}

Finally, using events where the left over jets are b-tagged, it was found that the invariant mass distribution of $(bW)$ exhibits a peak at the top mass,
as seen in Fig. \ref{fig.simul.12jet.Minv12jWb}. Such identification of top quarks in the events may open the door for a potential measurement of the 
mass of the top squark $\sTop$.
However, it is evident from Fig. \ref{fig.simul.12jet.Minv12jWb} that the available statistics is insufficient to analyze these events much further, and
such an analysis will have to wait for the HL-LHC. Another opportunity to place a (possibly rather weak) constraint on the $\sTop$ mass 
is provided by the measurement of the endpoint of the $(bbWW)$ invariant mass,
which relates the masses of $\sGlu$, $\sTop_1$, $\chipm_1$ and $\chiz_1$. 
This, together with the measurement of the $\sGlu$, $\chipm_1$ and $\chiz_1$ masses in previous analyses 
would allow the mass of the $\sTop$ to be determined.

In conclusion, we realize that analyses involving a large number of jets may be an idealization of the real ones, 
in which boosted objects, bosons and even taus, may be treated in a different, more optimal, way, 
e.g. as single jets of high mass and large radius. 
Given the masses adopted in the mixed scenario, 
the decay products from $\sQua$ and the highest neutralinos/charginos are the most likely to be boosted and prone to such different procedures.

\subsection{Summary}

\subsubsection{The mixed scenario}

The mass determination is summarized in Table \ref{tab.simul.summary.masses}.
\begin{table}[htb]
\begin{center}
\begin{tabular}{|c|ccccccc|} \hline
       				& $\sGlu$ & $\sQua_R$ & $\sQua_L$	& $\sTop_1$ 	& $\chiz_1$	& $\chipm_1$ 	& $\chipm_2$     \\
\hline
$W / Z / h$ incl 		&       	&       		&       		&       		&    X  		&    X  		&    X  	       \\
2 $W / Z / h$ incl 	&       	&       		&       		&       		&    V  		&    V  		&    V  	       \\
2j 				&       	&    V  		&       		&       		&    V  		&       		&       	       \\
4j 				&    X  	&    X  		&       		&       		&    X  		&       		&       	       \\
10j 				&       	&       		&    X  		&       		&    X  		&    X  		&    X  	       \\
12j no b 			&    V  	&       		&    V  		&       		&    V  		&    V  		&    V  	       \\
12j $\geq$ 1 b 		&    V  	&       		&    V  		&       		&    V  		&    V  		&    V  	       \\
 12j t 			&       	&       		&       		&    V 		&    V  		&    V  		&       	       \\
\hline
\end{tabular}
\caption{A summary of the mass determination in the Mixed scenario.
The "V" symbol indicates that relations between the masses are obtained, but their determination requires extra information.
The "X" symbol implies that the masses can be determined.
The heavier neutralinos are not explicitly included because they are nearly degenerate with the charginos.
}
\label{tab.simul.summary.masses}
\end{center}
\end{table}
We observe that  if SUSY is discovered in a Mixed scenario, the purely hadronic final states discussed in this section will
enable us to determine the masses of 
$\sGlu$, $\sQua_R$, $\sQua_L$, $\sTop_1$, $\chipm_2$, $\chipm_1$ and $\chiz_1$.
However, in several cases the analyses were limited by the available statistics for an integrated luminosity of $300\ {\rm fb}^{-1}$,
thus they would benefit considerably from collecting the HL-LHC luminosity of $3000\ {\rm fb}^{-1}$
and/or to increasing the energy like for the HE-LHC of 30 TeV.

\subsubsection{The other three scenarios}

\begin{figure}[!htb]
 \begin{center}
 \includegraphics[width=0.49\textwidth]{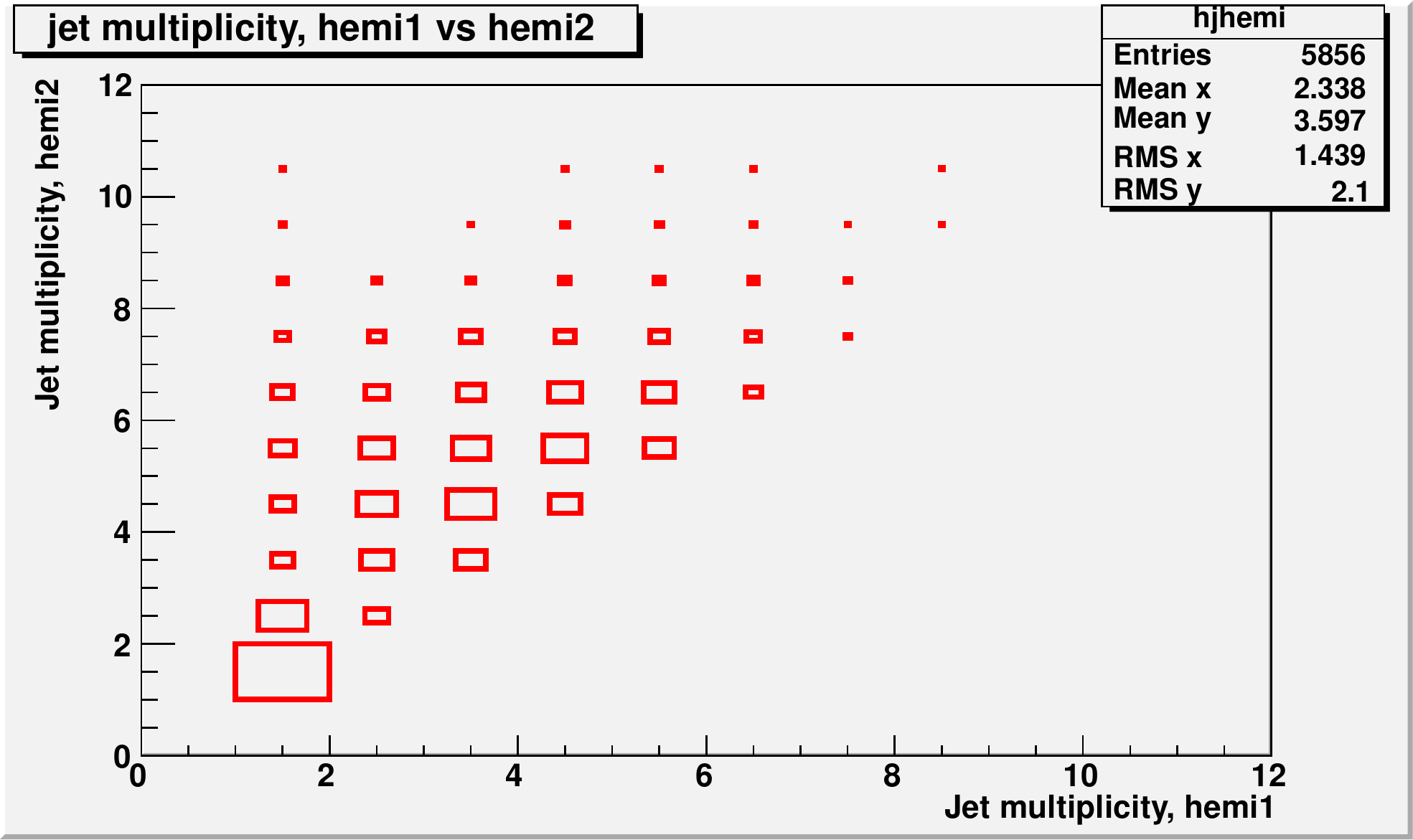}
 \includegraphics[width=0.49\textwidth]{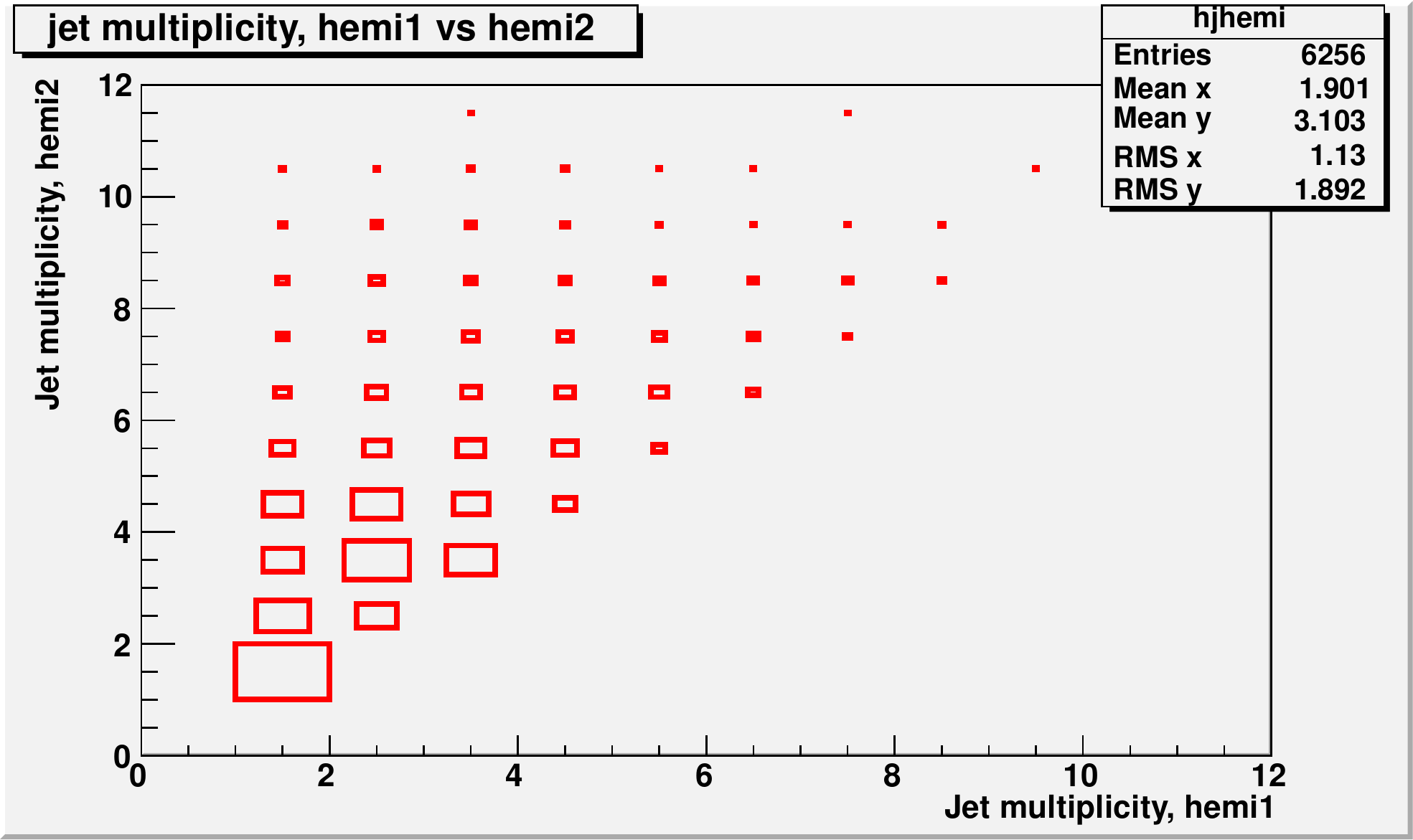}\\
 \includegraphics[width=0.49\textwidth]{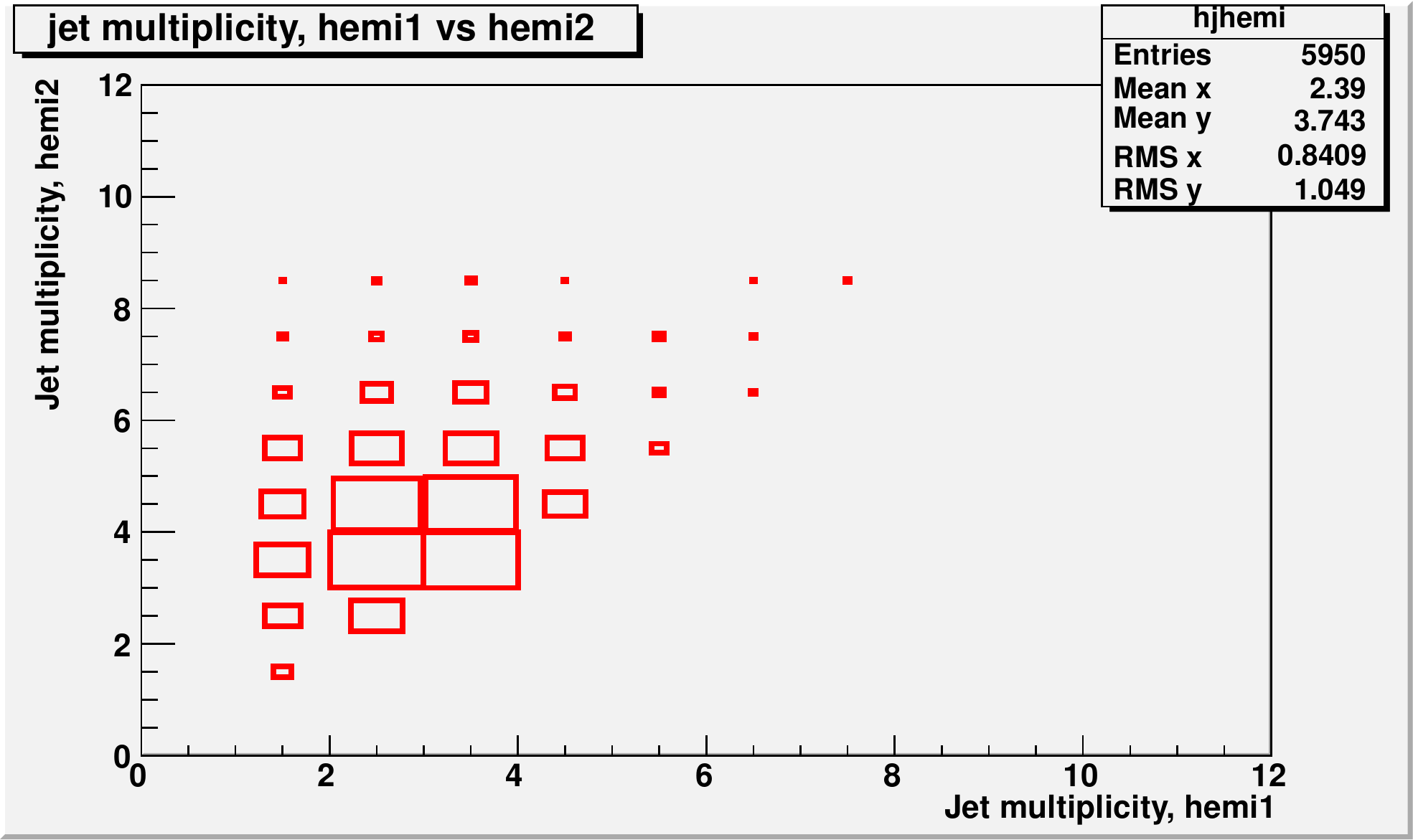}
 \includegraphics[width=0.49\textwidth]{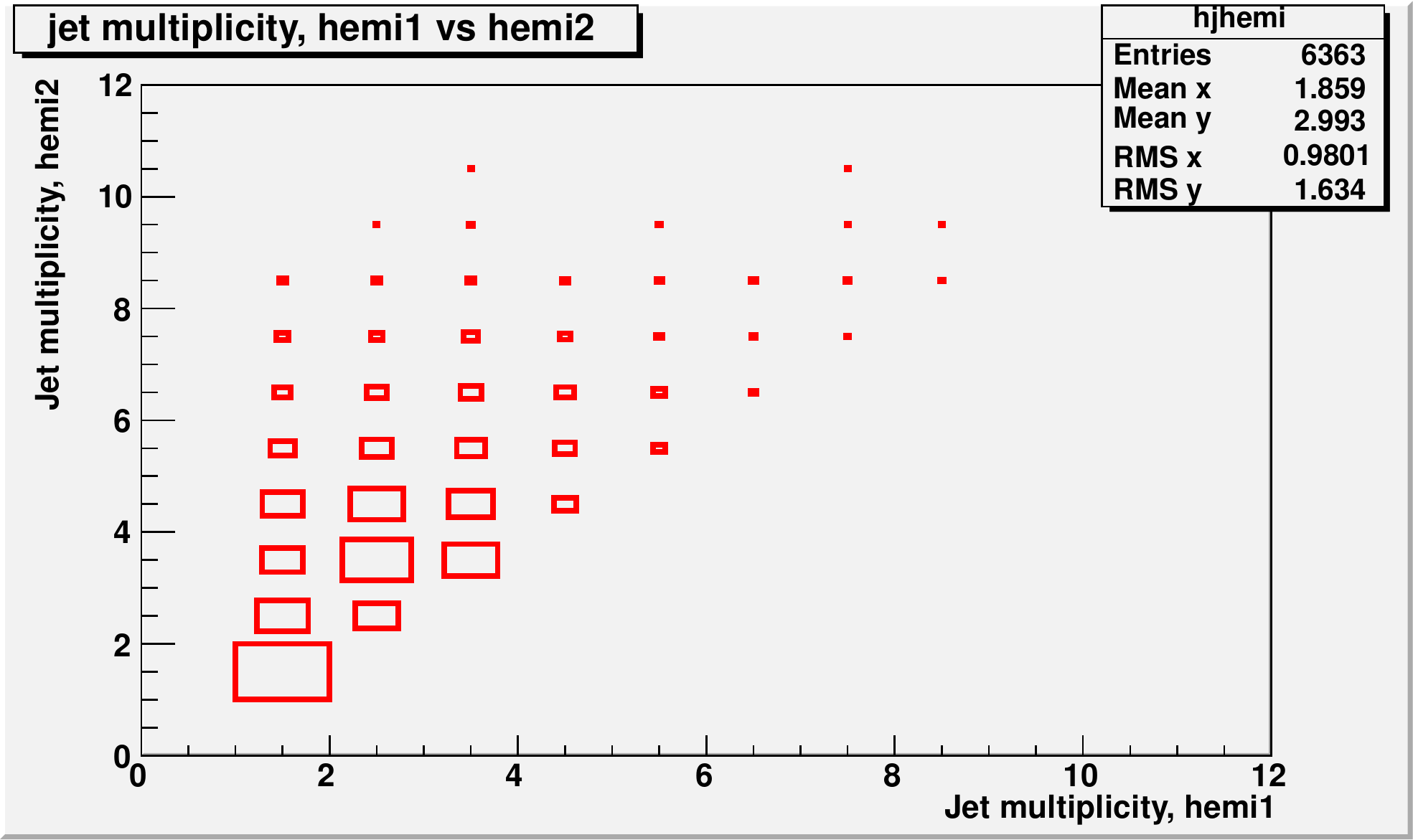}
 \caption{Jet multiplicities in the two hemispheres for the different scenarios: (clockwise from upper left) the mixed scenario, the Bino scenario, the Wino scenario and the Higgsino scenario.}
 \label{fig.simul.stats.hemimult}
 \end{center}
\end{figure}

So far, we have analyzed the "Mixed" scenario from Sec.~\ref{sec:mixed}.
We may also wonder to what extent the conclusions reached for the Mixed scenario would hold for the other three scenarios
considered in Sec.~\ref{sect:scenarios}.
In principle, the techniques used to analyze the various topologies will apply to the other scenarios as well,
and the results will be expected to be similar, as long as the same statistics is available.
To get a feeling of what would be achievable in the other scenarios, we can compare the available statistics for the different event topologies.
\begin{table}[htb]
\begin{center}
\begin{tabular}{|c|c|c|c|c|} \hline
       				& Mixed	& Bino 	& Wino	& Higgsino 	\\
\hline
2j 				& 1643 	& 1669	& 1613  	& (34)   		\\
4j 				& 104	& 299  	& 332 	& (292) 		\\ 
6j 				& (166)	& 443	& 523  	& 958 		\\
8j 				& (179) 	& 79  	& 88   	& 178  		\\
10j 				& 167	& (28)   	& (28)    	& (25)   		\\
12j 				& 48  	& 22   	& 6	   	& 3     		\\
\hline
\end{tabular}
\caption{Available signal statistics for $300 \; {\rm fb}^{-1}$ in various multijet event topologies for the different scenarios from Sec.~\ref{sect:scenarios}.
The first column ($nj$) counts the total number of jets in symmetric event topologies with $n/2$ jets per hemisphere after experimental cuts and hemisphere reconstruction.
Numbers in parentheses correspond to event topologies where most of the events have migrated from other topologies, and are
therefore unreliable.
}
\label{tab.simul.summary.others}
\end{center}
\end{table}
Table~\ref{tab.simul.summary.others} lists the available signal statistics for $300 \; {\rm fb}^{-1}$ in various multijet event topologies for the different scenarios from Sec.~\ref{sect:scenarios}.
As the mass determination is often obtained from the $M_{T2}$ distributions, in Table~\ref{tab.simul.summary.others} we continue to focus on symmetric events, 
with $n/2$ jets per hemisphere after experimental cuts and hemisphere reconstruction (the complete classification of events by 
jet multiplicity per hemisphere is pictorially shown in Fig.~\ref{fig.simul.stats.hemimult}).
As expected from our discussion in Sec.~\ref{sec:wino}, the Bino and the Wino scenarios have very similar statistics to each other
(compare the right panels in Fig.~\ref{fig.simul.stats.hemimult}).
They also have comparable statistics to the Mixed scenario, but differ strikingly from the Higgsino scenario.

The following quick conclusions can be drawn from Table~\ref{tab.simul.summary.others} 
and Fig.~\ref{fig.simul.stats.hemimult} (further information can be found in Appendix \ref{sect:app:simul.others}).
\begin{itemize}
\item \textbf{Bino scenario.} This is the scenario which is most studied in the literature, albeit rarely for purely hadronic final states as we are doing here.
There is  sizable statistics in 2j and 4j events, which will allow a measurement of the masses of $\sGlu$, $\sQua_R$ and $\chiz_1$. In addition, there is
reasonable statistics in 6j events, which can be interpreted as $\sQua_L \rightarrow q \chi_2 \rightarrow q V \chiz_1$, with
$\chi_2 = \left\{\chipm_1, \chiz_2\right\}$ and $V = \{W, Z^0, h^0\}$.
From this, we can measure the masses of $\sQua_L$, $\chipm_1$ ($\chiz_2$) and $\chiz_1$.
Altogether, in this scenario we measure the masses of $\sGlu$, $\sQua_L$, $\sQua_R$, $\chipm_1$ ($\chiz_2$) and $\chiz_1$.
However, neither the 10j nor the 12j samples are usable.
Furthermore, there is no sign of the existence of $\chipm_2$, $\chiz_3$ and $\chiz_4$, and the mass of $\sTop_1$ is unlikely to be measurable.
\item \textbf{Wino scenario.}
It also has sizable statistics in 2j and 4j events, for which the decays are similar to those of the Bino scenario, 
except that now $\sQua_L$ plays the role of $\sQua_R$. This will allow a measurement of the masses of $\sGlu$, $\sQua_L$ and $\chiz_1$.
As in the Bino scenario, neither the 10j nor the 12j samples are usable.
However, there is reasonable statistics in 6j events, which can be interpreted as $\sQua_R \rightarrow q \chiz_2 \rightarrow q V \chi_1$, with
$V = \{W, Z^0, h^0\}$ and $\chi_1 = \{\chipm_1, \chiz_1\}$.
From this decay chain one could measure the masses of $\sQua_R$, $\chiz_2$ and $\chiz_1$ ($\chipm_1$).
However, again there will be no sign of the existence of $\chipm_2$, $\chiz_3$ and $\chiz_4$.
There is also some marginal statistics in 8j events, which
may originate from $\sGlu \rightarrow q \sQua_R$ followed by the same decay of $\sQua_R$ as above.
These events could provide some further constraints, helping the 6j analysis, but cannot clarify the ambiguity encountered above.
Finally, the mass of $\sTop_1$ is again unlikely to be measurable.
\item \textbf{Higgsino scenario.} As seen in Fig.~\ref{fig.distscenarios.Hino}
the Higgsino scenario is characterised by the fact that the decay chains for both 
$\sQua_L$  and $\sQua_R$ result in a boson, $V=\{W, Z^0, h^0\}$, which increases the jet multiplicity. 
As a result, 2j and 4j events are not expected to be produced in fully hadronic decays in this scenario,
as can be easily verified from Tables \ref{tab.scenarios.BR} and \ref{tab.scenarios.BRstop}.
The surviving events in the 2j and 4j categories are due to one or both bosons decaying leptonically,
where the lepton was lost due to cuts, and are thus not usable
for mass reconstruction in a fully hadronic analysis.
The largest available statistics is in 6j events, which are most likely due to squark pair production and decay,
either 
$\sQua_L \rightarrow q \chi_4 \rightarrow q V \chi_1$, with 
$\chi_4=\{\chipm_2, \chiz_4\}$ and $\chi_1=\{\chipm_1, \chiz_2, \chiz_1\}$, 
or
$\sQua_R \rightarrow q \chiz_3 \rightarrow q V \chi_1$.
These two types of decays should produce different endpoints, as the masses of $\sQua_L$ and $\sQua_R$,
as well as of $\chi_4$ and $\chiz_3$, are different. Even if all endpoints can be observed, their interpretation will remain ambiguous,
as we will not be able to resolve the twofold ambiguity
$\sQua_L \leftrightarrow \sQua_R$ and $\chiz_4\leftrightarrow \chiz_3$.
(This is indeed supported by the analysis summarized in Appendix \ref{sect:app.simul.Hino}.)
Similarly to the Wino scenario, the 8j events, which are due to $\sGlu \rightarrow q \sQua_L (\sQua_R)$,
may bring some additional support, but they do not allow this ambiguity to be solved.
Once again, the mass of $\sTop_1$ is unlikely to be measurable.
\end{itemize}
We may conclude that if physics is realized in one of these alternative scenarios,
it will still be possible to measure the masses of $\sGlu$ and $\sQua$,
and to unravel the lower part of the electroweakino spectrum, but the higher electroweakino states remain unnoticed.
Moreover, the mass of $\sTop_1$ remains difficult to determine due to the lack of statistics, 
as it is only measurable in events with 12 or more jets.
The availability of the HL-LHC or HE-LHC may possibly allow a measurement of the $\sTop_1$ mass.

\section{Leptonic Final States with Decays through Sleptons}
\label{sect:leptslept}

Several final states with leptons originating from SUSY cascades with decays through on-shell sleptons
have been extensively studied in the literature. However, one should keep in mind that 
most existing studies have been done in SUSY models leading to a Bino-like scenario, with
the sleptons appearing as intermediate states between the Bino and the Winos, allowing for 
Wino-to-Slepton-to-Bino transitions.
Here we will summarize the important results as they pertain to our scenarios,
including some channels which previously have not been sufficiently investigated.

\subsection{Single lepton final states}
\label{sect:1lept}

Events with single leptons are expected to be the most copiously produced among all leptonic final states.
Their  discovery reach in mass may be comparable to the one of hadronic final states, depending of the branching fraction to sleptons.
The type of decay targeted by leptonic analyses is typically one where a squark decays to a chargino, followed by a decay to an on-shell charged
slepton $\sLep$ or sneutrino $\sNu$
\begin{eqnarray}
 \chipm_1 \rightarrow \{ l \sNu  , \nu \sLep \} \rightarrow l  \nu \chiz_1.
\end{eqnarray}
The squark from the other decay chain would then decay hadronically.

The single lepton channel provides an interesting test of lepton universality.
Non-universality might point to different masses for $\sEl$ and $\sMu$.
Unfortunately, there does not seem to be other interesting physics to extract.

\subsection{Same-flavor opposite-sign di-lepton final states, looking for $\chiz_2$}
\label{sect:OSdilchiz}

The Same Flavour Opposite Sign (SFOS) di-lepton events are targeting the decay chain
\begin{eqnarray}
\sQua \rightarrow q \chiz_2 \rightarrow q (l_1 \sLep )  \rightarrow q (l_1 l_2 \chiz_1),
\label{eq:OSdilchiz.chizdec}
\end{eqnarray}
in which the flavor and the charges of the two leptons $l_1$ and $l_2$ are fully correlated.
For a true di-lepton event, the other decay chain should result in a hadronic final state.)
This decay sequence has been extensively studied and was at the origin of the first pioneering studies
of mass reconstruction, see e.g. \cite{Bachacou:1999zb}.
This channel also enables us to make a test of lepton universality in the $\chiz_2$ decay   \cite{Goto:2004cpa}.

The measurement of the edge of the di-lepton invariant mass distribution 
provides an algebraic relation between the masses of the $\chiz_2$, the $\sLep$ and the $\chiz_1$
and is thus insufficient to determine all the masses.
To proceed further, two difficulties are encountered.
The first is to which jet the di-lepton system is associated,
the second is to identify who is the first lepton ($l_1$) and who is the second lepton ($l_2$) in the decay chain.
Both difficulties have been solved in principle \cite{Allanach:2000kt,Georgia08},
producing three more endpoints, for $M(l_1 q)$, $M(l_2 q)$ and $M(l l q)$.
These 4 endpoints allow us to determine the masses of $\sQua$, $\chiz_2$, $\sLep$ and $\chiz_1$ \cite{Gjelsten:2004ki}.

\subsection{Different-flavor opposite-sign di-lepton final states, looking for $\chipm_1$}
\label{sect:OSdilchipm}

In this case, each decay leg corresponds to a squark or antisquark decaying leptonically via a chargino $\chipm_1$:
\begin{eqnarray}
 \sQua &\rightarrow& q \chipm_1 \rightarrow q (l \sNu )  \rightarrow q (l \nu \chiz_1) \nonumber \\
 	 &\rightarrow& q \chipm_1 \rightarrow q (\nu \sLep ) \rightarrow q (\nu l \chiz_1) 
\label{eq:OSdilchipm.chipmdec}
\end{eqnarray}

Here the lepton flavors are uncorrelated, so by focusing on the Different Flavor Opposite Sign (DFOS) di-lepton events, 
one can be sure that there is no contribution from (\ref{eq:OSdilchiz.chizdec}).

\subsubsection{Mass reconstruction in decays with $\sNu$}
\label{sect:OSdilchipm.snu}

The first decay mode in (\ref{eq:OSdilchipm.chipmdec}), through the sneutrino, 
is a two-step decay with available endpoints for the overall $M_{T2}(q l)$, $M_{T2}(q)$ and $M_{T2}(l)$.
Taking the $M_{T2}(q l)$ endpoint, all masses can be determined from 3 values of the test mass.
There is, however, a two-fold ambiguity in the solution \cite{Burns:2008va}, which can be solved by using another constraint on the masses,
for example from the endpoint of $M_{T2}(q )$.
Finally, we can use the $M_{T2}(l)$ endpoints and again make three measurements, as in the case for $M_{T2}(q l)$.
This again determines all three masses,
demonstrating that the masses of $\sQua$, $\chipm_1$ and $\sNu$ are over-determined \cite{Burns:2008va}.

\subsubsection{Mass reconstruction in decays with $\sLep$}
\label{sect:OSdilchipm.slep}

The case of the second decay mode in (\ref{eq:OSdilchipm.chipmdec}) has not been extensively studied so far and here we largely 
report results from \cite{Lept}.
This case is special as it contains two unseen neutral particles in each decay chain.
Under these circumstances, it is observed that the $M_{T2}(q l)$ endpoint at the correct test mass value underestimates the parent mass.
The correct endpoint is recovered, however, by computing instead the $M_2$ with additional mass constraints imposed,
namely the equality of the two $\sQua$, $\chipm_1$ and $\sLep$ masses between the two decay chains.
There exist several ways to determine the masses.
One of them would be to determine the endpoints of the $M_{T2}(q l)$, $M_{T2}(q)$ and $M_{T2}(l)$ distributions,
which provide three constraints on the masses.
A fourth one may be obtained from using the $\chiz_1$ momentum resulting from the minimization of  $M_{2}(q l)$
and computing the invariant mass $M(l \chiz_1)$, which exhibits a peak at the $\sLep$ mass.
An alternative might be to use the relation between masses from the endpoints of the invariant mass $M(q l)$ endpoint.
All four masses, of $\sQua$, $\chipm_1$, $\sLep$ and $\chiz_1$, can thus be determined.

\subsection{Same-sign di-lepton final states}
\label{sect:SSdilept}

Same Sign (SS) di-lepton events can be produced by the decays of two same-sign charginos as in (\ref{eq:OSdilchipm.chipmdec}).
In order for this to be the case, the charginos must originate from the production of either two squarks, $\sQua$, or two anti-squarks,  $\bar{\sQua}$,
which may occur by a $t$-channel exchange of a gluino or from the pair production of gluinos.
The same-sign di-lepton channel has long been viewed as endowed with an excellent discovery potential given its small SM backgrounds
\cite{Nachtman:1999ua},
but it was only later realized that it also has an excellent potential for mass reconstruction \cite{Matchev:2009fh}.

The mass reconstruction in this channel can be treated in the same way as the OFOS events from charginos in Sec.~\ref{sect:OSdilchipm}.
The study in reference \cite{Matchev:2009fh} concentrated on the first decay of eq. (\ref{eq:OSdilchipm.chipmdec}) where the unseen neutral is a $\sNu$.
The interest of this approach is that a full mass determination is possible using only the leptons,
but exploiting the dependence on the upstream $P_T$.
Three different methods were proposed for the reconstruction of the individual masses of $\chipm_1$ and $\sNu$.
Two of these methods use the UTM, made of the ISR jets and potential decays of squarks and gluinos.
An advantage of this approach is that there is no need to identify ISR jets as such.
Obviously, the same approach can be applied to the  OFOS ($e^\pm \mu^\mp$) events of Sec.~\ref{sect:OSdilchipm}.

\subsection{Tri-lepton final states}
\label{sect:3lept}

In most cases studied in the literature, tri-lepton final states will typically originate from the inclusive production of 
one charged and one neutral electroweakino, e.g., $\chiz_2 \chipm_1$, one in each decay chain.
The $\chiz_2$ yields a SFOS lepton pair from the decays listed in (\ref{eq:OSdilchiz.chizdec}), while
the $\chipm_1$ produces a single lepton as in (\ref{eq:OSdilchipm.chipmdec}).
(In principle, for long enough decay chains, it may also be possible to obtain same-sign tri-lepton events, but their 
production rate tends to be extremely small.)

The clean final state should allow the masses to be reconstructed using invariant mass and $M_{T2}$ techniques as
in the previous sections, as the mass difference between $\chipm_1$ and $\chiz_2$ can typically be neglected.
In order to avoid additional combinatorial ambiguities, it is worth concentrating on the 
$e^+ e^- \mu^{\pm}$ and $\mu^+ \mu^- e^{\pm}$ final states.
However, these events have not been analyzed in detail so far 
(see \cite{Konar:2009qr} for a study employing an asymmetric $M_{T2}$ variable \cite{Barr:2009jv}).

\subsection{Four-lepton final states}
\label{sect:4lept}

Four-lepton final states should originate from the production of $\chiz_2 \chiz_2$, one in each decay chain,
where both $\chiz_2$ decay via $\sLep l$ as in (\ref{eq:OSdilchiz.chizdec}).

Here again, charge rates could be investigated and lepton universality can be tested.
Mass reconstruction would be a straightforward application of the SFOS analysis described in Sec.~\ref{sect:OSdilchiz},
but here again the main challenge is related to the production rates.

\subsection{Leptons from stop decays}
\label{sect:leptstop}

With the stop being a mixed state and $\sTop_1$ being by definition lighter than $\sTop_2$,
it is most likely that the first observation will be of $\sTop_1$.
The most relevant decay modes are then
\begin{eqnarray}
 \sTop_1 &\rightarrow& b \chipm_1 \rightarrow b (l \sNu )  \rightarrow b (l \nu \chiz_1) \nonumber \\
	 &\rightarrow& b \chipm_1 \rightarrow b (\nu \sLep ) \rightarrow b (\nu l \chiz_1) \nonumber \\
 	 &\rightarrow& t \chiz_1 \rightarrow (b W) \chiz_1 \rightarrow (b l \nu) \chiz_1
\label{eq:leptstop.chipmdec}
\end{eqnarray}
all leading to the same final state topology (the leptons can be either flavor).

The first decay mode can be analyzed as in Sec.~\ref{sect:OSdilchipm.snu}
and its mass reconstruction is straightforward.
It enables us to determine the masses of $\sTop_1$, $\chipm_1$ and $\sNu$.

The second decay mode is analogous to the case considered in
Sec.~\ref{sect:OSdilchipm.slep}.
Due to the presence of two unseen neutrals on each side of the event, $M_{T2}(bl)$ does not reproduce the correct $\sTop_1$ mass.
Fortunately,  using $M_2(bl)$, with equality constraints on the masses of $\sTop_1$, $\chipm_1$ and $\sLep$, 
the observed kinematic endpoint reaches the correct value \cite{Lept}.
Moreover, the invariant mass $M(l \chiz_1)$, where the $\chiz_1$ momentum is obtained from the $M_2$ minimization,
peaks at the correct value of the $\sLep$ mass, see Appendix~\ref{sect:MT2.M2}.
Two more relations between the masses can be obtained from $M_2(b)$ with the $\chipm_1$ as ``unseen neutral",
and from the endpoint of the invariant mass distribution $M(bl)$. All these constraints
in principle allow the masses of $\sTop_1$, $\chipm_1$, $\sLep$ and $\chiz_1$ to be determined \cite{Lept}.

The analysis described in the previous paragraph is largely based on the $M_2$ variables, but it
is still possible to find a useful application of the $M_{T2}$ variables \cite{Mahbubani:2012kx}.
It is well known that the left ($F_L$) and right ($F_R$) branches of the $M_{T2}(bl)$ endpoint functions 
cross at a special value of the $\chiz_1$ mass, where their $P_T$ dependence disappears, see, e.g., eq.~(\ref{eq:ptindep}).
The endpoint value at this mass reproduces the correct parent mass, and the 
location of this point is given by $M_{\chipm_1} M_{\chiz_1} / M_{\sLep}$.
Together with the endpoint of $M_{T2}(b)$ and the invariant mass $M(bl)$, there are in principle 
enough constraints to determine all masses, but
it is clear that the usage of $M_{T2}$ becomes more involved than simply switching to $M_2$.

The last decay mode $\sTop \rightarrow t \chiz_1$ is special as we know the masses of $t$ and $W$.
An interesting distribution is the invariant mass $M(b l)$, which endpoint is completely determined by the top decay,
and amounts to 153 GeV for a top mass  of 173 GeV.
Its observation is a smoking gun for this type of decay mode.

\subsection{Leptons from sbottom decays}
\label{sect:leptsbot}

Like for the case of stop, the lightest sbottom state (and likely the first one to be discovered) is the $\sBot_1$.
The most relevant decay modes are then
\begin{eqnarray}
 \sBot_1 &\rightarrow& b \chiz_2 \rightarrow b (l \sLep )  \rightarrow b (l l \chiz_1) \nonumber \\
	 &\rightarrow& b \chiz_2 \rightarrow b (Z \chiz_1) \rightarrow b (l l \chiz_1) 
\label{eq:leptsbot.chizdec}
\end{eqnarray}
which are the same types of decays as in (\ref{eq:OSdilchiz.chizdec}).
The mass reconstruction can thus proceed in the same way as for the SFOS events in Sec.~\ref{sect:OSdilchiz}.

\subsection{Summary of the mass determinations}

The potential mass determination from leptonic final states is summarized in Table \ref{tab.lept:simul.summary.massesl}
\begin{table}[htb]
\begin{center}
\begin{tabular}{|c|ccccccc|} \hline
       				& $\sQua$	& $\sTop_1$ 	& $\chiz_1$	& $\chiz_2$	& $\chipm_1$ 	& $\sLep$		& $ \sNu$     \\
\hline
SFOS 2l edge		&     X  		&       		&    X   		&    X   		&       		&    X  		&			\\
OS 2l $\sNu$		&     X 		&      			&       		&       		&    X  		&			&    X  	       \\
OS 2l $\sLep$		&     X 		&      			&    X  		&       		&    X  		&    X  		&			\\
SS 2l $\sNu$		&     X 		&      			&       		&       		&    X  		&			&    X  	       \\
SS 2l $\sLep$		&     X  		&      			&    X  		&       		&    X  		&    X  		&			\\
3l 				&     X  		&       		&    X 		&   X    		&    X  		&    X  		&			\\
2l $\sTop$ 		&       		&     X		&    X  		&       		&    X  		&    X  		&	X		\\
\hline
\end{tabular}
\caption{Masses which can be determined in the light $\sLep$ scenario.
An ``X" indicates that the mass can be determined.
}
\label{tab.lept:simul.summary.massesl}
\end{center}
\end{table}
for the case where the slepton $\sLep$ ($\sEl$ or $\sMu$) and the sneutrino $\sNu$ appear in the decays of $\chiz_2$ or $\chipm_1$.
It is seen that many masses can be determined in this scenario.
Needless to say, the projected power of these analyses depends strongly on the assumed decay branching fractions to sleptons.
Note that, as we can measure separately the mass of the $\sLep^{\pm}$ and of the $\sNu$,
a consistency test can be performed, as they should satisfy the model independent relation $m_{\sLep_L}^2 - m_{\sNu}^2 = - M_W^2 cos 2 \beta$.

\section{Leptonic Final States with Sleptons out of Reach}
\label{sect:lept}

When the sleptons are heavier than the electroweakinos, the leptonic branching fractions are 
typically suppressed and jetty events tend to dominate the signal yield.
This is why this case is rarely discussed in the literature in relation to leptonic final states.
Here a simple simulation was performed in the different scenarios, just like for the fully hadronic decays.
Leptons were selected if they have $p_T \geq 10$ GeV and $|\eta | \leq 2.4$.
The statistics accumulated with $300 \; {\rm fb}^{-1}$ for the different scenarios are listed in Table \ref{tab.lept:simul.summary.scenarios}.
\begin{table}[!htb]
\begin{center}
\begin{tabular}{|c|c|c|c|c|} \hline
       				& Mixed		& Bino		& Wino		& Higgsino	\\
\hline
1l 				&   1497		& 1343		&  1308		 &  1731		\\
SFOS 2l $Z^0$		&   189 		&  72  		&  44 		 &  116	      	 \\
OS 2l $e \mu$		&   111		&  52 		&  41			 &  66		\\
SS 2l			&   98		& 68  		&  50 		 &  62		\\
3l 				&   50		& 22  		& 16 			 &  21		\\
$\geq$4l 			&   6    		& 3    		& 1    		&   0 			\\
\hline
\end{tabular}
\caption{Event statistics for the different scenarios for the case of sleptons out of reach.
}
\label{tab.lept:simul.summary.scenarios}
\end{center}
\end{table}
We observe that the single lepton events dominate the statistics. They would allow a test of 
lepton universality to be performed, however, this should only be viewed as a sanity check since the decays of $W$ and $Z$ are known to satisfy lepton universality.
The number of multi-lepton events is significantly less, with the Mixed scenario having the largest statistics.

\subsection{Single lepton final states}
\label{sect:lept:simul.sinlep}

Single lepton final states are the most abundant in Table \ref{tab.lept:simul.summary.scenarios}.
One may try to estimate whether there is any contribution from direct slepton decays as in Sec.~\ref{sect:leptslept} as follows.
In Sec.~\ref{sect:had} we saw that the fully hadronic analysis, by looking for inclusive $W / Z / h$, can estimate 
the total number of observed $W / Z$s decaying hadronically, as well as their kinematics.
Then, from this kinematics, it should be possible to estimate the acceptances of $W / Z$ 
for both the hadronic, $A_h$, and leptonic, $A_l$, decays.
The number of observed events  from the decay $W \rightarrow l \nu$, $N_{W, lept }^{obs}$,
 can then be computed starting from the observed number of hadronic events, $N_{W, had}^{obs}$,
 \begin{eqnarray*}
 N_{W, lept }^{obs} = \frac{A_l}{A_h} N_{W, had}^{obs}.
\end{eqnarray*}
After subtracting this number from the total number of observed single lepton events,
the remaining events are a measure of the contribution from $\sLep \rightarrow l \chiz_1$.
As the single lepton events are expected to be detected before any multi-lepton final states,
it shows that we may know whether or not there is a contribution from sleptons in the events
well before their direct observation.

\subsection{Di-lepton final states}
\label{sect:lept:simul.dilep}

The di-lepton final states exhibit a clean $Z^0$ peak in the invariant mass distribution of the OSSF sample.
After selecting events inside the $Z^0$ peak, the invariant mass $M(q Z)$ was analyzed, 
but no clear endpoint could be observed, due to the marginal statistics and the difficulty of 
correctly assigning the jet accompanying the $Z^0$.

After vetoing events with dilepton mass in the $Z^0$ mass window, it is expected that the two leptons belong to different hemispheres
and reflect the production of two $\chipm_1$ decaying via a $W$.
The $M_2(q l)$ distribution was computed, but no clear endpoint is visible,
due to the difficulty to correctly pair the jets and leptons.
A clearer endpoint becomes, however, apparent in the SS di-lepton events, but the statistics is again marginal.

Stop production will lead to decay chains of the type $\sTop \rightarrow b \chipm$, $\chipm \rightarrow \chiz_1 W$ and $W \rightarrow l \nu$
with two unseen neutrals in each decay leg. Such events can be targeted by requiring two b-tagged jets and two OS leptons.
We can then study the kinematic endpoint of the distribution of $M_2(bl)$ with equal-mass requirements on the masses of $\sTop_1$ and $\chipm_1$,
and additionally enforcing the constraint $m(W) = 80$ GeV.
Adding the measurements of the kinematic endpoints of the $M_2(b)$ and $M(bl)$ distributions,
the masses of $\sTop_1$, $\chipm_1$ and $\chiz_1$ can be determined. However, 
it is difficult to distinguish this decay mode from the one with on-shell sleptons $\sLep$ in Sec.~\ref{sect:leptstop}.

\subsection{Final states with three or more leptons}
\label{sect:lept:simul.trilep}

As seen in Table \ref{tab.lept:simul.summary.scenarios},
the statistics for events with 3 or more leptons at $300 \; {\rm fb}^{-1}$ is too small to allow a meaningful analysis to be performed.

\subsection{Summary of mass determinations in leptonic events}

If the $\sLep$ are too heavy, the potential mass determination from leptonic final states is summarized in Table \ref{tab.lept:simul.summary.masses} assuming a Mixed scenario. 
Several masses can still be measured. 
The main interest of the leptonic channels, compared to the hadronic ones, is that they allow the $\chipm_1$ and the $\chiz_2$ to be distinguished.
\begin{table}[htb]
\begin{center}
\begin{tabular}{|c|ccccc|} \hline
       				& $\sQua$	& $\sTop_1$ 	& $\chiz_1$	& $\chiz_2$	& $\chipm_1$		\\
\hline
SFOS 2l $Z^0$		&     ? 		&       		&    ?   		&    ?   		&      		     		  \\
2l $W$			&     ?		&      			&    ?   		&    	  		&    ?  			\\
3l 				&       		&       		&     			&       		&      				\\
2l $\sTop$ $W$ 	&       		&     X		&    X  		&       		&    X  			\\
\hline
\end{tabular}
\caption{Masses which can be determined in the Mixed scenario.
A ``?" indicates that the masses could in principle perhaps be determined, but the endpoint was not very clear.
An ``X" means that the masses can be determined.
}
\label{tab.lept:simul.summary.masses}
\end{center}
\end{table}

\section{Correlations between Different Final States}
\label{sect:correl}

The preceding discussion in Sections~\ref{sect:had}-\ref{sect:lept} shows
that a considerable amount of information on the underlying physics scenario can be extracted after a discovery is made.
It is also clear that to perform this task, a huge amount of work of a non-negligible team of physicists will be required,
given the numerous channels to be studied.
Below, we outline in more detail the physics that can be extracted and the potential limitations.

It is of utmost importance to check whether a consistent picture of the physics emerges from the previous analyses.
The fully hadronic final states enable us to obtain several measurements of the mass of the same sparticle.
They should obviously be compatible, unless the interpretation of the events assumed for at least one of the measurement turns out to be wrong.
Another obviously very important point will be the comparison of the masses and sparticle identities obtained from the leptonic final states
with the ones from the fully hadronic analysis.

It was seen, and is supported by a simple simulation test, that we can in principle measure the masses of most sparticles from the purely hadronic final states.
From the two--jet and four-jet events, we may determine the masses of the $\sGlu$, $\sQua_R$, $\chiz_2$, $\chipm_1$, $\chiz_1$,
even with redundancy of the constraints.
They do not give access to the masses of the heavier chargino/neutralinos.
However, it was seen that some simple observations, 
like the presence of two bosons in the same decay chain or the existence of same charge tri- and/or four-lepton events,
provides at least evidence for the presence of higher chargino/neutralino states in the decays.
For a measurement of the masses of $\sQua_L$ and $\chipm_2$ in the Mixed scenario, we have to resort to 10-jet and/or 12-jet events, making the task more difficult.
In the 12-jet events with b-tags, there is a possibility to determine the mass of the $\sTop_1$ and/or $\sBot_1$.
Moreover, for these analyses we would profit from the HL-LHC with $3000 \; {\rm fb}^{-1}$ or of the HE-LHC.

The leptonic final states may bring evidence or exclude the presence of $\sLep$ in the electroweakino decays.
If a slepton is present, they also enable us to measure the masses of the $\sQua_L$, $\chipm_1$ and the $\chiz_1$.
Tests of lepton universality allow us to check whether the $\sEl$ and $\sMu$ are degenerate in mass or not.
Events with di-leptons and b-tags will also allow a determination of $\sTop_1$ and/or $\sBot_1$ masses.
All these masses can be compared to the values obtained in the hadronic analyses.
If no slepton is produced, a simple simulation test shows that we might still measure the masses of $\chipm_1$ and perhaps the $\chiz_2$.
Given the small statistics available, the leptonic analyses would also profit considerably from a HL-LHC or a HE-LHC.

The measurement of several sparticle masses will allow to start testing the high scale SUSY framework,
especially in the case of restricted models like MSUGRA which have only a handful of parameters.
A convenient method for performing these tests without going through a numerical RGE analysis is to 
test the predicted relations among the low-energy mass spectrum, often referred to as ``sum rules" 
\cite{Martin:1993ft}.

\section{Distinguishing the Different Scenarios}
\label{sect:distscenarios}

In all of our previous discussion, we have given explicit names to the sparticles.
However, we cannot with certainty know whether this interpretation is correct.
One remaining fundamental uncertainty has to do with the nature of the observed ``LSP" ---
it could be a standalone $\chiz_1$ state, as in the Bino and Mixed scenarios, but the $\chiz_1$ could also be mass degenerate with a $\chipm_1$ state
(Wino-LSP scenario), or even with both a $\chipm_1$ and a $\chiz_2$ state (higgsino-LSP scenario).
The only possibility to detect directly such a situation would be to observe events with a soft lepton or pion, or to identify the scenario indirectly.
As a consequence, we do not really know how to name the various observed electroweakino states.
Moreover, as was seen above, the identification of $\sQua_L$ and $\sQua_R$ also remains ambiguous at this stage.

In this section we shall discuss how one might attempt to solve some of these ambiguities.
It will be difficult, if not impossible, to identify the sparticles in a totally model independent way.
We will assume that the LSP is a $\chiz_1$-like object, but we will not introduce an assumption on the identity and mass ordering of the other sparticles.
We assume that the sparticles have been produced through strong interactions,
and conservatively ignore the electroweak direct production of charginos and/or neutralinos,
since the sensitivity for these channels will be very limited.

\subsection{Mass ordering and identification of $\sGlu$ and $\sQua$}

In full generality, several questions come up, which need to be answered in turn.
The first question is whether the $\sGlu$ is heavier or lighter than the light flavor squarks $\sQua$.
The hadronic analysis of four-jet events showed that it is possible to check whether the squark $\sQua$ in the decay $\sGlu \rightarrow q \sQua$
is on-shell or off-shell, i.e., whether the squark $\sQua$ is lighter or heavier than the gluino $\Glu$, respectively.
However, if a particular squark flavor $\sQua$ is found to be on-shell, 
this does not exclude the possibility that there could be other, heavier, squarks, called $\tilde{Q}$,
giving rise to the decay chain $\tilde{Q} \rightarrow q \sGlu \rightarrow q q \sQua$.
The heavy squarks $\tilde{Q}$ are also strongly produced, albeit at a lower rate, 
and would lead to decays with 3 additional jets, none of which are expected to reconstruct to a $W / Z / h$. 
Thus, a study of the jet multiplicity in those events (which would necessarily involve understanding the pattern of ISR) 
could reveal the presence of the heavier squarks $\tilde{Q}$.

Squarks are primarily produced by strong interactions --- either directly, or in the decays of gluinos.
Strong interactions tend to distribute the squark chiralities and flavors democratically, apart from pdf effects and 
phase space suppressions due to mass differences between the different squark states.
For the first two generations, $L-R$ mixing is negligible and squarks appear as purely $\sQua_L$ or $\sQua_R$ states.
The $\sQua_R$ decays $100 \%$ to a $\sBino$, which is the LSP in the Bino and the Mixed scenarios.
These would be observable as di-jet events.
The $\sQua_L$ will decay to Winos, with rates of 2/3 to the chargino and 1/3 to the neutralino.
The chirality of the $\sQua$ is thus correlated with the composition of the gauginos to which they decay,
i.e. from the scenario.
We have seen above that in all cases the masses can be determined.
Moreover, the $\sQua_L$ and $\sQua_R$ will most likely have different masses and can thus be identified as distinct states.
Finally, as the different $\sQua$ flavors may differ in mass, this would lead to several endpoints in the same invariant mass and $M_{T2}$ distribution.

If top squarks $\sTop$ are directly produced, the simplest event topology will be 6-jet events from $t \chiz_1$ or $b \chipm_1$ decays.
Therefore, the production of top squarks can be distinguished from that of light flavor squarks, thanks to the presence of a b-jet.
It was seen that the various decay modes can usually be distinguished from each other and that the masses involved can be reconstructed.
This would allow the identification and mass measurement of $\chiz_1$, $\chipm_1$ and $\sTop$.
As the $L$ and $R$ components of stop can be strongly mixed,
this mass probably refers to the lighter state, the $\sTop_1$, which is kinematically easier to produce.
Note that the decay mode $\sTop \rightarrow t \chiz_1$ clearly identifies the parent as a stop,
but it is not immediately clear if the measured mass is that of $\sTop_1$ or $\sTop_2$,
as both stop states may contribute.

There is a fair chance to observe the decay $\sBot \rightarrow b \chiz_1$ in di-jet events,
which unambiguously signals the presence of $\sBot$ quarks.
This motivates us to also look for events with higher multiplicities which contain b-tagged jets.
A large number of sbottom decays produce 6-jet events and, if taken alone, are indistinguishable from the stop decays due to the $W / Z$ ambiguity.
We might also have 6-jet events from $\sBot  \rightarrow b \chiz_2  \rightarrow b (Z / h^0 \chiz_1)$ with decays of $Z / h^0  \rightarrow  b b$.
These would be different from the $\sTop$ decays and may allow measuring the sbottom mass(es).
If both decay modes are observed, they might or might not correspond to the same sbottom squark ($\sBot_1$ or $\sBot_2$),
but they do provide an unambiguous identification of sbottom.
Then, by comparing to the masses from the other 6-jet topologies, one may figure out whether those also correspond to sbottom or include stop.

Needless to say, if a reliable charm-tagging method is implemented, this might be used to determine the mass of $\tilde{c}$, e.g. in di-jet events,
which can then be compared to the masses of the other first two generation squarks.
Compatibility of these masses would lend support to the universality hypothesis of the first two squark generations at the GUT scale.

\subsection{Mass ordering of the charginos and neutralinos}

The next question is the ordering in mass between the Bino, Winos and higgsinos.
It was seen in Sec.~\ref{sect:scenarios} that the absence of di-jet events singles out the Higgsino scenario.
However, it was also observed that in real life some Higgsino events will still populate the di-jet categories, 
but only when one or both squarks decay to leptons.
Using the leptonic analyses, these events can be identified, provided the leptons are accepted by the experimental cuts.
The events with a rejected lepton, called lost lepton events, can be evaluated by an extrapolation, giving an estimate of the total number of leptonic decays.
The comparison between this extrapolated number and the number of observed di-jet events then enables us to decide 
whether there are real hadronic di-jet events or not.

A further distinction may come from counting the bosons, $W / Z / h$, in each decay chain.
We could distinguish the decays via light flavor squarks from those via stop by selecting or vetoing b-tagged jets which are not part of a $Z / h$.
In decay chains involving only the first two generation squarks, it is seen from Table \ref{tab.scenarios.BR} 
that in the Bino-like, the Wino-like and the Higgsino-like scenarios only up to one boson per hemisphere is expected,
whereas for the Mixed scenario some decays will involve two bosons per hemisphere.
In decay chains involving stop, Table \ref{tab.scenarios.BRstop} shows that in the Bino-like, 
Wino-like or Mixed scenarios up to three bosons per hemisphere can be produced (one coming from a top decay);
in the Higgsino-like scenario, only two bosons per hemisphere may exist and one of them should come from a top decay.
In all cases, one more boson may possibly exist if the $\sTop_1$ originates from $\sGlu \rightarrow t \sTop_1$.
Hence, by counting the number of bosons in the decay chains, one could, in principle, attempt to distinguish among the different scenarios.

To check these statements, we again use a simple simulation as in Sec.~\ref{sect:had}.
Hemispheres were considered if they did not contain an identified lepton.
The results for the different scenarios are summarized in Table \ref{tab.simul.distscenarios.WZh}.
\begin{table}[htb]
\begin{center}
\begin{tabular}{|c|c|c|c|c||c|c|c|c|} \hline
       				& Mixed	& Bino 	& Wino	& Higgsino & Mixed & Bino 	& Wino	& Higgsino 	\\
\hline
 $\#$Hemis		& 11804	& 12430	& 12662	& 11978	& 1.		& 1.		& 1.		& 1.			\\
\hline
  \multicolumn{9}{|c|}{ Events with no b's }          \\
\hline
$0 \; W/Z/h$		& 5996 	& 7396	& 7430   	& 4006	& 0.508 	& 0.595	& 0.587	& 0.334   		\\
$1 \; W/Z/h$		& 2525	& 2756 	& 2897  	& 5281	& 0.214	& 0.222 	& 0.229  	& 0.441		\\ 
$2 \; W/Z/h$		& 1198	& 392	& 371  	& 440 	& 0.101	& 0.032	& 0.029  	& 0.037		\\
$3 \; W/Z/h$		& 146 	& 110  	& 55   	& 18  	& 0.012	& 0.009  	& 0.004	& 0.002   		\\
$4 \; W/Z/h$		& 7 		& 25	  	& 7   		&  1	 	& 0.001 	& 0.002	& 0.0006	&  0.0001	 	\\
\hline
  \multicolumn{9}{|c|}{ Events with b's }          \\
\hline
$0 \; W/Z/h$		& 484 	& 750	& 828  	& 961  	& 0.041 	& 0.060	& 0.065  	& 0.080  		\\
$1 \; W/Z/h$		& 901	& 568  	& 769 	& 1046 	& 0.076	& 0.046  	& 0.061 	& 0.087 		\\ 
$2 \; W/Z/h$		& 457	& 316	& 257  	& 215 	& 0.039	& 0.025	& 0.020  	& 0.018 		\\
$3 \; W/Z/h$		& 86 		& 91  	& 46   	& 10 		& 0.007 	& 0.007  	& 0.004	& 0.001 		\\
$4 \; W/Z/h$		& 4	 	& 24	  	& 1	   	& 0	  	& 0.0003	& 0.002  	& 0.0001	& 0.	  		\\
\hline
\end{tabular}
\caption{Statistics on $W / Z / h$ per hemisphere in the various scenarios for $300 \; {\rm fb}^{-1}$.
Columns 2 to 5 (6 to 9) list the number (fraction) of hemispheres with a given number of $W / Z / h$ per hemisphere.
}
\label{tab.simul.distscenarios.WZh}
\end{center}
\end{table}
The first line lists the number of hemispheres in total for each scenario, while the following lines categorize those hemispheres based on the 
number of reconstructed $W / Z / h$, in terms of total number (columns 2 to 5) or fraction (columns 6 to 9).
The categorization is done separately for hemispheres without b-tagged jets (top rows)
and with at least one b-tagged jet for jets not included in the bosons (bottom rows).

In the case with no b-tags, it is seen that:
\begin{itemize}
\item In the case of the Mixed scenario, as expected, the fraction of hemispheres with 2 $W / Z / h$ bosons is much larger than for the other scenarios.
\item In the case of the Higgsino scenario, again as expected, the fraction of hemispheres with 1 $W / Z / h$ boson compared to 0 $W / Z / h$ bosons 
is much larger than for the other scenarios.
This is due to the fact that in the Higgsino scenario all decay chains give rise to exactly one boson, and there are no direct squark decays to the LSP.
\item The results for the Bino and the Wino scenarios are very similar and do not allow these scenarios to be distinguished based on this counting.
\end{itemize}
Thus, from the non-b events, three classes of scenarios can be identified:
the indistinguishable Bino and Wino scenarios, the distinct Mixed scenario and the distinct Higgsino scenario.

It the $\sTop_1$ is sufficiently lighter than the $\sGlu$, there will also be events with b-tagged jets. The corresponding statistics 
is listed in the lower part of Table~\ref{tab.simul.distscenarios.WZh}. We observe that
the Mixed scenario stands out with with a larger number of hemispheres with 1 $W / Z / h$ boson 
compared to the number of hemispheres without any $W / Z / h$ bosons.
Another striking feature is that the Mixed and Bino scenarios have a non-negligible number of hemispheres with 3 $W / Z / h$.
That number is reduced for the Wino scenario and even more so for the Higgsino scenario.
However, this is not a robust distinction between the Bino and the Wino scenarios, 
because it relies on the existence of a certain $\sTop_1$ (and not $\sTop_2$) state in the data.
As the mass parameters of $\sTop_L$ and $\sTop_R$ at the GUT scale are arbitrary,
the composition of the $\sTop_1$ at the electroweak scale could very well be different
and the result seen in Table~\ref{tab.simul.distscenarios.WZh} is therefore somewhat ``accidental".

For completeness, we also consider the two inverted (twin) scenarios mentioned in Sec.~\ref{sect:scenarios}.
These are obtained by exchanging the masses of the Bino and Wino in the Mixed and in the Higgsino scenarios.
The corresponding results are listed in Table \ref{tab.simul.distscenarios.WZhprime}.
\begin{table}[htb]
\begin{center}
\begin{tabular}{|c|c|c||c|c|c|c|} \hline
       				& Inverted Mixed	& Inverted Higgsino & Inverted Mixed & Inverted Higgsino 	\\
\hline
 $\#$Hemis		& 11790	& 12030	& 1.		& 1.			\\
\hline
  \multicolumn{5}{|c|}{ Events with no b's }          \\
\hline
$0 \; W/Z/h$		& 5818 	& 3905	& 0.493 	& 0.325   		\\
$1 \; W/Z/h$		& 2507	& 5167	& 0.213	& 0.430 		\\ 
$2 \; W/Z/h$		& 1273	& 496 	& 0.108	& 0.041		\\
$3 \; W/Z/h$		& 110 	& 28  	& 0.009	& 0.002   		\\
$4 \; W/Z/h$		& 2 		&  1	 	& 0.0002 	&  0.0001	 	\\
\hline
  \multicolumn{5}{|c|}{ Events with b's }          \\
\hline
$0 \; W/Z/h$		& 625 	& 1072  	& 0.053 	& 0.089  		\\
$1 \; W/Z/h$		& 1003	& 1084 	& 0.085	& 0.090 		\\ 
$2 \; W/Z/h$		& 385	& 265 	& 0.033	& 0.022 		\\
$3 \; W/Z/h$		& 65	 	& 12 		& 0.006 	& 0.001 		\\
$4 \; W/Z/h$		& 2	 	& 0	  	& 0.0002	& 0.	  		\\
\hline
\end{tabular}
\caption{The same as Table~\ref{tab.simul.distscenarios.WZh}, but for the Inverted Mixed and Inverted Higgsino scenarios.
}
\label{tab.simul.distscenarios.WZhprime}
\end{center}
\end{table}
Comparing the results for the Mixed and Inverted Mixed scenarios from Tables~\ref{tab.simul.distscenarios.WZh} and \ref{tab.simul.distscenarios.WZhprime}
respectively, we see that the two scenarios are fully compatible, as already explained in Sec.~\ref{sect:scenarios}.
A similar match is observed between the Higgsino and Inverted Higgsino scenarios in the two tables.
This implies that for those two pairs of twin scenarios it will not be possible to disambiguate the chirality of the squarks 
($\sQua_L$ from $\sQua_R$)
on the basis of $W / Z / h$ multiplicities measured in the fully hadronic events.

Note that in the Wino and Inverted Mixed scenarios, the lightest chargino is Wino-like and almost degenerate with the Wino-like neutral LSP.
Similarly, in the Higgsino and Inverted Higgsino scenarios, the lightest three states in the electroweakino spectrum are the Higgsinos, which are 
also quite degenerate in mass. Therefore, another handle on distinguishing the twin scenarios is provided by the possibility 
to detect the soft decay products (electron or pion) from the $\chipm_1\to \chiz_1$ transition, which would indicate a Wino (as opposed to a Bino) 
or an Inverted  Mixed (as opposed to a Mixed) scenario. Correspondingly, this will also resolve the ambiguity regarding the squark chirality in those scenarios.
Unfortunately, the Higgsino and Inverted Higgsino scenarios cannot be distinguished by this method, 
as they both have a triplet of almost degenerate Higgsinos at the bottom of the mass spectrum.

The above discussion shows that most scenarios can indeed be distinguished on the basis of the $W / Z / h$ multiplicities per hemisphere,
possibly complemented with additional information.
This allows to know whether the LSP-like particle is a gaugino (either Bino or Wino) or a Higgsino.
If the soft decay products from $\chipm_1\to \chiz_1$ can be identified, it is also possible to distinguish the Mixed from the Inverted Mixed scenario,
and the Bino from the Wino scenario.
This also allows the identification of $\sQua_L$ and $\sQua_R$ for the Bino, the Wino and the Mixed scenarios,
but not for the Higgsino ones.

\section{Determination of Branching Ratios and Cross Sections}
\label{sect:BRs}

So far, the only information which we used in determining the sparticle masses
were the locations of endpoints of kinematic distributions. However, there is a whole lot more information in the data.
Ideally, once the masses are measured, one would like to introduce the new particles in an event generator and simulate their production
and decays, including the detector response.
However, in order to generate events, we also need to know the production modes, the decay sequences and the associated branching fractions
$B_{\tilde \psi_2\tilde\psi_1}$ for the transitions between two SUSY states $\tilde\psi_2$ and $\tilde\psi_1$ with $m_{\tilde\psi_2}>m_{\tilde\psi_1}$.
The main purpose of this section is to obtain a first rough estimate of the cross-sections for the individual strong production subprocesses, 
and the relevant branching fractions of colored sparticles. This is only intended as a starting point for any detailed Monte Carlo simulation
which can be used later on to fine-tune the initial naive estimates.

There are several ingredients to such an analysis which will have to be performed ahead of time:
\begin{itemize}
\item The measurement of the mass spectrum as already discussed above --- in particular, the determination of the 
masses of the colored sparticles (gluino and two types of squark states, $\sQua_L$ and $\sQua_R$) as well as the accessible electroweakinos, 
see Tables~\ref{tab.simul.summary.masses}, \ref{tab.lept:simul.summary.massesl} and \ref{tab.lept:simul.summary.masses}. 
\item In the case of hadronic channels, understanding the contributions from ISR jets, as well as the migration of events between different
multijet categories, see Fig.~\ref{fig.simul.stats.hemimult}.
\item Determination of the geometric acceptances\footnote{More precisely, $a_j$ is the product of the geometrical acceptance and the reconstruction and identification efficiencies.} 
$a_j$ for various objects $j$: $a_{q_R}$ ($a_{q_L}$, $a_q$) for jets resulting from the decay of a right-handed squark (left-handed squark, gluino)
and $a_V$ for hadronically decaying bosons $W$, $Z^0$ or $h^0$.
These can be estimated reliably, once we know from the data the kinematical properties of the final state objects in specific topologies.
We may then input the kinematics in a simple generator and obtain the acceptances.
\item Theoretical input and assumptions. Finally, given the mass spectrum, one can predict theoretically the ratio 
of branching ratios
\begin{eqnarray}
 \frac{B_{\sGlu \sQua_R}}{B_{\sGlu \sQua_L}} \equiv  \tan\alpha.
 \label{alphadef}
\end{eqnarray}
In addition, for the dominant decay modes, which do not compete much with other SUSY decay channels, the branching ratios can be safely assumed to be equal to 1.
\end{itemize}

With those preliminaries, we can now follow a ``piecewise inference" of the modes, as illustrated below.
The exact strategy would, of course, depend on which final state topologies have been observed (and hence on the scenario).
As an illustration, here we shall focus on the case of the Mixed scenario, and to further simplify the analysis, 
assume that we are only interested in the contribution from the light flavor squarks (first and second generation).
Let us define $N_i$ as the observed number of events in channel $i$.
\begin{itemize}
\item The number of di-jet events ($N_{2j}$) for a given integrated luminosity ${\mathcal L}$ 
will allow us to estimate the contribution from the pair production of $\sQua_R$.
The acceptance $a_{q_R}$ is obtained from the kinematics of the two jets.
The cross section can be written as\footnote{Here and below our notation for the cross-sections also implies the conjugated processes,
e.g., $\sigma_{\sQua_R \sQua_R}$ is the cross-section for producing a squark pair, an anti-squark pair and a squark-anti-squark pair.}
\begin{eqnarray}
 \sigma_{\sQua_R \sQua_R} = \frac{N_{2j}}{a_{q_R}^2 B_{\sQua_R \sBino}^2 \mathcal{L}}. 
\end{eqnarray}
The decay branching fraction $B_{\sQua_R \sBino}\equiv B(\sQua_R\to q \sBino)$ can in this case 
safely be assumed to be 1, so that $N_{2j}$ in principle provides a measurement of the cross section $ \sigma_{\sQua_R \sQua_R}$:
\begin{equation}
N_{2j}  \longrightarrow  \sigma_{\sQua_R \sQua_R}. 
\label{N2jtosigma}
\end{equation}
\item In a similar fashion, the observed number of 10-jet events $N_{10j}$
allows us to estimate the contribution from the pair production of $\sQua_L$,
followed by the decay $\sQua_L \rightarrow q \sWino \rightarrow q V_1 \sHino \rightarrow q V_1 V_2  \sBino$, 
where $V_1$ and $V_2$ stand for all three bosons, $W$,  $Z^0$ and $h^0$, summed over all charges.
Here, we need to know the kinematics of the individual quark jets and of the two bosons.
We also need to identify the first ($V_1$) and the second ($V_2$) boson in the decay sequence.
In Sec.~\ref{sec:10jet} we saw that this can be achieved by choosing the lower and higher invariant mass $M(qV)$,
which in turn allows to determine the acceptances $a_{V_1}$ and $a_{V_2}$, and from these, the cross section\footnote{Here and below 
we shall assume that the hadronic branching fractions of the three bosons $W$,  $Z^0$ and $h^0$ are similar (to a very good approximation), and 
furthermore, that these hadronic branching fractions are already incorporated into the acceptance $a_V$.}
\begin{eqnarray}
 \sigma_{\sQua_L \sQua_L} = \frac{N_{10j}}{a_{q_L}^2 a_{V_1}^2 a_{V_2}^2 B_{\sQua_L \sWino}^2 B_{\sWino \sHino}^2 B_{\sHino \sBino}^2 \mathcal{L}}. 
 \label{sigmaLLN10}
\end{eqnarray}
Since we are summing over all intermediate electroweakino states (and hence over all final states), we can 
safely assume that $B_{\sQua_L \sWino} = 1$ and $B_{\sWino \sHino} = B_{\sHino \sBino} = 1$, so that using eq.~(\ref{sigmaLLN10})
$N_{10j}$ provides a measurement of $\sigma_{\sQua_L \sQua_L} $, in analogy to (\ref{N2jtosigma}):
\begin{equation}
N_{10j}  \longrightarrow  \sigma_{\sQua_L \sQua_L}. 
\label{N10jtosigma}
\end{equation}
\item As discussed in Sec.~\ref{sec:4jet}, the 4-jet events can be interpreted in terms of the decay
$\sGlu \rightarrow q_1 \sQua_R \rightarrow q_1 q_2 \sBino$, see eq.~(\ref{eq:fourjet.sgtosq}).
In this case, it would be necessary to distinguish the first from the second jet in the decay chain, 
and, as discussed in Sec.~\ref{sec:4jet}, 
cuts can be designed to make the correct choice on average, and from there to obtain the kinematical properties and compute 
the corresponding acceptances $a_q$ and $a_{q_R}$.
As a starting point, we have the relation
\begin{eqnarray}
 B_{\sGlu \sQua_R}^2 \cdot \sigma_{\sGlu \sGlu} = \frac{N_{4j}}{a_{q}^2 a_{q_R}^2 B_{\sQua_R \sBino}^2 \mathcal{L}}. 
\end{eqnarray}
As before, we can safely assume $B_{\sQua_R \sBino}=1$, so that the number of 4-jet events measures 
the rate $R_{4j}$ as the product of the cross section times branching fraction squared for this process:
\begin{equation}
N_{4j}  \longrightarrow   R_{4j} \equiv \sigma_{\sGlu \sGlu}\ B_{\sGlu \sQua_R}^2 \equiv  \sigma_{\sGlu \sGlu}\ B_{\sGlu\sQua}^2\ \sin^2\alpha,
\label{N4jtosigma}
\end{equation}
where in the last relation we used the alternative parametrization (\ref{alphadef}) and introduced
\begin{equation}
B_{\sGlu\sQua}\equiv \sqrt{ B_{\sGlu \sQua_R}^2 + B_{\sGlu \sQua_L}^2 }.
\label{Bdef}
\end{equation}
\item We might use the number of events $N_{3j}$ in the $(2, 1)$ topologies to obtain the mixed decay from the $\sGlu \sQua_R$ production,
using the acceptances determined in the previous case.
The expression is now
\begin{eqnarray}
 B_{\sGlu \sQua_R} \cdot \sigma_{\sGlu \sQua_R} = \frac{N_{3j}}{a_q a^2_{q_R} B_{\sQua_R \sBino}^2 \mathcal{L}}, 
\end{eqnarray}
so that $N_{3j}$ determines the product of the cross section times the branching fraction:
\begin{equation}
N_{3j}  \longrightarrow    R_{3j} \equiv \sigma_{\sGlu \sQua_R} \ B_{\sGlu \sQua_R} \equiv   \sigma_{\sGlu \sQua_R} \ B_{\sGlu\sQua}\ \sin\alpha,
\label{N3jtosigma}
\end{equation}
\item The number of 12-jet events $N_{12j}$ enable us to estimate the pair production of $\sGlu$, followed by decays to $\sQua_L$:
\begin{eqnarray}
 B_{\sGlu \sQua_L}^2 \cdot \sigma_{\sGlu \sGlu} = \frac{N_{12j}}{a_{q}^2 a_{q_L}^2  a_{V_1}^2 a_{V_2}^2 B_{\sQua_L \sWino}^2 B_{\sWino \sHino}^2 B_{\sHino \sBino}^2 \mathcal{L}},  
\end{eqnarray}
from where
\begin{equation}
N_{12j}  \longrightarrow   R_{12j} \equiv  \sigma_{\sGlu \sGlu}\ B_{\sGlu \sQua_L}^2 \equiv  \sigma_{\sGlu \sGlu}\ B_{\sGlu\sQua}^2\ \cos^2\alpha.
\label{N12jtosigma}
\end{equation}
\item The only remaining strong production mode is $\sGlu \sQua_L$, which is
related to the number of events $N_{11j}$ in the $(6, 5)$ topology in case of the decay $\sGlu \rightarrow \sQua_L$
\begin{eqnarray}
 B_{\sGlu \sQua_L} \cdot \sigma_{\sGlu \sQua_L} = \frac{N_{11j}}{a_{q} a^2_{q_L}  a_{V_1}^2 a_{V_2}^2 B_{\sQua_L \sWino}^2 B_{\sWino \sHino}^2 B_{\sHino \sBino}^2 \mathcal{L}},
\end{eqnarray}
or the number of events $N_{7j}$ in the $(2, 5)$ topology in case of the decay $\sGlu \rightarrow \sQua_R$:
\begin{eqnarray}
 B_{\sGlu \sQua_R} \cdot \sigma_{\sGlu \sQua_L} = \frac{N_{7j}}{a_{q} a_{q_R} a_{q_L}  a_{V_1} a_{V_2} B_{\sQua_R \sBino} B_{\sQua_L \sWino} B_{\sWino \sHino} B_{\sHino \sBino} \mathcal{L}}.
\end{eqnarray}
Once again, setting $B_{\sQua_R \sBino} =B_{\sQua_L \sWino} =B_{\sWino \sHino} =B_{\sHino \sBino} =1$ leads to
\begin{eqnarray}
N_{11j}  &\longrightarrow&  R_{11j}\equiv  \sigma_{\sGlu \sQua_L} \ B_{\sGlu \sQua_L} \equiv   \sigma_{\sGlu \sQua_L} \ B_{\sGlu\sQua}\ \cos\alpha, \label{N11jtosigma}\\
N_{ 7j}  &\longrightarrow&  R_{7j}\equiv  \sigma_{\sGlu \sQua_L} \ B_{\sGlu \sQua_R} \equiv   \sigma_{\sGlu \sQua_L} \ B_{\sGlu\sQua}\ \sin\alpha. \label{N7jtosigma}
\end{eqnarray}
\end{itemize}

We can summarize the above results (\ref{N4jtosigma}), (\ref{N3jtosigma}),  (\ref{N12jtosigma}),  (\ref{N11jtosigma})  and (\ref{N7jtosigma}) for the rates $R_{nj}$ for events from gluino-induced processes as
\begin{eqnarray}
R_{4j} &=&  \sigma_{\sGlu \sGlu}\ B_{\sGlu\sQua}^2\ \sin^2\alpha, \label{R4} \\
R_{3j} &=&   \sigma_{\sGlu \sQua_R} \ B_{\sGlu\sQua}\ \sin\alpha,  \label{R3} \\
R_{12j} &=&  \sigma_{\sGlu \sGlu}\ B_{\sGlu\sQua}^2\ \cos^2\alpha,  \label{R12} \\
R_{11j}&=&   \sigma_{\sGlu \sQua_L} \ B_{\sGlu\sQua}\ \cos\alpha,  \label{R11}\\
R_{ 7j}  &=&   \sigma_{\sGlu \sQua_L} \ B_{\sGlu\sQua}\ \sin\alpha.  \label{R7}
\end{eqnarray}
These are 5 relations for 5 unknown parameters, namely, three cross-sections, 
$\sigma_{\sGlu \sGlu}$, $\sigma_{\sGlu \sQua_R}$ and $ \sigma_{\sGlu \sQua_L} $, and
two branching ratios parameterized in terms of (\ref{alphadef}) and (\ref{Bdef}).
However, the system of equations (\ref{R4}-\ref{R7}) is invariant under the following rescaling by an arbitrary constant $C$:
\begin{eqnarray}
B_{\sGlu\sQua} &\to& C \times B_{\sGlu\sQua} , \\
\sigma_{\sGlu \sGlu} &\to& \frac{1}{C^2} \times \sigma_{\sGlu \sGlu}, \\
\sigma_{\sGlu \sQua_R} &\to& \frac{1}{C} \times \sigma_{\sGlu \sQua_R}, \\
\sigma_{\sGlu \sQua_L} &\to& \frac{1}{C} \times \sigma_{\sGlu \sQua_L}.
\end{eqnarray}
Therefore, we need additional information in order to fix this ambiguity. For example, one can 
try to estimate the contribution from gluino decays to third generation squarks 
from the presence of b-tagged jets in the sample.
If this contribution turns out to be negligible, then we must have $B_{\sGlu\sQua}=1$, since the gluino must
necessarily decay through squarks. If, on the other hand, this contribution is measurable, 
then it would provide an estimate of $B_{\sGlu\sQua}$ itself.

Returning to the system (\ref{R4}-\ref{R7}), there are several useful relations which can be further derived from it.
For example, one can {\em measure} the ratio of branching ratios (\ref{alphadef}) by taking the 
ratio of (\ref{R11}) and (\ref{R7}):
\begin{equation}
\tan\alpha = \frac{R_{7j}}{R_{11j}}
\end{equation}
and compare the obtained result with the theory prediction.
The gluino pair-production cross section $\sigma_{\sGlu \sGlu}$ is related to the combination 
\begin{equation}
R_{4j} + R_{12j} = \sigma_{\sGlu \sGlu}\ B_{\sGlu\sQua}^2
\end{equation}
and can be readily extracted once $B_{\sGlu\sQua}$  is known.

Obviously, information from the leptonic channels discussed in Secs.~\ref{sect:leptslept} and \ref{sect:lept} 
could also be used in the calculation. In the Mixed scenario the leptons originate solely from the decays of $W /Z / h$ whose leptonic branching fractions are known.
The data from the leptonic channels can then be used to corroborate the results from the hadronic channels, and also to
disentangle the individual contributions from $W$, $Z^0$ and $h^0$.

The analysis described above will provide an initial determination of the sparticle masses, cross sections and branching ratios and their uncertainties.
As we have identified the scenario, the masses of the electroweakinos enable us to determine the MSSM parameters $M_1$, $M_2$, $M_3$ and $\mu$,
while the value of $tan \beta$ remains largely unconstrained. In order to set limits on the allowed range of $\tan\beta$,
one can pursue several possible approaches, typically involving detailed studies of the third generation sfermions  \cite{Golling:2016gvc}.

To make further progress, one must assume a complete model, such as the MSSM, 
and feed these initial estimates for the model parameters in a suitable package (or combination of packages) like 
SFITTER \cite{Lafaye:2004cn}, Fittino \cite{Bechtle:2004pc}, SModelS \cite{Kraml:2013mwa}, FastLim \cite{Papucci:2014rja}, MadAnalysis \cite{Conte:2014zja},
MasterCode \cite{deVries:2014vaa}
or GAMBIT \cite{Balazs:2017moi,Workgroup:2017bkh}, which can provide adequate simulation of the data including the detector response.
The output from this simulation can then be fitted to the multiplicity distributions of jets, b-jets and leptons,  
as well as the kinematic distributions (of $M_{T2}$ and other variables) for the events surviving the experimental cuts.
The simulation will also account correctly for the MSSM backgrounds due to migrations in each topology.
Moreover, some decay chains are repeated in different channels, 
e.g. the decay sequence of $\sQua_L$ is present not only in events of direct $\sQua_L$ production, 
but also appears in events initiated by a $\sGlu$ decaying as $\sGlu \rightarrow q \sQua_L$.
Hence, there is a gain when combining all channels in an overall fit. By performing several iterations
of the fit, one would also improve the precision on the production cross sections for squarks and gluinos.
At this point, it will also be important to test for the spins of the particles, as there can be several
possible spin assignments for the new particles, all leading to the same final state\footnote{A well-studied example
of this duality is provided by the models of supersymmetry and universal extra dimensions \cite{Cheng:2002ab,Smillie:2005ar,Datta:2005zs,Barr:2005dz,Alves:2006df}.}.
Given the observed mass spectrum, one can compare the measured cross sections to the theoretically calculated ones\footnote{The cross-sections 
for the most popular new physics models are already known quite accurately, at least at the level of NLO \cite{Kramer:2012bx}.} for different spin assignments, 
taking advantage of the dependence of the cross-section on the particle spins, see e.g. \cite{Datta:2005vx,Meade:2006dw,Hubisz:2008gg}.
Since the cross-sections for different spins may vary by as much as an order of magnitude, this
comparison can already exclude some of the alternative scenarios.
On the other hand, one may also use the steep dependence of the cross sections on the sparticle masses themselves
to obtain an even more accurate measurement of the masses of $\sGlu$ and $\sQua$.
Of course, the most general approach is to do a global fit of the cross-sections and angular distributions to 
the masses, spins, couplings and their chiralities.

Once the low-energy parameters of the model (in this case the MSSM) are extracted, one may 
perform an RGE analysis in order to test the high-scale boundary conditions and possibly determine the
mechanism of SUSY breaking and its mediation to the observable sector \cite{Zerwas:2002as,Adam:2010uz,Athron:2017qdc}.
This is clearly the most fundamental physics output from all the analyses outlined in the previous sections.
However, this is a very long-term program and it will take a significant amount of time and effort to reach this stage,
and one should be prepared for surprises along the way as well.

\section{Conclusions}
\label{sect:concl}

In the assumption that new physics in the form of an MSSM-like model with R-parity conservation has been discovered at the LHC,
a strategy was worked out to determine the mass spectrum from the kinematic endpoints of only invariant mass distributions and/or from $M_{T2}$ or $M_2$ distributions.
It was found that in a scenario where the $\sGlu$ and $\sQua$ masses are slightly below 2 TeV
(which would enable the SUSY discovery to be made with less than $100 \; {\rm fb}^{-1}$)
{\em most of the sparticle masses can potentially be measured in the fully hadronic final states}, after collecting about $300 \; {\rm fb}^{-1}$.
In CMS, this is thanks to the excellent jet energy resolution obtained in the particle flow reconstruction.
In ATLAS, a similar resolution can be reached, due to its superior hadron calorimetry.
However, several potential measurements are limited or impossible due to statistical limitations, which 
provides a strong motivation for the HL-LHC with $3000 \; {\rm fb}^{-1}$ (assuming that the first hints for a discovery appear below $300 \; {\rm fb}^{-1}$)
and/or a HE-LHC of 30 TeV.

In the \textbf{fully hadronic final state}, it was found that the hadronic decays of the $W$ and $Z$ bosons can be distinguished from the hadronic decays of the Higgs boson $h$,
but the jet energy resolution does not allow a clear separation between the peaks of the $W$ and the $Z$.
The set of sparticles whose masses can be measured from the hadronic channels depends on the electroweakino scenario, 
i.e., on the composition of the charginos and neutralinos. The six main electroweakino scenarios were considered, as described in Section \ref{sect:scenarios}.
The masses of the gluino $\sGlu$ and the first two generation squarks $\sQua$ are always measurable.
In the ``Mixed" scenario, where the LSP is a Bino, the next-to-lightest charginos and neutralinos are Higgsinos and the heaviest ones are Winos,
the masses of all charginos/neutralinos can be determined.
But in scenarios where the LSP is a Bino or a Wino and the Higgsinos are the heaviest electroweakinos, only the masses of the LSP and the 
next-to-lightest $\chiz_2$ are measurable.
If, on the other hand, the LSP is a Higgsino, the decays do not allow the $\sQua_L$ and $\sQua_R$ to be distinguished.
As a consequence, it will be difficult to determine the masses of the Bino-like and Wino-ike chargino/neutralinos.
This scenario deserves a more detailed analysis to establish what exactly is feasible in this case.
Except for the Mixed scenario, the mass of $\sTop_1$ is unlikely to be measured, due to lack of statistics.

The \textbf{leptonic final states} give the unique opportunity to detect a slepton, if the latter is light enough to appear in the decays of electroweakinos.
They may also provide additional measurements of the sparticle mass spectrum
and allow a better identification of the nature of the sparticles thanks to the ability to make charge measurements.
But if the sleptons are too heavy, then 
the leptonic channels are severely limited by statistics for an integrated luminosity amounting to $300 \; {\rm fb}^{-1}$.
Nevertheless, the lepton channels differentiate between the $W$ and $Z$ bosons, which helps in identifying the scenario.

The above observations lead to a \textbf{complete reversal of the usual paradigm}, which was primarily concerned with leptonic channels and tended to ignore the fully hadronic decays.
Here we find that the hadronic channels, which were previously not considered or not believed to be useful,
may allow to extract a significant amount of new physics. For example, it was 
found that the multiplicities of bosons, $W$, $Z^0$ and $h^0$, enable us to identify the type of scenario of the underlying physics,
i.e. whether the ``LSP" is a gaugino or a Higgsino, and whether we are dealing with a Mixed scenario.
This is very important in order to correctly reveal the identity (quantum numbers) of the measured sparticles.

Depending on what Nature prepares for us, we should be in a position to identify and to measure the masses
of at least the lowest lying MSSM states.
Whether and how many of them will be found is unpredictable and depends on the sparticle spectrum.
It should be borne in mind that this program can only be undertaken after sufficient statistics is accumulated.

Apart from the measurement of the masses of the sparticles, this analysis also shows that it is possible to determine the decay branching fractions
and the cross sections of the various production modes.
These can be compared to the theoretical expectations, which provides a powerful test of the spin assignment of the sparticles
from the measured mass and cross section of the produced $\sGlu$ and $\sQua$. In turn, this allows,
at least in principle, the possibility to distinguish between the competing models of
R-parity conserving MSSM (or Littlest Higgs with T-parity) and UED with KK-parity.

A final caveat: the simulation tests presented in this document are meant to provide only a proof of principle for what could be achieved.
The production of ISR and FSR jets is not included in the simulation and the SM backgrounds are ignored.
To develop a more realistic analysis, it will be necessary to design an efficient method to distinguish quark from gluon jets
(in order to identify ISR jets), to analyze the jet substructure (the present analysis was done at the parton level) and to control the SM backgrounds with high precision.
Progress on these topics is currently being made within the LHC collaborations, often taking advantage of machine learning techniques.
Moreover, the endpoint values were simply read off the histograms, whereas in real life a more objective method would be required \cite{Curtin:2011ng,Debnath:2015wra,Debnath:2016mwb}.
Therefore, several of the conclusions reached here should be considered as tentative and the results need to be confirmed by detailed analyses in the future.
It is hoped that the obtained results are sufficiently promising for such further work to be undertaken.
Last but not least, this whole approach only makes sense if New Physics is discovered in the data.

\section*{ACKNOWLEDGEMENTS}

This work is supported in part by the United States Department
of Energy DE-SC0010296.

\clearpage

\appendix

\section{Mass Determination from Invariant Mass Endpoints}
\label{sect:invmass}

\newcommand{\oneMrat}[2]{\ensuremath{\left(1 - \frac{{#1}^2}{{#2}^2}\right)}}
\newcommand{\onePrat}[2]{\ensuremath{\left(1 + \frac{{#1}^2}{{#2}^2}\right)}}

In this Appendix we collect the necessary formulae for the kinematic endpoints of 
the invariant mass distributions for the relevant combinations of visible particles 
which are emitted within a single decay chain.
The method was first proposed in Ref.~\cite{Bachacou:1999zb},
which also listed some formulae relating endpoints to sparticle masses,
assuming that the visible particles are massless.
However, the extraction of the sparticle masses from the measured endpoints 
was plagued by ambiguities in this approach.
The endpoints have been derived for many additional final states in \cite{Luc06}, which is unpublished,
and an approach was proposed to avoid these ambiguities (see also \cite{Allanach:2000kt}).
We will, therefore, summarize them here for some of the main decay channels.
Some part of this additional material has previously appeared in \cite{Matchev:2009iw}.

\subsection{Two-step decays}
\label{sect:invmass.2step}

Two-step decays are of the generic type
\begin{eqnarray}
 X \rightarrow f_1 R, \qquad R \rightarrow f_2 O
\label{eq:invmass.2step.general}
\end{eqnarray}
where $f_1$ and $f_2$ are observed particles (quarks or leptons) and $O$ is considered as the LSP.
A typical SUSY example is the neutralino decay, $\chiz_2 \rightarrow l_1 \sLep, \; \sLep \rightarrow l_2 \chiz_1$.

If the observed particle masses are negligible, the endpoint $M_{ff}^{max}$ of the invariant mass distribution 
$d\Gamma/dM_{ff}$ is given by 
\begin{eqnarray}
 M_{ff}^{max} = M_X \sqrt{\left(1-\frac{M_R^2}{M_X^2}\right)\left(1-\frac{M_O^2}{M_R^2}\right) }
\label{eq.invmass.2step.Mffmax}
\end{eqnarray}
for an on-shell intermediate particle $R$.
It occurs in a configuration where $f_1$ and $f_2$ are back-to-back in the rest frame of $X$.
A two-step decay thus provides one constraint (\ref{eq.invmass.2step.Mffmax}) 
on three unknown masses, $M_X$, $M_R$ and $M_O$,
and is therefore insufficient to determine all three of them.

The invariant mass distribution is given by
\begin{eqnarray}
 \frac{1}{\Gamma} \frac{d\Gamma}{dM_{ff}} = \frac{2}{(M_{ff}^{max})^2} M_{ff} 
 \label{eq:triangleshape}
\end{eqnarray}
and thus has a triangular shape, which is a direct consequence of the isotropic decay of particle $R$.

If $R$ is off-shell, the kinematic endpoint becomes
\begin{eqnarray}
 M_{ff}^{max} = M_X - M_O
\label{eq.invmass.2step.Mffmax1}
\end{eqnarray}
which is reached for a configuration where particle $O$ is at rest in the rest frame of $X$.
The shape of the invariant mass distribution $d\Gamma/dM_{ff}$ in this case is different from (\ref{eq:triangleshape}),
which allows the on-shell and off-shell cases to be distinguished \cite{Cho:2012er}.

\subsection{Three-step decays with on-shell particles}
\label{sect:invmass.3step}

\begin{figure}[t]
\large
\vspace*{0.5cm} 
\makebox[5cm][c]{${\mathbf C}_1$}
\makebox[5cm][c]{${\mathbf C}_2$}
\makebox[5cm][c]{${\mathbf C}_3$} \\
\makebox[5cm]{%
\includegraphics[width=0.3\textwidth]{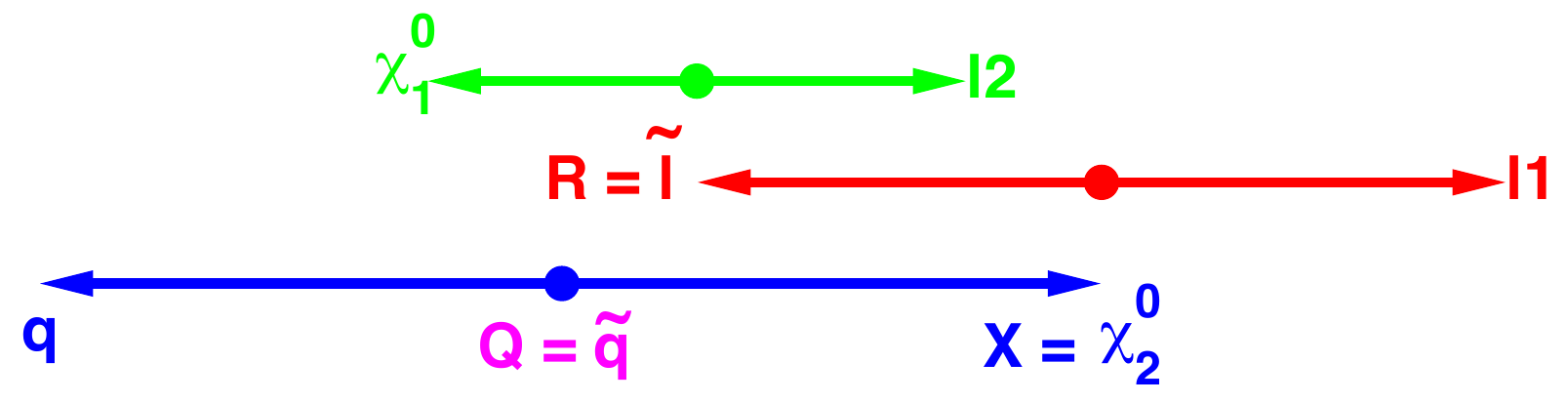}}
\makebox[5cm]{%
\includegraphics[width=0.3\textwidth]{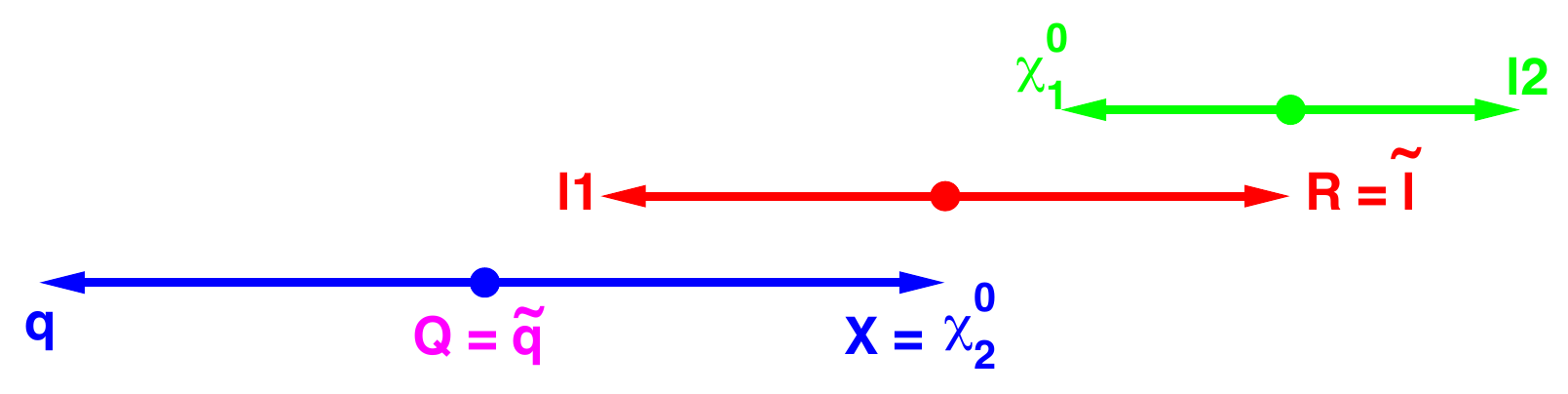}}
\makebox[5cm]{%
\includegraphics[width=0.3\textwidth]{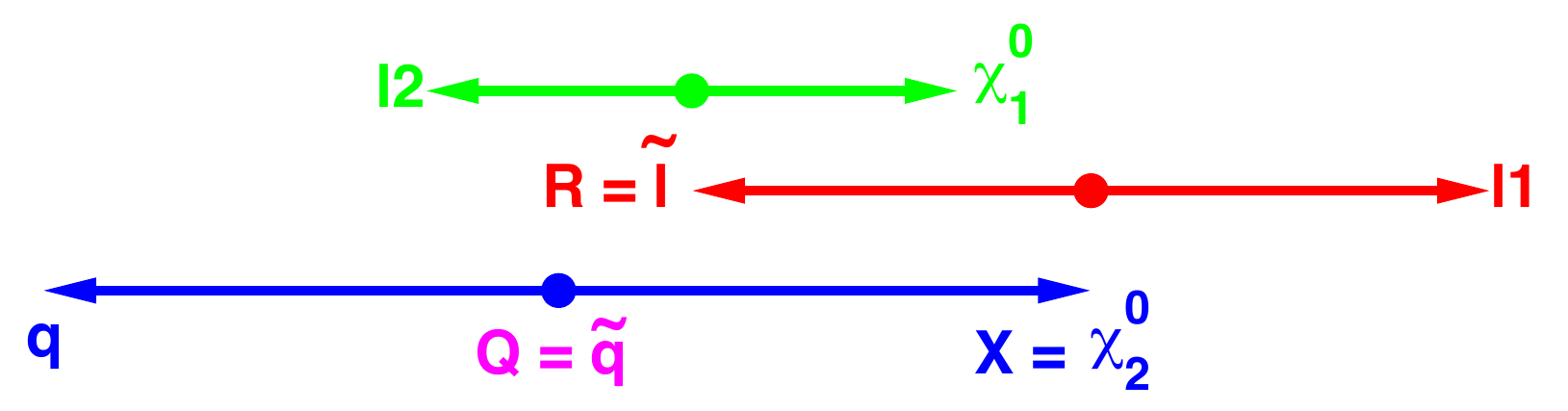}}
\vspace*{0.1cm} 
\makebox[5cm][c]{$M_{ff}^{min}$}   \makebox[5cm][c]{$M_{ff}^{max}$}  \makebox[5cm][c]{$M_{ff}^{max}$}  \\
\makebox[5cm][c]{$M_{qf_1}^{max}$} \makebox[5cm][c]{$M_{qf_1}^{min}$}  \makebox[5cm][c]{$M_{qf_1}^{max}$}  \\
\makebox[5cm][c]{$M_{qf_2}^{max,1}$} \makebox[5cm][c]{$M_{qf_2}^{max,2}$}  \makebox[5cm][c]{$M_{qf_2}^{min}$}  \\
\makebox[5cm][c]{$M_{qff}^{max,1}$} \makebox[5cm][c]{$M_{qff}^{max,2}$}  \makebox[5cm][c]{$M_{qff}^{max,3}$} 
\normalsize
\caption{The three relevant collinear configurations, $\mathbf{C}_1$, $\mathbf{C}_2$ and $\mathbf{C}_3$, for the three-step decay (\ref{eq:invmass.3step.general}),
illustrated for the case of a squark decay through a heavier neutralino $\tilde\chi^0_2$ and a charged
slepton $\tilde l$: $Q=\tilde q$, $X=\tilde\chi^0_2$, $R=\tilde l$, $O=\tilde\chi^0_1$,
$f_1=l_1$ and $f_2=l_2$.}
\label{fig.invmass.3step.configs}
\end{figure}

The next longest decay chain is the three-step decay of the generic type
\begin{eqnarray}
 Q \rightarrow q X, \qquad X \rightarrow f_1 R, \qquad R \rightarrow f_2 O,
\label{eq:invmass.3step.general}
\end{eqnarray}
as in the decay of a squark, $\sQua \rightarrow q \chiz_2$, followed by the same neutralino decay as in Sec.~\ref{sect:invmass.2step}.\footnote{Four-step 
decays could be encountered if a gluino decays to a squark which then gives rise to the three-step decay (\ref{eq:invmass.3step.general}).
A very large number of endpoints are then available. The corresponding analytical results are found in \cite{Luc06} and will not be repeated here.} 
In what follows, we shall assume that both intermediate particles $X$ and $R$ are on-shell\footnote{The case 
of a direct four-body decay can be distinguished from the one with on-shell intermediate sparticles.
It was demonstrated in \cite{Kim:2015bnd}  that for pure phase space decays, the endpoints of the invariant mass distributions
have to satisfy specific relations, which are summarized in App. \ref{sect:invmass.phasespace}.}.
In addition to the endpoint (\ref{eq.invmass.2step.Mffmax}), several other endpoints are available.
They are reached in the collinear configurations\footnote{For a pictorial three-dimensional view of the 
allowed parameter space for the decay chain (\ref{eq:invmass.3step.general}) in terms of the invariant masses
$M_{ff}$, $M_{qf_1}$ and $M_{qf_2}$, see Fig.~9 in Ref.~\cite{Kim:2015bnd}. 
The points $P_1$, $P_2$ and $P_3$ in Fig.~9 correspond to configurations $\mathbf{C}_1$, $\mathbf{C}_2$ and $\mathbf{C}_3$, respectively.} 
illustrated in Fig.~\ref{fig.invmass.3step.configs}.
(There is an additional configuration where all visible particles are parallel, but
in that case all invariant masses are zero and hence it can be ignored.)

The kinematic endpoint $M_{ff}^{max}$ for the dilepton mass $M_{ff}$ (reached in the collinear configurations $\mathbf{C}_2$ and $\mathbf{C}_3$, among many others)
is still given by (\ref{eq.invmass.2step.Mffmax}).
The formulas for the endpoints of the invariant mass distributions of the $(q f)$ systems are a bit more involved.
When the quark is paired with the ``near" lepton $(qf_1)$, the endpoint configuration can be $\mathbf{C}_1$ or $\mathbf{C}_3$ (among many others).
The formula is unique and is given in analogy to (\ref{eq.invmass.2step.Mffmax}) as
\begin{equation}
 M_{q f_1}^{max} = M_Q \sqrt{\left(1-\frac{M_X^2}{M_Q^2}\right)\left(1-\frac{M_R^2}{M_X^2}\right)}. 
 \label{Mqf1maxformula}
\end{equation}
On the other hand, when the quark is paired with the ``far" lepton, $(qf_2)$, the candidate endpoint configurations are $\mathbf{C}_1$ and $\mathbf{C}_2$, 
giving the respective endpoint formulas
\begin{eqnarray}
 M_{q f_2}^{max,1} &=&  M_Q \sqrt{\left( 1- \frac{M_X^2}{M_Q^2}\right)\frac{M_R^2}{M_X^2}\left(1-\frac{M_O^2}{M_R^2}\right)}, 
 \label{eq.invmass.3step.Mqfmax1}\\
 M_{q f_2}^{max,2} &=& M_Q \sqrt{\left( 1- \frac{M_X^2}{M_Q^2}\right)\left(1-\frac{M_O^2}{M_R^2}\right)}.
\label{eq.invmass.3step.Mqfmax}
\end{eqnarray}
From these two equations, it is easy to see that since $M_R<M_X$, the endpoint candidate value $M_{q f_2}^{max,1}$ is always less than $M_{q f_2}^{max,2}$ and therefore cannot be the true endpoint for the $(qf_2)$ case, implying that
\begin{equation}
 M_{q f_2}^{max}\equiv  M_{q f_2}^{max,2} .
\end{equation}

A possible complication arises when $f_1$ and $f_2$ cannot be distinguished on a case by case basis
which is precisely the situation with the ``near" and ``far" lepton of the SUSY cascade (\ref{eq:invmass.3step.general}).
One way to deal with the resulting two-fold ambiguity is to plot both combinations, $M_{q f_1}$ and $M_{q f_2}$,
and measure the kinematic endpoint of the resulting combined distribution $M_{q f_1}\cup M_{q f_2}$ \cite{Matchev:2009iw}.
Its value will be given by the larger of the two formulas, (\ref{Mqf1maxformula}) or (\ref{eq.invmass.3step.Mqfmax}),  depending on the sparticle spectrum.
(Having $M_{q f_2}^{max,2} > M_{q f_1}^{max}$ requires that $M_R^2 > M_X  M_O$.)

It is worth observing from eqs.~(\ref{eq.invmass.3step.Mqfmax1},\ref{eq.invmass.3step.Mqfmax}) and Fig.~\ref{fig.invmass.3step.configs},
that the maximum mass for the $(q f_2)$ system is 
reached for the same configuration, $\mathbf{C}_2$, which also maximizes the mass of the $(f f)$ system.
On the contrary, the maximum $(q f_1)$ mass is reached for any $M_{f f}$.
This allows to distinguish the two $(q f)$ endpoints,
by plotting, for example, $M_{f f}$ versus $M_{q f}$ 
for both combinations of the $(q f)$ system,
Although $f_1$ and $f_2$ cannot be distinguished on an event by event basis, their endpoints can thus unambiguously be identified \cite{Costanzo:2009mq,Burns:2009zi,Matchev:2009iw}.

Given that the $\chiz_2$ has spin $1/2$, the $(q f_1)$ mass distribution
is affected by spin effects.
However, the differences induced are between the $(q f^+)$ and $(q f^-)$
distributions and, if no sign selection is made, 
they cancel to yield a pure phase space distribution.
The invariant mass distribution for $M_{q f_1}$ is thus again triangular
\begin{eqnarray}
 \frac{1}{\Gamma} \frac{d\Gamma}{dM_{q f_1}} = \frac{2}{(M_{q f_1}^{max})^2} M_{q f_1}.
\label{eq.invmass.3step.Minvqf1}
\end{eqnarray}
The differential decay width in $M_{q f_2}$ has been computed
in \cite{Miller:2005zp} to be (neglecting spin correlations)
\begin{equation}
 \frac{1}{\Gamma} \frac{d \Gamma}{d M_{q f_2}}
 =
\left\{
 \begin{array}{l}
  \frac{4 M_{q f_2}}{(M_{q f_2}^{max})^2 \left(1-\frac{M_R^2}{M_X^2}\right)}
 \ln \frac{M_X}{M_R}
 \qquad \mathrm{for}\;\; 
  0 \leq M_{q f_2} \leq M_{q f_2}^{max,1};
 \\ 
 \frac{4 M_{q f_2}}{(M_{q f_2}^{max})^2 \left( 1-\frac{M_R^2}{M_X^2} \right)}
 \ln \frac{M_{q f_2}^{max}}{M_{q f_2}}
 \qquad \mathrm{for}\;\; 
 M_{q f_2}^{max,1} \leq M_{q f_2} \leq M_{q f_2}^{max}.
\end{array} 
\right.
\label{shape.invmass.3step.Minvqf2}
\end{equation}
It is made up of two pieces, the first one being again triangular in shape (i.e., simply proportional to $M_{q f_2}$), 
the second one first increasing then decreasing eventually to zero at the endpoint.
As $M_R \rightarrow M_X$, the second piece will be difficult to observe due to the small lever arm.
On the other hand, the shape of the combined distribution $M_{q f_1}\cup M_{q f_2}$ is given by the sum of
(\ref{eq.invmass.3step.Minvqf1}) and (\ref{shape.invmass.3step.Minvqf2}).
If the mass spectrum is such that $M_R^2 > M_X  M_O$, the true upper kinematic endpoint of the combined distribution is 
given by $M_{q f_2}^{max}$, but the ``hidden" endpoint $M_{q f_1}^{max}$ may also be observable, 
due to the characteristic triangular shape (\ref{eq.invmass.3step.Minvqf1}).

The endpoint of the invariant mass distribution of $M_{q f f}$ is more complicated.
As indicated in Fig. \ref{fig.invmass.3step.configs}, 
there are three collinear configurations which can determine the endpoint,
depending on the mass spectrum:
\begin{eqnarray}
 M_{q f f}^{max, 1} &=& M_Q \sqrt{\left(1-\frac{M_X^2}{M_Q^2}\right)\left(1-\frac{M_O^2}{M_X^2}\right)} \qquad  \mathrm{if} \qquad \frac{M_X}{M_Q} \leq \frac{M_R}{M_X}\frac{M_O}{M_R} ,\label{eq.invmass.3step.Mqffmax1}  \\
 M_{q f f}^{max, 2} &=& M_Q \sqrt{\left(1-\frac{M_R^2}{M_Q^2}\right)\left(1-\frac{M_O^2}{M_R^2}\right)} \qquad \mathrm{if} \qquad \frac{M_O}{M_R} \leq \frac{M_R}{M_X}\frac{M_X}{M_Q}  ,
 \label{eq.invmass.3step.Mqffmax2} \\
 M_{q f f}^{max, 3} &=& M_Q \sqrt{\left(1-\frac{M_R^2}{M_X^2}\right)\oneMrat{(M_X M_O)}{(M_Q M_R)}} \qquad \mathrm{if} \qquad \frac{M_R}{M_X} \leq \frac{M_O}{M_R}\frac{M_X}{M_Q},
\label{eq.invmass.3step.Mqffmax3}
\end{eqnarray}
where ``if" represents the condition under which the endpoint is the true endpoint.
Finally, there is yet another candidate configuration, which is {\bf acollinear} --- it describes the case when
the unseen neutral $O$ is at rest in the $Q$ rest frame\footnote{This situation can 
also be reached in the collinear configurations of Fig.~\ref{fig.invmass.3step.configs}, but only
for specific relations among the sparticle masses: 
$M_Q M_O = M_X^2$ for ${\mathbf C}_1$,
$M_Q M_O = M_R^2$ for ${\mathbf C}_2$
and
$M_Q M_R^2 = M_O M_X^2$ for ${\mathbf C}_3$.}. 
The corresponding candidate value
for the endpoint is given by
\begin{equation}
 M_{q f f}^{max, 4} = M_Q - M_O.
\label{eq.invmass.3step.Mqffmaxacol}
\end{equation}
The actual kinematic endpoint for $M_{q f f}$ is then given by
\begin{equation}
 M_{q f f}^{max} = \max \left\{  M_{q f f}^{max, 1},  M_{q f f}^{max, 2},  M_{q f f}^{max, 3},  M_{q f f}^{max, 4} \right\}.
 \label{truemqffmax}
\end{equation}

The shape of the distribution for $M_{q f f}$ is also known from \cite{Miller:2005zp}, but the formula is rather complicated and will not be quoted here.

In summary, for the three-step decay chain (\ref{eq:invmass.3step.general}) 
we already have four measurable endpoints, $M_{ff}^{max}$, $M_{q f_1}^{max}$, $M_{q f_2}^{max}$ and $M_{q f f}^{max}$,
which in principle should be sufficient to determine the four unknown masses involved.
However, the formula for $M_{q f f}^{max}$ corresponding to the true endpoint depends on the mass hierarchy,
which introduces an ambiguity (see \cite{Burns:2009zi} and references therein).
An additional problem is that $M_{q f f}^{max, 2}$ and $M_{q f f}^{max, 3}$ are not independent of the other endpoints, as
\begin{eqnarray}
 \left( M_{q f f}^{max, 2} \right)^2 = (M_{q f_2}^{max})^2 + (M_{f f}^{max})^2 \nonumber \\
  \left( M_{q f f}^{max, 3} \right)^2 = (M_{q f_1}^{max})^2 + (M_{f f}^{max})^2
\end{eqnarray}
so that if one of these gives the true $M_{q f f}^{max}$ endpoint (\ref{truemqffmax}), the $M_{q f f}^{max}$ measurement 
does not introduce an independent constraint and the mass values cannot be completely determined \cite{Gjelsten:2004ki}.
This situation can be detected by correlations,
as $M_{q f f}^{max, 2}$ and $M_{q f f}^{max, 3}$ occur in configurations $\mathbf{C}_2$ and $\mathbf{C}_3$, respectively,
for which $M_{f f}$ is maximum,
whereas $M_{q f f}^{max, 1}$ is obtained in configuration $\mathbf{C}_1$ where $M_{f f}$ is minimum.

In fact, there may also be a lower endpoint for $M_{q f f}$.
The minimum value of $M_{q f f}$ is zero in general.
But it is possible to obtain a non-zero lower endpoint value by selecting events where
$M_{f f}$ is above some minimal threshold.
For example, taking events with $cos\theta^* > 0$ in the rest frame of
$R$, i.e. with $M_{f f} > M_{f f}^{max} / \sqrt{2}$, the following lower endpoint
is produced in the $M_{q f f}$ distribution:
\begin{eqnarray}
 \left( M_{q f f}^{min} \right)^2 &=& M_Q^2 + M_O^2 - \frac{1}{4 M_X^2 M_R^2}
   \left[ (M_Q^2 + M_X^2) (M_X^2 + M_R^2) (M_R^2 + M_O^2) \right. \nonumber \\
   &+& \left. (M_Q^2 - M_X^2)
   \sqrt{(M_X^2 - M_R^2)^2(M_R^2 + M_O^2)^2 + 4 M_X^2 M_R^2 (M_R^2 - M_O^2)^2}
   \right]
\end{eqnarray}
which was used as additional mass constraint in \cite{Bachacou:1999zb}.

Perhaps a better approach would be to compute the mass distribution of $M_{q f_1}^2 + M_{q f_2}^2$ \cite{Matchev:2009iw},
for which the endpoint is given by
\begin{eqnarray}
 \left( M_{q f_1}^2 + M_{q f_2}^2 \right)^{max} = \left( M_{q f f}^{max, 1} \right)^2.
\label{eq.invmass.3step.Mqf1f2sqmax}
\end{eqnarray}
This additional measurement can be used to verify whether the observed $M_{q f f}^{max}$ indeed corresponds to $M_{q f f}^{max, 1}$,
and if not, use $M_{q f f}^{max, 1}$ as an additional measurement. 

To summarize, in spite of all the different challenges, as shown in \cite{Matchev:2009iw,Debnath:2016gwz},
in principle one can always determine all four masses involved in the three-step decays.

Finally, a possibility which was not used so far would be to determine the lower and upper endpoints of $M_{q f f}$
for a fixed value of $M_{f f}$.
These are reached in a collinear configuration with both fermions parallel
and their system respectively along or opposite to the direction of $Q$.
Then,
\begin{eqnarray}
 \left( M_{q f f}^{min, max} \right)^2 &=& M_Q^2 + M_O^2 - \frac{1}{2} M_Q^2 \left(1+\frac{M_X^2}{M_Q^2}\right) \left(1+ \frac{M_O^2}{M_X^2}\right)
  + \frac{M_Q^2}{2 M_X^2} \left(1+\frac{M_X^2}{M_Q^2}\right) M_{f f}^2 \nonumber \\
  & \mp& \frac{M_Q^2}{2} \left(1- \frac{M_X^2}{M_Q^2}\right)
       \sqrt{ \left( 1 - \frac{M_O^2}{M_X^2} - \frac{M_{f f}^2}{M_X^2} \right)^2 
       - 4 \frac{M_O^2}{M_X^2} \frac{M_{f f}^2}{M_X^2}}.
\end{eqnarray}
Hence, this dependence on $M_{f f}$ could be fitted, which  might provide more accurate values of the masses $M_Q$, $M_X$ and $M_O$ than from endpoint
measurements only.
The mass $M_R$ remains however undetermined and requires the use of other endpoints,
in principle $M_{q f_1}^{max}$ and/or $M_{q f_2}^{max}$.

So far, the formulae were derived under the assumption that the masses of all visible particles can be neglected.
There are, however, final states with top or $W / Z / h$ where the masses cannot be neglected.
The formulae in these cases are rather cumbersome, but some examples are collected in Appendices \ref{sect:invmass.hquark} and \ref{sect:invmass.h0Z0}.

\subsection{Three-step decays with an off-shell particle $R$}
\label{sect:invmass.3stepoff}

In the three-step decay sequence (\ref{eq:invmass.3step.general}), one of the intermediate particles, e.g., $R$, can be off-shell ($M_R>M_X$),
leading to the generic 2-body plus 3-body decay chain
\begin{eqnarray}
 Q \rightarrow q X, \qquad X \rightarrow f_1 f_2 O.
\label{eq:invmass.3stepoff.general}
\end{eqnarray}
This situation may occur when a squark decay $\sQua \rightarrow q \chiz_2$ is followed by a three-body neutralino decay
$\chiz_2 \rightarrow l_1 l_2 \chiz_1$, which is the case when the sleptons are too heavy.
As the two fermions are indistinguishable, we can form the invariant mass distributions for $M_{ff}$, $M_{qf}$ and $M_{qff}$.
Their kinematic endpoints are given by \cite{Lester:2006cf}
\begin{eqnarray}
 M_{f f}^{max} &=& M_X - M_O,  \label{eq.invmass.offshell.Mffmax}  \\
 M_{q f}^{max} &=& M_Q \sqrt{\left(1-\frac{M_X^2}{M_Q^2}\right)\left(1-\frac{M_O^2}{M_X^2}\right)}, \label{eq.invmass.offshell.Mqfmax} \\
 M_{q f f}^{max} &=& 
 \left\{
 \begin{array}{ll}
 M_Q \sqrt{\left(1-\frac{M_X^2}{M_Q^2}\right)\left(1-\frac{M_O^2}{M_X^2}\right)}, &  \mathrm{if\ } M_Q M_O \geq M_X^2; \\ [2mm]
 M_Q-M_O,  &  \mathrm{otherwise}. 
 \end{array}
 \right.
\label{eq.invmass.3stepoff.Mmax}
\end{eqnarray}
The upper line in (\ref{eq.invmass.3stepoff.Mmax}) is realized when
one fermion is at rest in the rest frame of $X$, leading to the same expression as in (\ref{eq.invmass.offshell.Mqfmax}).
Therefore, in the case of $M_Q M_O \geq M_X^2$, eqs.~(\ref{eq.invmass.offshell.Mffmax}-\ref{eq.invmass.3stepoff.Mmax})
provide only two constraints for three unknowns $M_Q$, $M_X$ and $M_O$, and the masses cannot be fully determined.
On the other hand, when $M_Q M_O \leq M_X^2$,
the $M_{q f f}$ endpoint is given by the lower line in (\ref{eq.invmass.3stepoff.Mmax}), which arises from a non-collinear configuration 
(with $O$ at rest in the rest frame of $Q$).
In that case, all masses can be reconstructed.

\subsection{Endpoints in squark cascade decays to heavy quarks}
\label{sect:invmass.hquark}

In this section, we reconsider the three-step decay chain (\ref{eq:invmass.3step.general}),
only this time we allow the quark $q$ to be massive, $m_q>0$, as exemplified by the top quark.
A typical example for such decay would be
\begin{eqnarray}
\sTop \rightarrow t \chiz_2 , \qquad \chiz_2 \rightarrow f_1 + \sFer , \qquad \sFer \rightarrow f_2 + \chiz_1.
\end{eqnarray}
Given the large top mass, the expressions from Appendix~\ref{sect:invmass.3step} are not applicable and need to be generalized.
To simplify the notation, let us define
\begin{eqnarray}
 E_q &=& \frac{M_Q^2 + m_q^2 - M_X^2}{2 M_Q} , \qquad p_q = \sqrt{E_q^2 - m_q^2},  \\
 E_X &=& \frac{M_Q^2 + M_X^2 - m_q^2}{2 M_Q} , \qquad p_X = \sqrt{E_X^2 - M_X^2}, 
\end{eqnarray}
with the energy and momentum expressed in the squark rest frame.
Since we treat the quark as massive, we will have both lower and upper endpoints.
For $M_{q f_1}$ they are
\begin{eqnarray}
  (M_{q f_1}^{max})^2 &=& m_q^2 + \frac{1}{2} M_Q^2  \left( 1 - \frac{M_X^2}{M_Q^2} + \frac{m_q^2}{M_Q^2} \right)
   \left( 1 - \frac{M_R^2}{M_X^2} \right) \frac{E_X + p_X}{M_Q} \left( 1 + \sqrt{1 - \frac{m_q^2}{E_q^2}}\right),~~~ \\
  (M_{q f_1}^{min})^2 &=& m_q^2 + \frac{1}{2} M_Q^2  \left( 1 - \frac{M_X^2}{M_Q^2} + \frac{m_q^2}{M_Q^2} \right)
   \left( 1 - \frac{M_R^2}{M_X^2} \right) \frac{E_X - p_X}{M_Q} \left( 1 - \sqrt{1 - \frac{m_q^2}{E_q^2}}\right).
\end{eqnarray}
Similarly, the two endpoints for $M_{q f_2}$ are
\begin{eqnarray}
  (M_{q f_2}^{max})^2 &=& m_q^2 + \frac{1}{2} M_Q^2  \left( 1 - \frac{M_X^2}{M_Q^2} + \frac{m_q^2}{M_Q^2} \right)
   \left( 1 - \frac{M_O^2}{M_R^2} \right) \frac{E_X + p_X}{M_Q} \left( 1 + \sqrt{1 - \frac{m_q^2}{E_q^2}}\right), \\
  (M_{q f_2}^{min})^2 &=& m_q^2 + \frac{1}{2} M_Q^2  \left( 1 - \frac{M_X^2}{M_Q^2} + \frac{m_q^2}{M_Q^2} \right)
   \left( 1 - \frac{M_O^2}{M_R^2} \right) \frac{M_R^2}{M_X^2} \frac{E_X - p_X}{M_Q} \left( 1 - \sqrt{1 - \frac{m_q^2}{E_q^2}}\right).~~~~
\end{eqnarray}
Finally, the $M_{q f f}$ endpoints in configuration ${\mathbf C}_1$ are
\begin{eqnarray}
  (M_{q f f}^{max})^2 &=& m_q^2 + \frac{1}{2} M_Q^2  \left( 1 - \frac{M_X^2}{M_Q^2} + \frac{m_q^2}{M_Q^2} \right)
   \left( 1 - \frac{M_O^2}{M_X^2} \right) \frac{E_X + p_X}{M_Q} \left( 1 + \sqrt{1 - \frac{m_q^2}{E_q^2}}\right),  \\
  (M_{q f f}^{min})^2 &=& m_q^2 + \frac{1}{2} M_Q^2  \left( 1 - \frac{M_X^2}{M_Q^2} + \frac{m_q^2}{M_Q^2} \right)
   \left( 1 - \frac{M_O^2}{M_X^2} \right) \frac{E_X - p_X}{M_Q} \left( 1 - \sqrt{1 - \frac{m_q^2}{E_q^2}}\right).
\end{eqnarray}
Depending on the relations among the sparticle masses, other configurations may sometimes yield the true $M_{q f f}$ endpoint.
The corresponding formulas can be derived analogously and will not be listed here.

Used together with the $M_{f f}$ endpoint, the above endpoints in
$M_{q f}$ and $M_{q f f}$ in principle provide sufficient information to reconstruct 
the sparticle masses.
However, the formulas are clearly more cumbersome than for massless visible systems.

\subsection{Endpoints in squark or gluino cascade decays to heavy SM bosons}
\label{sect:invmass.h0Z0}

The discussion of the benchmark scenarios in Section~\ref{sect:scenarios}
revealed that at higher SUSY mass scales, it is quite possible that the 
electroweakinos decay to on-shell SM bosons,
$W$, $Z$ or $h$. For concreteness, consider a heavy neutralino decay to a Higgs boson, e.g.,
$\chiz_2\to h\chiz_1$.
The full decay chain starting from the squark is 
\begin{eqnarray}
 \sQua \rightarrow q \chiz_2 \rightarrow q h \chiz_1.
\label{squarktoHiggs}
\end{eqnarray}
As discussed in Section~\ref{sec:hadronicselection}, the Higgs boson $h$ can be identified through its decay into $b \bar{b}$
and its mass will therefore  be reconstructed.
Note that the formulas derived below for the case of $h$ from $\chiz_2\to h\chiz_1$
are equally applicable to the analogous decays $\chiz_2\to Z\chiz_1$ and $\chipm_1\to W^\pm \chiz_1$.

For the decay chain (\ref{squarktoHiggs}), the only available invariant mass is $M_{q h}$, whose endpoints are
\begin{eqnarray}
 (M_{qh}^{max})^2 &=& M_h^2 + \frac{M_Q^2 - M_X^2}{2 M_X^2}
 \left( M_X^2 - M_O^2 + M_h^2 + 
 \sqrt{(M_X^2 - M_O^2 - M_h^2)^2 - 4 M_O^2 M_h^2} \right), \label{Mqhmax} \\
 (M_{qh}^{min})^2 &=& M_h^2 + \frac{M_Q^2 - M_X^2}{2 M_X^2}
 \left( M_X^2 - M_O^2 + M_h^2 - 
 \sqrt{(M_X^2 - M_O^2 - M_h^2)^2 - 4 M_O^2 M_h^2} \right). ~~~~  \label{Mqhmin} 
\end{eqnarray}
(Similar formulas are obtained when considering the decays to $Z$ or $W^\pm$.)
If it is assumed that the decay follows approximately phase space,
the upper kinematic endpoint should be well visible, as the 
distribution should drop steeply near the upper end.

Even if we assume that both endpoints (\ref{Mqhmax}) and (\ref{Mqhmin})
are measurable, they lead to only two constraints on the three unknown masses.
Thus they do not, on their own, allow a determination of all masses involved.
They may, however, bring valuable information if used in conjunction 
with another decay mode, e.g. the $Z$ or di-leptons,
provided the latter has a sufficient branching ratio. 
An interesting, but rather infrequent, situation may arise,
where the two decay modes of the squark,
$\sQua \rightarrow q \chiz_2 \rightarrow q \sLep l \to q l l \chiz_1$ and 
$\sQua \rightarrow q \chiz_4 \rightarrow q h \chiz_1$,
compete, allowing the $h$ decay channel (taken together with the di-lepton channel)
to determine the mass of the $\chiz_4$.

On the other hand, if the squark itself originates from the decay of a gluino,
more constraints become available, which would allow an unambiguous determination of the masses.
This motivates us to extend the analysis to gluino decays of the type
$ \sGlu \rightarrow q_1 \sQua  \rightarrow q_1 q_2 \chiz_2 \rightarrow q_1 q_2 h \chiz_1$,
or in terms of our generic notation,
\begin{eqnarray}
 G \rightarrow q_1 Q, \qquad  Q \rightarrow q_2 X , \qquad X \rightarrow h O,
\end{eqnarray}
where the additional kinematical quantities available are $M_{q_1 q_2}$, $M_{q_1 h}$
and $M_{q_1 q_2 h}$.
The $M_{q_2 h}$ endpoints are the same as the ones given above in eqs.~(\ref{Mqhmax}) and (\ref{Mqhmin}).

The $M_{q_1q_2}$ mass distribution is triangular and its endpoint is given by a formula analogous to (\ref{eq.invmass.2step.Mffmax}):
\begin{eqnarray}
 M_{q_1q_2}^{max} = M_G \sqrt{\left(1-\frac{M_Q^2}{M_G^2}\right)\left(1-\frac{M_X^2}{M_Q^2}\right)}.
\label{eq.higgsgdec.q1q2.q1q2max}
\end{eqnarray}
The upper and lower endpoints of $M_{q_1 h}$ are
\begin{eqnarray}
 (M_{q_1 h}^{max})^2 &=& M_h^2 + \frac{M_G^2 - M_Q^2}{2 M_X^2}
 \left( M_X^2 - M_O^2 + M_h^2 + 
 \sqrt{(M_X^2 - M_O^2 - M_h^2)^2 - 4 M_O^2 M_h^2} \right),  \\
 (M_{q_1 h}^{min})^2 &=& M_h^2 + \frac{M_G^2 - M_Q^2}{2 M_X^2}
 \left( M_X^2 - M_O^2 + M_h^2 - 
 \sqrt{(M_X^2 - M_O^2 - M_h^2)^2 - 4 M_O^2 M_h^2} \right). ~~~
\end{eqnarray}
The configuration in which the upper endpoint of $M_{q_2 h}$
is reached is independent of the orientation with respect to $q_1$,
hence independent of the mass $M_{q_1 q_2}$.
On the other hand, the upper endpoint of $M_{q_1 h}$ requires $M_{q_1 q_2}$
to be minimal.
This allows the two types of $M_{qh}$ endpoints to be distinguished.

It is also possible to compute the expressions for the endpoint of $M_{q_1 q_2 h}$.
This leads to the same type of ambiguities depending on the mass hierarchy as observed in Appendix~\ref{sect:invmass.3step}.
As we already have a sufficient number of endpoints to determine all masses involved,
we will not discuss them here.

Note that the above expressions can also be applied to the decay chain of $\sQua \rightarrow q \chipm_1$, $\chipm_1 \rightarrow W^\pm \chiz_1$,
where the $W$ would decay hadronically.
The mass $M_X$ is then the chargino mass and $M_h$ is to be replaced by $M_W$.

\section{Mass Determination from $M_{T2}$ or $M_2$ Endpoints}
\label{sect:MT2}

The invariant mass techniques discussed in the previous Appendix~\ref{sect:invmass} were focused on 
a single SUSY decay chain. At the same time, in the MSSM with R-parity conservation the production of sparticles occurs in pairs,
so that there are {\em two} SUSY decay chains per event. Each of them
may proceed through one or several steps via intermediate sparticles and end with the production of the LSP.
The LSP is a weakly interacting particle that remains undetected and will leave a signature of missing momentum (MET).
Rather than analyzing the sparticle contents in the same decay sequence, as discussed above,
it is also possible to analyze both decay chains simultaneously by using the variables $M_{T2}$ or $M_2$.
$M_{T2}$ was originally developed \cite{MT2variable} for di-jet and di-lepton events, but could also
be used for multi-jet events provided the event is viewed as two ``pseudo-jets" associated with the original sparticles produced \cite{Barr:2003rg}.
The formation of ``pseudo-jets" can be accomplished by the ``hemisphere" reconstruction method \cite{hemisphere}.

\subsection{Definition of $M_{T2}$}
\label{sect:MT2.MT2def}

As the next technique of mass reconstruction, we consider the variable $M_{T2}$, introduced by the Cambridge group in \cite{MT2variable} and defined as
\begin{eqnarray}
 M_{T2} (\Mtil_0) = \mathrm{min}_{\vec{p}_T^{\;\chi (1)} + \vec{p}_T^{\;\chi (2)} = \vec{p}_T ^{\;miss}} \left[ \mathrm{max} \left(M_T^{(1)}, M_T^{(2)}\right) \right],
 \label{eq:MT2basicdef}
\end{eqnarray}
where the numerical subscript $(i)$ identifies the hemisphere, $i=1$ or $i=2$.
$M_T^{(i)}$ is the transverse mass of the $i$-th hemisphere, including the unseen neutral particle $\chi(i)$, and is defined by
\begin{eqnarray}
 \left( M_{T}^{(i)} \right)^2 &=& \left( E_{T}^{(i)} +  E_{T}^{\chi(i)} \right)^2 - \left( \vec{p}_{T}^{\; (i)} + \vec{p}_{T}^{\; \chi(i)} \right)^2 \nonumber \\
 &=& \left( m^{(i)} \right)^2 +  \left( m^{\chi(i)} \right)^2 + 2 \left(  E_{T}^{(i)} E_{T}^{\chi(i)} - \vec{p}_{T}^{\; (i)} \cdot \vec{p}_{T}^{\; \chi(i)} \right),
\end{eqnarray}
where $E_{T}^{(i)}$ and $\vec{p}_{T}^{\; (i)}$ are respectively the transverse energy and transverse momentum of the visible pseudo-particle 
in the $i$-th hemisphere, while $E_{T}^{\chi(i)}$ and $\vec{p}_{T}^{\; \chi(i)}$ are the corresponding transverse energy and transverse momentum of 
the unseen neutral particle $\chi(i)$:
\begin{eqnarray}
\left( E_{T}^{(i)} \right)^2-\left( \vec{p}_{T}^{\; (i)} \right)^2 &=& \left( m^{(i)} \right)^2 \\ [2mm]
\left(E_{T}^{\chi(i)} \right)^2 - \left(\vec{p}_{T}^{\; \chi(i)} \right)^2 &=& \left( m^{\chi(i)} \right)^2 \equiv \tilde M_0^2.
\end{eqnarray}
To compute $M_{T2}$, one chooses the larger of the two $M_T^{(i)}$ and minimizes its value by varying the transverse momenta $\vec{p}_{T}^{\; \chi(i)}$
for the unseen neutrals $\chi(i)$, keeping their vector sum equal to the missing transverse momentum $\vec{p}_T^{\; miss}$.
The mass of the unseen neutral, $\Mtil_0$, assumed to be the same in the two decay legs, remains a free variable.
This minimization ensures the resulting $M_{T2}$ to remain always lower than the true mass of the parent particle,
provided that the true value $M_0$ of the mass for the unseen neutral is used.
This means that the true mass of the parent particle of the decay chain can be measured as the 
endpoint of the $M_{T2}$ distribution.\footnote{The above discussion still leaves open the question whether the 
parent mass bound is saturated or not --- for details and some counterexamples, see \cite{Barr:2011xt,Mahbubani:2012kx}.}

\subsection{Mass determination from $M_{T2}$ in a one-step decay}
\label{sect:MT2.MT2}

The variable $M_{T2}$ can be computed for the complete event analytically \cite{Cho:2007qv}, 
after subdividing the latter into two hemispheres \cite{hemisphere}.
As the mass of the LSP is a free parameter (the test mass $\Mtil_0$ in the $M_{T2}$ calculation),
we can vary the LSP test mass $\Mtil_0$ and measure the endpoint $M_{T2}^{max}$ of the $M_{T2}$ distribution for each mass value.
This procedure results in a function, $M_{T2}^{max}(\Mtil_0)$, which is indicative of the mass splitting between the parent and the LSP.
In order to see this explicitly, consider the generic one-step decay without any intermediate sparticles,
\begin{equation}
X_1 \rightarrow x_0 X_0,
\label{1stepgeneric}
\end{equation}
where $x_0$ is a visible system and the capital letters correspond to sparticles, with $X_1$ ($X_0$) having mass $M_1$ ($M_0$).
Under those circumstances, the function $M_{T2}^{max}(\Mtil_0)$ is known analytically.
For events without ISR jets, it is given by \cite{Cho:2007dh,Burns:2008va}
\begin{eqnarray}
 M_{T2}^{max} ( \Mtil_0 ) = \mu + \sqrt{\mu^2 + \Mtil_0^2},
 \label{mt2max110}
\end{eqnarray}
where 
\begin{eqnarray}
 \mu = \frac{M_1}{2} \left( 1 - \frac{M_0^2}{M_1^2} \right) = \frac{1}{2} \frac{M_1^2 - M_0^2}{M_1}.
 \label{mu110definition}
\end{eqnarray}
Eq.~(\ref{mt2max110}) shows that $\mu$ is the only mass parameter which can be extracted from these repeated $M_{T2}$ endpoint measurements\footnote{Measurements of 
$M_{T2}$  endpoints for several test mass values $\tilde M_0$ do not help, as all of them measure the same value of $\mu$. In fact,
in order to extract the value of $\mu$, a single measurement of the $M_{T2}$
endpoint for an arbitrarily chosen value of the test mass is sufficient.}.
Of course, for the correct value of the LSP mass, $\tilde M_0=M_0$, the function (\ref{mt2max110}) will provide the
correct value $M_1$ of the parent mass:
\begin{equation}
M_1= M_{T2}^{max} ( M_0 ),
\end{equation}
however, without additional considerations, one would not know what is the proper value of $\tilde M_0$ to use in the above equation.
For example, a convenient choice would be to take $\Mtil_0 = 0$, but then
\begin{eqnarray}
 M_{T2}^{max} ( \Mtil_0 = 0 ) = 2 \mu = \frac{M_1^2 - M_0^2}{M_1}
\end{eqnarray}
would only lead to a lower limit on $M_1$ (since $M_0$ cannot be negative).

In summary, the measurement of the $\mu$ parameter from the $M_{T2}$ endpoint (\ref{mt2max110})
provides a relation between $M_1$ and $M_0$, but does not allow both masses to be obtained.\footnote{This difficulty 
only exists for one-step decays, like $\sQua \rightarrow q \chiz_1$. 
For longer decay chains, it will be seen below that full mass determination from $M_{T2}$ endpoint measurements is possible.}
To make further progress, three different approaches have been proposed, and we now discuss each one of them in turn.

\subsubsection{Kink method}

If the $M_{T2}$ variable is computed for the whole event, as originally proposed in \cite{MT2variable},
any potential ISR jets are a priori included in the calculation of $M_{T2}$.
This will distort the $M_{T2}$ distribution, which can now extend beyond the true endpoint.
One possibility would be to identify by some method\footnote{For leptonic channels, this identification is rather trivial
\cite{Matchev:2009fh}.} the ISR jets, and treat them 
as an upstream system with some Upstream Transverse Momentum (UTM) $\vec{P}_T$.
There exists no analytical form for the calculation of the event by event $M_{T2}$ in the presence of UTM,
but it can be calculated numerically with the codes of \cite{Cheng:2008hk,Walker:2013uxa,Lester:2014yga,Lally:2015xfa}.

The $M_{T2}$ endpoint formula for a one-step decay with UTM is in full generality built out of two pieces  \cite{Burns:2008va}
\begin{eqnarray}
 M_{T2}^{max} ( \Mtil_0, P_T ) = F_{L, R} ( \Mtil_0, P_T )
  \label{mt2max110PT}
\end{eqnarray}
where the function $F_L$ applies to the ``left" branch with $\Mtil_0 \leq M_0$,
while the function $F_R$ gives the ``right" branch with $\Mtil_0 \geq M_0$.
These functions are given by
\begin{eqnarray}
 F_L ( \Mtil_0, P_T ) &=& \left\{ \left[ \mu (P_T) + \sqrt{ \left( \mu (P_T) + \frac{P_T}{2} \right)^2  + \Mtil_0^2} \right]^2 - \frac{P_T^2}{4} \right\}^{1/2}, \label{FL110}\\
 F_R ( \Mtil_0, P_T ) &=& \left\{ \left[ \mu (-P_T) + \sqrt{ \left( \mu (-P_T) - \frac{P_T}{2} \right)^2  + \Mtil_0^2} \right]^2 - \frac{P_T^2}{4} \right\}^{1/2}  \label{FR110}
\end{eqnarray}
in terms of the $P_T$-dependent $\mu$ parameter
\begin{eqnarray}
 \mu ( P_T ) = \mu \cdot \left( \sqrt{1 + \left( \frac{P_T}{2 M_1} \right)^2} - \frac{P_T}{2 M_1} \right).
\end{eqnarray}
Hence, the two branches are related by
\begin{eqnarray}
 F_R ( \Mtil_0, P_T ) = F_L ( \Mtil_0, -P_T )
\end{eqnarray}
It can be easily verified that for $\Mtil_0 = M_0$, the two branches join together.
The two branches correspond to special momentum configurations in which all three transverse momentum vectors are collinear:
the transverse momenta of the two visible systems are parallel to each other, and then they are either parallel (for $F_L$) or anti-parallel (for $F_R$) to the UTM $\vec{P}_T$,
see Appendix A1 of \cite{Burns:2008va}.

\begin{figure}[htb]
 \begin{center}
 \includegraphics[width=0.49\textwidth]{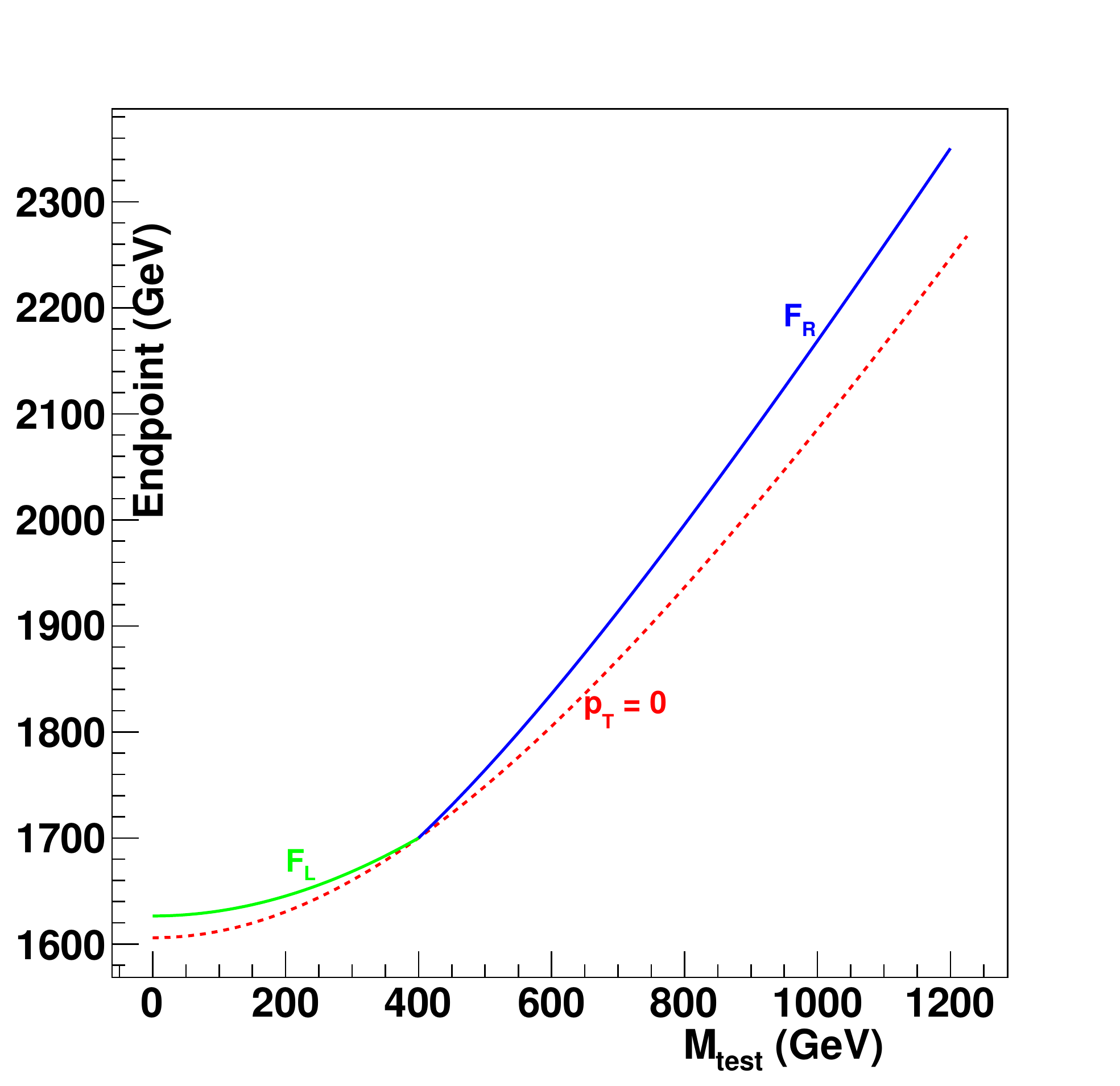}
 \caption{The endpoint value $M_{T2}^{max}$ versus the test mass $\tilde M_0$,
for the case of $M_0 = 400$ GeV and $M_1 = 1700$ GeV.
 The red dotted line shows the case of $P_T = 0$, i.e., eq.~(\ref{mt2max110})
 The green (blue) solid line shows $F_L$ ($F_R$) for a fixed $P_T = 1$ TeV.
}
 \label{fig.MT2.MT2.endpoint}
 \end{center}
\end{figure}

The dependence of the $M_{T2}$ endpoint (\ref{mt2max110PT}) on the test mass $\tilde M_0$ is illustrated in Fig.~\ref{fig.MT2.MT2.endpoint}.
The case without any upstream momentum, $P_T = 0$, is shown with the red dotted line, which is simply the function from eq.~(\ref{mt2max110}).
The solid lines in Fig.~\ref{fig.MT2.MT2.endpoint} then illustrate the two branches (\ref{FL110}) and (\ref{FR110})
for  the case when the UTM is significant, $P_T = 1000$ GeV.

It can be shown mathematically that when the test mass $\tilde M_0$ equals the true LSP mass $M_0$, we have
\begin{eqnarray}
 M_{T2}^{max} ( \Mtil_0 = M_0, P_T ) = M_1
 \label{eq:ptindep}
\end{eqnarray}
regardless of the value of $P_T$.  Hence, all branches cross at a single point, $(M_0 , M_1)$, for {\em any} value of the $P_T$
(this property is also visible in Fig.~\ref{fig.MT2.MT2.endpoint}).
But, since the expressions (\ref{FL110}) and (\ref{FR110}) for $F_L$ and $F_R$ are different, 
this also means that at the point $\Mtil_0 = M_0$ there is a kink in the function (\ref{mt2max110PT})
when viewed as a function of $\tilde M_0$ \cite{Cho:2007dh}.
This property would allow to locate the true value of the LSP mass.
Unfortunately, the kink is not always visible, which limits the applicability of this method.

\subsubsection{Using the dependence on the UTM }

An alternative approach would be to fit the $P_T$-dependent endpoint function (\ref{mt2max110PT}) to the data.
As now there is an explicit dependence of $\mu ( P_T )$ on $P_T$ and $M_1$, this provides an additional 
handle for the determination of both $M_0$ and $M_1$.
Thus, {\em for a one-step decay with UTM, $M_{T2}$ enables us to determine both sparticle masses $M_0$ and $M_1$.}
In principle, it is sufficient to measure the endpoints for two $p_T$ bins at a fixed value of the test mass,
or alternatively, to measure the endpoints at a fixed (non-zero) value of $P_T$ at two points in $\Mtil_0$ ---
one below, the other above the true LSP mass (in other words, one point on each of the branches $F_L$ and $F_R$).
An example application of this method to the same-sign di-lepton final states has been presented in \cite{Matchev:2009fh}.

\subsubsection{Distributions of $M_{T2\perp}$}
\label{sect:MT2.MT2perp}

Another variation of the previous method would be to search for the value of the test mass for which the difference
\begin{eqnarray}
 \Delta M_{T2} ( \Mtil_0, P_T ) \equiv M_{T2}^{max} ( \Mtil_0, P_T ) - M_{T2}^{max} ( \Mtil_0, P_T = 0 )
 \label{eq:MT2.MT2.DeltaM}
\end{eqnarray}
is minimized, the idea being that for the true value $\Mtil_0 = M_0$, this difference should be zero \cite{Konar:2009wn}.
This method was studied, again for the same-sign di-lepton final states, in \cite{Konar:2009wn},
where it was proposed to estimate $M_{T2}^{max} ( \Mtil_0, P_T = 0 )$ as the endpoint of the $M_{T2\perp}(\tilde M_0)$ distribution.
The variable $M_{T2\perp}$ was introduced in \cite{Konar:2009wn} as follows.
Instead of computing the 2D variable $M_{T2}$ for the rather small subset of events with $P_T=0$, 
it was proposed to use all events, and then compute the 1D components of $M_{T2}$  along ($M_{T2\parallel}$) and perpendicular to ($M_{T2\perp}$) the UTM.
Of particular interest is $M_{T2\perp}$, which is independent of the UTM. 
On an event by event basis, it is given by the analytical formula
\begin{eqnarray}
 M_{T2\perp} (\Mtil_0)  &=& \sqrt{A_{T \perp}} + \sqrt{A_{T \perp} + \Mtil_0^2}, 
 \label{eq:mt2perpdef}
\end{eqnarray}
where
\begin{eqnarray}
  A_{T \perp} &\equiv& \frac{1}{2} \left( | \vec{p}^{\;(1)}_{T \perp} | | \vec{p}^{\;(2)}_{T \perp} | + \vec{p}^{\;(1)}_{T \perp} \cdot \vec{p}^{\;(2)}_{T \perp} \right)
\end{eqnarray}
and the momenta $\vec{p}^{\;(i)}_{T \perp}$ ($i = 1, 2$) are the projections of the transverse momenta $\vec{p}^{\;(i)}_{T}$ 
of the visible systems onto the direction orthogonal to the UTM $\vec{P}_T$:
\begin{eqnarray}
  \vec{p}^{\;(i)}_{T \perp} &=& \vec{p}^{\;(i)}_{T} - \frac{1}{P_T^2} \left( \vec{p}^{\;(i)}_{T} \cdot \vec{P}_T \right) \vec{P}_T.
\end{eqnarray}

The endpoint of the $M_{T2\perp}$ distribution is given by
\begin{eqnarray}
 M_{T2\perp}^{max} (\Mtil_0) &=& \mu + \sqrt{\mu^2 + \Mtil_0^2}  
 \label{eq:MT2perpmaxmu}
\end{eqnarray}
with $\mu$ again given by (\ref{mu110definition}).

The shape of the $M_{T2\perp}$ distribution is also known analytically: 
\begin{eqnarray}
 \frac{d N}{d M_{T2\perp}} &=& N_0 \; \delta \left( M_{T2\perp} - \Mtil_0 \right) + (1 - N_0) \frac{M_{T2\perp}^4 - \Mtil_0^4}{\mu^2 M_{T2\perp}^3} \ln \left( \frac{2 \mu M_{T2\perp}}{M_{T2\perp}^2 - \Mtil_0^2}  \right).
 \label{eq:MT2perpshape}
\end{eqnarray}
The first term is a Dirac delta function peaking at the value of the test mass $\Mtil_0$. It arises from events for which
$\vec{p}^{\;(1)}_{T \perp}$ and $\vec{p}^{\;(2)}_{T \perp}$ are back-to-back and as a result, $A_{T \perp}=0$ in eq.~(\ref{eq:mt2perpdef}).
The second term provides the relevant shape of the distribution, and contains the dependence on the 
measurable parameter $\mu$. Note that the shape (\ref{eq:MT2perpshape})
does not contain any unknown kinematic parameters, such as the unknown center-of-mass energy 
or the longitudinal momentum of the initial hard scattering. 
It is also insensitive to spin correlation effects, whenever the upstream momentum results from production 
and/or decay processes involving scalar particles (e.g., squarks) 
or vectorlike couplings (e.g., the QCD gauge coupling). 
By design, it is also independent of the actual value of the upstream momentum $P_T$; as a result, 
we are not restricted to a particular $P_T$ range and can use the whole event sample in the $M_{T2\perp}$ analysis. 
For any given choice of $\Mtil_0$, eq.~(\ref{eq:MT2perpshape}) 
is a one-parameter curve which can be fitted to the data to obtain the parameter $\mu$ and then from (\ref{eq:MT2perpmaxmu}), 
compute the value of
$$
M_{T2\perp}^{max} ( \Mtil_0)=M_{T2}^{max} ( \Mtil_0, P_T = 0 ),
$$
which is needed for eq.~(\ref{eq:MT2.MT2.DeltaM}).
The knowledge of the shape (\ref{eq:MT2perpshape}) is of primary importance, given the potentially soft logarithmic decrease near the endpoint,
which may make the latter difficult to locate with great precision.
However, care should be exercised for distortions due to the experimental cuts and the smearing from the energy resolution.

\subsection{Mass determination from $M_{T2}$ in a two-step decay}
\label{sect:MT2.MT2.2step}

One can generalize the $M_{T2}$ discussion above to the case of two-step decay chains
as in Sec.~\ref{sect:invmass.2step}: 
\begin{equation}
X_2 \rightarrow x_1X_1 \rightarrow x_1 x_0 X_0,
\label{2stepgeneric}
\end{equation}
This generic chain may represent gluino decay, $\sGlu \rightarrow q_1 \sQua \rightarrow q_1 q_2 \chiz_1$,
or neutralino decay, $\tilde\chi^0_2\rightarrow l\tilde l\rightarrow ll\tilde\chi^0_1$, in SUSY.
The presence of the intermediate state $X_1$ constrains the kinematics, thus affecting
the endpoint value of the overall $M_{T2}$. Of course, in order to determine all  three masses,
$M_2$, $M_1$ and $M_0$, additional independent measurements are needed.
For this purpose, Ref.~\cite{Burns:2008va} introduced the concept of the ``subsystem $M_{T2}$",
by applying $M_{T2}$ to a suitable subset of the full event.
For the example in (\ref{2stepgeneric}), there are three different subsystems, and correspondingly, 
three different $M_{T2}$ variables.
The first, $M_{T2}(x_1x_0)$, is obtained by considering the full event of $X_2$ pair production, 
where the $X_0$'s are the invisible particles.
The second, $M_{T2}(x_1)$, again starts from $X_2$ pair production, but effectively treats the $X_1$'s as the
missing particles, by adding the $x_0$ transverse momenta to the measured $\vec{p}_T^{\;miss}$.
Finally, $M_{T2}(x_0)$ starts from $X_1$ pair production, adding the $x_1$ transverse momenta to the UTM.
This procedure allows for three $M_{T2}$ kinematic endpoint measurements, which, when taken together with 
the kinematic endpoint of the $M_{x_0x_1}$ invariant mass distribution, are more than sufficient to determine
the three unknown masses $M_2$, $M_1$ and $M_0$ \cite{Burns:2008va}.

The endpoint formulae for the various $M_{T2}$ subsystems have been derived in \cite{Burns:2008va}.
The calculation of $M_{T2}({x_1 x_0})$ and $M_{T2}({x_1})$ is easier, since there is no UTM.
To simplify the notation, it is convenient to generalize the definition of $\mu$ used in Sec.~\ref{sect:MT2.MT2} as follows
\begin{eqnarray}
 \mu_{(n, p, c)} = \frac{M_n}{2} \left( 1 - \frac{M_c^2}{M_p^2} \right),
\end{eqnarray}
where $n$ is the total number of decay steps, $p$ labels the parent of the subsystem and $c$ the child of the subsystem.
Then, the kinematic endpoint for the overall $M_{T2}(x_1x_0)$ is again defined in terms of two branches \cite{Burns:2008va}
\begin{equation}
 M_{T2}^{max} (x_1x_0) = 
 \left\{
\begin{array}{ll}
 F_{L} ( \Mtil_0), & {\rm for\ } \tilde M_0\le M_0  \\ [2mm]
 F_{R} ( \Mtil_0), & {\rm for\ } \tilde M_0\ge M_0
\end{array}
\right.
\label{MT2x1x0branches}
\end{equation}
where 
\begin{eqnarray}
 F_L ( \Mtil_0) &=& \mu_{(2, 2, 0)} + \sqrt{  \mu_{(2, 2, 0)}^2 + \Mtil_0^2}, \label{FL2step} \\ [2mm]
 F_R ( \Mtil_0) &=& \mu_{(2, 2, 1)} + \mu_{(2, 1, 0)} + \sqrt{  \left( \mu_{(2, 2, 1)} - \mu_{(2, 1, 0)} \right)^2 + \Mtil_0^2}.
 \label{FR2step}
\end{eqnarray}
The two branches exhibit different dependences on the underlying masses
and thus {\em allow the determination of all three masses $M_2$, $M_1$ and $M_0$},
as shown in \cite{Burns:2008va}. For instance, this could be achieved by choosing 
a low test mass, say $\Mtil_0 = 0$, which necessarily belongs to the left branch,
and then choosing two sufficiently high values of $\Mtil_0$ on the right branch.

The subsystem $M_{T2}(x_1)$ also has no UTM and is a one-step decay.
Hence, as seen in Sec.~\ref{sect:MT2.MT2},
we will only obtain a relation between the masses of the parent particle $X_2$ and the daughter particle $X_1$. 

For $M_{T2}(x_0)$, there is UTM provided by the visible particles $x_1$.
This allows both the mass of the parent $X_1$ and the mass of the daughter $X_0$ to be extracted,
following the procedure discussed in Sec.~\ref{sect:MT2.MT2}.

In conclusion, using subsystem $M_{T2}$ kinematic endpoints, we obtain 
enough constraints to overdetermine all three masses involved in the decay chain (\ref{2stepgeneric}).

\subsection{Mass determination from $M_{T2}$ in a direct 3-body decay}
\label{sect:MT2.MT2.3body}

Next, consider the case of a direct three-body decay without an intermediate resonance $X_1$:
\begin{equation}
X_2 \rightarrow x_1 x_0 X_0.
\label{3bodygeneric}
\end{equation}
Now, the only two sparticles involved in the process are $X_2$ and $X_0$ 
and the only meaningful variable is the overall $M_{T2}(x_1x_0)$.
This case is treated in \cite{Cho:2007qv}, where the endpoint formulas are also listed.
Once again, the kinematic endpoint $M^{max}_{T2}(x_1x_0)$ is given by two branches 
as in eq.~(\ref{MT2x1x0branches}), only now
\begin{eqnarray}
 F_L ( \Mtil_0) &=&  \mu_{(2, 2, 0)}  + \sqrt{  \mu_{(2, 2, 0)}^2 + \Mtil_0^2}, \label{FL3body}\\ [2mm]
 F_R ( \Mtil_0) &=& \left(M_2 - M_0 \right) + \Mtil_0.  \label{FR3body}
\end{eqnarray}
We see that the two branches involve different combinations of sparticle masses
and hence both masses $M_2$ and $M_0$ can be determined from two values of the test mass, one on each branch.

Note that the left branches in (\ref{FL2step}) and (\ref{FL3body}) are given by the same expression, so that the left branch alone
will not be able to distinguish between the sequential two-body decays (\ref{2stepgeneric}) 
with an on-shell particle $X_1$ and the direct three-body decays (\ref{3bodygeneric}).
The right branch is, however, different for those two cases, and will in general allow them to be 
distinguished: for the three-body decay (\ref{3bodygeneric}), the dependence of $F_R$ on $\Mtil_0$ is linear, see eq.~(\ref{FR3body}).
For the sequential two-body decay (\ref{2stepgeneric}) the corresponding behavior (\ref{FR2step}) is in general different, 
but may fake a linear relationship whenever the quantity
$$\mu_{(2, 2, 1)} - \mu_{(2, 1, 0)} = \frac{M_2}{2}  \left( \frac{M_0^2}{M_1^2}  - \frac{M_1^2}{M_2^2} \right)$$
is
negligible compared to $\Mtil_0$. Even then, the shape of the $M_{T2}$ distribution can be used to identify the type of decay.

\subsection{Mass determination from $M_2$}
\label{sect:MT2.M2}

$M_{T2}$ is a {\em transverse} kinematical variable which makes crucial use
of momentum conservation in the transverse plane in order to reduce the number of 
independent unknown degrees of freedom (the transverse components $\vec{p}_T^{\; \chi(i)}$ of the invisible momenta).
The remaining independent transverse components are subsequently fixed by the minimization in eq.~(\ref{eq:MT2basicdef}).
However, the same chain of thought which led to eq.~(\ref{eq:MT2basicdef}) can be extended to $3+1$ dimensions
and one can define an analogous class of invariant mass variables known as $M_2$ \cite{Barr:2011xt,Mahbubani:2012kx}\footnote{The 
logic behind this notation is that the $M_2$ variables are 3+1 dimensional and the transverse index ``$T$" is unnecessary.
Unfortunately, this may lead to confusion with the Wino mass parameter or the mass of the parent particle $X_2$
in (\ref{2stepgeneric}) or (\ref{3bodygeneric}). We hope that the meaning of $M_2$ throughout this report is clear from the context.}
\begin{eqnarray}
 M_{2} (\Mtil_0) = \mathrm{min}_{\vec{p}_T^{\; \chi (1)} + \vec{p}_T^{\; \chi (2)} = \vec{p}_T ^{\; miss}} \left[ \mathrm{max} \left(M^{(1)}, M^{(2)}\right) \right],
 \label{eq:M2basicdef}
\end{eqnarray}
where now the minimization applies to the full invariant masses $M^{(i)}$ of the hypothesized parent particles, rather than their transverse masses
$M_T^{(i)}$ used in (\ref{eq:MT2basicdef}).
Since the minimization in (\ref{eq:M2basicdef}) is performed by varying the full 4-momenta of the unseen neutral particles $\chi(i)$
(including the longitudinal components $p_z^{\; \chi(i)}$), as a byproduct of the minimization one also obtains a specific ansatz for the invisible 4-momenta.

The $3+1$ dimensional nature of the $M_2$ variables opens the door for several new opportunities 
which are not present in the case of purely transverse variables like $M_{T2}$:
\begin{itemize}
\item {\em Enforcing invariant mass constraints during the minimization.} 
For a given event topology, there may be extra constraints which can placed on the invisible 
momenta, in addition to the MET constraint  $\vec{p}_T^{\; \chi (1)} + \vec{p}_T^{\; \chi (2)} = \vec{p}_T ^{\; miss}$. 
For example, consider symmetric events of $X_2$ pair production, followed by the decay (\ref{2stepgeneric}).
Focusing on the $(x_1x_0)$ subsystem for the moment, 
one can define four different variants of $M_2(x_1x_0)$ \cite{Mahbubani:2012kx,Cho:2014naa}:
\begin{itemize}
\item $M_{2CX}$, additionally demanding that the masses of the parent particles $X_2$ in the two decay chains are the same.
\item $M_{2XC}$, additionally demanding that the masses of the intermediate particles $X_1$ in the two decay chains are the same.
\item $M_{2CC}$, enforcing both of the above constraints.
\item $M_{2XX}$, adding no additional invariant mass constraints to eq.~(\ref{eq:M2basicdef}).  
\end{itemize}
The constrained $M_2$ variables for the other two subsystems, $(x_1)$ and $(x_0)$, can be defined analogously.
It can be shown that in any subsystem, the following hierarchy holds true \cite{Cho:2014naa}
\begin{eqnarray}
 M_{T2}= M_{2XX} = M_{2CX} \leq M_{2XC} \leq M_{2CC}.
 \label{MT2M2hierarchy}
\end{eqnarray}
This means that the unconstrained $M_2$ variable $M_{2XX}$ and the singly constrained $M_{2CX}$ version are both equivalent to $M_{T2}$ event by event,
providing the link between the two approaches. More importantly, eq.~(\ref{MT2M2hierarchy}) shows that the addition of constraints generally  
leads to {\em higher} event by event values for $M_2$ in comparison to $M_{T2}$. At the same time, the upper kinematic endpoint is unchanged, 
since it is bounded by the actual parent mass. This implies that in the case of the constrained $M_2$'s, 
the events are being pushed closer to the endpoint, making it more pronounced and
therefore more easily measurable experimentally \cite{Cho:2014naa}.
\item {\em Recognizing the correct event topology.} 
The capability to optionally enforce different on-shell constraints
allows us to perform a consistency check on the assumed event topology.
In general, any given event topology is accompanied with a specific set of on-shell conditions,
either due to symmetry between the two decay chains, or because some of the intermediate
resonances have known masses (for example, the top quark and the $W$-boson).
In Ref.~\cite{Cho:2014naa} it was demonstrated that if one uses the wrong 
constraints\footnote{That is, the constraints corresponding to the wrong event topology.} 
in computing $M_2$, there are sizeable tails in the $M_2$ distribution, which would indicate that 
the hypothesis was wrong. In particular, it was shown that this method can 
distinguish the sequence (\ref{2stepgeneric}) from the sequence (\ref{3bodygeneric}),
as well as from an asymmetric event topology.\footnote{For an alternative approach 
which does not use the $M_2$ variables but focuses instead on the
invariant masses of various combinations of visible particles in the event, see \cite{Bai:2010hd}.}
\item {\em 4-dimensional ansatz for the invisible momenta.} As already mentioned, the $M_2$ minimization is based on 
varying the full 4-momentum vectors of the unseen neutral particles, including the longitudinal spatial component.  
Hence, after convergence, the full 4-momenta are available as a viable ansatz. In the case of $M_{T2}$
the analogous ansatz (named MAOS \cite{Cho:2008tj}) for the {\em transverse} components $\vec{p}_T^{\; \chi (i)}$
has already been shown to be of great value --- as it turns out, the MAOS momenta are not too different from the true invisible momenta in the event, 
especially for events near the $M_{T2}$ kinematic endpoint \cite{Cho:2008tj,Choi:2009hn,Park:2011uz}.
\begin{itemize}
\item {\em The 4D ansatz can be used to determine the masses of accompanying particles.}
The minimization in (\ref{eq:M2basicdef}) provides a lower bound on the masses of the parent particles within a given subsystem.
However, except for the simplest event topology (\ref{1stepgeneric}), there are typically several other new particles in the event 
--- they can be intermediate, upstream or downstream along the decay chain.
Once we have an ansatz for the invisible momenta, the kinematics of the event is fully determined, and we can readily compute the 
masses of any additional particles in the event.
It was found in \cite{Cho:2014naa}  that the computed invariant mass distributions for the additional sparticles indeed 
peak at the correct mass values.\footnote{Of course, this is only true if the test mass $\tilde M_0$ for the 
invisible particle is chosen correctly. In general, the sparticle masses are determined as functions of $\tilde M_0$,
just like the case in eqs.~(\ref{mt2max110}), (\ref{mt2max110PT}), (\ref{eq:MT2perpmaxmu}) and (\ref{MT2x1x0branches}).}
Furthermore, the shape of the reconstructed invariant mass distribution provides an indication of the reliability of the method ---
when the peak value correctly identifies the mass, the mass distribution is very narrow and symmetric.
On the other hand, when the mass distribution has an abrupt end and a long tail towards low masses, 
the peak value may not be reliable and one should instead extract the mass from the 2D correlation plot  
of the reconstructed invariant mass versus $M_2$, focusing on the region near the $M_2$ endpoint \cite{Kim:2017awi}.
\item {\em The 4D ansatz can be used to resolve combinatorial ambiguities.} 
A typical SUSY-like event at the LHC suffers from the combinatorial problem of
partitioning the visible particles in the event into two groups, one for each decay chain. 
This problem is traditionally handled with the ``hemisphere" algorithm \cite{hemisphere}. 
Proposed improvements to it introduce
suitable cuts on the jet $p_T$ \cite{Rajaraman:2010hy} or on $M_{T2}$ \cite{Baringer:2011nh}, 
exclude certain jets from the clustering algorithm \cite{Nojiri:2008vq,Alwall:2009zu}, or 
attempt partial event reconstruction \cite{Jackson:2017gcy}.   
The MAOS invisible momenta were used to further refine the algorithm \cite{Choi:2011ys}. 
Ref.~\cite{Debnath:2017ktz} showed that the selection efficiency is further improved 
if one instead uses the 4D ansatz for the invisible momenta provided by
the $M_2$ variables.
\end{itemize}
\end{itemize}

The addition of invariant mass constraints turns the minimization procedure in eq.~(\ref{eq:M2basicdef}) 
into a complex mathematical problem for which no analytical solutions exist, unlike the case of $M_{T2}$ 
where at least some special cases can be treated analytically \cite{Lester:2011nj}.
Fortunately, a general code \cite{Cho:2015laa} has been developed and it can compute any constrained $M_2$ variable numerically.
It is based on an augmented Lagrange multiplier method, which provides a large flexibility for optionally imposing different types of 
kinematic constraints. As a downside, the $M_2$ code runs considerably slower than analogous $M_{T2}$ codes \cite{Cheng:2008hk,Walker:2013uxa,Lester:2014yga,Lally:2015xfa}.

\section{Identification of Point-like (Off-Shell) Decays}
\label{sect:invmass.phasespace}

In \cite{Kim:2015bnd}, general relations were derived for the endpoints of invariant mass distributions in the decay of a 
massive parent $D$ to $N$ visible particles and one invisible particle $A$ via point-like interactions (eqs.~(\ref{1stepgeneric} and (\ref{3bodygeneric})
illustrate the two simplest cases with $N=1$ and $N=2$, respectively). 
Such relations remain valid even when the decays are mediated by heavy intermediate particles (perhaps only slightly above the kinematic limit)
which are off-shell --- the presence of a heavy intermediate particles will modify the shape of the invariant mass distribution, but leave its endpoint intact.
An interesting aspect of this study is that one does not need to know the exact order in which the visible particles were emitted along the decay chain.

Assume that the decay results in $N$ indistinguishable massless particles and that we are interested in $n$-particle invariant masses, with $n<N$.
For a fixed $n$, he number of such $n$-particle invariant masses is given by the combinatoric factor $C_n^N$.
Therefore, the total number of invariant mass variables which can be studied is given by
\begin{eqnarray}
 \sum_{n = 2}^N C_n^N = 2^N - N -1.
\end{eqnarray}
The variable $n$ is called the order of the invariant mass.
Next, we can sort the invariant masses of a given order from the largest to the smallest value event by event.
Call the position of an invariant mass in the sorted list $r$, the rank of the variable, so that we can label the invariant masses as $m_{(n, r)}$.
It is clear that the endpoint of any invariant mass distribution for the decay of a particle $D$ into $N$ visible particles and an unseen neutral $A$
can only depend on the masses $m_D$ and $m_A$.
Defining the ratio $R_{AD} = m_A^2 / m_D^2$, so that $R_{AD}<1$,
it was shown that the general formula for the endpoint of the invariant mass distribution of $m_{(2, r)}$ is given by
\begin{eqnarray}
 m_{(2, r)}^{max} = \frac{1}{\sqrt{r}} (m_D - m_A) = \frac{m_D}{\sqrt{r}} (1 - \sqrt{R_{AD}}),
\label{eq:invmass.phasespace.endpt2}
\end{eqnarray}
provided $r$ is not too large.
It is found that if $r$ is larger than a certain value, the factor $1 / \sqrt{r}$ overestimates the endpoint value,
hence the formula provides an upper bound on the endpoint position.
The range of applicability of this formula is restricted to
\begin{equation}
 r \leq
 \left\{
 \begin{array}{ll}
  6 k^2 			& {\rm for\ } N = 4 k, \\ [2mm]
  6 k^2 + 3 k 		& {\rm for\ } N = 4 k +1, \\ [2mm]
  6 k^2 + 6 k + 1 	& {\rm for\ } N = 4 k +2, \\ [2mm]
  6 k^2 + 9 k + 3 	& {\rm for\ } N = 4 k +3,
\end{array}
\right.
\end{equation}
where $k$ is an integer $\geq 1$.

In full generality, the endpoint is located at
\begin{eqnarray}
 m_{(n, r)}^{max} = \left\{ \begin{array}{ll}
 	m_D (1 - \sqrt{R_{AD}}) & \mathrm{for} \; r < C_{n-2}^{N-2} ,\\ [2mm]
	 \sqrt{\frac{C_{n-2}^{N-2}}{r}}m_D (1 - \sqrt{R_{AD}}) & \mathrm{for} \; r \geq C_{n-2}^{N-2}.
 \end{array} \right.
\label{eq:invmass.phasespace.endptn3}
\end{eqnarray}

An interesting example is $N = 3$, for which the MSSM decay chains could be
\begin{eqnarray}
 \sQua &\rightarrow& q \chipm_1 \rightarrow q W \chiz_1 \rightarrow q q q \chiz_1 \nonumber \\
 	&\rightarrow& q \chiz_2 \rightarrow q (Z^0 / h^0) \chiz_1 \rightarrow q q q \chiz_1 \nonumber \\
	&\rightarrow& q \chiz_2 \rightarrow q l \sLep \rightarrow q l l \chiz_1
\end{eqnarray}
or direct four-body decays.
From equations (\ref{eq:invmass.phasespace.endpt2}) and (\ref{eq:invmass.phasespace.endptn3}),
we expect the following relations among the endpoints to hold:
\begin{eqnarray}
 \frac{m_{(3, 1)}^{max}}{m_{(2, 1)}^{max}} = 1, \qquad
 \frac{m_{(2, 2)}^{max}}{m_{(2, 1)}^{max}} = \frac{1}{\sqrt{2}}, \qquad
 \frac{m_{(2, 3)}^{max}}{m_{(2, 1)}^{max}} = \frac{1}{\sqrt{3}}.
\end{eqnarray}
Finding out experimentally that these equations are satisfied would be evidence for point-like decays with no intermediate on-shell particles,
whereas a significant deviation from these values would signal the presence of intermediate on-shell states.

\section{Jet Multiplicities in Other Scenarios}
\label{sect:app:simul.others}

Tables~\ref{tab.simul.summary.others.Bino}-\ref{tab.simul.summary.others.Hino} 
provide some more detailed information on the cutflow of the events in the various scenarios:
the Bino scenario (Table~\ref{tab.simul.summary.others.Bino}), 
the Wino scenario (Table~\ref{tab.simul.summary.others.Wino})
and the Higgsino scenario (Table~\ref{tab.simul.summary.others.Hino}).
The numbers in the tables list the number of events with equal number of jets in the two hemispheres.
The column ``Generated" gives the number of events at generation level before experimental cuts are applied.
Those events are then divided into ``Rejected", i.e., those which were rejected by the experimental cuts,
and ``Left", i.e., those which passed the cuts.
The last column labelled ``Hemispheres" gives the number of events after applying the hemisphere reconstruction algorithm \cite{hemisphere}.
\begin{table}[htb]
\begin{center}
\begin{tabular}{|c|c|c|c|c|} \hline
\textbf{Bino Scenario}	& Generated	& Rejected 	& Left	& Hemispheres 	\\
\hline
2j 					& 1788 		& 49		& 1739  	& 1669   		\\
4j 					& 215		& 45  	& 170 	& 299 		\\ 
6j 					& 1085		& 50		& 1035 	& 443 		\\
8j 					& 117 		& 31  	& 86   	& 79  		\\
10j 					& 1			& 0	   	& 1    	& (28)   		\\
12j 					& 37  		& 12     	& 25  	& 22     		\\
14j 					& 0  			& 0	     	& 0	   	& (7)     		\\
16j 					& 35  		& 16     	& 19   	& 5     		\\
18j 					& 0	  		& 0	     	& 0	   	& 0     		\\
20j 					& 3	  		& 1	     	& 2	   	& 0     		\\
\hline
\end{tabular}
\caption{Statistics in the various jet multiplicities in the Bino scenario.
Numbers in parentheses correspond to topologies where most of the events are migrations from other topologies,
hence unreliable.
}
\label{tab.simul.summary.others.Bino}
\end{center}
\end{table}

\begin{table}[h!]
\begin{center}
\begin{tabular}{|c|c|c|c|c|} \hline
\textbf{Wino Scenario}	& Generated	& Rejected 	& Left	& Hemispheres 	\\
\hline
2j 					& 1738 		& 56		& 1682  	& 1613   		\\
4j 					& 160		& 55  	& 105 	& 332 		\\ 
6j 					& 1104		& 40		& 1064 	& 523 		\\
8j 					& 274 		& 60  	& 214   	& 88  		\\
10j 					& 3			& 0	   	& 3    	& (28)   		\\
12j 					& 39  		& 20     	& 19	  	& 6     		\\
14j 					& 0  			& 0	     	& 0   		& 2     		\\
16j 					& 14	  		& 0     	& 14   	& 2     		\\
\hline
\end{tabular}
\caption{The same as Table~\ref{tab.simul.summary.others.Bino}, but for the Wino scenario.
}
\label{tab.simul.summary.others.Wino}
\end{center}
\end{table}

\begin{table}[h!]
\begin{center}
\begin{tabular}{|c|c|c|c|c|} \hline
\textbf{Higgsino Scenario}& Generated	& Rejected 	& Left	& Hemispheres 	\\
\hline
2j 					& 121 		& 97		& 24  	& (34)   		\\
4j 					& 33			& 31  	& 2	 	& (292) 		\\ 
6j 					& 2076		& 72		& 2004  	& 958 		\\
8j 					& 678 		& 122  	& 556   	& 178  		\\
10j 					& 4			& 0	   	& 4    	& (25)   		\\
12j 					& 42  		& 3     	& 39  	& 3     		\\
14j 					& 0  			& 0	     	& 0  	 	& 0     		\\
16j 					& 1	  		& 0     	& 1  	 	& 0     		\\
\hline
\end{tabular}
\caption{The same as Table~\ref{tab.simul.summary.others.Bino}, but for the Higgsino scenario.
In this case the 2j and 4j numbers are in parentheses because they originate from events with leptonic decays.
}
\label{tab.simul.summary.others.Hino}
\end{center}
\end{table}

Table~\ref{tab.simul.summary.others.Bino} reveals that in the Bino scenario, 2j, 4j, 6j and 8j events are usable.
Higher jet multiplicities are either too contaminated or have too low statistics. The same conclusions hold for the 
Wino scenario in Table~\ref{tab.simul.summary.others.Wino}. This is expected, as the two scenarios are twins to each other, 
see Fig.~\ref{fig.distscenarios.BWino} and Table~\ref{tab:binowinoBR}.

In the Higgsino scenario of Table~\ref{tab.simul.summary.others.Hino}, only the 6j and 8j events are usable.
Higher jet multiplicities are either too contaminated or have too low statistics.
The 2j and the 4j events originate from leptonic decays of one or both $W$s and so cannot be reliably 
used for the hadronic analyses of Sec.~\ref{sect:had}.

\section{Six-jet Events in the Higgsino Scenario}
\label{sect:app.simul.Hino}

In the Higgsino scenario we do not expect many fully hadronic events in the 2j and 4j final states, thus
the lowest relevant jet multiplicity is 6j events. Table~\ref{tab.simul.summary.others.Hino} confirms that 
the most common events are indeed of the 6j type. It is therefore worthwhile studying in some more
detail what can be measured in this topology (the complete analysis is presented in \cite{Hadr}).
Selecting events with three jets in each hemisphere, the decay chains can be interpreted as being one of the following
\begin{eqnarray}
 \sQua_L &\rightarrow&  q' \; \chipm_2 \rightarrow q' \; (W/Z/h) \; \chiz_1, \nonumber \\
	 &\rightarrow&  q \; \chiz_4 \rightarrow q \; (W/Z/h) \; \chiz_1, \label{higgsinochains}\\
 \sQua_R &\rightarrow&  q \; \chiz_3 \rightarrow q \; (W/Z/h) \; \chiz_1, \nonumber 
\end{eqnarray}
where $\chiz_1$ could also stand for a $\chipm_1$ or a $\chiz_2$, since all three states are nearly mass degenerate.
The study of the decay chains (\ref{higgsinochains}) might potentially allow the determination of the masses 
of the two types of squarks, $\sQua_{L}$ and $\sQua_{R}$, the two nearly mass-degenerate Wino-like states $\chipm_2$ and $\chiz_4$,
and the Bino-like $\chiz_3$. In general, the squarks $\sQua_{L}$ and $\sQua_{R}$ will have different masses, and 
similarly the masses of $\chiz_4$ and $\chiz_3$ can be different as well.
This complicates the analysis, since the kinematic distributions may exhibit several
distinct kinematic endpoints, which will need to be appropriately interpreted.

\begin{figure}[htb]
 \begin{center}
 \includegraphics[width=0.49\textwidth]{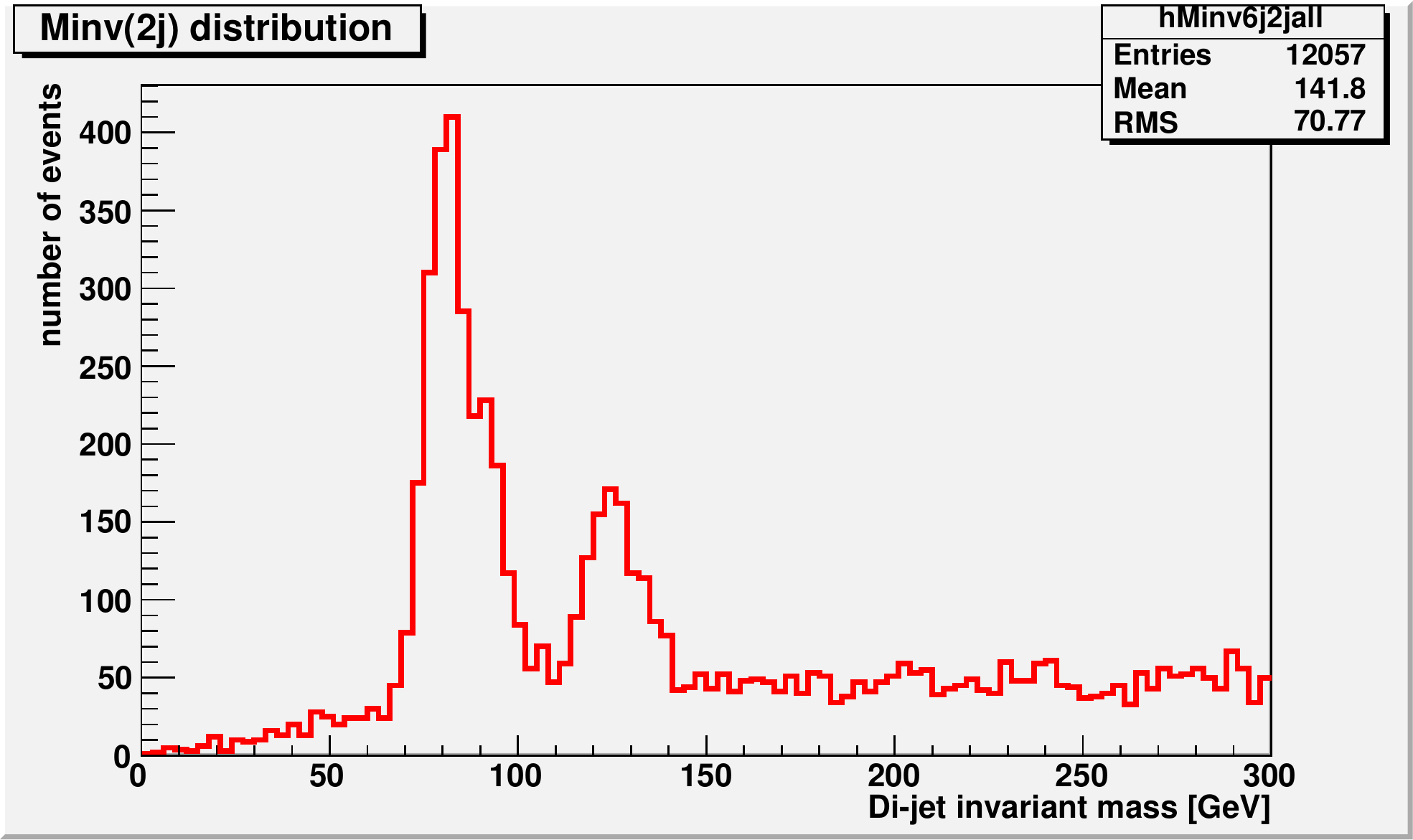}
 \includegraphics[width=0.49\textwidth]{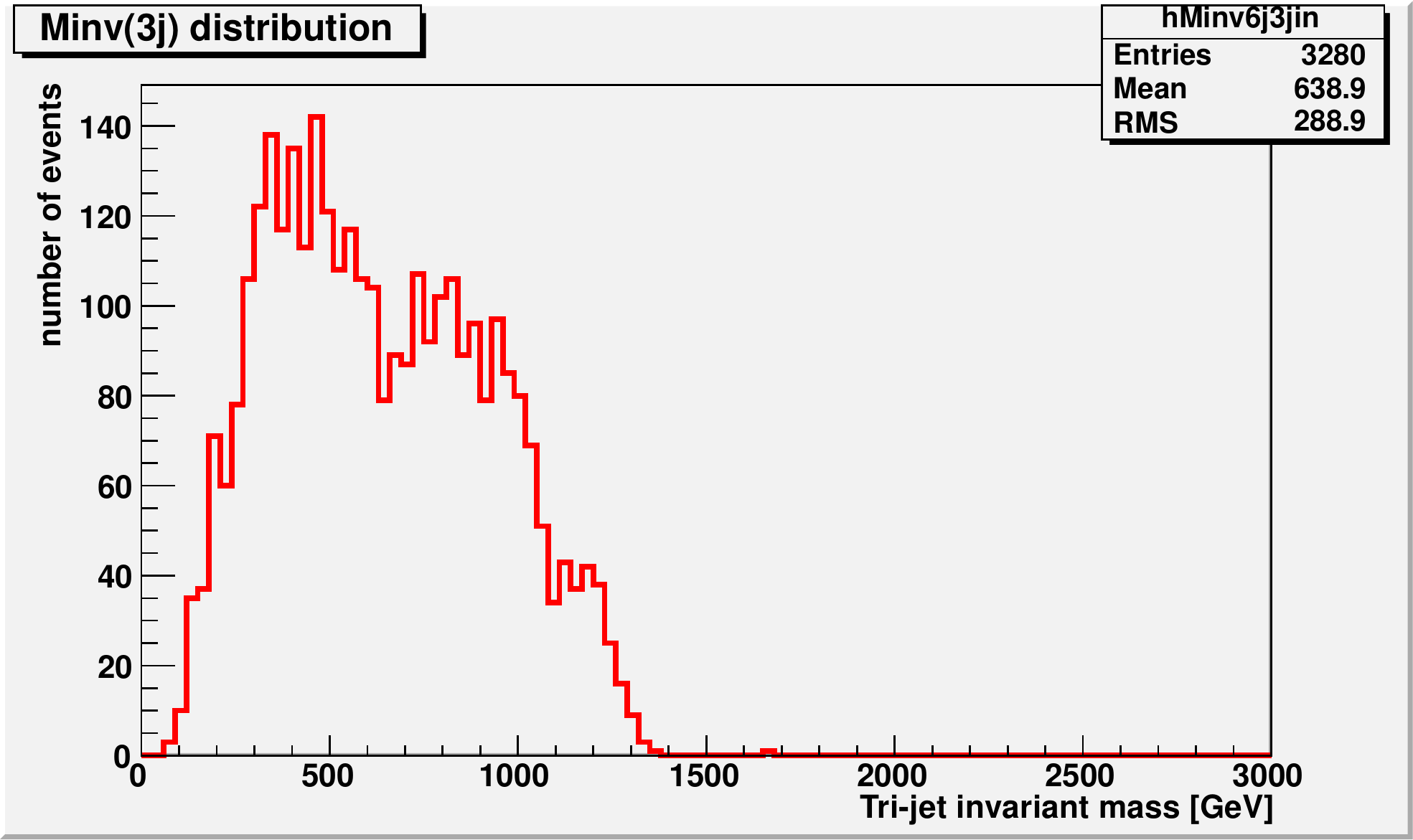}
\caption{ Left: di-jet invariant mass distribution in 6-jet events, including all 3 di-jet combinations from each hemisphere.
Right: invariant mass distribution of the three jets in a hemisphere, provided that one di-jet combination is in the $W / Z / h$ mass window (from 70 to 140 GeV).
}
 \label{fig.simul.6jet.Hino.Minv2jall}
 \end{center}
\end{figure}

As in Sec.~\ref{sec:hadronicselection}, let us start by analyzing the di-jet invariant mass
distributions, including all three combinations per hemisphere, in order to check for the presence of massive SM bosons in the final state.
The low mass range of the resulting distribution is shown in the left panel of Fig.~\ref{fig.simul.6jet.Hino.Minv2jall}.
The $W / Z$ peak is clearly observed, together with a well separated $h$ peak.
Selecting events where at least one di-jet combination is inside the $W / Z / h$ mass window ranging from 70 to 140 GeV,
the invariant mass distribution of the three jets in the corresponding hemisphere is displayed 
in the right panel of Fig.~\ref{fig.simul.6jet.Hino.Minv2jall}.
The expected endpoint value for the $\sQua_L$ decay in (\ref{higgsinochains}), computed numerically, is 1053 GeV,
while for the $\sQua_R$ decay it is 1260 GeV.
These values are in reasonable agreement with the observed distribution in the right panel of Fig.~\ref{fig.simul.6jet.Hino.Minv2jall}.
The measurement of these two kinematic endpoints will provide two relations between the masses of the sparticles,
however, we will not know which endpoint corresponds to the $\sQua_L$ decay and which 
to the $\sQua_R$ decay in (\ref{higgsinochains}).

\begin{figure}[htb]
 \begin{center}
 \includegraphics[width=0.49\textwidth]{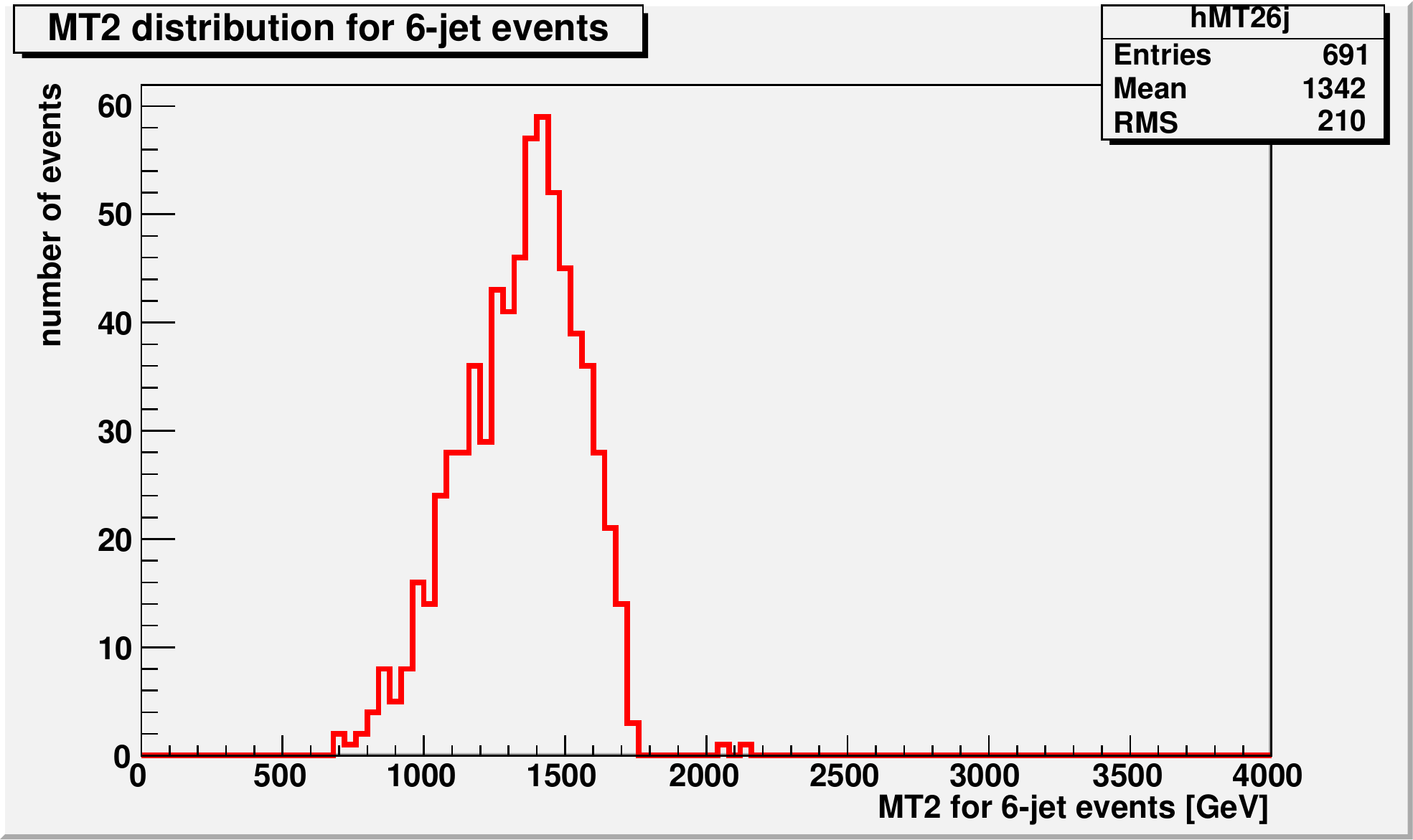}
\caption{Distribution of $M_{T2}(qV)$, $V=\{W,Z,h\}$, for 6-jet events in which
each hemisphere contains an identified $V$ candidate with a dijet mass between 70 and 140 GeV.  }
 \label{fig.simul.Hino.MT26j}
 \end{center}
\end{figure}

Following Sec.~\ref{WZhInDifferentHemispheres}, we proceed with the analysis of $M_{T2}$-type distributions for events 
with an identified $W/Z/h$ candidate (di-jet invariant mass between 70 and 140 GeV) in each hemisphere.
In Fig. \ref{fig.simul.Hino.MT26j} we show the distribution of $M_{T2}(qV)$, $V=\{W, Z, h\}$,
when the true value of the LSP test mass is used. The distribution 
exhibits an endpoint at around 1700 GeV, near the value of the squark masses (1656 GeV).
Since in our simulation test the masses of $\sQua_L$ and $\sQua_R$ were taken to be the same,
a separate endpoint is not visible, as the two endpoints coincide.

\begin{figure}[htb]
 \begin{center}
 \includegraphics[width=0.49\textwidth]{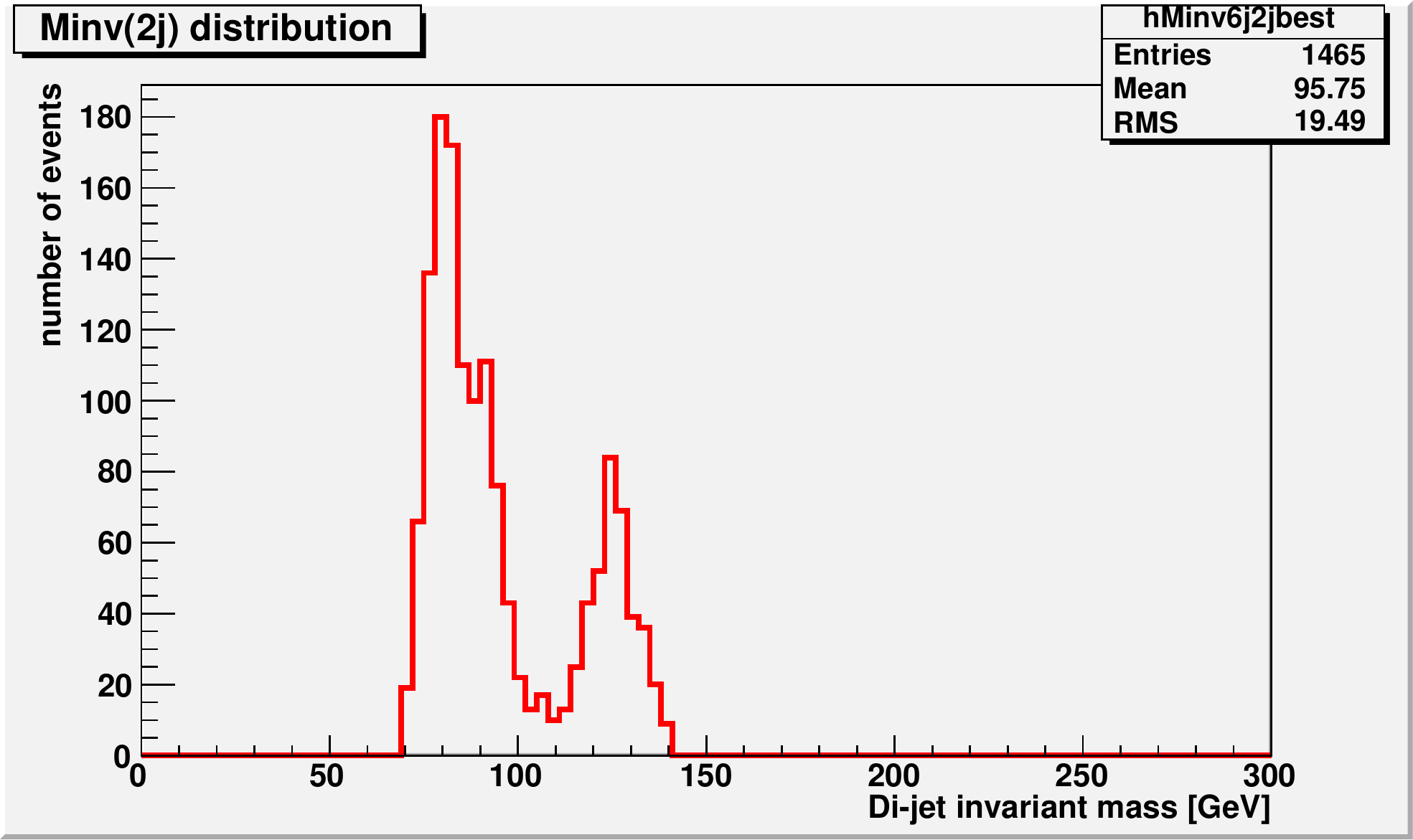}
\caption{Di-jet invariant mass distributions of the best $V=\{W, Z, h\}$ candidates
for events with at least one $V$ candidate in each of the two hemispheres.
}
 \label{fig.simul.Hino.Minvbest}
 \end{center}
\end{figure}

\begin{figure}[!htb]
 \begin{center}
 \includegraphics[width=0.49\textwidth]{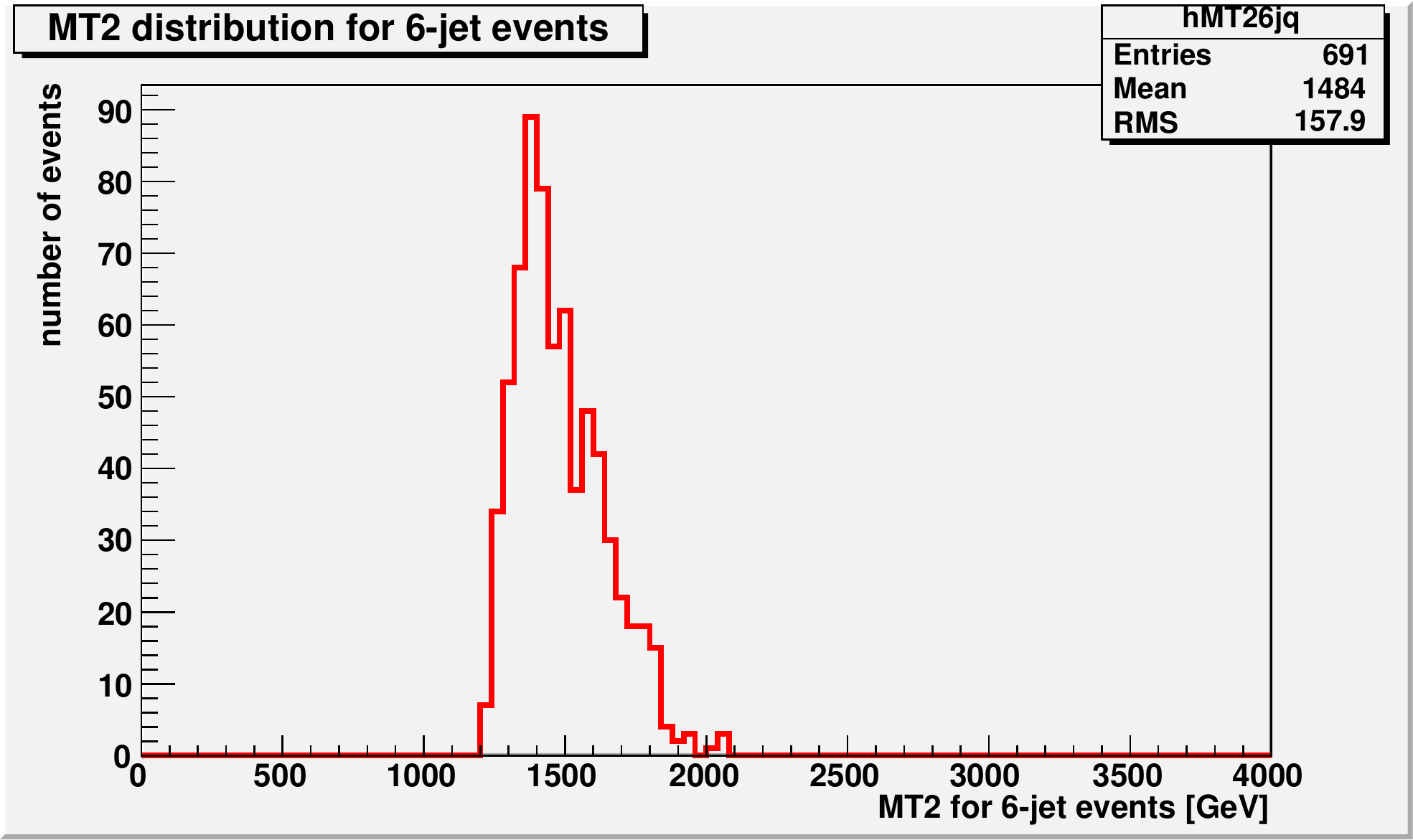}
 \includegraphics[width=0.49\textwidth]{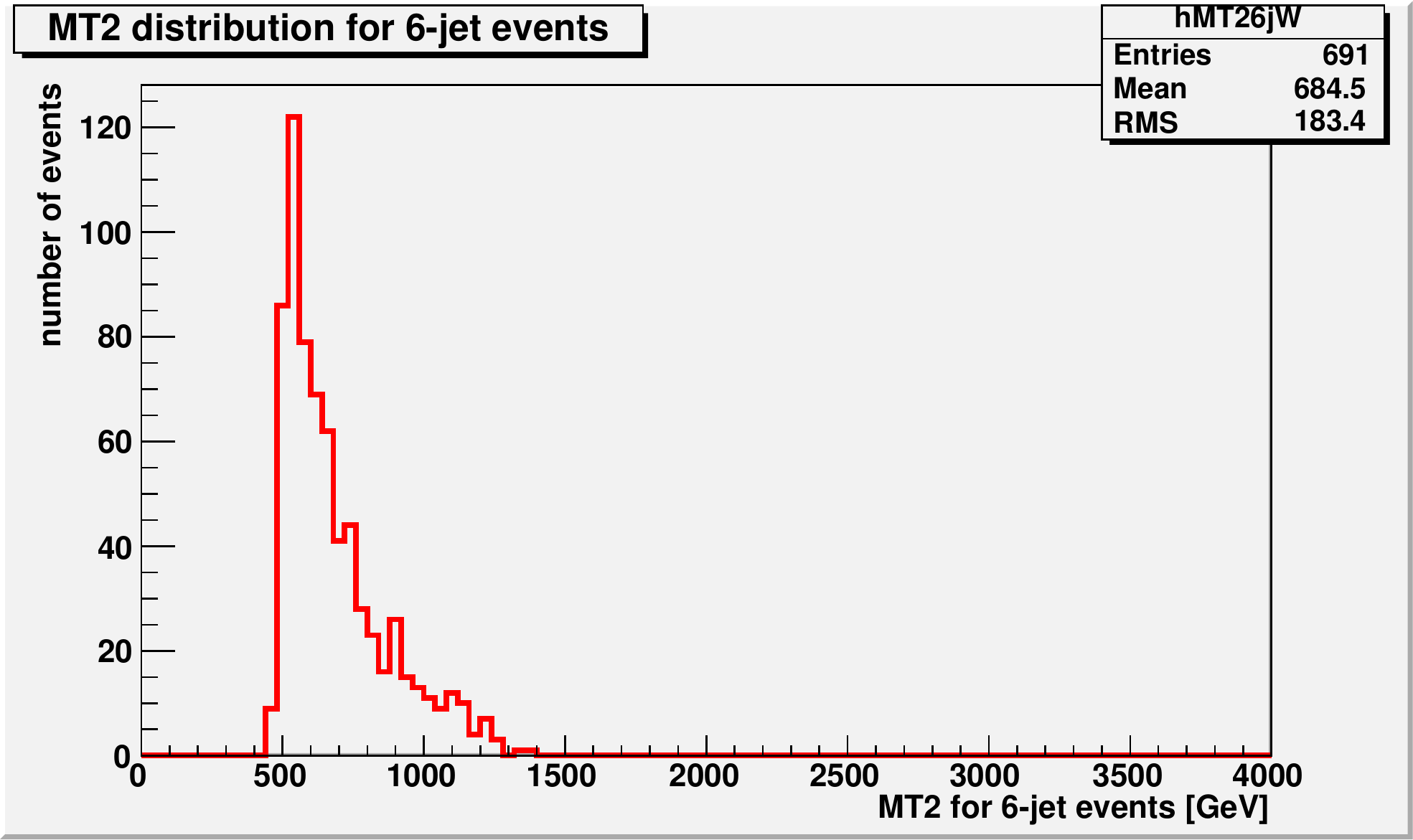}
\caption{Subsystem $M_{T2}$ distributions,  $M_{T2}(q)$ (left panel) and $M_{T2}(V)$ (right panel),
for 6-jet events with a $V=\{W, Z, h\}$ candidate
in each hemisphere, and after choosing the best $V$ candidate.
}
 \label{fig.simul.Hino.MT26jq}
 \end{center}
\end{figure}

Next, we want to build the distribution of the subsystem $M_{T2}(q)$, where $q$ is the leftover jet.
As there may be several acceptable $V=\{W, Z, h\}$ candidates in the same hemisphere,
only the solution closest to the mass of either $W$, $Z$ or $h$ is kept.
The resulting di-jet invariant mass is shown in Fig. \ref{fig.simul.Hino.Minvbest}.
The $M_{T2}(q)$ distribution is plotted in the left panel of Fig.~\ref{fig.simul.Hino.MT26jq},
where the jets from the $V$ are added to the missing momentum
and the $\chipm_1$ is thus treated as a missing neutral particle (its test mass is taken to be the true mass of 1200 GeV).
The distribution is negatively skewed and the endpoint will be more difficult to extract --- it appears to be between 1800 and 1950 GeV,
in reasonable agreement with the estimate from Fig.~\ref{fig.simul.Hino.MT26j}.

Finally, we can compute the subsystem $M_{T2}(V)$, treating the leftover jet in each hemisphere as part of the upstream $P_T$
and treating $\chiz_1$ as the missing particle.
This  distribution is seen in the right panel of Fig.~\ref{fig.simul.Hino.MT26jq} (the test mass was taken to be the true LSP mass).
The kinematic endpoint is around 1250 GeV, close to the mass of 1230 GeV of the $\chipm_2$.
Hence, the correct masses are obtained, but the $\sQua_L \leftrightarrow \sQua_R$ ambiguity could not be resolved.

The same masses can also be measured in 8j events, starting from a $\sGlu$.
However, this does not solve the above two-fold ambiguity either.
In summary, in the Higgsino scenario the sparticle masses can be determined,
but it is not known whether they correspond to a $\sQua_L$ or a $\sQua_R$ decay.

\bibliographystyle{mystylefile}
\bibliography{mybibliography}

\end{document}